\documentclass[a4paper,fleqn,usenatbib]{mnras}
\usepackage{newtxtext,newtxmath}

\usepackage[T1]{fontenc}
\usepackage{ae,aecompl}

\usepackage{graphicx}
\usepackage{amsmath}
\usepackage{amssymb}
\usepackage{float}
\usepackage{enumerate}  
\usepackage{lipsum}
\usepackage{gensymb}
\usepackage{graphicx}
\usepackage{longtable}
\usepackage{rotate}
\usepackage{rotating}
\usepackage{mathtools}
\usepackage{pdflscape}
\usepackage{scalerel}
\usepackage{times}
\usepackage{verbatim}
\usepackage{tablefootnote}
\usepackage{subfig}
\usepackage{lscape}
\newif\ifAMStwofonts

\def\suzaku{{\it Suzaku}}

\def\rosat{{\it ROSAT}}

\def\asca{{\it ASCA}}

\def\swift{{\it Swift}}
\def\xmm{{\it XMM-Newton}}

\def\mrk335{{Mrk~335}}

\def\feka{{Fe~K$\alpha$}}

\def\oiii{{[O~\textsc{iii}]}}

\def\feii{{Fe~\textsc{ii}}}

\def\arcs{{\hbox{$^{\prime\prime}$}}}

\def\cm{{\rm\thinspace cm}}
\def\erg{{\rm\thinspace erg}}
\def\eV{{\rm\thinspace eV}}

\def\kev{{\rm\thinspace keV}}
\def\km{{\rm\thinspace km}}

\def\Mpc{{\rm\thinspace Mpc}}

\def\s{{\rm\thinspace s}}

\def\ps{{\rm\thinspace s^{-1}}}

\def\ergps{\hbox{$\erg\s^{-1}\,$}}

\def\kmps{\hbox{$\km\ps\,$}}
\def\kmpspMpc{\hbox{$\kmps\Mpc^{-1}\,$}}

\def\pscm{\hbox{$\cm^{-2}\,$}}

\title[X-ray spectral properties of \suzaku\ Seyfert 1s]{A \suzaku\ sample of unabsorbed narrow-line and broad-line Seyfert 1 galaxies: I. X-ray spectral properties}

\author[S. G. H. Waddell \& L. C. Gallo]{
S. G. H. Waddell$^{1}$\thanks{E-mail: swaddell@mpe.mpg.de}
and L. C. Gallo$^{1}$
\\
$^{1}$Department of Astronomy \& Physics, Saint Mary's University, 923 Robie Street, Halifax, Nova Scotia, B3H 3C3, Canada \\
}

\date{Accepted XXX. Received YYY; in original form ZZZ}

\pubyear{2020}

\begin{document}
\label{firstpage}
\pagerange{\pageref{firstpage}--\pageref{lastpage}}
\maketitle

\begin{abstract}
A sample of narrow-line (NLS1) and broad-line Seyfert 1 (BLS1) galaxies observed with \suzaku\ is presented. The final sample consists of 22 NLS1s and 47 BLS1s, for a total of 69 AGN that are all at low redshift ($z<0.5$) and exhibit low host galaxy column densities ($<10^{22}$ \pscm). The average spectrum for each object is fit with a toy model to characterise important parameters, including the photon index, soft excess, Compton hump (or hard excess), narrow iron line strength, luminosity and X-ray Eddington ratio (L$_x$/L$_{\rm Edd}$). We confirm previous findings that NLS1s have steeper power laws and higher X-ray Eddington ratios, but also find that NLS1 galaxies  have stronger soft and hard excesses than their BLS1 counterparts. Studying the correlations between parameters shows that the soft and hard excesses are correlated for NLS1 galaxies, while no such correlation is observed for BLS1s. Performing a principal component analysis (PCA) on the measured X-ray parameters shows that while the X-ray Eddington ratio is the main source of variations within our sample (PC1), variations in the soft and hard excesses form the second principal component (PC2) and it is dominated by the NLS1s. The correlation between the soft and hard excess in NLS1 galaxies may suggest a common origin for the two components, such as a blurred reflection model. The presented \suzaku\ sample of Seyfert 1 galaxies is a useful tool for analysis of the X-ray properties of AGN, and for the study of the soft and hard excesses observed in AGN.
\end{abstract}




\section{Introduction}
\label{sect:intro}
Active Galactic Nuclei (AGN) are accreting supermassive black holes at the centres of galaxies. These objects are responsible for some of the most energetic events in the Universe, and in many cases, can even outshine their host galaxies. By studying the X-ray emission from these objects, we are able to probe some of the most extreme processes occurring at the innermost regions of the AGN.

The X-ray spectra of Seyfert 1 galaxies (optical type-1), where we have a direct view of the central region, show key similarities. The primary component exhibits a power law shape across the entire X-ray band and is attributed to the Comptonisation of accretion disc photons in the hot corona surrounding the inner disc (e.g. \citealt{Haardt+1991,Haardt+1993}). Below $2\kev$, many AGN exhibit a surplus of photons, called the soft excess, above the primary power law. The origin of the soft excess is highly disputed, and many possible physical interpretations exist to explain this feature, including a partial covering or obscuration scenario (e.g. \citealt{Gierlinski+2004,Tanaka+2004}), blurred reflection originating in the inner accretion disc (e.g. \citealt{Ballantyne+2001,RossFabian+2005}), and soft Comptonisation of X-ray photons by a secondary, cooler corona (e.g. \citealt{Magdziarz+1998,Done+2012}). These models have all been used successfully to explain the spectra of Seyfert 1 AGN. 

At higher energies, a second excess, called the hard excess, peaking at $\sim20\kev$ is often detected. When hard X-ray photons produced in the corona are incident on optically thick material, they are scattered to lower energies through Compton down-scattering resulting in the so called Compton hump. 

In the $4-7\kev$ band, AGN spectra show other important features, mainly those associated with iron. Interaction of  the primary X-rays with iron in the surround medium can produce a strong X-ray fluorescent emission line at $6.4\kev$ if the material is neutral. This line may be narrow when coming from the distant torus (e.g. \citealt{Nandra+2007}), or relativistically broadened due to the proximity to the supermassive black hole if it originates in the accretion disc (e.g. \citealt{Fabian+1989}). Typically, we see a superimposition of these two lines, where both the torus and inner accretion disc are illuminated by the corona and thus a narrow core and extended line profile are seen simultaneously (e.g. \citealt{Pounds+1990,Nandra+1994,Reeves+2006}).

Type-1 AGN are classified based on their optical spectra. NLS1s are those AGN with the full-width-half-maximum (FWHM) of the H$\beta$ lines less than $2000~\textrm{km}\ps$, while BLS1s have FWHM greater than $2000~\textrm{km}\ps$ (\citealt{Osterbrock+1985}; \citealt{Goodrich+1989}). NLS1 galaxies also show strong \feii\ emission lines, and weak emission from forbidden \oiii\ lines.

The differences in the spectra and properties of BLS1 and NLS1 AGN beyond their optical properties is an important area of study. In general, it is observed that NLS1 AGN have lower mass black holes than their BLS1 counterparts. These smaller AGN have weaker gravitational fields, which may be partially responsible for the narrower optical lines produced in the broad line region of NLS1 galaxies. Additionally, it has been shown that NLS1 AGN are more rapidly accreting than their broad-line counterparts \citep{Pounds+1995,Grupe+2004,Grupe+2010}. NLS1s are often found to be accreting at a significant fraction of their Eddington limit. Together with the lower black hole mass, these pieces of information suggest that NLS1s may be a younger population of AGN (see for example, \citealt{Grupe1996,Mathur+2000}). If this is indeed the case, the study of NLS1 galaxies and the comparison of their spectral properties to other Seyfert 1 AGN may lead to important clues on the growth and evolution of supermassive black holes throughout cosmic time. 

In the X-ray regime, many NLS1 and BLS1 galaxies show large amplitude variability in their luminosities on very short timescales of hours (e.g. \citealt{Nik+2009}; \citealt{Gallo+2018}). The variability in NLS1 galaxies is more extreme (e.g. \citealt{Boller+1996}), and flux changes by factors of $\sim100$ within days have been observed. The reason for this variability is not well understood, but may be correlated with the mass and/or accretion rates of the black holes (e.g. \citealt{Ponti+2012,Nik+2009}, and references therein). However, the very short timescales (hours) on which this variability is observed implies that X-rays originate from a very small and compact region, much smaller than the radius of the outer accretion disc or the BLR. The driving mechanism for this variability is therefore often associated with changes in the temperature, density and/or shape of the X-ray emitting corona (e.g \citealt{Wilkins+2015}; \citealt{Gallo+2019m335}; \citealt{Alston+2019}).

\cite{Boller+1996} examined the soft X-ray spectra of a sample of NLS1 AGN observed with \rosat. They found that NLS1 galaxies have very steep spectral slopes of $\Gamma\approx3.5$ at low energies ($0.1-2.4\kev$), where $\Gamma$ is the power law photon index. This is much steeper than typical BLS1 galaxies, which have indices of $\Gamma\approx2.3$ in this energy range (\citealt{Walter+1993}). These results are confirmed by \cite{Leighly1999}, who perform a similar analysis on a sample of NLS1 AGN observed with \asca, and other subsequent works. Additionally, NLS1 galaxies have been shown to show stronger soft excesses, or increased excess emission below $\sim2\kev$ using a variety of toy models (e.g. \citealt{Vaughan+1999,Grupe+2010,Gliozzi+2020,Ohja+2020}), and some NLS1 galaxies also show extreme soft excesses (e.g. \citealt{Gallo+2004,Middleton+2007}). These properties suggest that the soft excesses of NLS1 AGN may differ from those of BLS1 AGN. 

The X-ray spectral slopes of NLS1 galaxies in the $2-10\kev$ energy range have also been shown to be steeper than those of BLS1 galaxies (e.g. \citealt{Brandt+1997,Grupe+2004}). This range is expected to be dominated by emission from the corona, which takes the form of a power law. Like the rapid variability, this may point to differences in the temperature of the corona \citep{Gallo2006,Gallo+2018}, where in particular, cooler coronae produce steeper X-ray spectral slopes (\citealt{Pounds+1995}). Clearly, NLS1 galaxies occupy an extreme parameter space in many respects and are worthy of further analysis. 

Many previous works have taken advantage of the broad-band observing capabilities of \suzaku\ for the analyses of samples of AGN. These range significantly in scope, from \feka\ line profile studies (e.g. \citealt{Fukazawa+2011,Fukazawa+2016,Mantovani+2016}) and physical spectral modelling (e.g. \citealt{Noda+2013,Walton+2013,Iso+2016}) to spectral variability studies (e.g. \citealt{Miyazawa+2009}).

In this work, we present a sample of NLS1 and BLS1 AGN observed with the \suzaku\ satellite in order to characterise their X-ray spectral properties, and investigate differences between the two classes. This work will expand upon previous samples by including a larger number of optical type-1 AGN, and studying the properties of NLS1 and BLS1 galaxies. The goal is to examine a range of parameters from the soft to hard band to search for similarities and differences between the two classes. In Section~\ref{sect:sample}, we describe the sample selection and present literature parameters for each AGN. In Section~\ref{sect:data}, we describe the data analysis. Section~\ref{sect:spectra} describes the spectral model used for analysis and discusses parameter distributions. In Section~\ref{sect:5}, we present correlations between parameters and compare the two samples. A discussion of results is given in Section~\ref{sect:disc}, and conclusions are presented in Section~\ref{sect:conclusion}.

\section{Sample selection}
\label{sect:sample}
The X-ray observatory used for this work is the \suzaku\ satellite (\citealt{suzaku}), operational between 2005 and 2015. The observatory collected data using two main detectors; the four X-ray Imaging Spectrometers (XIS) and the Hard X-ray Detector (HXD). The HXD instrument includes the PIN and GSO detectors intended for the capture of high energy X-rays. By combining data from the XIS and PIN detectors, \suzaku\ is sensitive between $\sim0.5-100\kev$, allowing for the simultaneous observation of the X-ray soft excess, power law emission from the corona, \feka\ emission and Compton hump. The simultaneous detection of all these components allows for detailed, broad-band X-ray spectral characterisation.

The sample is assembled to include objects that exhibit relatively small levels of absorption so that the low-energy emission (i.e. the soft excess) can be constrained.
All \suzaku\ observations used in this work are publicly available on the \suzaku\ DARTS{\footnote{https://darts.isas.jaxa.jp/astro/suzaku/}} website. These observations are divided into several sub-categories based on object class. AGN are typically classified as extra-galactic compact sources, and we thus begin our search for Seyfert 1 galaxies in this sub-section. In total, 518 observations of approximately 350 sources were made over the mission lifetime. Of these, 225 were found to be Seyfert galaxies, with 106 being type-1 AGN. 

We first selected all Seyfert 1, 1.2, and 1.5 galaxies, as these objects are expected to show little X-ray absorption due to obscuration by the torus. We include only those galaxies with intrinsic (host-galaxy) column densities of less than $10^{22}\pscm$, allowing for some constraint to be placed on the soft excess component. Objects with redshifts $z>0.5$ are also excluded to ensure that the intrinsic spectrum down to at least $\sim1\kev$ could be measured. 

The remaining 72 objects were then divided into sub-samples of NLS1 and BLS1 galaxies based on classifications found in literature. All the NLS1s have FWHM(H$\beta$) $<2000~\textrm{km}\ps$, relatively strong \feii\ and weak \oiii\ according to previous works. One source, MCG+04-22-042, has several different measurements for the FWHM(H$\beta$), with the smallest width recorded as $1946\pm$

\onecolumn
\begin{longtable}{lllll}
		\hline
		(1) & (2) & (3) & (4) & (5) \\
		Object & Redshift & N$_{\rm H}$  & FWHM (H$\beta$) & M$_{\rm BH}$ \\
		& & [$\times 10^{22}$~\pscm] & [\kmps] & [$\times$ 10$^7$ M$_{\rm sun}$] \\
		\hline
		\endhead
		\multicolumn{5}{c}{NLS1} \\
		\hline
		1H0323+342         & 0.061   & 0.127   & 1520$^{(1)}$ & 2.00$^{(2)}$ \\
		1H0707-495         & 0.04057 & 0.0431  & 1000$^{(3)}$ & 0.204$^{(4)}$ \\
		Ark 564            & 0.02468 & 0.0534  & 750$^{(3)}$  & 0.186$^{(5)}$ \\
		IRAS 05262+4432    & 0.03217 & 0.321   & 700$^{(3)}$  & 2.06$^{(6)}$ \\
		IRAS 13224-3809    & 0.0658  & 0.0476  & 650$^{(3)}$  & 0.631$^{(7)}$ \\
		MCG-6-30-15        & 0.00775 & 0.0392  & 1700$^{(4)}$ & 0.200$^{(5)}$ \\
		Mrk 1040           & 0.01665 & 0.0673  & 1830$^{(6)}$ & 4.37$^{(7)}$ \\
		Mrk 110            & 0.03529 & 0.013   & 1760$^{(8)}$ & 1.96$^{(9)}$ \\
		Mrk 335            & 0.02578 & 0.0356  & 1710$^{(8)}$ & 1.70$^{(9)}$ \\
		Mrk 359            & 0.01739 & 0.0426  & 900$^{(6)}$  & 0.170$^{(6)}$ \\
		Mrk 478            & 0.07906 & 0.0105  & 1630$^{(8)}$ & 1.99$^{(10)}$ \\
		Mrk 766            & 0.01293 & 0.0178  & 1100$^{(8)}$ & 0.664$^{(9)}$ \\
		NGC 4051           & 0.00234 & 0.0115  & 1170$^{(8)}$ & 0.135$^{(9)}$ \\
		PG 1211+143        & 0.0809  & 0.0274  & 1900$^{(8)}$ & 4.07$^{(5)}$ \\
		PG 1404+226        & 0.098   & 0.0222  & 880$^{(3)}$  & 0.450$^{(11)}$ \\
		PKS 0558-504       & 0.1372  & 0.036   & 1250$^{(12)}$ & 4.50$^{(12)}$ \\
		RE J1034+396       & 0.04244 & 0.04244 & 1500$^{(4)}$  & 0.245$^{(13)}$ \\
		RX J0134.2-4258    & 0.237   & 0.0167  & 900$^{(8)}$  & 2.00$^{(14)}$ \\
		RX J1633.3+4718    & 0.1158  & 0.0174  & 900$^{(15)}$  & 0.300$^{(15)}$ \\
		SWIFT J2127.4+5654 & 0.0144  & 0.765   & 2000$^{(16)}$ & 1.51$^{(16)}$ \\
		TON S180           & 0.06198 & 0.0136  & 970$^{(8)}$  & 0.710$^{(9)}$ \\
		WKK 4438           & 0.016   & 0.291   & 1700$^{(16)}$ & 0.200$^{(6)}$ \\
		\hline
		\multicolumn{5}{c}{BLS1} \\
		\hline
		1H0419-577         & 0.104   & 0.0116  & 2580$^{(8)}$ & 13.0$^{(8)}$ \\
		3C 111             & 0.0485  & 0.285   & 4800$^{(17)}$ & 360$^{(17)}$ \\
		3C 120             & 0.03301 & 0.194   & 3711$^{(18)}$  & 5.50$^{(9)}$ \\
		3C 382             & 0.05787 & 0.0619  & ...  & 61.7$^{(19)}$ \\
		3C 390.3           & 0.0561  & 0.0441  & 10000$^{(4)}$  & 43.5$^{(9)}$ \\
		3C 78              & 0.02865 & 0.104   & ...  & 39.8$^{(20)}$ \\
		4C+74.26           & 0.104   & 0.114   & 11000$^{(21)}$ & 398$^{(22)}$ \\
		Ark 120            & 0.03271 & 0.0998  & 5800$^{(4)}$ & 11.7$^{(9)}$ \\
		B3 0309+411        & 0.136   & 0.101   & ...  & ... \\
		CBS 126            & 0.0791  & 0.0131  & 2980$^{(8)}$ & 7.09$^{(8)}$ \\
		ESO 511-G030       & 0.02239 & 0.0434  & 3335$^{(23)}$ & 6.92$^{(19)}$ \\
		ESO 548-G081       & 0.01448 & 0.0237  & ...  & 5.50$^{(19)}$ \\
		Fairall 9          & 0.04702 & 0.0286  & 5780$^{(4)}$ & 19.9$^{(9)}$ \\
		IGR J16185-5928    & 0.03463 & 0.207   & 2918$^{(23)}$ & 2.80$^{(24)}$ \\
		IGR J19378-0617    & 0.01025 & 0.191   & 2700$^{(24)}$  & ... \\
		III Zw 2           & 0.0898  & 0.0592  & 5295$^{(23)}$ & 11.7$^{(9)}$ \\
		KAZ 102            & 0.136   & 0.0475  & ...  & 10.0$^{(25)}$ \\
		LEDA 168563        & 0.029   & 0.437   & 7400$^{(26)}$ & 11.0$^{(26)}$ \\
		MCG-02-14-009      & 0.02845 & 0.1     & ...  & 1.35$^{(5)}$ \\
		MCG-02-58-22       & 0.04686 & 0.0334  & 6360$^{(4)}$  & 30.0$^{(27)}$ \\
		MCG+04-22-042*     & 0.03235 & 0.035   & 2600$^{(23)}$ & 31.6$^{(22)}$ \\
		MCG+08-11-11       & 0.02048 & 0.297   & 3630$^{(4)}$  & 12.0$^{(28)}$ \\
		MR 2251-178        & 0.06398 & 0.0264  & 4617$^{(23)}$ & 19.0$^{(29)}$ \\
		Mrk 1018           & 0.04244 & 0.0267  & 5858$^{(30)}$  & 6.92$^{(31)}$ \\
		Mrk 1148           & 0.064   & 0.0501  & ... & ... \\
		Mrk 1320           & 0.103   & 0.024   & ... & ... \\
		Mrk 205            & 0.07085 & 0.029   & 4560$^{(23)}$ & 20.9$^{(5)}$ \\
		Mrk 279            & 0.03045 & 0.0159  & 5360$^{(4)}$  & 2.72$^{(9)}$ \\
		Mrk 352            & 0.01486 & 0.0531  & 4214$^{(23)}$ & 0.85$^{(32)}$ \\
		Mrk 509            & 0.0344  & 0.0393  & 2270$^{(4)}$ & 11.2$^{(9)}$ \\
		Mrk 530            & 0.02952 & 0.0442  & ...  & 11.5$^{(33)}$ \\
		Mrk 79             & 0.02219 & 0.0543  & 4000$^{(23)}$ & 4.09$^{(9)}$ \\
		Mrk 841            & 0.03642 & 0.0246  & 5470$^{(4)}$  & 7.59$^{(34)}$ \\
		NGC 4593           & 0.009   & 0.0167  & 4910$^{(8)}$ & 0.950$^{(8)}$ \\
		NGC 6814           & 0.00521 & 0.148   & ...  & 1.09$^{(9)}$ \\
		NGC 7213           & 0.00584 & 0.0111  & 13000$^{(35)}$ & 10.00$^{(35)}$ \\
		NGC 7469           & 0.01632 & 0.0539  & 3388$^{(4)}$  & 0.904$^{(9)}$ \\
		NGC 985            & 0.04314 & 0.0346  & ...  & ... \\
		PG 1322+659        & 0.168   & 0.017   & ...  & 19.5$^{(36)}$ \\
		PG 1626+554        & 0.133   & 0.0108  & ...  & 53.7$^{(37)}$ \\
		RBS 1124           & 0.208   & 0.0156  & ...  & 18.0$^{(38)}$ \\
		SBS 1301+540       & 0.0299  & 0.0173  & 7174$^{(23)}$ & 3.16$^{(22)}$ \\
		SWIFT J0501.9-3239 & 0.01244 & 0.0184  & ...  & ... \\
		SWIFT J0904.3+5538 & 0.03714 & 0.0232  & 3022$^{(23)}$ & ... \\
		SWIFT J1310.9-5553 & 0.104   & 0.208   & ...  & ... \\
		UGC 6728           & 0.00652 & 0.0446  & 2308$^{(30)}$  & 0.513$^{(9)}$ \\
		ZW 229-15          & 0.02788 & 0.0534  & 3360$^{(39)}$ & 0.818$^{(9)}$ \\
		\hline
		\caption{AGN parameters for all NLS1s and BLS1s in the \suzaku\ sample. Column (1) gives the object name, and columns (2) and (3) show the redshift and Galactic column density, obtained from NED and \protect\cite{Willingale+2013}, respectively. Columns (4) and (5) show the FWHM and black hole mass for each source.  One source, MCG+04-22-042, is included in the BLS1 sample but has been alternatively classified as an NLS1 and is marked with an asterisk. Masses and FWHM values are taken from; (1) \protect\cite{Paliya+2014}; (2) \protect\cite{Landt+2017}; (3) \protect\cite{Vaughan+1999}; (4) \protect\cite{Bian+2003}; (5) \protect\cite{Ponti+2012}. (6) \protect\cite{Wang+2001}; (7) \protect\cite{GM+2012}; (8) \protect\cite{Grupe+2010}; (9) \protect\cite{AGNmass}; (10) \protect\cite{Porquet+2004}; (11) \protect\cite{Mallick+2018}; (12) \protect\cite{Gliozzi+2007}; (13) \protect\cite{Czerny+2016};  (14) \protect\cite{Grupe+2000}; (15) \protect\cite{Yuan+2010}; (16) \protect\cite{Malizia+2008}; (17) \protect\cite{3c111}; (18) \protect\cite{Du+2018}; (19) \protect\cite{Vasudevan+2009}; (20) \protect\cite{Woo+2002}; (21) \protect\cite{Bal+2005}; (22) \protect\cite{Soldi+2014}; (23) \protect\cite{swift70}; (24) \protect\cite{Malizia+2008}; (25) \protect\cite{kaz102}; (26) \protect\cite{Liebmann+2018}; (27) \protect\cite{Rivers+2011}; (28) \protect\cite{mcg08}; (29) \protect\cite{Wang+2009}; (30) \protect\cite{Winter+2010}; (31) \protect\cite{m1018}; (32) \protect\cite{swift9}; (33) \protect\cite{m530}; (34) \protect\cite{m841}; (35) \protect\cite{ngc7213}; (36) \protect\cite{pg13}; (37); \protect\cite{pg16}; (38) \protect\cite{rbs1124}; (39) \protect\cite{zw229hb}. }
		\label{tab:samp}
\end{longtable}
\twocolumn

\noindent $170~\textrm{km}\ps$ (\citealt{LaMura+2011}). This borderline object is included in the BLS1 sample and marked with an asterisk in Table~\ref{tab:samp}. It appears comparable to other BLS1s in our analysis.

Two objects, Mrk 1239 (typically classified as an NLS1) and ESO 323-G077 (BLS1), were rejected from the sample as they exhibited atypical spectra (see \citealt{m1239} and \citealt{m1239+2020} for Mrk 1239, and \citealt{Miniutti+2014} for ESO 323-G077). Both AGN show significant optical polarization and ESO 323-G077 has been described as a borderline Seyfert 1/2 (e.g. \citealt{Schmid+2003}).  Mrk 590, typically classified as  BLS1, was also removed. This source has been shown to change look (i.e., show dramatic optical emission line variability), and although typically classified as a Seyfert 1, did not show a typical type-1 optical spectrum during the \suzaku\ observation (\citealt{m590}). Additionally, only two out of the six observations of the BLS1 galaxy 3C 120 are included here due to calibration uncertainties during the first four observations, which were taken during the  calibration phase. The final sample consists of 22 narrow-line Seyfert 1 AGN, and 47 broad-line Seyfert 1 AGN, for a total of 69 AGN suitable for analysis. 

To characterise our sample, we first perform a literature search to determine the black hole mass and H$\beta$ FWHM of each object. Redshifts for each object are taken from NED{\footnote{https://ned.ipac.caltech.edu/}}, and column densities viewed through the Galaxy are taken from \cite{Willingale+2013}. The results, where available, are presented in Table~\ref{tab:samp}. The sample covers a redshift range of $z=0.00234$ to $z=0.237$, with most AGN having $z<0.1$. NLS1 galaxies appear to have lower black hole masses on average than their BLS1 counterparts, in agreement with previous observations (e.g. \citealt{Pounds+1995,Wang+1996,Grupe+2004,Grupe+2010}).

\section{Data reduction}
\label{sect:data}
\subsection{\suzaku\ XIS}
\label{sect:xis}
The \suzaku\ XIS detectors feature four CCD cameras; XIS0, XIS1, XIS2 and XIS3. In late 2006, the XIS2 detector was struck by a small meteorite and became unsuitable for use in scientific analysis, so observations taken after November 2006 have data from only three XIS detectors. The XIS1 detector differs from the others in that it is back-illuminated (BI). The detector response is significantly different from the other three front-illuminated (FI) CCDs, and can not be co-added with the other detectors. XIS1 has very high background at high energies (above $\sim7\kev$), which complicates analysis of the \feka\ line profile. For the remainder of this work, we therefore focus only on the FI detectors XIS0 and XIS3 (and XIS2 when available). 

Cleaned event files from the FI CCDs were used for the extraction of data products in {\sc xselect v2.4g}. For each instrument, source photons were extracted using a $240\arcs$ region centred around the source, while background photons were extracted from a $180\arcs$ off-source region. Calibration zones located in the corners of the CCDs were avoided in the background extraction. Response matrices for each detector were generated using the tasks {\sc xisrmfgen} and {\sc xissimarfgen}. The data and responses from each of the FI detectors were then merged to improve signal-to-noise. 

In this work we are comparing overall average properties of the sample. The data from AGN observed at multiple epochs are merged to form one single spectrum for each unique source.  Source and background spectra, as well as responses are merged for each object, with spectra and responses weighted based on exposure time. The merged spectrum for each object is then used in spectral modelling for the remainder of this work. For each merged spectrum, we model the $0.6-10\kev$ energy band, excluding the calibration uncertainty regions between $1.72-1.88\kev$ and $2.19-2.37\kev$ (\citealt{Nowak+2011}). Observations are summarized in Table~\ref{tab:obs}.

\subsection{\suzaku\ HXD}
\label{sect:hxd}
Data from the \suzaku\ HXD-PIN detector are also used in the analysis. In general, the $15-40\kev$ energy band is modelled, but for some objects, this band is smaller to improve the detection signal. Data are processed using the tool {\sc hxdpinxbpi}. Both the non-X-ray background (NXB) and the cosmic X-ray background (CXB) are considered to determine the detection threshold for each observation, where a detection of 5 per cent of background is considered significant (\citealt{Fukazawa+2009}). As with the XIS, data from individual observations of each source are merged to create one averaged spectrum for each source. A cross calibration factor between the PIN and the XIS of 1.16 or 1.18 is applied, depending on whether the observation was taken in XIS or HXD nominal pointing mode, respectively. The exception for this is NGC 6814, where a much lower calibration constant of 0.5 is required (\citealt{ngc6814}). We also neglect the PIN data for Zw 229-15, as these data were shown to diverge significantly from all models and other high energy data sets, possibly due to low data quality or calibration uncertainties (\citealt{zw229}).

\section{Spectral modelling}
\label{sect:spectra}
\subsection{Modelling procedure}
\label{sect:xspec}

All spectral fits are performed using {\sc xspec v12.9.0n} (\citealt{xspec}) from {\sc heasoft v6.26}. Abundances are taken from \citealt{Wilms+2000}, and values for the Galactic column density for each source are taken from \cite{Willingale+2013}. Spectra from the \suzaku\ XIS detectors have been grouped using optimal binning (\citealt{optbin}), using the {\sc ftgrouppha} tool, while data from the PIN have been grouped to a minimum of 20 counts per bin. Model fitting was done using C-statistic (\citealt{Cash1979}), to account for the fact that the data do not follow a Gaussian distribution, but rather a Poisson distribution. All errors are quoted at the 90 per cent confidence level. For ease of comparison between objects, luminosities free of Galactic and host-galaxy absorption, rather than fluxes or model normalisations, are presented. These luminosities are calculated using the {\sc lumin} command in {\sc xspec}. We adopt a value for the Hubble constant of ${\rm H}_0 = 70$\kmpspMpc, and $\Lambda_0 = 0.73$. 

\begin{figure}
	\centering
	\includegraphics[width=0.95\columnwidth]{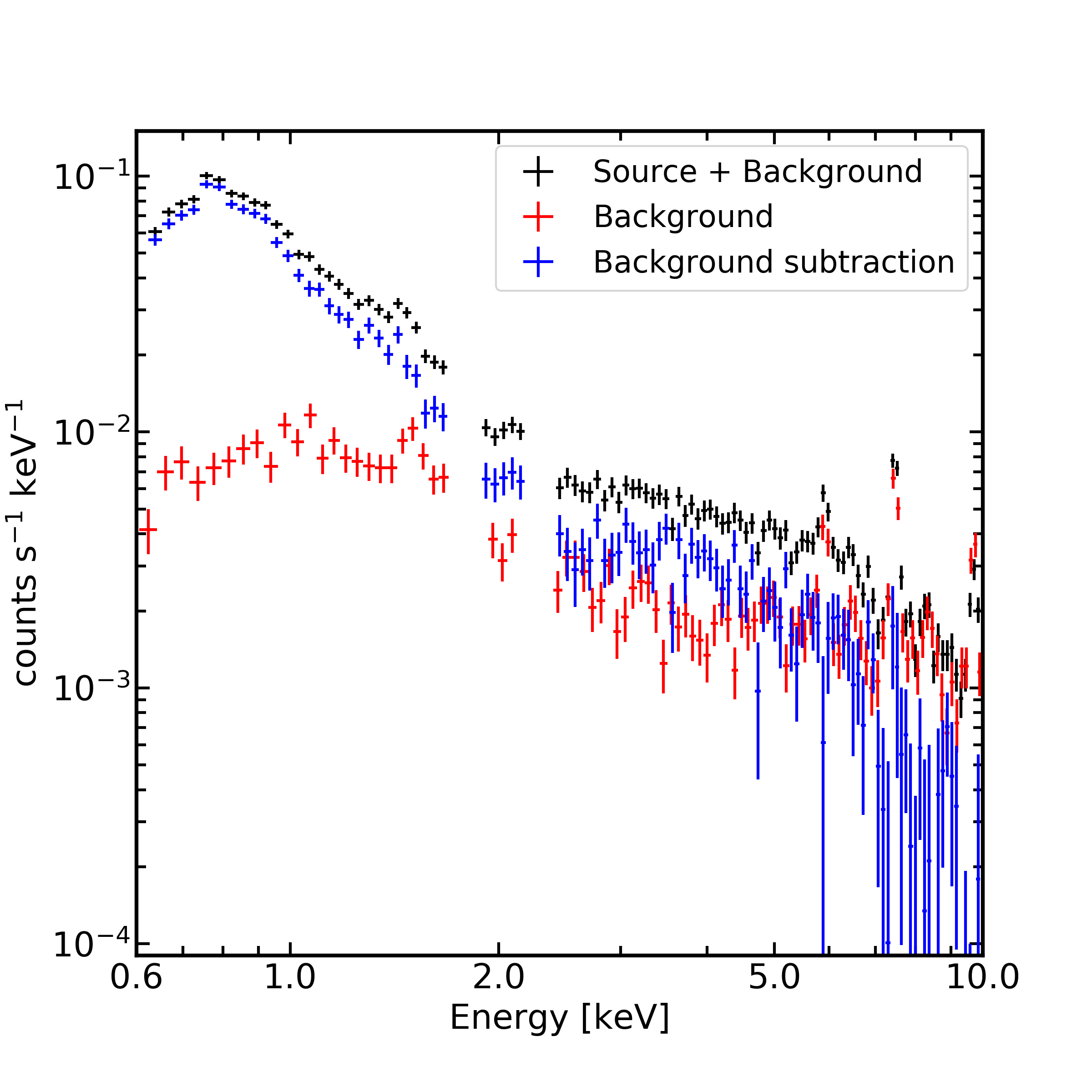}
	\caption{Source+background (black), background (red) and background subtracted spectrum (blue) for a NLS1, 1H0707-495. The background emission lines begin to dominate the spectrum above $6\kev$, but the source contribution is still detectable.  The source contribution above $6\kev$ can be better constrained by modelling the background rather than subtracting it or cutting the spectrum.}
	\label{fig:back}
\end{figure}

To maximize the spectral information from each observation, we model the XIS background spectrum obtained from the off-source regions rather than simply subtracting it. This is required when using C-statistics, and allows for improved analysis of dim sources which are background dominated at high energies, such as 1H0707-495 (Fig.~\ref{fig:back}). This source becomes dominated by emission from the background above $\sim6\kev$ (see the background subtracted spectrum in blue), but by modelling the background, we can consider up to $10\kev$. 

\onecolumn
\renewcommand{\arraystretch}{1.0}
	\begin{longtable}{lllllllll}
		\hline
		(1) & (2) & (3) & (4) & (5) & (6) & (7) & (8) & (9) \\
		Object & Obs ID & Date & Detector & Exposure (s) & SC cts & BG cts & PIN cts  & PIN \% \\
		& & YYYY-MM-DD & & (all XIS) & (all XIS) & (all XIS) & & \\
		\hline
		\endhead
		\multicolumn{9}{c}{NLS1} \\
		\hline
		1H0323+342         & 704034010 & 2009-07-26  & XIS 0+3   & 371000    & 232762  & 5491  & 40370   & 9.0\\
		& 707015010 & 2013-03-01  & XIS 0+3   &         &       &         &       &        \\
		1H0707-495         & 700008010 & 2005-12-03  & XIS 0+2+3 & 292900       & 23750   & 3938  & 16720    & 2.9*  \\
		Ark 564            & 702117010 & 2007-06-26  & XIS 0+3   & 306400     & 886516  & 8138  & 27061 & 8.4 \\
		& 710018010 & 2015-05-25  & XIS 0+3   &         &       &         &       &        \\
		IRAS 05262+4432    & 703019010 & 2008-09-12  & XIS 0+3   & 164200      & 16817   & 1907  & 12503    & 2.1*   \\
		IRAS 13224-3809    & 701003010 & 2007-01-26  & XIS 0+3   & 395900     & 49337   & 5192  & -   & -   \\
		MCG-6-30-15        & 700007010 & 2006-01-09  & XIS 0+2+3 & 1015000   & 2797730 & 23791 & 115952 & 23.3 \\
		& 700007020 & 2006-01-23 & XIS 0+2+3 &         &       &         &       &        \\
		& 700007030 & 2006-01-27 & XIS 0+2+3 &         &       &         &       &        \\
		Mrk 1040           & 707046010 & 2013-08-11  & XIS 0+3   & 274900    & 533369  & 8471  & 36897 & 31.1 \\
		Mrk 110            & 702124010 & 2007-11-02  & XIS 0+3   & 181800    & 229892  & 3684  & 27197 & 15.8 \\
		Mrk 335            & 701031010 & 2006-06-21  & XIS 0+2+3 & 1052000    & 702650  & 16100 & 43096  & 9.2\\
		& 708016020 & 2013-06-14  & XIS 0+3   &         &       &         &       &        \\
		& 708016010 & 2013-06-11  & XIS 0+3   &         &       &         &       &        \\
		Mrk 359            & 701082010 & 2007-02-06  & XIS 0+3   & 215000    & 79777   & 2776  & 19905  & 4.1*\\
		Mrk 478            & 706041010 & 2011-07-14  & XIS 0+3   & 170600    & 43759   & 2070  & 13877 & 3.1* \\
		Mrk 766            & 701035010 & 2006-11-16  & XIS 0+3   & 315100   & 277229  & 3996  & 45721   & 9.0  \\
		& 701035020 & 2007-11-17  & XIS 0+3   &         &       &         &       &        \\
		NGC 4051           & 703023010 & 2008-11-06  & XIS 0+3   & 1065000   & 1526650 & 6667  & 129761 & 15.3\\
		& 700004010 & 2005-11-10 & XIS 0+2+3 &         &       &         &       &        \\
		& 703023020 & 2008-11-13  & XIS 0+3   &         &       &         &       &        \\
		PG 1211+143        & 700009010 & 2005-11-24  & XIS 0+2+3 & 289200    & 86466   & 4345  & 17679  & 4.0*\\
		PG 1404+226        & 708045010 & 2013-07-13  & XIS 0+3   & 365400    & 12833   & 3818  & 24168  & 7.7\\
		& 707026010 & 2012-12-23  & XIS 0+3   &         &       &         &       &        \\
		PKS 0558-504       & 701011030 & 2007-01-19  & XIS 0+3   & 200100   & 310836  & 2806  & 16297  & 8.2 \\
		& 701011010 & 2007-01-17  & XIS 0+3   &         &       &         &       &        \\
		& 701011040 & 2007-01-20  & XIS 0+3   &         &       &         &       &        \\
		& 701011050 & 2007-01-21  & XIS 0+3   &         &       &         &       &        \\
		& 701011020 & 2007-01-18  & XIS 0+3   &         &       &         &       &        \\
		RE J1034+396        & 707039010 & 2012-11-14  & XIS 0+3   & 199800    & 34194   & 2098  & 14164  & 5.5\\
		RX J0134.2-4258    & 707014010 & 2012-12-29  & XIS 0+3   & 162900    & 34665   & 1627  & 1519   & 5.8 \\
		RX J1633.3+4718    & 706027040 & 2012-02-05  & XIS 0+3   & 334200   & 35951   & 4043  & 21475   & 2.8*\\
		& 706027030 & 2012-01-13  & XIS 0+3   &         &       &         &       &        \\
		& 706027010 & 2011-07-01  & XIS 0+3   &         &       &         &       &        \\
		& 706027020 & 2011-07-08  & XIS 0+3   &         &       &         &       &        \\
		SWIFT J2127.4+5654 & 702122010 & 2007-12-09  & XIS 3     & 91730     & 131319  & 2227  & 30599  & 20.6\\
		TON S180           & 701021010 & 2006-12-09  & XIS 0+3   & 241400  & 176949  & 3602  & 23364  & 4.8*  \\
		WKK 4438           & 706011010 & 2012-01-22  & XIS 0+3   & 140600     & 91783   & 1446  & 13314  & 6.2\\
		\hline
		\multicolumn{9}{c}{BLS1} \\
		\hline
		1H0419-577         & 702041010 & 2007-07-25   & XIS 0+3 & 658000 & 770926 & 13936 & 48285 & 15.5 \\
		& 704064010 & 2010-01-16   & XIS 0+3 & & & & & \\
		3C 111             & 703034010 & 2008-08-22   & XIS 0+3 & 725700 & 1153400 & 15114 & 27277 & 37.4 \\
		& 705040010 & 2010-09-02   & XIS 0+3 & & & & & \\
		& 705040020 & 2010-09-09   & XIS 0+3 & & & & & \\
		& 705040030 & 2010-09-14   & XIS 0+3 & & & & & \\
		3C 120 & 706042010 & 2012-02-09 & XIS 0+3 & 602200 & 1665920 & 16153 & 89641 & 36.0 \\
		& 706042020 & 2012-02-14 & XIS 0+3 & & & & & \\
		3C 382             & 702125010 & 2007-04-27   & XIS 0+3 & 261200 & 722194 & 5800 & 43401 & 24.7 \\
		3C 390.3 & 708034010 & 2013-05-24 & XIS 0+3 & 200700 & 481139 & 3355 & 32140 & 32.1 \\
		3C 78              & 706013010 & 2011-08-20   & XIS 0+3 & 194000 & 23330 & 2318 & - & - \\
		4C+74.26           & 702057010 & 2007-10-28   & XIS 0+3 & 386000 & 562383 & 7614 &  32440 & 22.2 \\
		& 706028010 & 2011-11-23   & XIS 0+3 & & & & & \\
		Ark 120            & 702014010 & 2007-04-01   & XIS 0+3 & 201700 & 410487 & 6401 & 32393 & 21.2 \\
		B3 0309+411        & 706036010 & 2012-02-19   & XIS 0+3 & 209200 & 55068 & 3994 & 16910 & 2.4* \\
		CBS 126            & 705042010 & 2010-10-18   & XIS 0+3 & 203100 & 70422 & 2704 & 22703 & 4.2* \\
		ESO 511-G030       & 707023020 & 2012-07-22   & XIS 0+3 & 551900 & 725334 & 11944 & 13447 & 20.4 \\
		& 707023030 & 2012-08-17   & XIS 0+3 & & & & & \\
		ESO 548-G081       & 704026010 & 2009-08-03   & XIS 0+3 & 78830 & 58808 & 1495 & 10054 & 13.8 \\
		Fairall 9          & 705063010 & 2010-05-19   & XIS 0+3 & 794200 & 1350110 & 13064 & 103887 & 20.4 \\
		& 702043010 & 2007-06-07   & XIS 0+3 & & & & & \\
		IGR J16185-5928    & 702123010 & 2008-02-09   & XIS 0+3 & 153200 & 73459 & 4219 & 22482 & 9.7  \\
		IGR J19378-0617 & 705055010 & 2010-10-16 & XIS 0+3 & 155600 & 274166 & 4215 & 19433 & 11.0 \\
		III Zw 2           & 706031010 & 2011-06-14   & XIS 0+3 & 162900 & 85023 & 2827 & 21272 & 10.4 \\
		KAZ 102            & 701012010 & 2006-06-09   & XIS 0+3 & 82920 & 10915 & 1022 & 3298 & 3.1* \\
		LEDA 168563        & 708017010 & 2013-09-01   & XIS 0+3 & 200100 & 149668 & 3884 & 22645 & 19.2 \\
		MCG-02-14-009      & 703060010 & 2008-08-28   & XIS 0+3 & 284300 & 73480 & 3786 & 24412 & 6.0 \\
		MCG-02-58-22 & 704032010 & 2009-12-02 & XIS 0+3 & 278000 & 1009640 & 5994 & 39753 & 38.0 \\
		MCG+04-22-042      & 704028010 & 2009-11-22   & XIS 0+3 & 81980 & 74668 & 1439 & 10480 & 12.3 \\
		MCG+08-11-11 & 702112010 & 2009-09-17 & XIS 0+3 & 197500 & 572367 & 6899 & 39533 & 39.6 \\
		MR 2251-178        & 704055010 & 2009-05-07   & XIS 0+3 & 273900 & 623692 & 4969 & 37765 & 26.7 \\
		Mrk 1018 & 704044010 & 2009-07-03 & XIS 0+3 & 87820 & 55075 & 1551 & 10700 & 13.4 \\
		Mrk 1148 & 708033010 & 2013-12-30 & XIS 0+3 & 56650 & 64418 & 835 & 5676 & 17.9 \\
		Mrk 1320 & 704006010 & 2009-06-28 & XIS 0+3 & 41560 & 3448 & 530 & - & - \\
		Mrk 205            & 705062010 & 2010-05-22   & XIS 0+3 & 201900 & 114219 & 4128 & 27039 & 14.1 \\
		Mrk 279 & 704031010 & 2009-05-14 & XIS 0+3 & 320700 & 92008 & 3426 & 40011 & 9.2 \\
		Mrk 352            & 704025010 & 2010-01-06   & XIS 0+3 & 91370 & 61034 & 1893 & 6303 & 8.0 \\
		Mrk 509            & 705025010 & 2010-11-21   & XIS 0+3 & 471000 & 1689290 & 11724 & 70667 & 31.9 \\
		& 701093040 & 2006-11-27   & XIS 0+3 & & & & & \\
		& 701093020 & 2006-10-14   & XIS 0+2+3 & & & & & \\
		& 701093010 & 2006-04-25   & XIS 0+2+3 & & & & & \\
		& 701093030 & 2006-11-15   & XIS 0+3 & & & & & \\
		Mrk 530 & 707003010 & 2012-06-02 & XIS 0+3 & 202500 & 313042 & 2667 & 25044 & 17.3 \\
		Mrk 79             & 702044010 & 2007-04-03   & XIS 0+3 & 167400 & 131341 & 1941 & 25288 & 9.6 \\
		Mrk 841 & 701084010 & 2007-01-22 & XIS 0+3 & 914700 & 686420 & 20200 & 119074 & 11.3 \\
		& 701084020 & 2007-07-23 & XIS 0+3 & & & & & \\
	    & 706029010 & 2012-01-05 & XIS 0+3 & & & & & \\
		& 706029020 & 2012-01-18 & XIS 0+3 & & & & & \\
		NGC 4593           & 702040010 & 2007-12-15   & XIS 0+3 & 971600 & 965252 & 15277 & 47703 & 16.0 \\
		& 709014050 & 2014-12-30   & XIS 0+3 & & & & & \\
		& 709014040 & 2014-12-26   & XIS 0+3 & & & & & \\
		& 709014020 & 2014-06-22   & XIS 0+3 & & & & & \\
		& 709014030 & 2014-12-15   & XIS 0+3 & & & & & \\
		& 709014010 & 2014-06-16   & XIS 0+3 & & & & & \\
		NGC 6814 & 706032010 & 2011-11-02 & XIS 0+3 & 84260 & 36352 & 1046 & 10096 & 4.8* \\
		NGC 7213 & 701029010 & 2006-10-22 & XIS 0+2+3 & 272200 & 428912 & 4137 & 32255 & 16.2 \\
		NGC 7469 & 703028010 & 2008-06-24 & XIS 0+3 & 224200 & 254110 & 4818 & 29828 & 19.3 \\
		NGC 985 & 704042010 & 2009-07-15 & XIS 0+3 & 64000 & 47761 & 1263 & 7607 & 13.9 \\
		PG 1322+659        & 706018010 & 2011-11-27   & XIS 0+3 & 163400 & 11544 & 1641 & - & -  \\
		PG 1626+554        & 706017010 & 2011-11-11   & XIS 0+3 & 121700 & 38611 & 1788 & - & - \\
		RBS 1124           & 702114010 & 2007-04-14   & XIS 0+3 & 172500 & 48581 & 1816 & 27027 & 7.5 \\
		SBS 1301+540       & 709018010 & 2014-11-30   & XIS 0+3 & 189500 & 47297 & 2000 & 13525 & 6.9 \\
		SWIFT J0501.9-3239 & 703014010 & 2008-04-11   & XIS 0+3 & 82600 & 176307 & 2556 & 13042 & 24.3 \\
		SWIFT J0904.3+5538 & 704027010 & 2009-04-28   & XIS 0+3 & 83890 & 11307 & 989 & 6885 & 5.7 \\
		SWIFT J1310.9-5553 & 706009010 & 2011-07-19   & XIS 0+3 & 165600 & 41357 & 1993 & 17854 & 6.5 \\
		UGC 6728           & 704029010 & 2009-06-06   & XIS 0+3 & 98010 & 80844 & 1710 & 12202 & 13.5 \\
		ZW 229-15          & 706035010 & 2011-06-03   & XIS 0+3 & 337400 & 91424 & 5253 & - & - \\
		\hline
		\caption{Observation parameters for each object in the sample. Columns (1), (2) and (3) give the object name, observation ID and observation start date, respectively. The XIS detectors used in the observation are shown in column (4). Column (5) gives the combined exposure for all available XIS detectors. Source, background and PIN (source plus background) counts are given in columns (6), (7) and (8), respectively. Source and background counts are quoted for all combined XIS detectors. Column (9) gives the source detection level over the background (per cent) in the PIN, where values marked with an asterisk (*) are below the significant detection threshold of 5 per cent. AGN with PIN data with a significance less than 1 per cent, or AGN where no PIN data was available, are marked with a dash.}
		\label{tab:obs}\\
	\end{longtable}
\twocolumn

\noindent The background model for each source is constructed using narrow emission features to model the instrumental lines from the detectors and multiple power law components to match the shape of the background continuum. The resulting background models are then applied to the combined source+background spectrum before modeling the source contribution.

\subsection{Broad-band model}
\label{sect:model}

The goal of this work is not to assume and test a physical model, but instead to characterise the shape of the spectrum and compare important values within the sample. To characterise the shape of the X-ray spectrum, we use a simple toy model. Not all of the AGN spectra are perfectly fit with the toy model, but the spectral shape is sufficiently well measured to allow for comparison.

The model consists of a power law to account for emission from the corona, and a blackbody component to model the soft excess below $\sim2\kev$. We also model \feka\ emission with a Gaussian emission profile, with the energy frozen to $6.4\kev$ and the width frozen to $1\eV$ so that only the normalisation is free to vary. 

To approximate the hard excess emission that might arise from the Compton hump peaking at $\sim20\kev$, we use a second Gaussian emission profile, with the width fixed to $5\kev$ and the energy frozen to $20\kev$. We do so because the hard excess is constrained only by the PIN data, which has poorer statistics compared to the XIS data. More complicated models that extend over a broad energy band are dominated by the statistics in the XIS band, which results in the hard excess being improperly modelled for some sources. The Gaussian model is sufficient to approximate the shape of the hard excess and it improves spectral modelling. 

This model is modified by both Galactic absorption as well as an additional absorption component, redshifted to the host galaxy frame. In {\sc xspec}, the source model is employed as {\sc tbabs $\times$ ztbabs(zgauss + zgauss + blackbody + powerlaw)}. For each object, the individual background model is also applied to the spectrum to form the total model. For each AGN in our sample, model parameters are given in Table~\ref{tab:params}. 

\begin{figure}
	\centering
	\includegraphics[width=\columnwidth]{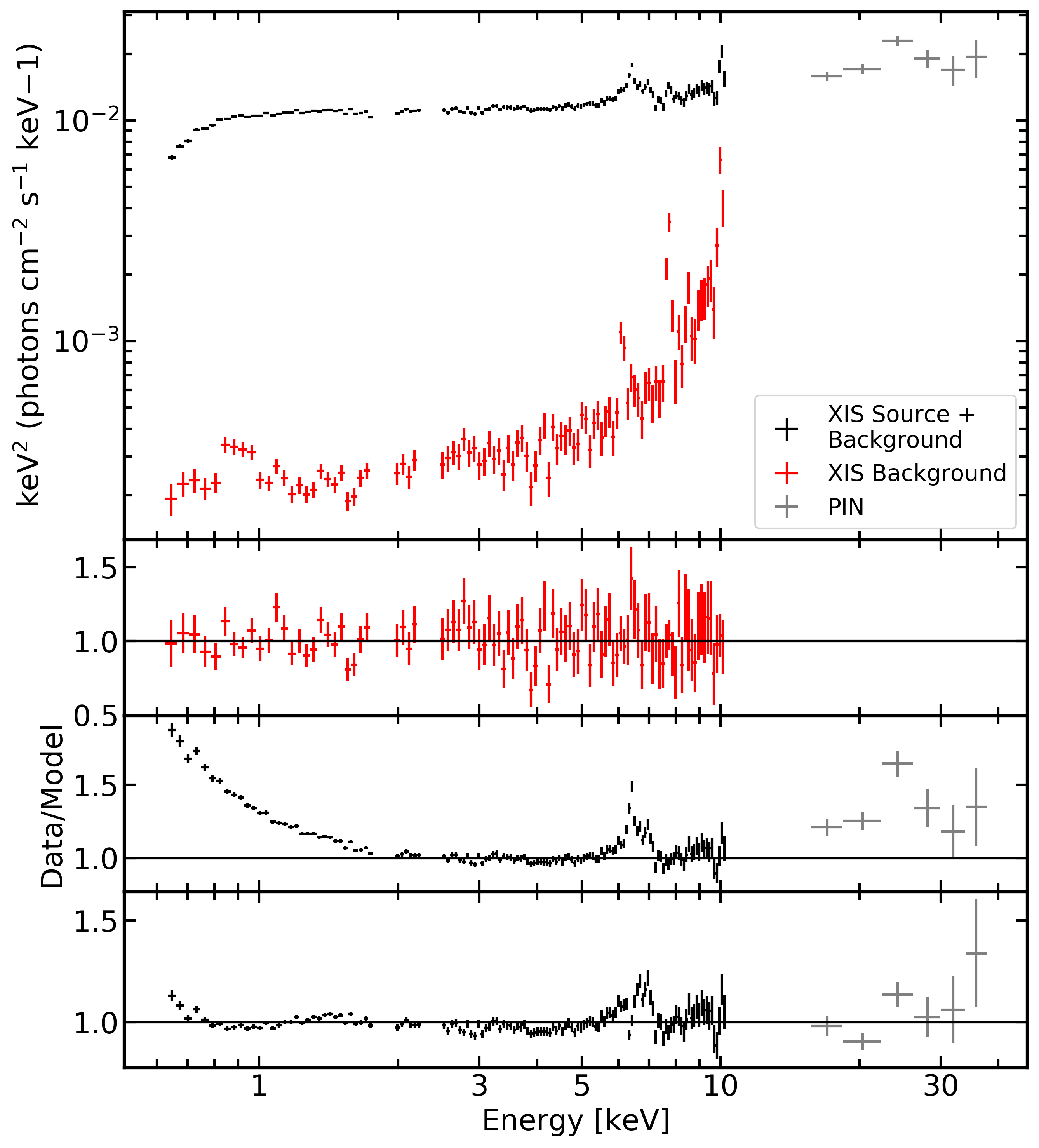}
	\caption{Top panel: XIS source+background (black), XIS background (red), and PIN (gray) spectra unfolded against a power law with $\Gamma$=0. The background increases toward higher energies. Second panel: residuals (data/model) for the background model. No clear residuals are present. Third panel: residuals for a power law model applied to the background modelled spectrum, fit between $2-4$ and $7-10\kev$ and extrapolated to lower and higher energies. A strong soft excess below $2\kev$, \feka\ emission from $5-7\kev$, and a hard excess peaking at $20-25\kev$ are visible. Bottom panel: residuals for the toy model applied to the data. Some residuals are present on either side of the narrow \feka\ line, but the spectrum is otherwise well approximated. All spectra have been redshift corrected.}
	\label{fig:sp}
\end{figure}

To illustrate this modelling process, the top panel of Fig.~\ref{fig:sp} shows the XIS source+background (black), XIS background (red), and PIN (gray) spectra unfolded against a power law with $\Gamma$=0 for the BLS1 galaxy Ark 120. A similar figure has been prepared for each source, and these are included as supplementary online material. The data have been redshift corrected for ease of interpretation. The background increases significantly at higher energies. Several instrumental lines are also seen in the background spectrum, including strong features at $6.5$, $7.5$ and $9.7\kev$ (see \citealt{suzbkg}). These features, especially the line at $9.7\kev$, are strong enough to be seen in the source+background spectrum.

The second panel shows the residuals (data/model) for the background model applied to the background spectrum. This model is comprised of multiple power law and Gaussian components to fit the observed curvature. No clear residuals can be seen, and the background data are well fit. In the third panel, we show the residuals for a power law model applied to the background modelled spectrum, fit between $2-4$ and $7-10\kev$ and extrapolated to all energies. A strong soft excess below $2\kev$, \feka\ emission between $5-7\kev$, and a hard excess peaking at $20-25\kev$ are visible. Finally, the bottom panel presents the residuals to our toy model applied to the background modelled XIS data and the PIN data. Aside from some remaining curvature in the $5-7\kev$ band on either side of the narrow \feka\ line at $6,4\kev$, the X-ray continuum appears to be reasonably approximated.

We select a number of parameters of interest for further study. These include the host galaxy absorption column density (znH), the blackbody temperature (kT), the photon index ($\Gamma$), and the $3-10\kev$ unabsorbed luminosity (L$_x$). We also approximate the Eddington ratio using the ratio of X-ray luminosity to the Eddington luminosity, L$_x$/L$_{\rm Edd}$, which we call the X-ray Eddington ratio (see Section.~\ref{sect:ledd}). 

Since \suzaku\ has the capability to simultaneously observe the soft and hard excesses, measuring these parameters is of great interest. We therefore define the soft excess strength, SE, as the ratio of the (unabsorbed) blackbody luminosity to the power law luminosity in the $0.6-1.5\kev$ band: 
\begin{equation}
\centering
{\rm SE} = \frac{\rm BB_{0.6-1.5\kev}}{\rm PL_{0.6-1.5\kev}}
\end{equation}

The hard excess, HE, is similarly defined as the ratio of the Compton hump luminosity in the $15-25\kev$ band to the power law luminosity in the same energy range:
\begin{equation}
\centering
{\rm HE} = \frac{\rm CH_{15-25\kev}}{\rm PL_{15-25\kev}}
\end{equation}

Finally, to assess the strength of the narrow \feka\ line relative to the continuum for each object, we define the iron line strength (Fe/L$_x$) as the ratio of the iron line luminosity to the $3-10\kev$ X-ray luminosity. All parameters will be measured and discussed throughout the remainder of this work.

\subsection{The Eddington luminosity}
\label{sect:ledd}

Previous works (e.g. \citealt{Pounds+1995,Wang+1996,Grupe+2004,Grupe+2010}) have shown that on average NLS1 exhibit higher L$_{\rm bol}$/L$_{\rm Edd}$ ratios than BLS1s. Studies with optical spectra of Seyfert 1 galaxies suggest that L$_{\rm bol}$/L$_{\rm Edd}$ may be the driving 

\onecolumn
\renewcommand{\arraystretch}{1.1}
\begin{landscape}
\begin{longtable}{llllllllll}
	\hline
	(1) & (2) & (3) & (4) & (5) & (6) & (7) & (8) & (9) & (10) \\
	Object & znH [$\times 10^{22}$\pscm] & kT [$\kev$] & $\Gamma$ & SE & HE & Fe/L$_x$ & L$_x$ [\ergps] & L$_x$/L$_{\rm Edd}$ & C-stat/dof \\
	\hline
	\endhead
	\multicolumn{10}{c}{NLS1} \\
	\hline
	1H0323+342 & < 0.0028 & 0.134$\pm0.003$ & 1.90$\pm0.01$ & 0.65$\pm0.04$ & 0.21$\pm0.11$ & 0.0032$\pm0.001$ & 6.58$\times10^{43}$ & 0.026 & 392/249 \\
	
	1H0707-495 & 0.18$_{-0.09}^{+0.68}$ & 0.093$\pm0.005$ & 1.75$\pm0.09$ & 8.2$_{-3.3}^{+5.5}$ & 4.8$\pm3.5$ & < 0.016 & 1.40$\times10^{42}$ & 0.005 & 280/218 \\
	
	Ark 564 & < 0.0006 & 0.135$\pm0.001$ & 2.542$\pm0.008$ & 0.59$\pm0.01$ & 0.75$\pm0.02$ & 0.0039$\pm0.0011$ & 1.94$\times10^{43}$ & 0.081 & 681/296 \\
	
	IRAS 05262+4432 & < 0.048 &  0.14$\pm0.01$ & 2.08$_{-0.06}^{+0.04}$ & 0.66$_{-0.08}^{+0.29}$ & 2.0$\pm1.8$ & < 0.00009 & 2.45$\times10^{42}$ & 0.001 & 240/208 \\
	
	IRAS 13224-3809 & 0.10$\pm0.01$ & 0.104$\pm0.001$ & 2.22$\pm0.03$ & 2.6$\pm0.2$ & - &  < 0.0059 & 5.99$\times10^{42}$ & 0.007 & 391/210 \\
	
	MCG-6-30-15 & 0.320$\pm0.002$ & 0.0539$\pm0.0004$ & 1.880$\pm0.002$ & 0.78$\pm0.05$ & 0.12$\pm0.02$ & 0.0067$\pm0.0004$ & 4.37$\times10^{42}$ & 0.017 & 2290/368 \\
	
	Mrk 1040 & 0.451$\pm0.004$ & 0.048$\pm0.002$ & 1.926$\pm0.004$ & 0.33$_{-0.14}^{+0.25}$ & 0.45$\pm0.04$ & 0.0115$\pm0.001$ & 1.73$\times10^{43}$ & 0.003 & 665/251 \\
	
	Mrk 110 & < 0.0030 & 0.14$\pm0.01$ & 1.79$\pm0.01$ & 0.17$\pm0.02$ & 0.223$_{-0.07}^{+0.004}$ & 0.0064$\pm0.0015$ & 4.66$\times10^{43}$ & 0.019 & 308/282 \\
	
	Mrk 335 & < 0.00054 & 0.129$\pm0.001$ & 1.897$\pm0.007$ & 0.91$\pm0.02$ & 0.98$\pm0.17$ & 0.019$\pm0.001$ & 9.35$\times10^{42}$ & 0.004 & 1445/287 \\
	
	Mrk 359 & < 0.0037 & 0.17$_{-0.02}^{+0.01}$ & 1.71$_{-0.02}^{+0.03}$ & 0.24$\pm0.03$ & 0.091$_{-0.03}^{+0.003}$ & 0.014$\pm0.003$ & 2.80$\times10^{42}$ & 0.013 & 270/236 \\
	
	Mrk 478 & < 0.0084 & 0.139$\pm0.007$ & 2.21$\pm0.04$ & 0.59$\pm0.06$ & 1.0$\pm0.9$ & < 0.000004 & 3.02$\times10^{43}$ & 0.012 & 242/210 \\
	
	Mrk 766 & 0.207$\pm0.006$ & 0.062$_{-0.001}^{+0.002}$ & 1.813$\pm0.006$ & 0.66$_{-0.11}^{+0.13}$ & 0.262$_{-0.08}^{+0.002}$ & 0.008$\pm0.0014$ & 3.92$\times10^{42}$ & 0.005 & 528/247 \\
	
	NGC 4051 & < 0.00037 & 0.103$\pm0.002$ & 1.910$\pm0.004$ & 0.23$\pm0.01$ & 0.52$\pm0.05$ & 0.0141$\pm0.0007$ & 1.73$\times10^{41}$ & 0.001 & 1771/295 \\
	
	PG 1211+143 & 0.06$\pm0.02$ & 0.080$\pm0.003$ & 1.79$\pm0.01$ & 0.74$_{-0.11}^{+0.13}$ & 0.05899$_{-0.06}^{+0.34}$ & 0.0095$_{-0.0027}^{+0.0028}$ & 5.08$\times10^{43}$ & 0.010 & 346/237 \\
	
	PG 1404+226 & 0.10$_{-0.07}^{+0.06}$ & 0.093$\pm0.006$ & 1.39$\pm0.08$ & 4.6$_{-1.2}^{+1.8}$ & 12.4$_{-2.4}^{+2.5}$ & < 0.00003 & 6.39$\times10^{42}$ & 0.011 & 252/182 \\
	
	PKS 0558-504 & < 0.0029 & 0.130$\pm0.003$ & 2.21$\pm0.01$ & 0.43$\pm0.02$ & 0.42$\pm0.18$ & < 0.0022 & 5.33$\times10^{44}$ & 0.066 & 299/256 \\
	
	RE J1034+396 & < 0.0070 & 0.113$\pm0.003$ & 2.34$\pm0.04$ & 1.7$\pm0.2$ & 10.7$\pm2.9$ & 0.0089$_{-0.0071}^{+0.0076}$ & 3.24$\times10^{42}$ & 0.010 & 248/187 \\
	
	RX J0134.2-4258 & 0.08$\pm0.02$ & 0.07$\pm0.02$ & 2.36$\pm0.02$ & 0.28$_{-0.22}^{+3.12}$ & 3.6$\pm2.9$ & < 0.0076 & 2.64$\times10^{44}$ & 0.100 & 187/171 \\
	
	RX J1633.3+4718 & 0.61$\pm0.05$ & 0.101$\pm0.008$ & 2.12$\pm0.03$ & 0.95$_{-0.26}^{+0.40}$ & 2.1$\pm1.2$ & < 0.00002 & 3.58$\times10^{43}$ & 0.093 & 261/202 \\
	
	SWIFT J2127.4+5654 & 0.27$\pm0.01$ & 0.21$\pm0.01$ & 1.96$\pm0.02$ & 0.22$\pm0.03$ & 0.21$\pm0.07$ & 0.0069$\pm0.0016$ & 1.22$\times10^{43}$ & 0.006 & 310/273 \\
	
	TON S180 & < 0.0030 & 0.135$_{-0.004}^{+0.003}$ & 2.31$\pm0.02$ & 0.48$\pm0.02$ & 1.5$\pm0.5$ & 0.0029$\pm0.0021$ & 3.75$\times10^{43}$ & 0.041 & 351/241 \\
	
	WKK 4438 & < 0.0032 & 0.25$\pm0.03$ & 1.90$\pm0.02$ & 0.091$_{-0.034}^{+0.033}$ & 0.029$_{-0.009}^{+0.001}$ & 0.0077$_{-0.0022}^{+0.0023}$ & 5.34$\times10^{42}$ & 0.021 & 273/240 \\
	\hline
	\multicolumn{10}{c}{BLS1} \\
	\hline
	1H0419-577 & < 0.00068 & 0.147$\pm0.003$ & 1.831$\pm0.007$ & 0.35$\pm0.01$ & 0.28$\pm0.05$ & 0.0040$\pm0.0007$ & 3.40$\times10^{44}$ & 0.021 & 1012/299 \\
	
	3C 111 & 0.816$\pm0.004$ & < 0.19 & 1.678$\pm0.003$ & 0.0024$_{-0.0002}^{+0.006}$ & 0.00226$_{-0.0007}^{+0.00001}$ & 0.0050$\pm0.0005$ & 1.97$\times10^{44}$ & 0.0004 & 645/303 \\
	
	3C 120 & < 0.00053 & 0.168$\pm0.006$ & 1.817$\pm0.005$ & 0.147$\pm0.007$ & 0.13$\pm0.02$ & 0.0080$\pm0.0005$ & 9.27$\times10^{43}$ & 0.014 & 1008/274 \\
	
	3C 382 & < 0.0017 & 0.091$\pm0.004$ & 1.808$\pm0.004$ & 0.16$\pm0.03$ & 0.0385$_{-0.01}^{+0.0002}$ & 0.0059$\pm0.0007$ & 2.61$\times10^{44}$ & 0.003 & 486/300 \\
	
	3C 390.3 & < 0.0028 & 0.152$\pm0.007$ & 1.709$\pm0.008$ & 0.19$\pm0.01$ & 0.0370$_{-0.01}^{+0.0004}$ & 0.0071$\pm0.0009$ & 2.39$\times10^{44}$ & 0.004 & 371/249 \\
	
	3C 78 & 0.12$_{-0.09}^{+0.12}$ & 0.16$\pm0.01$ & 2.35$\pm0.06$ & 0.40$\pm0.08$ & - & < 0.0079 & 1.45$\times10^{42}$ & 0.0001 & 241/169 \\
	
	4C+74.26 & 0.05$\pm0.01$ & 0.05$\pm0.01$ & 1.811$\pm0.007$ & 0.11$_{-0.09}^{+2.28}$ & 0.28$\pm0.04$ & 0.0070$\pm0.0008$ & 6.05$\times10^{44}$ & 0.001 & 489/258 \\
	
	Ark 120 & < 0.0014 & 0.134$\pm0.004$ & 1.977$\pm0.009$ & 0.37$\pm0.02$ & 0.38$\pm0.07$ & 0.013$\pm0.001$ & 5.62$\times10^{43}$ & 0.004 & 713/294 \\
	
	B3 0309+411 & 0.06$\pm0.02$ & 0.07$\pm0.01$ & 1.81$_{-0.02}^{+0.01}$ & 0.39$_{-0.26}^{+1.11}$ & 0.0201$_{-0.006}^{+0.0003}$ & 0.0056$\pm0.0028$ & 1.58$\times10^{44}$ & - & 202/200 \\
	
	CBS 126 & < 0.011 & 0.066$\pm0.004$ & 1.99$\pm0.01$ & 0.53$_{-0.17}^{+0.27}$ & 0.74$\pm0.42$ & 0.0085$\pm0.0031$ & 4.87$\times10^{43}$ & 0.006 & 325/249 \\
	
	ESO 511-G030 & < 0.0013 & 0.134$\pm0.005$ & 1.778$\pm0.006$ & 0.22$\pm0.01$ & 0.22$\pm0.07$ & 0.0106$\pm0.0009$ & 1.85$\times10^{43}$ & 0.002 & 574/258 \\
	
	ESO 548-G081 & < 0.012 & 0.14$\pm0.02$ & 1.66$\pm0.02$ & 0.16$_{-0.03}^{+0.06}$ & 0.151$_{-0.045}^{+0.004}$ & 0.016$\pm0.003$ & 5.15$\times10^{42}$ & 0.001 & 257/252 \\
	
	Fairall 9 & < 0.00047 & 0.150$\pm0.003$ & 1.909$\pm0.006$ & 0.270$\pm0.008$ & 0.42$\pm0.04$ & 0.0143$\pm0.0007$ & 9.60$\times10^{43}$ & 0.004 & 1389/283 \\
	
	IGR J16185-5928 & < 0.0089 & 0.16$_{-0.03}^{+0.02}$ & 1.88$\pm0.03$ & 0.206$_{-0.04}^{+0.08}$ & 0.80$\pm0.23$ & 0.0093$_{-0.0026}^{+0.0027}$ & 1.72$\times10^{43}$ & 0.006 & 340/271 \\
	
	IGR J19378-0617 & < 0.0016 & 0.135$\pm0.003$ & 2.20$\pm0.01$ & 0.44$\pm0.02$ & 0.125$_{-0.04}^{+0.002}$ & 0.0071$\pm0.0017$ & 4.06$\times10^{42}$ & - & 492/256 \\
	
	III Zw 2 & < 0.0063 & 0.18$_{-0.13}^{+0.05}$ & 1.58$_{-0.02}^{+0.03}$ & 0.07412$_{-0.03}^{+0.81}$ & 0.18$\pm0.13$ & 0.0070$_{-0.0021}^{+0.0022}$ & 1.49$\times10^{44}$ & 0.011 & 242/247 \\
	
	Kaz 102 & < 0.016 & 0.24$_{-0.05}^{+0.06}$ & 1.53$\pm0.03$ & 0.18$_{-0.08}^{+0.13}$ & 0.295$_{-0.09}^{+0.009}$ & 0.0066$_{-0.0062}^{+0.0068}$ & 8.61$\times10^{43}$ & 0.007 & 205/202 \\
	
	LEDA 168563 & < 0.024 & 0.12$\pm0.02$ & 1.63$\pm0.01$ & 0.19$_{-0.07}^{+0.18}$ & 0.0704$_{-0.02}^{+0.001}$ & 0.0091$\pm0.0016$ & 2.68$\times10^{43}$ & 0.002 & 308/244 \\
	
	MCG-02-14-009 & < 0.0069 & 0.15$\pm0.02$ & 1.79$\pm0.03$ & 0.28$_{-0.04}^{+0.06}$ & 0.56$_{-0.28}^{+0.29}$ & 0.014$\pm0.003$ & 6.09$\times10^{42}$ & 0.004 & 358/268 \\
	
	MCG-2-58-22 & < 0.00095 & 0.143$\pm0.008$ & 1.755$_{-0.007}^{+0.006}$ & 0.15$\pm0.01$ & 0.12$\pm0.03$ & 0.0064$\pm0.0006$ & 2.32$\times10^{44}$ & 0.006 & 689/281 \\
	
	MCG+04-22-042 & < 0.0044 & 0.13$\pm0.01$ & 1.88$\pm0.02$ & 0.24$_{-0.04}^{+0.06}$ & 0.30$_{-0.18}^{+0.19}$ & 0.012$\pm0.003$ & 2.72$\times10^{43}$ & 0.001 & 290/248 \\
	
	MCG+8-11-11 & 0.014$\pm0.003$ & 0.056$\pm0.008$ & 1.699$\pm0.004$ & 0.06$_{-0.04}^{+0.15}$ & 0.0261$_{-0.008}^{+0.0001}$ & 0.0111$\pm0.0008$ & 4.94$\times10^{43}$ & 0.003 & 497/296 \\
	
	MR 2251-178 & 0.11$\pm0.01$ & 0.059$\pm0.003$ & 1.649$\pm0.004$ & 0.29$_{-0.10}^{+0.16}$ & 0.00282$_{-0.0008}^{+0.00001}$ & 0.0036$\pm0.0007$ & 3.22$\times10^{44}$ & 0.013 & 626/283 \\
	
	Mrk 1018 & < 0.015 & 0.16$\pm0.03$ & 1.80$\pm0.03$ & 0.14$_{-0.03}^{+0.07}$ & 0.52$\pm0.22$ & 0.0088$_{-0.0031}^{+0.0032}$ & 3.39$\times10^{43}$ & 0.004 & 300/251 \\
	
	Mrk 1148 & < 0.013 & 0.12$\pm0.01$ & 1.82$\pm0.02$ & 0.27$_{-0.06}^{+0.10}$ & 0.39$\pm0.18$ & 0.0065$\pm0.0030$ & 1.26$\times10^{44}$ & - & 251/216 \\
	
	Mrk 1320 & < 0.043 & < 0.15 & 1.50$\pm0.06$ & < 0.02 & - & < 0.00053 & 2.82$\times10^{43}$ & - & 151/153 \\
	
	Mrk 205 & < 0.0071 & 0.12$\pm0.01$ & 1.89$\pm0.02$ & 0.27$_{-0.04}^{+0.06}$ & 1.08$_{-0.90}^{+0.17}$ & 0.013$\pm0.002$ & 8.57$\times10^{43}$ & 0.003 & 397/258 \\
	
	Mrk 279 & < 0.0039 & 0.19$_{-0.03}^{+0.02}$ & 1.55$\pm0.03$ & 0.16$\pm0.03$ & 0.68$\pm0.18$ & 0.032$\pm0.003$ & 8.56$\times10^{42}$ & 0.002 & 346/264 \\
	
	Mrk 352 & < 0.0068 & 0.11$_{-0.03}^{+0.02}$ & 1.79$_{-0.03}^{+0.02}$ & 0.19$_{-0.07}^{+0.18}$ & 0.102$_{-0.03}^{+0.003}$ & 0.011$\pm0.003$ & 4.90$\times10^{42}$ & 0.004 & 282/249 \\
	
	Mrk 509 & < 0.00019 & 0.152$\pm0.005$ & 1.875$\pm0.006$ & 0.187$_{-0.006}^{+0.007}$ & 0.34$\pm0.03$ & 0.0078$\pm0.0006$ & 1.06$\times10^{44}$ & 0.008 & 1980/279 \\
	
	Mrk 530 & < 0.0022 & 0.179$_{-0.01}^{+0.009}$ & 1.91$\pm0.01$ & 0.19$\pm0.02$ & 0.27$\pm0.09$ & 0.011$\pm0.001$ & 3.26$\times10^{43}$ & 0.002 & 387/242 \\
	
	Mrk 79 & 0.12$_{-0.02}^{+0.03}$ & 0.067$_{-0.005}^{+0.006}$ & 1.596$\pm0.008$ & 0.24$_{-0.09}^{+0.18}$ & 0.0134$_{-0.004}^{+0.0001}$ & 0.019$\pm0.002$ & 1.35$\times10^{43}$ & 0.003 & 429/276 \\
	
	Mrk 841 & < 0.00074 & 0.097$\pm0.004$ & 1.770$\pm0.005$ & 0.24$_{-0.03}^{+0.04}$ & 0.43$\pm0.06$ & 0.0145$\pm0.0009$ & 2.71$\times10^{43}$ & 0.003 & 1130/283 \\
	
	NGC 4593 & < 0.0012 & 0.063$\pm0.002$ & 1.695$\pm0.003$ & 0.18$_{-0.04}^{+0.05}$ & 0.13$\pm0.06$ & 0.0214$\pm0.0008$ & 2.45$\times10^{42}$ & 0.002 & 628/278 \\
	
	NGC 6814 & < 0.011 & < 0.082 & 1.51$\pm0.01$ & 0.0351$_{-0.01}^{+0.0006}$ & 0.4245$_{-0.13}^{+0.007}$ & 0.012$\pm0.003$ & 5.40$\times10^{41}$ & 0.0002 & 250/226 \\
	
	NGC 7213 & < 0.0065 & 0.119$\pm0.009$ & 1.733$\pm0.007$ & 0.095$_{-0.014}^{+0.019}$ & 0.0683$_{-0.02}^{+0.0005}$ & 0.012$\pm0.001$ & 1.42$\times10^{42}$ & 0.0001 & 456/286 \\
	
	NGC 7469 & < 0.0017 & 0.095$_{-0.006}^{+0.007}$ & 1.760$\pm0.008$ & 0.18$_{-0.04}^{+0.06}$ & 0.27$\pm0.07$ & 0.019$\pm0.002$ & 9.89$\times10^{42}$ & 0.009 & 506/282 \\
	
	NGC 985 & 0.13$\pm0.04$ & 0.057$_{-0.004}^{+0.005}$ & 1.78$\pm0.01$ & 0.53$_{-0.24}^{+0.51}$ & 0.28$\pm0.19$ & 0.011$\pm0.003$ & 4.66$\times10^{43}$ & - & 251/248 \\
	
	PG 1322+659 & < 0.074 & 0.08$\pm0.04$ & 1.95$\pm0.05$ & 0.29$_{-0.24}^{+9.7}$ & - & < 0.010 & 4.39$\times10^{43}$ & 0.002 & 195/161 \\
	
	PG 1626+554 & < 0.10 & 0.15$\pm0.02$ & 2.05$\pm0.04$ & 0.27$_{-0.05}^{+0.07}$ & - & 0.011$\pm0.004$ & 1.27$\times10^{44}$ & 0.002 & 188/168 \\
	
	RBS 1124 & < 0.093 & 0.12$\pm0.02$ & 1.9$\pm0.02$ & 0.25$_{-0.07}^{+0.15}$ & 0.53$\pm0.17$ & 0.0044$_{-0.0026}^{+0.0027}$ & 4.76$\times10^{44}$ & 0.021 & 369/267 \\
	
	SBS 1301+540 & < 0.069 & 0.16$\pm0.02$ & 1.62$_{-0.03}^{+0.02}$ & 0.21$_{-0.04}^{+0.05}$ & 0.144$_{-0.04}^{+0.005}$ & 0.010$\pm0.003$ & 6.94$\times10^{42}$ & 0.002 & 281/193 \\
	
	SWIFT J0501.9-3239 & < 0.0035 & 0.107$_{-0.008}^{+0.009}$ & 1.89$\pm0.01$ & 0.16$_{-0.03}^{+0.05}$ & 0.27$\pm0.07$ & 0.011$\pm0.002$ & 8.56$\times10^{42}$ & - & 371/271 \\
	
	SWIFT J0904.3+5538 & 0.18$_{-0.10}^{+0.12}$ & 0.07$_{-0.01}^{+0.02}$ & 1.66$\pm0.03$ & 0.27$_{-0.19}^{+1.30}$ & 0.39$_{-0.12}^{+0.01}$ & 0.017$_{-0.007}^{+0.008}$ & 1.13$\times10^{43}$ & - & 183/197 \\
	
	SWIFT J1310.9-5553 & < 0.15 & 0.12$_{-0.03}^{+0.04}$& 1.50$\pm0.02$ & 0.19$_{-0.10}^{+0.32}$ & 0.185$_{-0.06}^{+0.005}$ & < 0.0028 & 1.22$\times10^{44}$ & - & 257/232 \\
	
	UGC 6728 & < 0.0035 & 0.16$\pm0.01$ & 1.77$\pm0.02$ & 0.29$\pm0.03$ & 0.19$\pm0.16$ & 0.0091$_{-0.0026}^{+0.0027}$ & 1.06$\times10^{42}$ & 0.002 & 310/251 \\

	ZW 229-15 & < 0.0042 & 0.16$_{-0.02}^{+0.01}$ & 1.74$\pm0.03$ & 0.25$\pm0.03$ & - & 0.013$\pm0.003$ & 5.66$\times10^{42}$ & 0.006 & 480/224 \\
	\hline
	\caption{Toy model parameters and associated errors. Object names are given in column (1). Host galaxy column densities (znH) are given in column (2) in units of 10$^{22}$~\pscm. The blackbody temperature (kT) and photon index ($\Gamma$) are given in columns (3) and (4), respectively. Columns (5), (6), and (7) give the soft excess (SE), hard excess (HE) and iron line strength (Fe/L$_x$), respectively. Error bars on the soft and hard excesses have been computed using error propagation of the blackbody, powerlaw and Compton hump luminosities. The $3-10\kev$ luminosities are given in column (8), and the X-ray Eddington ratios (L$_x$/L$_{\rm Edd}$) are given in column (9). Column (10) gives the fit quality, in terms of the C-stat over the degrees of freedom (dof). }
	\label{tab:params}
\end{longtable}
\end{landscape}

\twocolumn

\noindent mechanism behind the so-called eigenvector 1, the primary axis of variance between the two classes (\citealt{Boroson+1992}). It is therefore of great interest to properly characterise this value. 

Many works suggest that the bolometric luminosity can be calculated using the X-ray luminosity with a scaling factor ranging from $\sim5$ to $\sim100$, depending on various host galaxy and AGN properties (e.g. \citealt{Lusso+2010,Lusso+2012}). There are also indications that the ratio of L$_{\rm bol}$/L$_{\rm Edd}$ may be closely linked with the photon index of the power law component, $\Gamma$ (e.g. \citealt{Shemmer+2008}; \citealt{Brightman+2013}). 

\begin{figure}
	\centering
	\includegraphics[width=0.9\columnwidth]{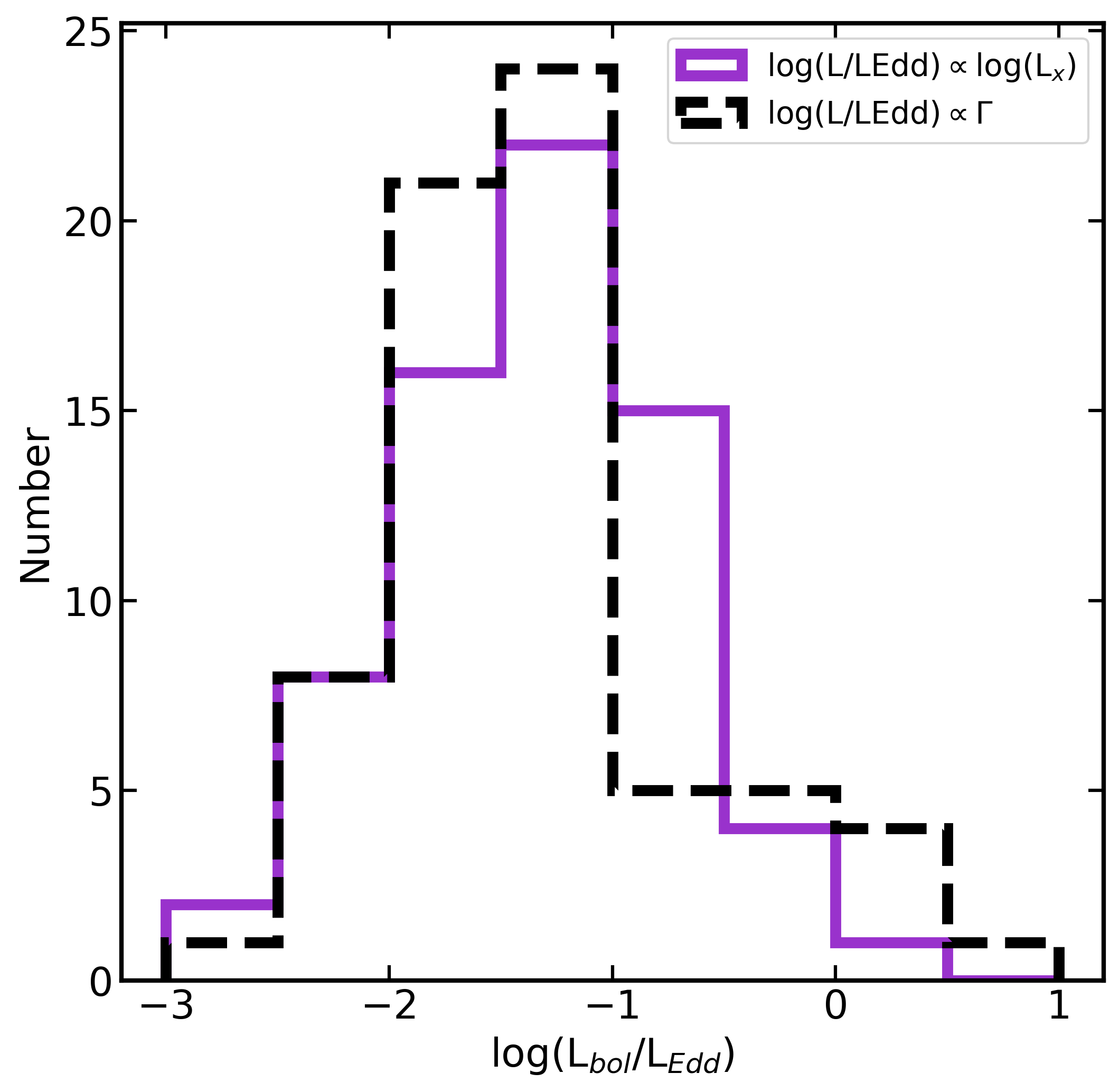}
	\caption{Comparison of methods for calculating the Eddington luminosity ratio, shown for the entire sample. The purple solid line shows the Eddington ratio estimated as $10~\times$ L$_x$/L$_{\rm Edd}$, while the black dashed line shows the calculation using the $\Gamma$--L$_{\rm bol}$/L$_{\rm Edd}$ relationship of \protect\cite{Brightman+2013}. The two distributions are in close agreement, supporting the use of L$_x$/L$_{\rm Edd}$ to approximate the Eddington luminosity ratio for further analysis.}
	\label{fig:ledd}
\end{figure}

To investigate this, we use two different methods to approximate the Eddington ratio. The histograms for these distributions, shown for the entire sample of NLS1 and BLS1 galaxies, is presented in Fig.~\ref{fig:ledd}. In the first we use the X-ray luminosity to approximate the bolometric luminosity, scaled by a factor of 10 (purple solid line). In the second, we use equation (2) of \cite{Brightman+2013}, which gives the relationship between $\Gamma$ and L$_{\rm bol}$/L$_{\rm Edd}$ (black dashed lines). For some individual sources, there can be large discrepancies, however the two methods give very similar distributions and median values, suggesting that L$_{x}$/L$_{\rm Edd}$ is a reasonable proxy for L$_{\rm bol}$/L$_{\rm Edd}$. Throughout the remainder of this work, we define this ratio of X-ray luminosity to Eddington luminosity, L$_{x}$/L$_{\rm Edd}$, as the X-ray Eddington ratio. This is advantageous so that we can continue to treat photon index and L$_{x}$/L$_{\rm Edd}$ independently and compare them in Section~\ref{sect:5}.

\begin{figure}
	\centering
	\includegraphics[width=0.9\columnwidth]{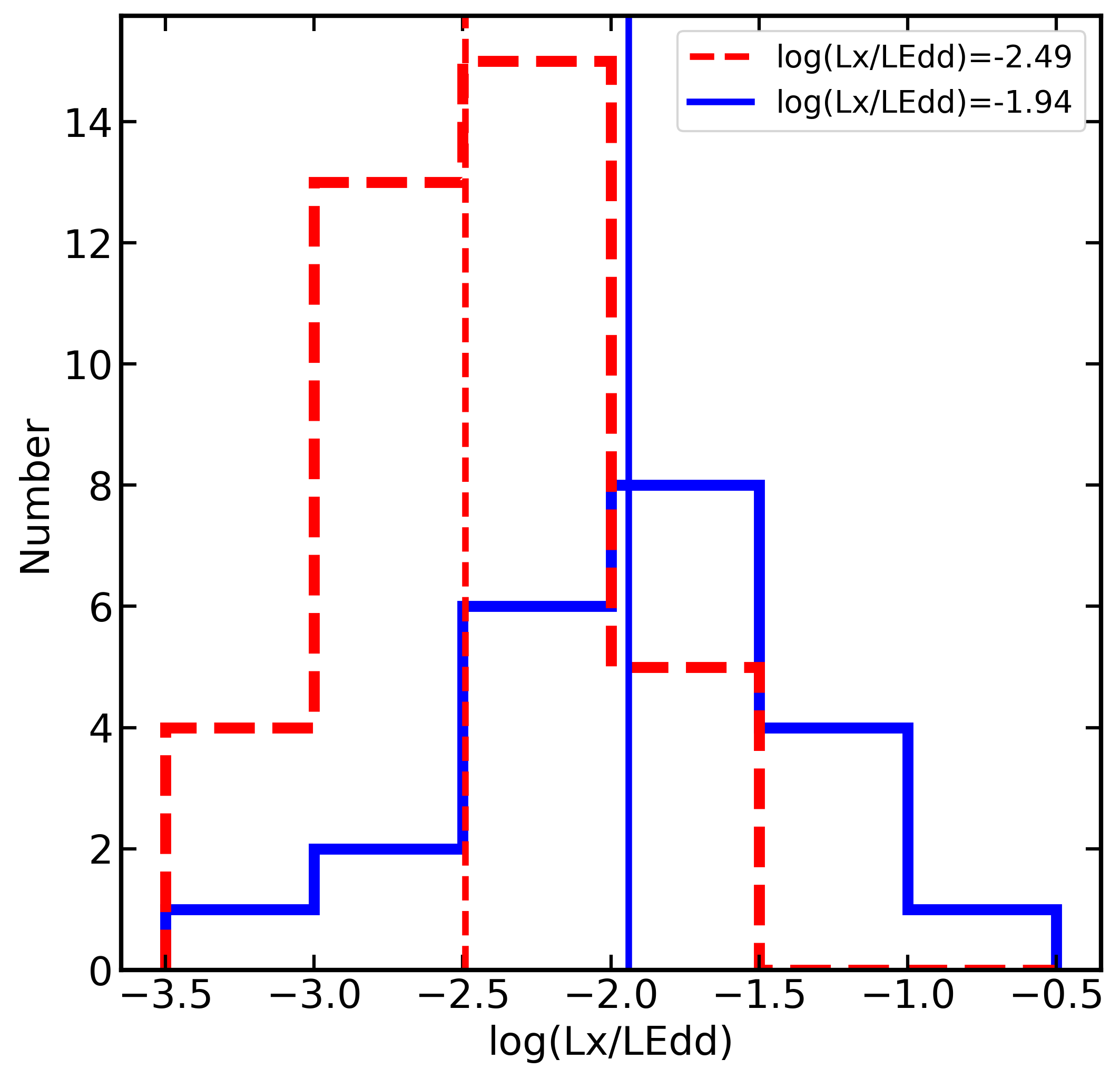}
	\caption{Distribution of log(L$_x$/L$_{\rm Edd}$) for NLS1 and BLS1 galaxies, where the NLS1 parameter values are shown in blue solid lines and BLS1 parameter values are shown as red dashed lines. The median values for each parameter are shown as vertical lines in corresponding colours and line styles. As expected, NLS1 galaxies show evidence for higher X-ray Eddington ratios (KS test p-value 0.00038).}
	\label{fig:lcomp}
\end{figure}

Fig.~\ref{fig:lcomp} shows the distributions of L$_x$/L$_{\rm Edd}$ for NLS1 and BLS1 galaxies, shown in log space for ease of interpretation. NLS1 galaxies are shown as blue solid lines and BLS1 galaxies are shown as red dashed lines. The median values for each parameter are shown as vertical lines in corresponding colours and line styles. We find that NLS1 galaxies show higher median L$_x$/L$_{\rm Edd}$, with a median value of log(L$_x$/L$_{\rm Edd}$) = -1.94 (corresponding to L$_x$/L$_{\rm Edd}$ = 0.01) for NLS1 galaxies compared to log(L$_x$/L$_{\rm Edd}$) = -2.49 (corresponding to L$_x$/L$_{\rm Edd}$=0.003) for BLS1 galaxies.

To assess the similarity of the distributions, we compute the Kolmogorov-Smirnov (KS) test on the two samples. This test compares two samples to compute the statistical likelihood that they are drawn from the same parent population. The KS test, implemented in {\sc python} using {\sc scipy.stats.ks\_2samp()}, returns a KS test statistic and a p-value. If the KS statistic is large (p-value is small), we can reject the null hypothesis that the two samples are drawn from the same distribution. For the ratio of L$_x$/L$_{\rm Edd}$, we compute a p-value of 0.00038, implying that the samples are different at above the 99.9 per cent confidence level.

\subsection{Parameter distributions}
\label{sect:histogram}

\begin{figure*}
	\centering
	\subfloat[]{
		\includegraphics[width=59mm]{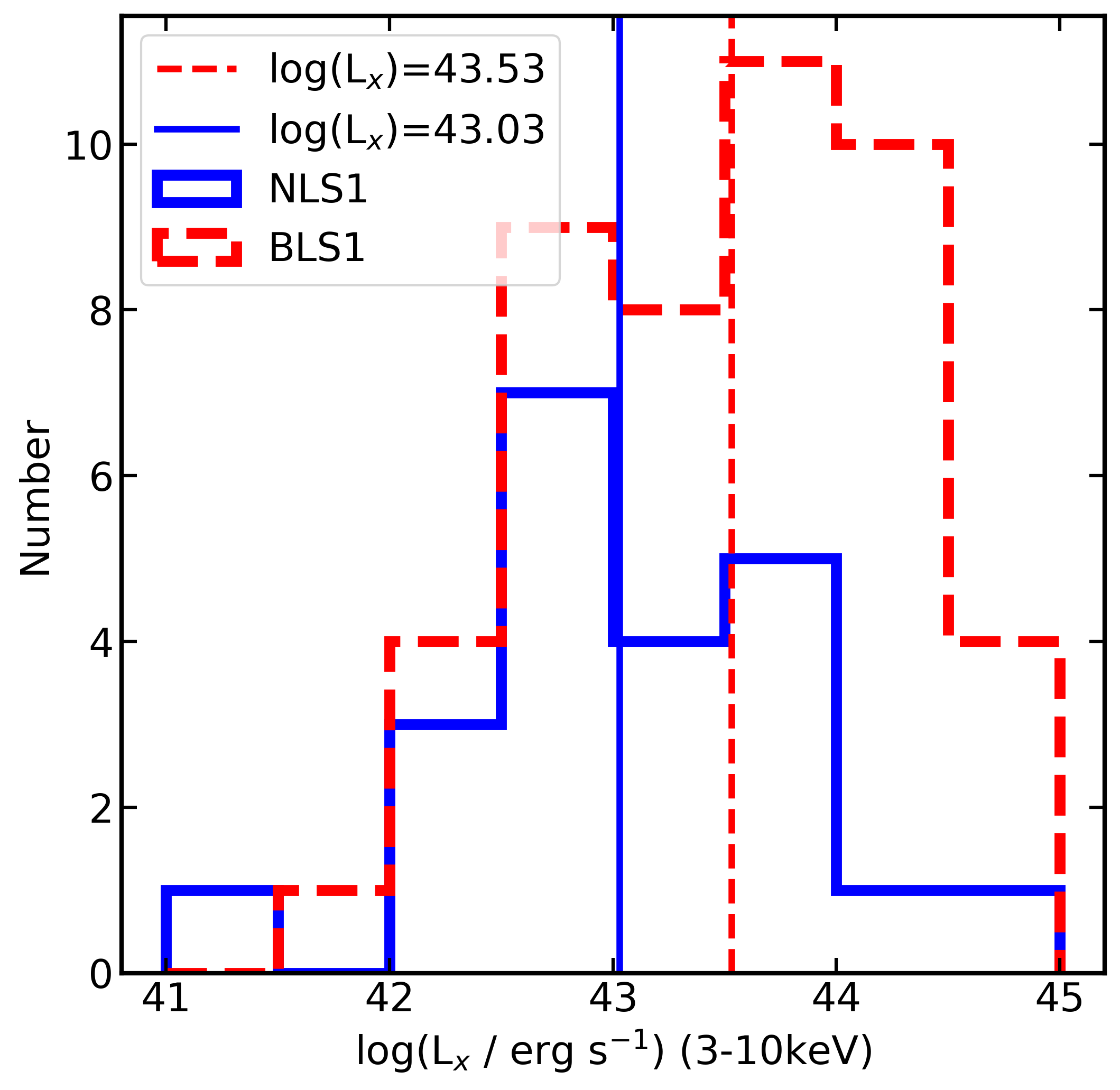}
	}
	\subfloat[]{
		\includegraphics[width=59mm]{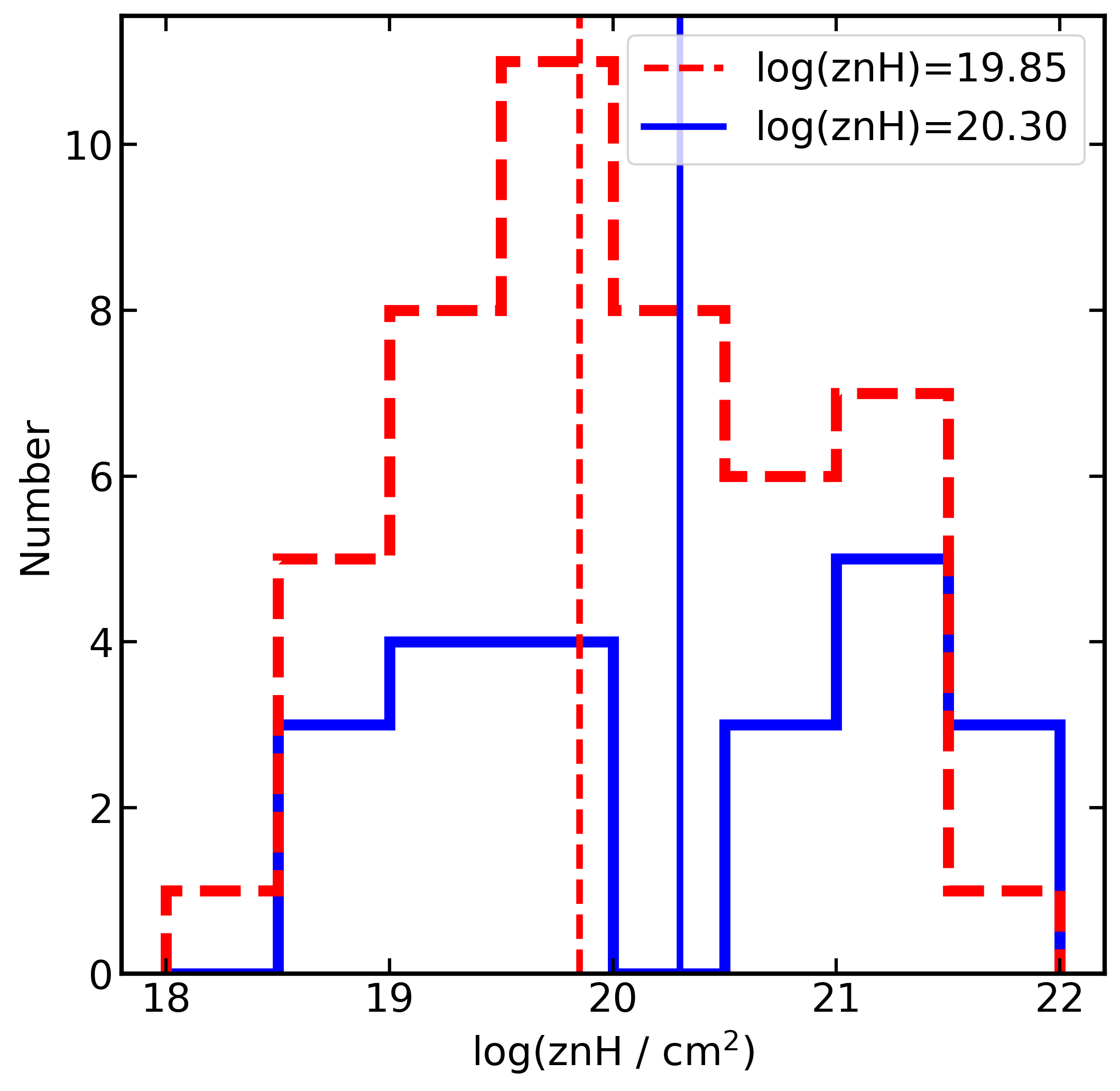}
	}
	\subfloat[]{
	\includegraphics[width=59mm]{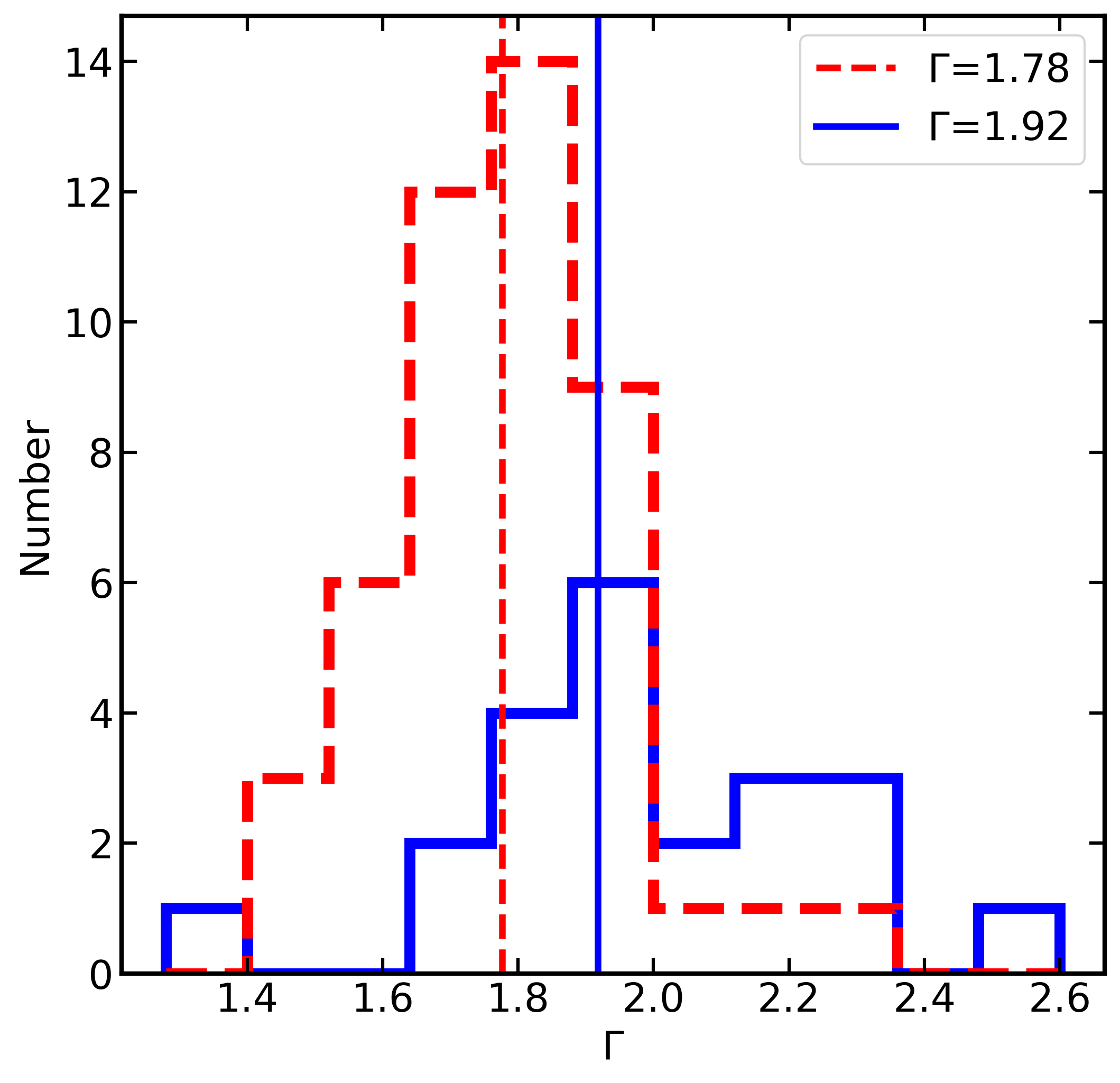}
	}
	\hspace{0mm}
	\subfloat[]{
		\includegraphics[width=59mm]{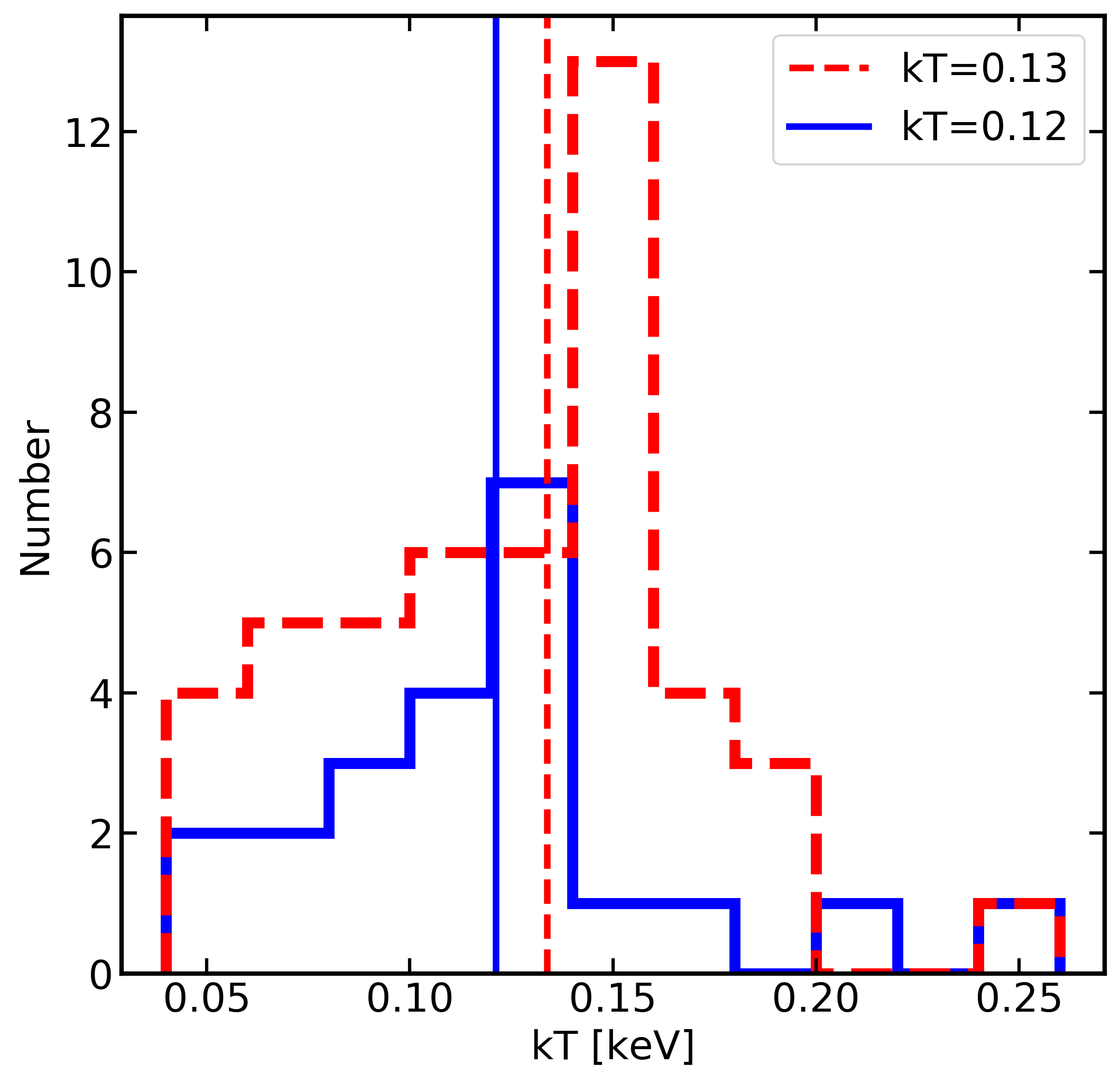}
	}
	\subfloat[]{
		\includegraphics[width=59mm]{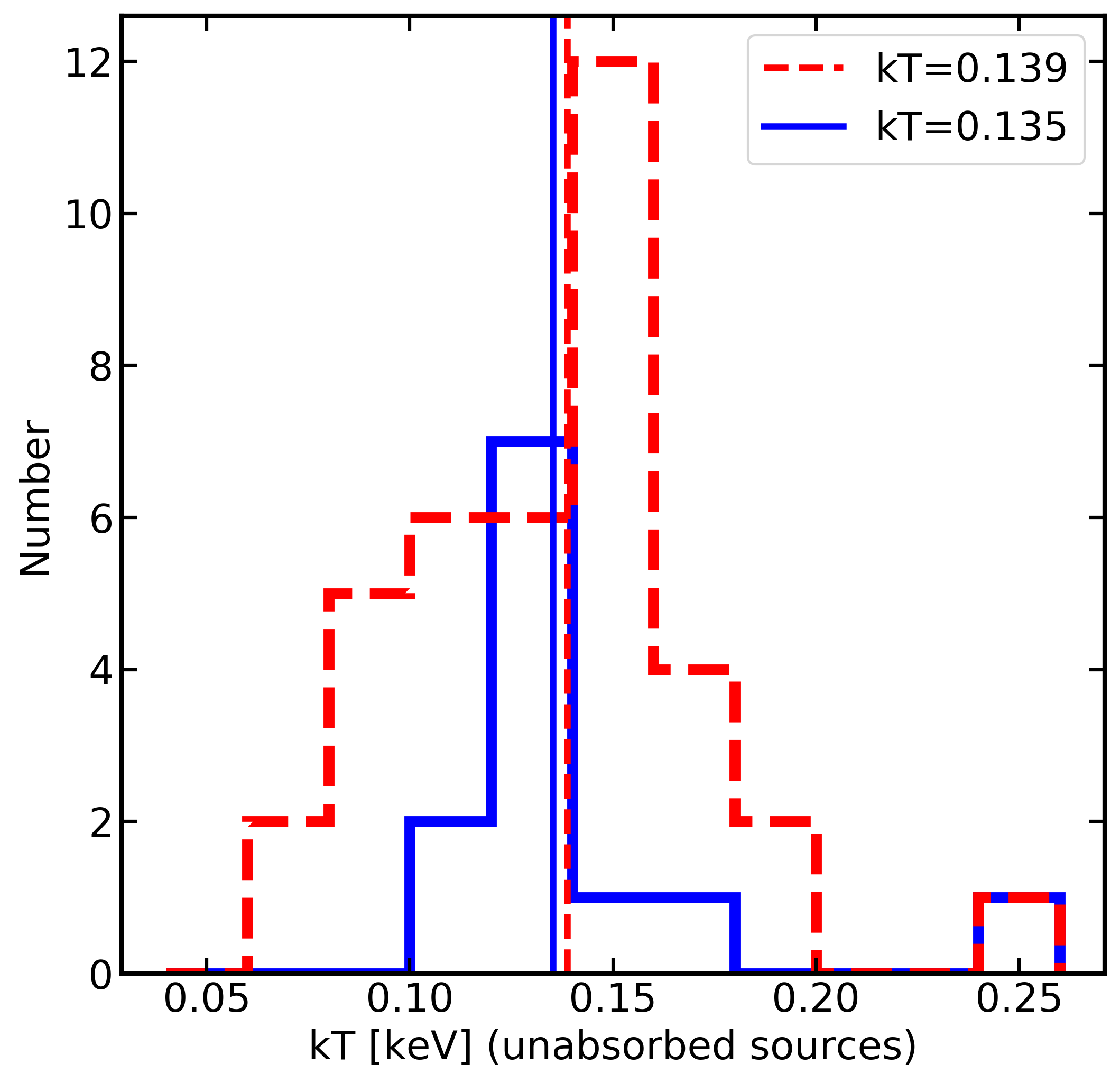}
	}
	\subfloat[]{
		\includegraphics[width=59mm]{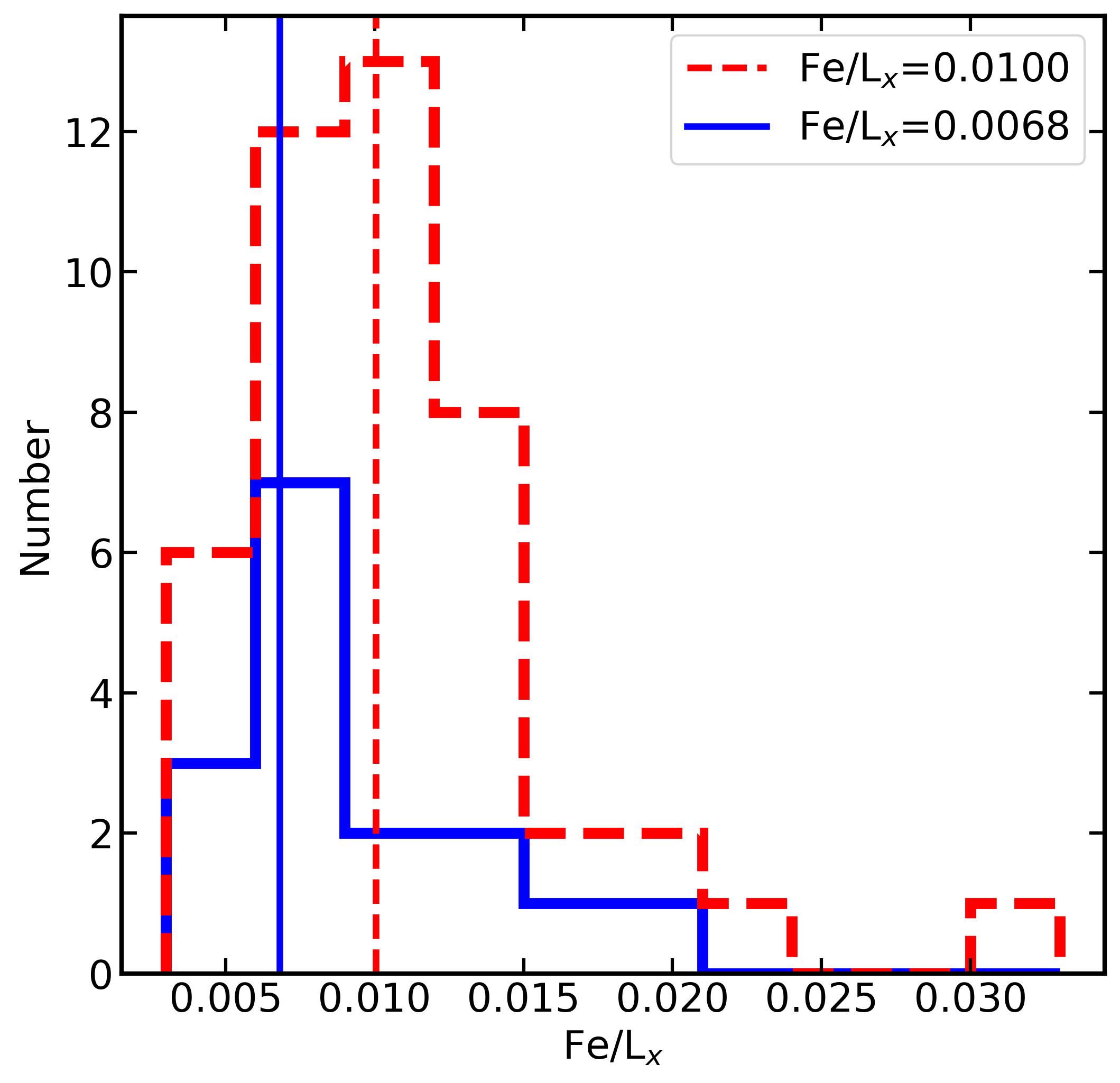}
	}
	\hspace{0mm}
	\subfloat[]{
		\includegraphics[width=59mm]{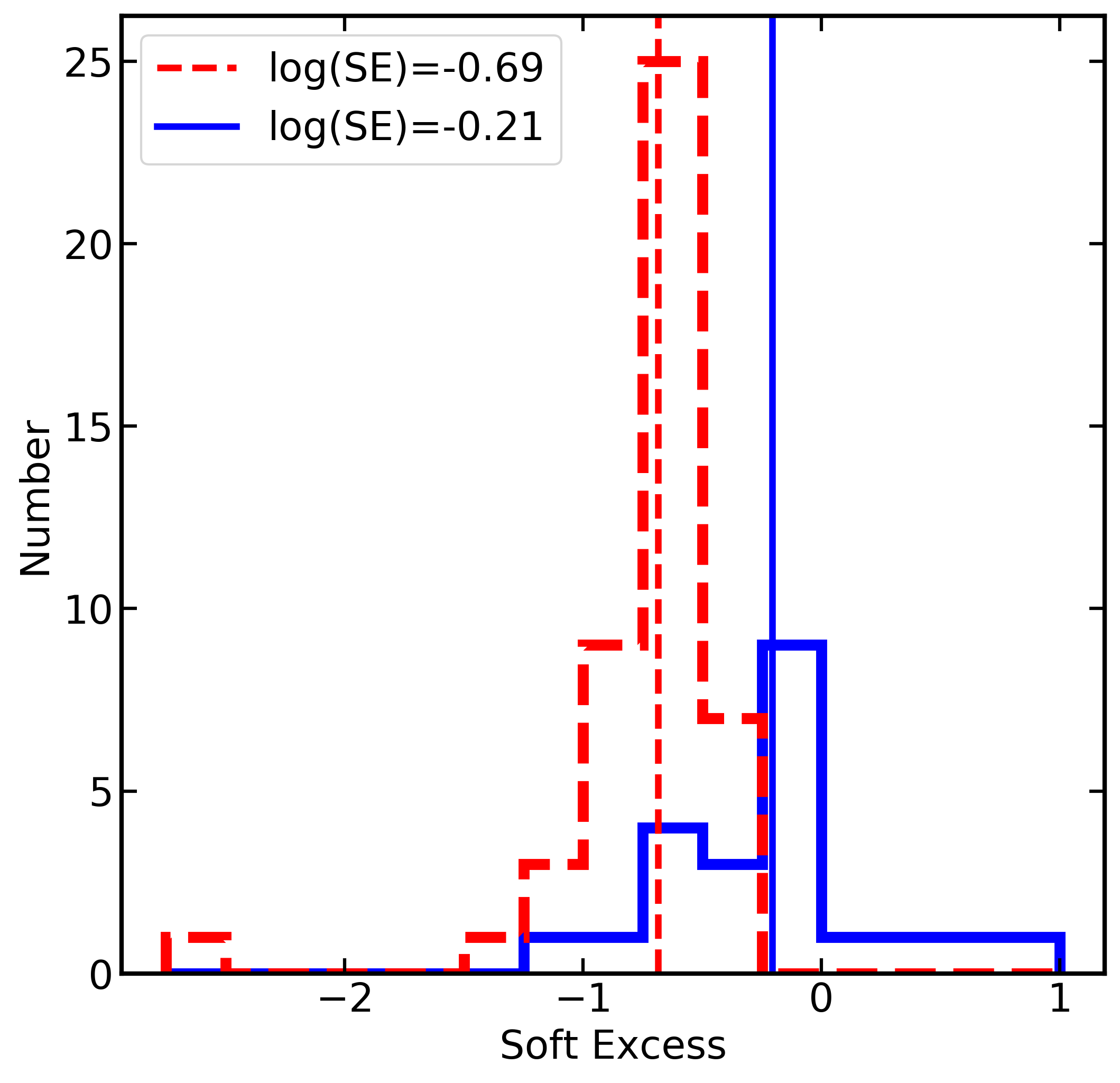}
	}
	\subfloat[]{
		\includegraphics[width=59mm]{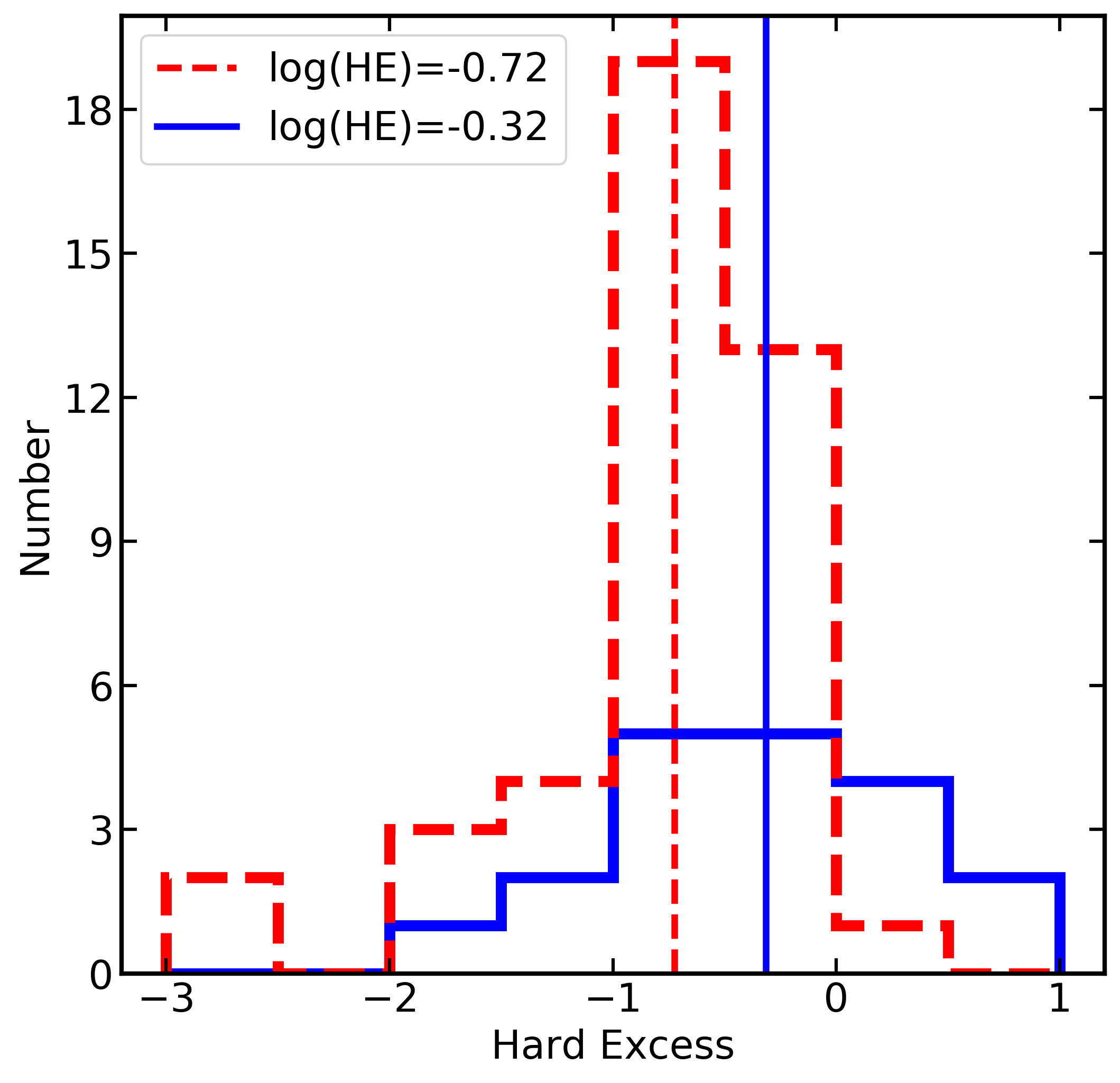}
	}
	\caption{Histograms showing parameter distributions, where the NLS1 parameter values are shown in blue solid lines and BLS1 parameter values are shown as red dashed lines. The median values for each parameter are shown as vertical lines in corresponding colours and line styles, and median values are given in the label for each plot. Panel (a) shows the X-ray luminosity, L$_x$ ($3-10\kev$) in log space (KS test p-value 0.12); (b) shows host-galaxy absorption in log values (p-value 0.27);  (c) shows photon index, $\Gamma$ (p-value 0.00040). Panel (d) shows black body temperature, kT (p-value 0.19). To show the influence of absorption on the temperature panel (e) shows the kT distribution only for sources that the absorption (znH) is an upper-limit consistent with zero. Many of the low kT objects are missing from this comparison, but the medians are very similar between samples (p-value 0.40). Panel (f) shows iron line strength, Fe/L$_x$ (p-value 0.028); (g) shows soft excess strength on a log-scale, log(SE) (p-value 8.0$\times$10$^{-6}$); and (h) shows the hard excess on a log-scale, log(HE) (p-value 0.0098).}
	\label{fig:hist}
\end{figure*}

Upon obtaining all the relevant parameters from the model, it is of interest to compare the distributions of each of these parameters between the NLS1 and BLS1 galaxies to search for similarities and differences between the two samples. All such distributions are presented in Fig.~\ref{fig:hist}. In each panel, the histogram for NLS1 galaxies is shown as a blue solid line, and the BLS1 histogram is shown as a red dashed line. The median values for each parameter are shown as vertical lines in corresponding colours and line styles, and median values are given in the label for each plot. 

First, panel (a) presents the luminosity distributions of the two samples, shown as log(L$_x$) in the $3-10\kev$ range. This band was selected to avoid any effects of absorption or contributions from the soft excess, which may have complicated a direct comparison between sources. The distributions show that a suitable range of luminosities are covered between the two samples. The NLS1 luminosities have a lower median value (log(L$_x$)=43.03 compared to log(L$_x$)=43.53 for BLS1 galaxies), likely owing to their lower black hole masses. We have also performed the KS test on the redshift distributions, and the measured p-value of 0.8 shows that no differences within the sample are attributable to redshift. The symmetry of the distributions suggests that the sample is largely unbiased in luminosity and covers a sufficient scope of type-1 AGN. From the KS test, we compute a p-value of 0.12, indicating that the distributions are similar.

Panel (b) shows the distributions of host-galaxy absorption (znH), again shown in log space. Many of these values are upper limits, as many galaxies showed little to no clear evidence for absorption within the \suzaku\ bands. When only an upper limit is present, the upper limit is shown in the histogram. We note also that very low absorption values of $10^{18}-10^{19}\pscm$ are very difficult to constrain given the limited \suzaku\ bandpass at low energies, and should thus be interpreted with caution. While the median level for absorption appears slightly higher for NLS1 galaxies, the KS test returns a p-value of 0.27, implying that the distributions are not statistically different. 

Many of these parameters include large errors, so we have attempted to quantify the confidence range for the KS test results. To do so, we take each parameter for each source and draw a pertubed value based on a normal distribution, centred on the parameter value and extending to the upper and lower error bars. This process is repeated 10000 times, and the p-value calculated for each perturbed sample. By taking the 5$^{\rm th}$ and 95$^{\rm th}$ percentile values, a range of p-values is obtained. For znH, we find a range of $0.16-0.37$. This shows that even when considering errors, no statistical differences in the distributions can be found. The same procedure is used to report a confidence range for the p-values for each of the remaining model parameters. This will be given along with measured value based on the best-fit.

Panel (c) shows the distribution of photon index ($\Gamma$) values. We note that some sources show very flat $\Gamma$ values of $1.4-1.6$; these may potentially be artificially flattened due to the presence of some underlying absorption or reflection component. The median $\Gamma$ value is significantly higher for NLS1 galaxies, with a p-value of 0.00040 ($0.00037-0.013$) implying $>99.9$ per cent confidence. More commonly, average values of the photon index are reported, so to ease comparison with other works, we determine the average values in our sample to be $\Gamma=1.78\pm0.02$ for the BLS1 galaxies and $\Gamma=2.00\pm0.06$ for the NLS1 sample. These are reasonably comparable to the canonical values of $\sim$1.9 and $\sim$2.1 (e.g. \citealt{Nandra+1994}; \citealt{Vaughan+1999}; \citealt{Reeves+2000}) that are often adopted. The difference could be attributed to systematic differences between telescopes (e.g. \citealt{Ishida+2011,Kettula+2013}); but more importantly, our sample confirms the well-known result that NLS1s possess steeper photon indices. 

In panel (d), we present the distribution of blackbody temperatures (kT), in units of keV. The distributions are statistically similar, with a p-value of 0.19 ($0.087-0.38$). However, we find that many AGN have kT values around $\sim0.05\kev$, compared to more typical values of $\sim0.10-0.15\kev$ (e.g. \citealt{Mateo+2010}). To check if this is an effect of the absorption component, panel (e) shows the same histogram, but including only the sources with upper limits on the absorption component (i.e., sources where the host-galaxy absorption agrees with 0). The median values are now nearly identical for NLS1 and BLS1 galaxies (p-value 0.40; $0.21-0.65$), and many of the lowest kT sources have been removed, suggesting some degeneracy between these components.

Panels (f), (g) and (h) present the distributions of the iron line strength (Fe/L$_x$), soft excess strength (SE; shown in log values), and hard excess strength (HE; shown in log values), respectively. For Fe/L$_x$, NLS1 galaxies exhibit lower median values than BLS1 galaxies (p-value 0.028; $0.0048-0.065$). By contrast, NLS1 galaxies have stronger median soft excesses and hard excesses than BLS1 galaxies. These differences appear significant in both cases, with p-values of 8.0$\times$10$^{-6}$ (7.9$\times$10$^{-7}$--1.5$\times$10$^{-5}$) and 0.0098 ($0.0018-0.013$) for the SE and HE distributions, respectively. The soft excess histogram (panel g) is particularly striking, with the median value for the NLS1 galaxies lying above the value of any BLS1 galaxy. The BLS1 galaxy with a very weak soft excess is 3C 111, a source which shows a very high level of host-galaxy absorption (e.g. \citealt{3c111}). Our analysis of this source gives log(znH)=21.9, just below our sample cutoff of 22.0. 

The KS test results and distributions show that NLS1 galaxies have steeper photon indices and higher X-ray Eddington ratios with confidence above the 99.9 per cent level, well known results from previous studies (e.g. \citealt{Pounds+1995,Boller+1996,Brandt+1997,Leighly1999,Gierlinski+2004,Grupe+2004,Crummy+2006,Ohja+2020}). The KS test also suggests that NLS1 galaxies exhibit stronger soft excesses than their BLS1 counterparts. Even when considering the p-values based on parameter uncertianties, this difference is significant at the 99.99 per cent confidence level and is the most statistically significant difference between NLS1 and BLS1 galaxies in our sample. This has been observed in various works, although all using slightly differing definitions of the soft excess strength (e.g. \citealt{Boller+1996,Vaughan+1999,Grupe+2010,Gliozzi+2020}). We also find that the hard excess is stronger in NLS1 galaxies, at the $>99$ per cent level. There is also some evidence that the iron strength is weaker in NLS1 galaxies, at the $95$ per cent confidence level. This suggests that the differences in the X-ray spectra in NLS1 and BLS1 galaxies go far beyond their power law slopes. 

\section{Sample analysis}
\label{sect:5}
\subsection{Correlation analysis}
\label{sect:corr}

The next step of our analysis is to search for correlations between the various parameters. We present this analysis in the form of a correlogram, constructed using the function {\sc corrplot} implemented in {\sc R}. This technique fits each pair of parameters and reports the Pearson correlation coefficient, $r$, obtained from the linear fit. Values of the correlation coefficient of larger than 0.5 indicate a significant positive correlation, while values less than -0.5 indicate a significant anti-correlation. 

\begin{figure}
	\centering
	\includegraphics[width=\columnwidth]{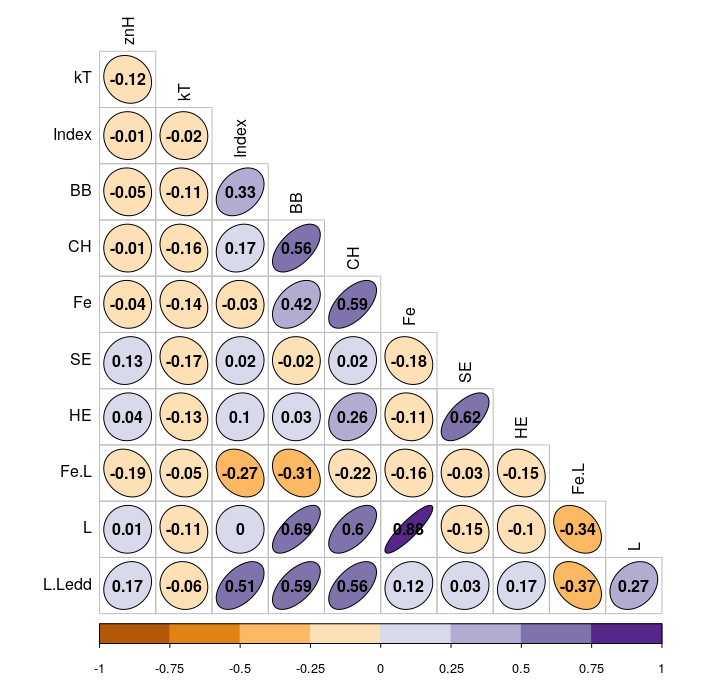}
	\caption{Correlation plots for the complete NLS1 and BLS1 sample. Yellow ellipses indicate anti-correlations, while purple indicates positive correlations. The Pearson correlation coefficient ($r$) values are written on the ellipses in black. We use eleven parameters in the analysis: host-galaxy column density (znH), blackbody temperature (kT), photon index (Index), blackbody luminosity in the $0.6-1.5\kev$ band (BB), Compton hump luminosity in the $15-25\kev$ band (CH), iron line luminosity (Fe), soft excess strength (SE), hard excess strength (HE), iron line strength (Fe.L), $3-10\kev$ luminosity (L), and X-ray Eddington ratio (L.Ledd). The analysis reveals nine significant positive correlations ($r > 0.5$).}
	\label{fig:quilt}
\end{figure}

The result of the correlation analysis for the entire Seyfert sample is shown in Fig.~\ref{fig:quilt}. The colour bar at the bottom shows the Pearson correlation coefficient, $r$, for each pair of parameters; yellow ellipses indicate anti-correlations, while purple indicates positive correlations. Higher ellipticity represents a tighter correlation. To help show which correlations are significant, the $r$ value is also printed on each ellipse. 

We use eleven parameters in the analysis; the eight parameters discussed in the previous Section (namely znH, kT, Index ($\Gamma$), SE, HE, Fe/L$_x$, L and  L$_{x}$/L$_{\rm Edd}$), as well as the blackbody luminosity in the $0.6-1.5\kev$ band (BB), the Compton hump luminosity in the $15-25\kev$ band (CH), and the iron line luminosity (Fe). The analysis reveals nine significant correlations including three with the X-ray Eddington ratio L$_{x}$/L$_{\rm Edd}$ (BB, CH, and Index); three with L$_{x}$ (BB, CH, and Fe); two others with CH (Fe and BB); and between SE and HE. Most of these correlations are associated with variations in luminosity of some component within the sample, thus it may be expected. For example, brighter AGN typically have higher blackbody, Compton hump, and iron line luminosities. Although not significant, there are also weak anti-correlations between the iron line strength and L, as well as the iron line strength and L$_{x}$/L$_{\rm Edd}$, in agreement with the X-ray Baldwin effect; an anti-correlation between the equivalent width of the \feka\ line and the luminosity of the AGN (e.g. \citealt{Baldwin}). 

The correlation between $\Gamma$ and the Eddington ratio is well known and has been studied in many works (e.g. \citealt{Shemmer+2008,Brightman+2013,Ohja+2020}). This relationship can be explained by considering a link between the Eddington ratio and the X-ray emitting corona. For systems accreting at a higher fraction of their Eddington limit, more emission from the disc will be incident upon the corona, resulting in more Compton up-scattering that causes the corona to cool more quickly. These cooler coronae then produce steeper photon indices (\citealt{Pounds+1995}). 

The tight correlation between the soft and hard excess components, SE and HE (defined as the ratio of blackbody and Compton hump luminosities, respectively, to the power law luminosity) is less well studied empirically. A hint at such a correlation was considered in \cite{Vasudevan+2014}, but using different definitions for the soft and hard excess strengths and non-simultaneous observations of the soft and hard energy bands. This result is therefore worthy of further consideration.

\begin{figure*}
	\centering
	\subfloat[]{
		\includegraphics[width=85mm]{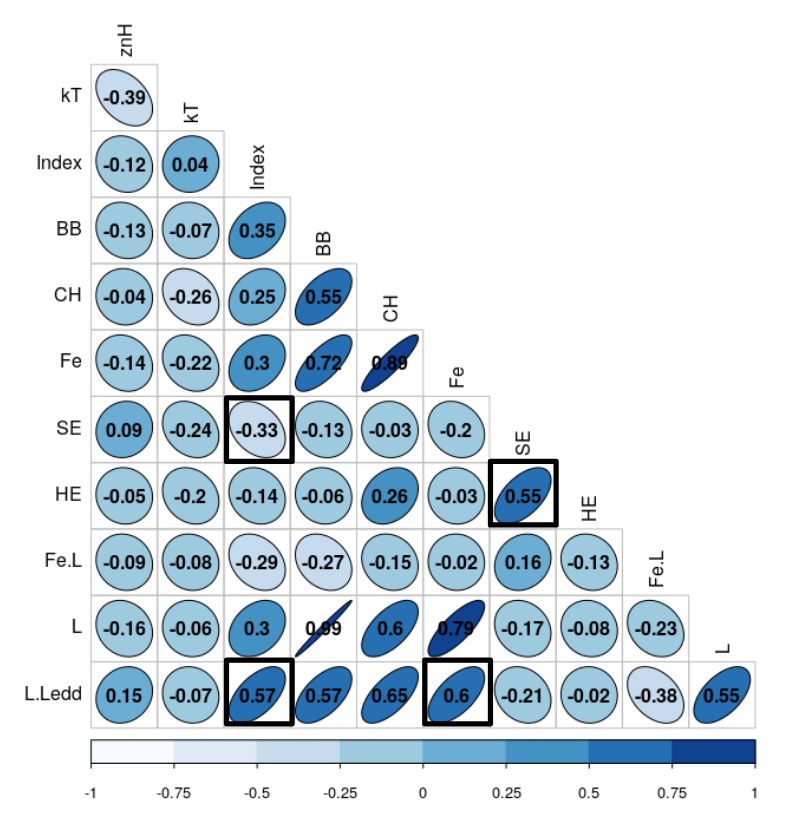}
	}
	\subfloat[]{
		\includegraphics[width=85mm]{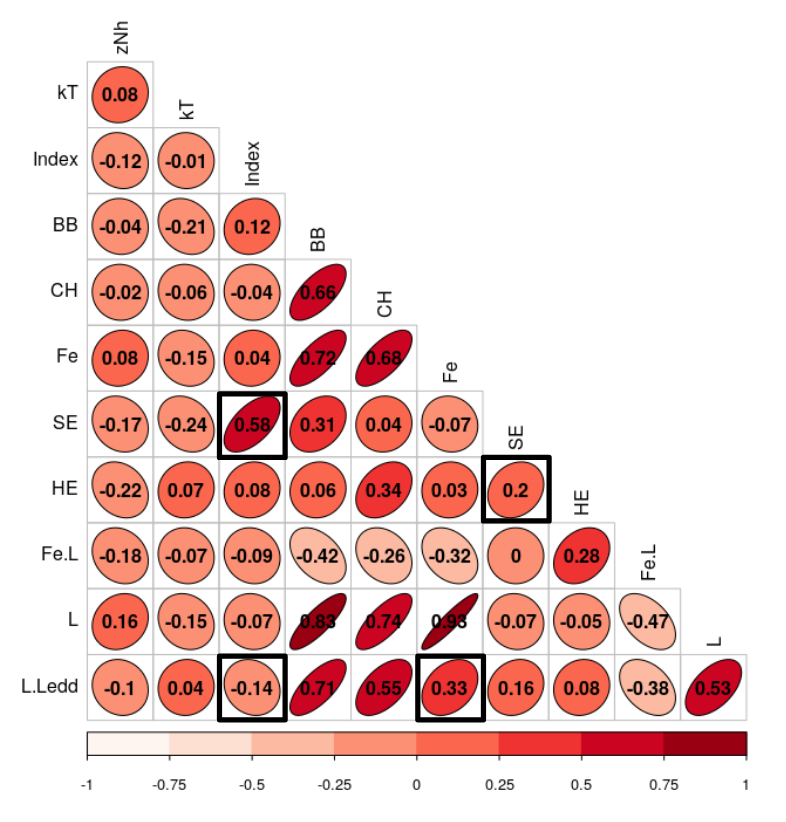}
	}
	\caption{Correlation analysis performed separately for NLS1 (left: blue) and BLS1 (right: red) galaxies. The colour bars at the bottom of each image show the corresponding $r$ values, and these are printed on each ellipse. Column and row names are the same as in Fig.~\ref{fig:quilt}. Two correlations, L with L$_{x}$/L$_{\rm Edd}$ and Fe with BB, are significant for the individual classes, but not for the full sample. Four correlations are significant in one sample, but not the other; these cells are outlines in black. BLS1 galaxies show a significant positive correlation between the soft excess strength and photon index (SE and $\Gamma$), while this correlation is weakly anti-correlated for NLS1 galaxies. The photon index $\Gamma$ and the iron line luminosity (Fe) are both significantly positively correlated with the X-ray Eddington ratio (L$_{x}$/L$_{\rm Edd}$) for NLS1 galaxies, but not for BLS1 galaxies. Finally, the soft excess and hard excess (SE and HE) are significantly positively correlated for NLS1 galaxies, but not for BLS1 galaxies.}
	\label{fig:2quilt}
\end{figure*}

To search for differences between NLS1 and BLS1 galaxies, we perform the same analysis separately on the NLS1 and BLS1 samples. The results are shown in Fig.~\ref{fig:2quilt}. The correlations for NLS1 galaxies are shown in blue, and those for BLS1 galaxies are shown in red. The colour bars below each image show the range of $r$ values for different colours, and the $r$ value for each correlation is printed over the corresponding ellipse. 

A comparison between the separate correlograms in Fig.~\ref{fig:2quilt} with the correlogram for the entire sample in Fig.~\ref{fig:quilt}, shows that some trends are stronger in the individual samples than in the total sample.  Specifically, the relation between iron line (Fe) and blackbody (BB) luminosities is rather significant in the individual samples ($r\approx0.72$) and only moderately important in the full sample analysis ($r=0.42$).  Likewise, L$_x$ and L$_{x}$/L$_{\rm Edd}$ are significantly correlated in the individual samples ($r\approx0.54$), but only mildly correlated in the full analysis ($r=0.27$). The straightforward explanation is that both the NLS1 and BLS1 samples exhibit a similar trend, but are slightly offset as their average luminosities are marginally different (Fig.~\ref{fig:hist}). When combining the two samples the slight differences degrade the overall correlation.

\begin{figure*}
	\centering
	\subfloat[]{
		\includegraphics[width=85mm]{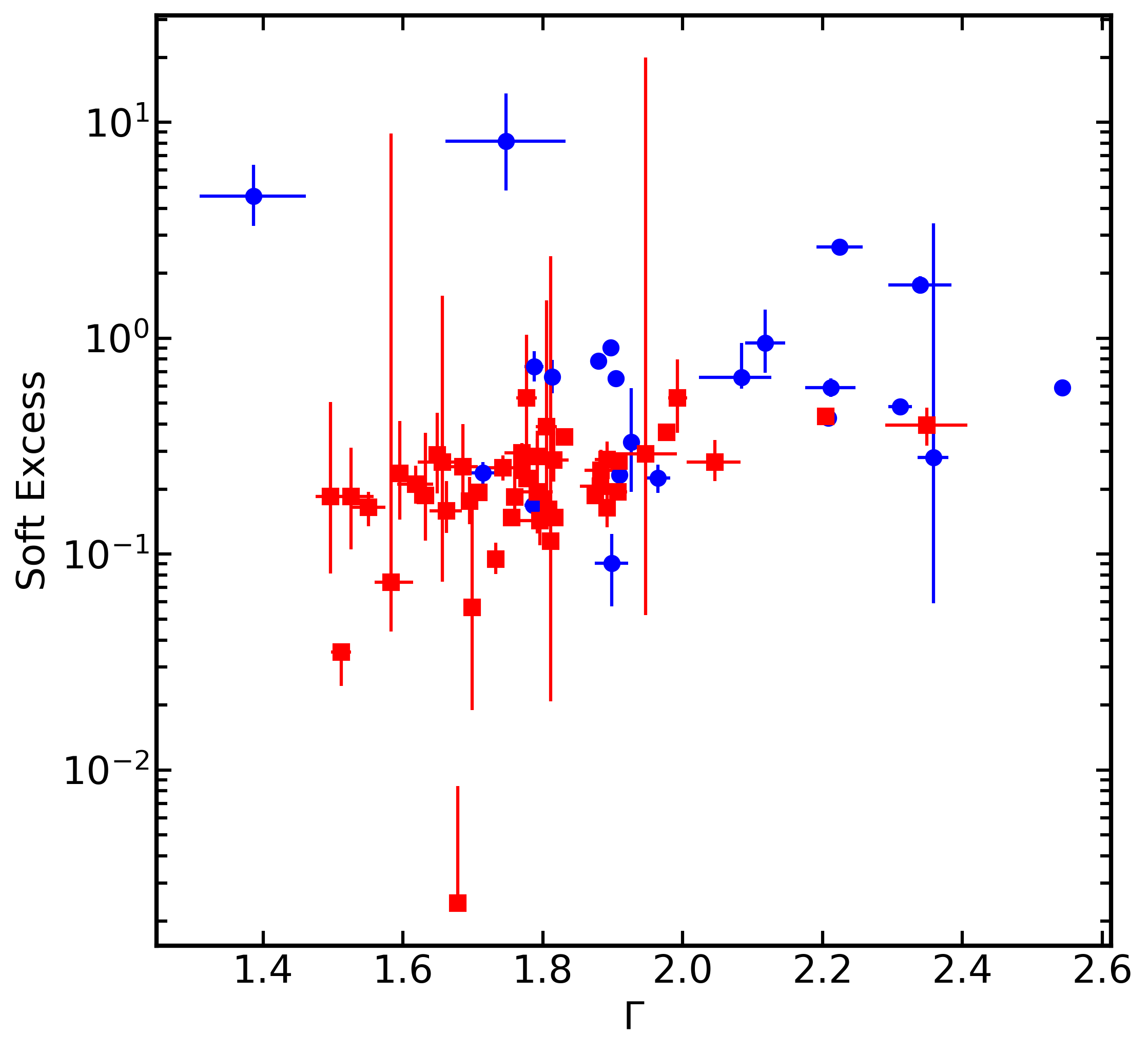}
	}
	\subfloat[]{
		\includegraphics[width=85mm]{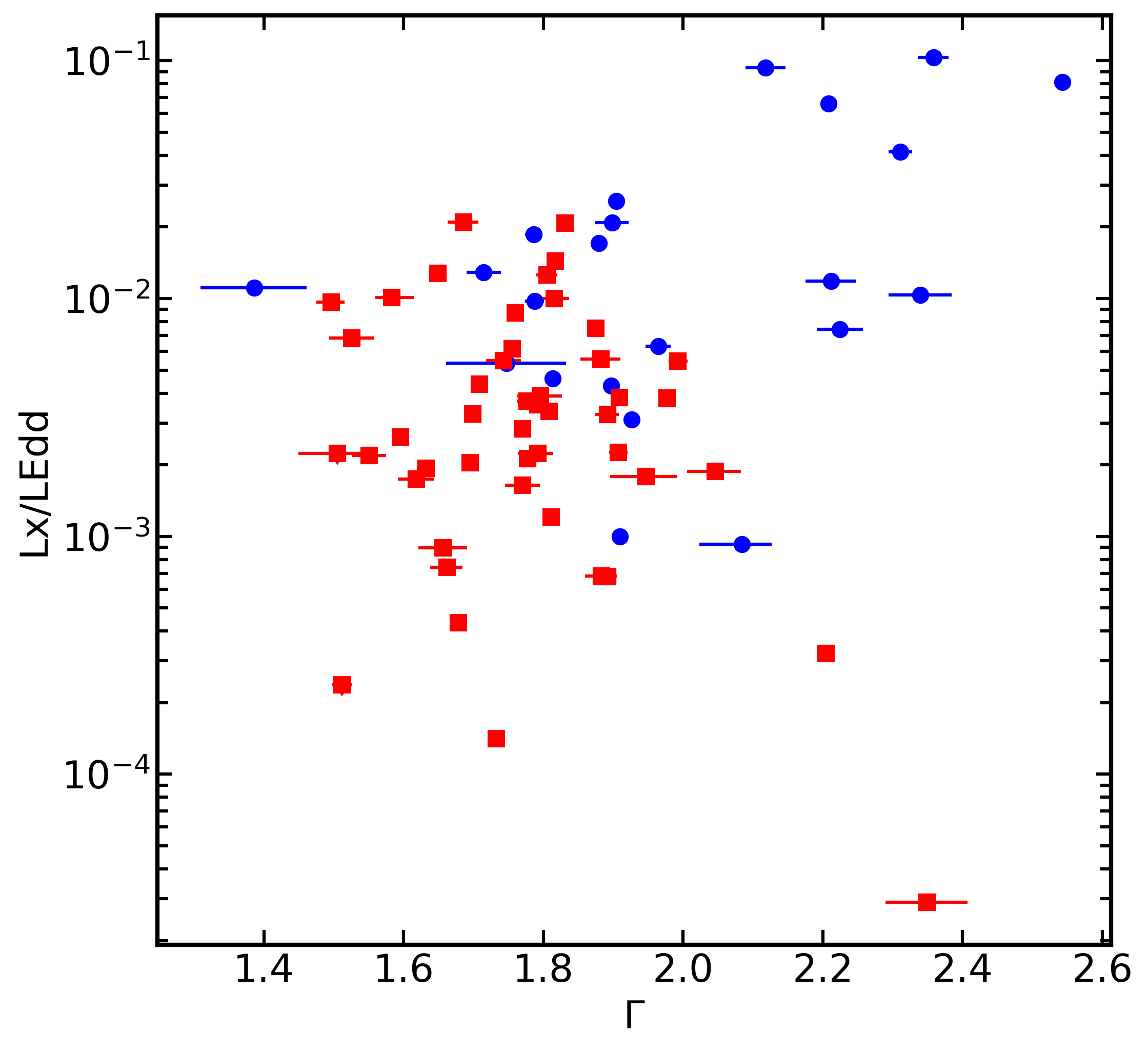}
	}
	\hspace{0mm}
	\subfloat[]{
		\includegraphics[width=85mm]{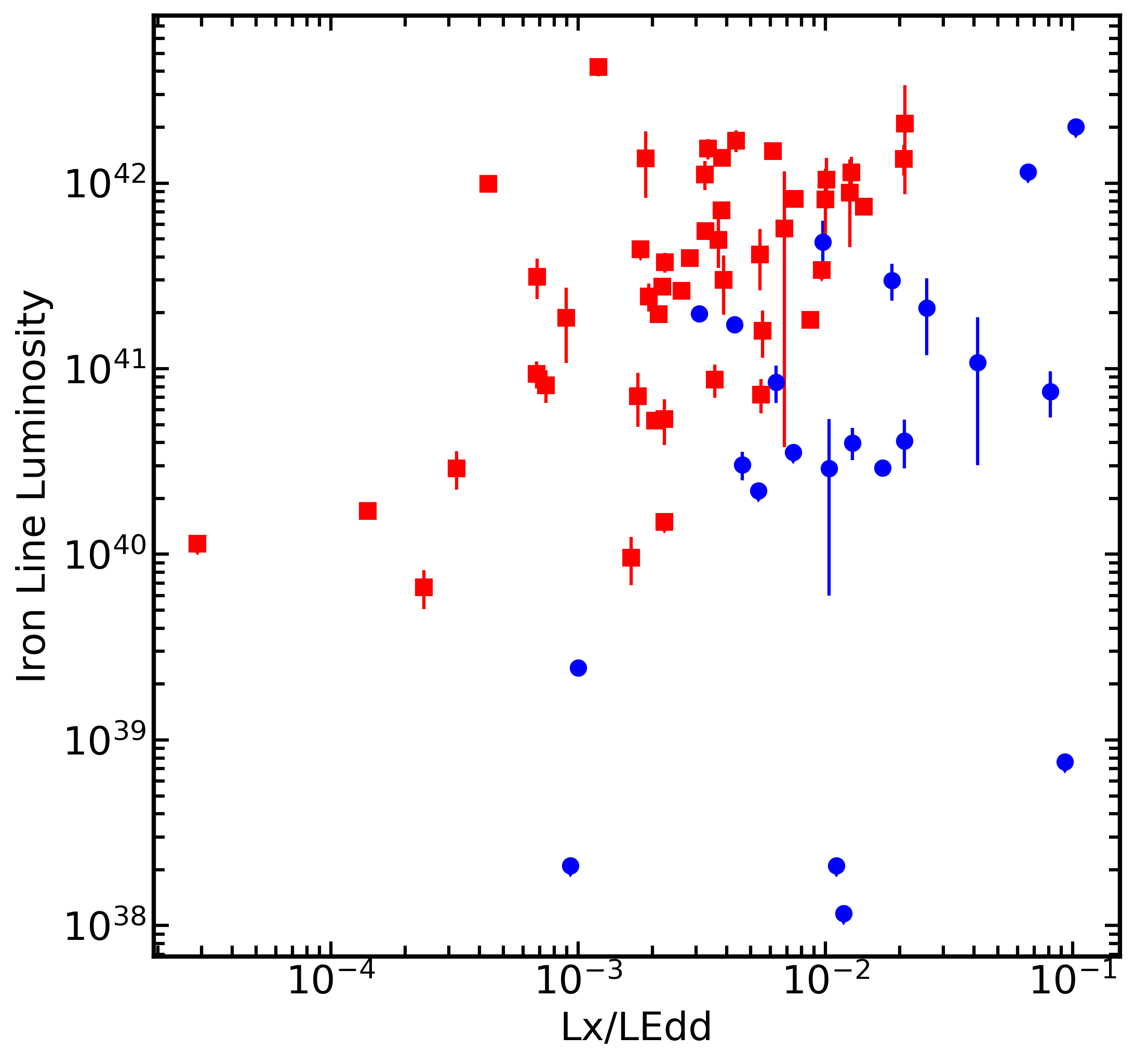}
	}
	\subfloat[]{
		\includegraphics[width=85mm]{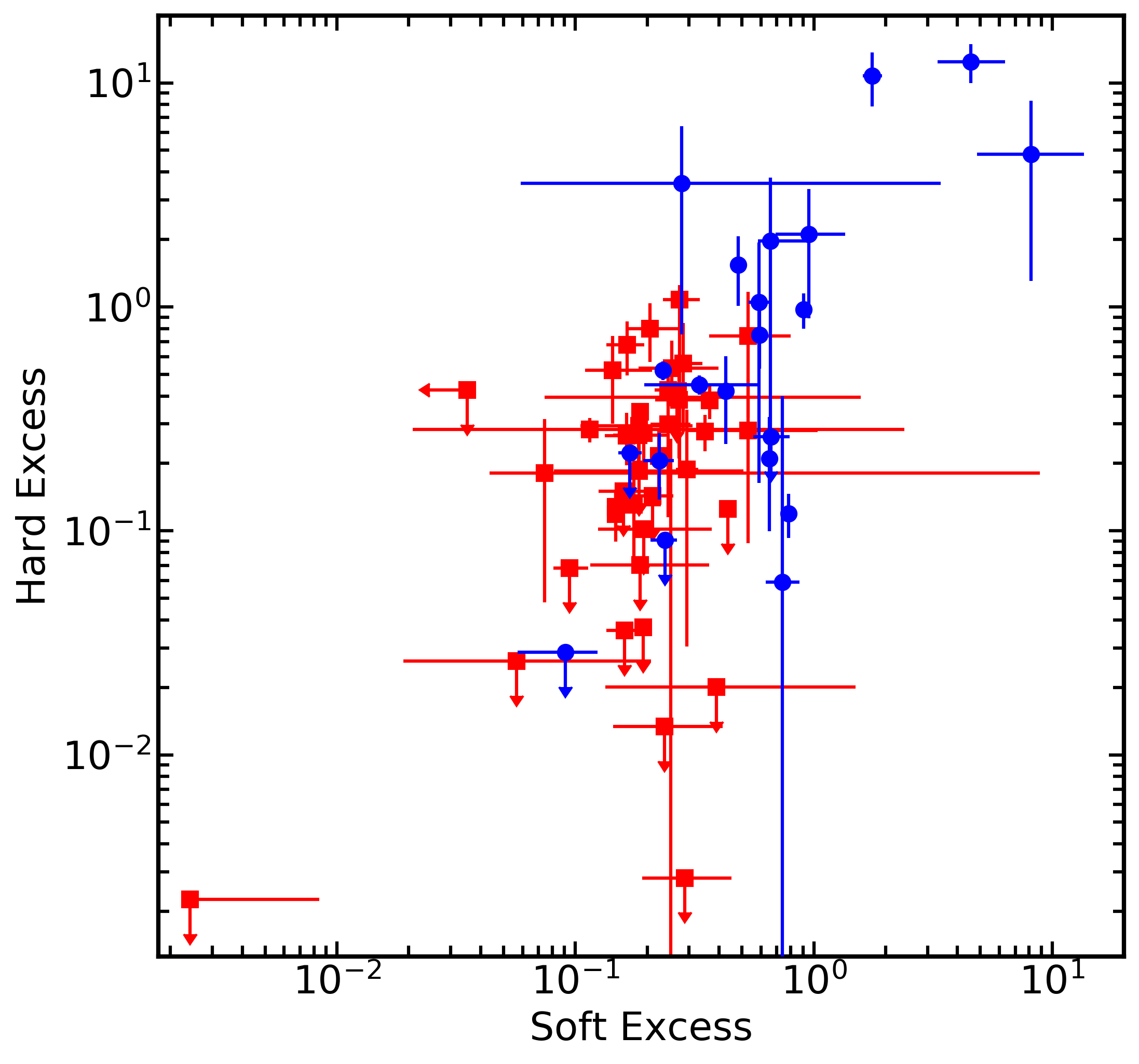}
	}
	\caption{Correlations for NLS1 (blue circles) and BLS1 (red squares) galaxies which are only significant for one sample. Upper limits on measurements are indicated with arrows. Panel (a) shows the distribution of $\Gamma$ with soft excess strength. The two extreme outliers for NLS1 galaxies are 1H0707-495 and PG 1404+226, two AGN whose extreme spectral curvature and poor data quality may contribute to poor measurement of these values. Panel (b) shows the correlation between $\Gamma$ and L$_{x}$/L$_{\rm Edd}$. A strong positive correlation is apparent for NLS1 galaxies, while outliers skew the distribution for BLS1 galaxies. Panel (c) shows the correlation between iron line luminosity and L$_{x}$/L$_{\rm Edd}$, with both axes shown in log scale for clarity. The correlation is strong for NLS1 galaxies, but weaker for BLS1 galaxies. Panel (d) shows the relationship between the soft excess and hard excess strengths, both shown in log scale. The correlation is strong for NLS1 galaxies, while BLS1 galaxies do not show any apparent correlation.}
\label{fig:scatter}
\end{figure*}

There are also four correlations which are significant in one sample, but not the other. These are highlighted in the correlograms (Fig.~\ref{fig:2quilt}), and shown in Fig.~\ref{fig:scatter}, where NLS1 galaxies are shown as blue circles and BLS1 galaxies are shown as red squares. Upper limits are indicated with arrows. First, in the top left panel (a), BLS1 galaxies show a significant positive correlation between the soft excess strength and photon index (SE and Index). The soft excess is shown on a log axis to show the large range of values.

For NLS1 galaxies, there is an anti-correlation between these parameters, although it is of lower significance with an $r$ value of -0.33. Examining the NLS1 points in panel (a) shows that this anti-correlation for NLS1 galaxies is driven by two sources with strong soft excesses and very flat $\Gamma$. These objects are 1H0707-495 and PG 1404+226. Both sources have poor data quality and extremely strong soft excesses. In particular, 1H0707-495 shows extreme spectral complexity in observations with \xmm\ (e.g. \citealt{Boller+2002,Gallo+2004,Fabian+2012}), likely indicating the presence of additional absorption or reflection. It is therefore possible that these measured $\Gamma$ values are not representative of the true underlying continuum. Removing these two outliers diminishes the negative trend but does not result in a significant correlation for the NLS1 sample. To preserve the size and scope of the sample, these objects are kept in the sample for the remainder of this work.

The other three panels show correlations which are significant for NLS1 galaxies, but not for the BLS1 sample. Panel (b) shows the relationship between $\Gamma$ and L$_{x}$/L$_{\rm Edd}$, where the X-ray Eddington ratio is shown on a log axis for clarity. NLS1 galaxies show a clear positive correlation, while the BLS1 galaxies do not show a significant trend. This is again likely affected by outliers, as both 3C 78 and PG 1626+554 have very low X-ray Eddington ratios but very steep photon indices. These spectra both have poor data quality, and 3C 78 also shows evidence for significant absorption, which may lead to poor measuring of these parameters. A positive correlation is more apparent, in agreement with previous works (e.g. \citealt{Shemmer+2008,Brightman+2013,Ohja+2020}), if these sources were removed. 

Panel (c) shows the relationship between the iron line luminosity and L$_{x}$/L$_{\rm Edd}$. Both values are shown on log scales for clarity. There appears to be more scatter in the distribution for BLS1 galaxies, which explains the slightly lower correlation coefficient for BLS1 galaxies ($r=0.33$) compared to NLS1 galaxies ($r=0.60$). The correlations appear rather similar in slope, but there is a clear shift in the measured luminosity of the narrow iron line with BLS1 galaxies exhibiting more luminous lines than NLS1 galaxies at a given L$_{x}$/L$_{\rm Edd}$. This also explains the insignificant correlation that was found between these parameters in the full sample (Fig.~\ref{fig:quilt}). 

Panel (d) shows the relationships between the soft excess (SE) and hard excess (HE) strengths over the power law. The AGN that do not have PIN data have not been included, as no hard excess could be measured. Both values are shown on log scales to better show the range of values, which span several orders of magnitude. For NLS1 galaxies, a very clear positive trend between these two values is observed. NLS1 galaxies with stronger soft excesses also show stronger hard excesses. The trend observed in the full sample seems therefore to be driven by the NLS1 sample. Neither the soft excess nor hard excess is highly correlated with L$_{x}$, which suggests that this relationship cannot be attributed to variations in luminosity. The correlation between these two parameters may instead point to a common origin between the soft excess and hard excess for NLS1 galaxies. 
	
This correlation between SE and HE is not observed for BLS1 galaxies, with a low correlation coefficient of $r=0.20$. As in the histograms presented in Fig.~\ref{fig:hist}, we observe that 3C 111 is an outlier, with extremely weak soft excess (likely due to large amounts of absorption) and only an upper limit placed on the hard excess. Aside from this source, the distribution of soft excesses is very narrow despite a wide range of hard excesses, spanning a factor of $\sim10$ in BLS1s compared to a factor of $\sim100$ in NLS1s. Further discussion is given in Section~\ref{sect:disc}. 

Finally, it is of interest to note that, although not highly significant in our sample, we do observe anti-correlations between the iron line strength (Fe/L$_x$) and the X-ray luminosity (L$_{x}$), as well as the X-ray Eddington ratio (L$_{x}$/L$_{\rm Edd}$) in both samples. In particular, for BLS1 galaxies, the $r$ value for the correlation between Fe/L$_x$ and L$_x$ is -0.47, very close to our definition of significant. This suggests that the X-ray Baldwin effect is an important feature in our sample, especially for BLS1 galaxies. 

\subsection{Principal Component Analysis}
\label{sect:pca}

To better understand the variations between objects in our sample, we employ principal component analysis (PCA). This technique uses eigenvalue decomposition to find the components that contribute to the maximum variations in a data set. We use the package {\sc prcomp} in {\sc R} to perform this analysis. As the correlation plots revealed many tight correlations between the luminosity of the AGN in different energy bands (as expected), we remove the blackbody, Compton hump and iron line luminosities, leaving only the $3-10\kev$ luminosity (L$_{x}$). We also remove the absorption parameter, znH. This leaves seven parameters for the analysis: kT, $\Gamma$, SE, HE, Fe/L$_x$, L$_{x}$ and L$_{x}$/L$_{\rm Edd}$, and thus our data has seven principal components.

\begin{figure*}
	\centering
	\subfloat[]{
		\includegraphics[width=90mm]{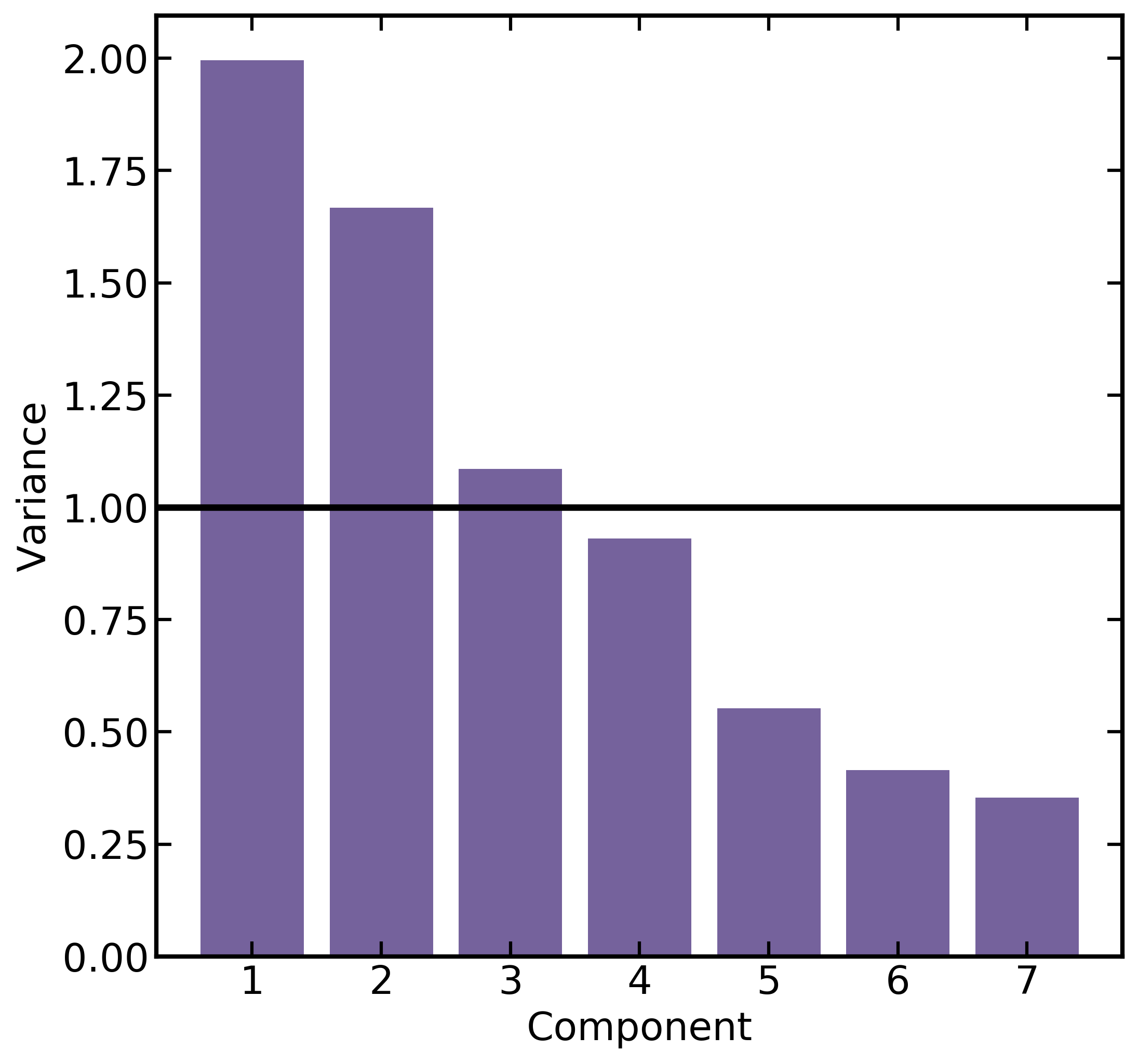}
	}
	\subfloat[]{
		\includegraphics[width=90mm]{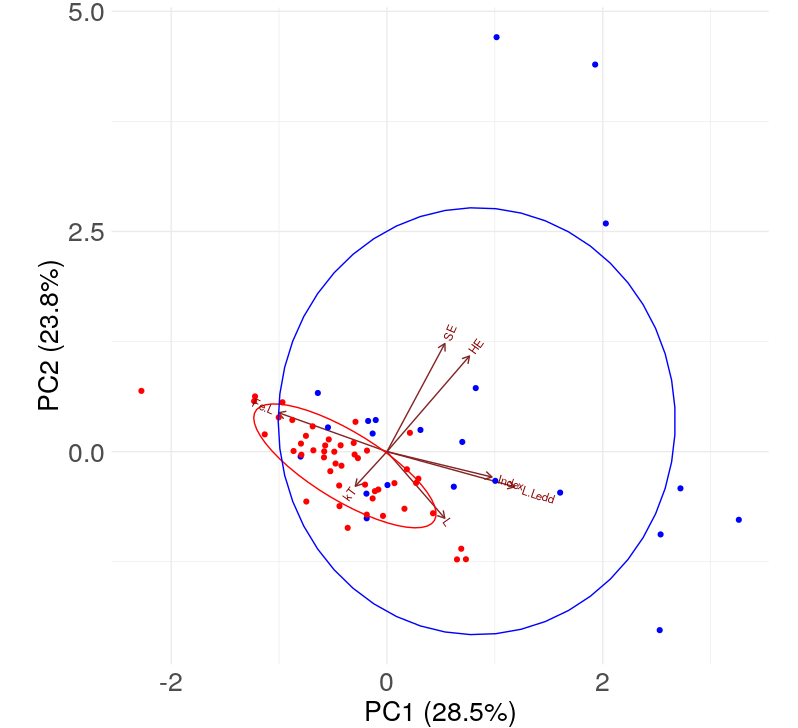}
	}
	\caption{Left: The variance produced by each principal component. A component is considered significant if its variance is greater than 1. PC1 and PC2 are clearly important. The variance in PC3 is only marginally greater than 1. Right: PC2 is plotted as a function of PC1 and the distribution of NLS1s (blue) and BLS1s (red) is shown.  The samples overlap, but do appear different. PC2 is largely dominated by the NLS1s.  For BLS1s, PC2 appears to be anti-correlated with PC1, but no such correlation is apparent for NLS1s. PC2 is largely composed of contribution for the soft excess (SE) and hard excess (HE) strengths, while significant contributions to PC1 are from the X-ray Eddington ratio (L$_{x}$/L$_{\rm Edd}$), photon index, and Fe/L$_x$ (see Table~\ref{tab:pca}).}
	\label{fig:pca}
\end{figure*}

First, we compute the variance for each principal component. The result is shown in panel (a) of Fig.~\ref{fig:pca}. For PCA, components with variances larger than 1 (shown with a solid black line) should be considered significant. Principal component 1 (PC1) and 2 (PC2) clearly meet this criteria. PC3 hovers just above the line ($\sim1.08$) and is likely not a significant component. 

The resulting PCA is shown in panel (b) of Fig.~\ref{fig:pca}. PC1 accounts for 28.5 per cent of the total variability, while PC2,  accounts for 23.8 per cent. NLS1 galaxies are shown as blue points, and BLS1 galaxies are shown as red points. Ellipses mark the parameter space occupied by each distribution. It is apparent that NLS1 galaxies are scattered in the parameter space and several occupy extremes of the parameter space. BLS1 galaxies are more consistent. Of interest is the tight anti-correlation between PC1 and PC2 that exists for BLS1 galaxies, but not for NLS1 galaxies. This behaviour may be highlighting intrinsic differences between NLS1 and BLS1 AGN.

The contributions of each measured parameter to each principal component are shown with dark red arrows. Longer arrows indicate higher contributions to the principal component. To assess the importance of these contributions to each PC, we compute the fractional contribution of each parameter to each of the first three principal components. The results are summarised in Table~\ref{tab:pca}. Parameters that contribute more than the average fractional contribution ($1/7 \simeq 0.15$) are considered important contributors to the PC. These values are shown in bold in the table.  

\begin{table}
	\resizebox{\columnwidth}{!}{%
		\begin{tabular}{lllll}
			\hline
			(1) & (2) & (3) & (4) & (5) \\
			Parameter & Name & PC1 & PC2 & PC3\\
			\hline
			kT & kT & 0.0187 & 0.0401 & \textbf{0.4214} \\
			Gamma ($\Gamma$) & Index & \textbf{0.2058} & 0.0217 & \textbf{0.1970} \\
			Soft excess (SE) & SE & 0.0625 & \textbf{0.3913} & 0.0026 \\
			Hard excess (HE) & HE & 0.1270 & \textbf{0.3062} & 0.0003\\
			Fe/L$_x$ & Fe.L & \textbf{0.2194} & 0.0507 & 0.0027 \\
			L$_x$ & L & 0.0623 & 0.1485 & \textbf{0.3605} \\
			L$_{x}$/L$_{\rm Edd}$ & L.Ledd & \textbf{0.3043} & 0.0416 & 0.0154 \\
			\hline
		\end{tabular}%
	}
	\caption{Fractional contributions of each parameter to the first three principal components. Column (1) gives the parameter name, column (2) gives the name as displayed  in panel (b) of Fig.~\ref{fig:pca}, column (3)  gives the fractional contribution to PC1, column (4) gives the contribution to PC2, and column (3) gives the contribution to PC3. Fractional contributions larger than $\sim0.15$ are considered important and are shown in bold.}
	\label{tab:pca}
\end{table}

Table~\ref{tab:pca} and panel (b) of Fig.~\ref{fig:pca} show that PC1 is mostly comprised of contributions from L$_{x}$/L$_{\rm Edd}$, along with weaker contributions from $\Gamma$ and the iron line strength, Fe/L$_x$. As we saw in the correlation analysis, and as discussed in previous works, both $\Gamma$ and the iron line emission are highly linked to the Eddington ratio. \cite{Shemmer+2008} and \cite{Brightman+2013}, among others, show a strong positive correlation between the Eddington ratio and $\Gamma$, which is also seen in this work (see Fig.~\ref{fig:quilt}). The X-ray Baldwin effect is also known to produce an anti-correlation between the accretion rate and the equivalent width of the \feka\ line (e.g. \citealt{Bianchi+2007}), which is seen also in our sample. In the PCA, the correlation between L$_{x}$/L$_{\rm Edd}$ and $\Gamma$ is indicated as both components contribution to PC1 in a correlated manner, while the anti-correlation between iron line strength and L$_{x}$/L$_{\rm Edd}$ is shown through the two components having opposite contributions to PC1 (L$_{x}$/L$_{\rm Edd}$ and Fe/L$_x$ point in opposite directions in panel (b) of Fig.~\ref{fig:pca}). It is therefore likely that PC1 is primarily due to variations in L$_{x}$/L$_{\rm Edd}$ within the sample.

Examining PC2, we see significant contributions from two components: the soft and hard excess strengths, SE and HE. These both contribute in the same way to PC2, as both arrows point in the same direction in Fig.\ref{fig:pca}. This is of interest, as the KS test analysis presented in Section~\ref{sect:histogram} indicated that the distributions of soft excess and hard excess strengths between NLS1 and BLS1 galaxies were different at the $>99.99$ per cent and $>99$ per cent levels, respectively, with NLS1 galaxies showing stronger soft and hard excesses. The correlation analysis also showed that these two parameters are significantly correlated, and that this correlation is driven by the NLS1 sample (where $r=0.55$), while BLS1 galaxies show little evidence for correlation between these parameters ($r=0.2$).

Although PC3 is likely not an important contributor to the observed variance, we show the fractional contributions of each parameter to this component in Table~\ref{tab:pca}. PC3 is mostly contributions from kT and L$_x$, with a weak contribution from $\Gamma$. Notably, kT and L$_x$ are the two components which were not important in PC1 or PC2. For this reason, and given that PC3 has a low variance, we do not consider this component in further discussion. 

As an additional check, we also compute the PCA separately for the NLS1 and BLS1 samples. For the NLS1 sample, only the first two principal components are significant. PC1 is dominated by contributions from the X-ray Eddington ratio, with weaker contributions from L$_x$ and $\Gamma$, while PC2 is dominated by the soft and hard excess strengths. This is largely in agreement with what we observe in the the PCA analysis for the full sample, and matches what we observe in the correlation analysis. For the BLS1 sample, PC1 has significant contributions from L$_{x}$/L$_{\rm Edd}$, L$_x$, and Fe/L$_x$. This is sensible, as all these parameters are highly correlated. However, for PC2, we see strong contributions from the soft excess strength and $\Gamma$, while there is no significant contribution from the hard excess. This again suggests that the relationship between the soft and hard excess, as well as the overall sample variance, are not the same between the NLS1 and BLS1 samples. These findings will come into play as we examine the markedly different behaviour of NLS1 and BLS1 galaxies seen in panel (b) of Fig.~\ref{fig:pca} in Section~\ref{sect:disc}.

\section{Discussion}
\label{sect:disc}
\subsection{The \suzaku\ sample}
\label{sect:discsample}
Our analysis of a sample of 22 NLS1 and 47 BLS1 galaxies reveals many interesting results. A key advantage of \suzaku\ is that it provides a simultaneous view of the soft excess, \feka\ line, and hard excess, which is advantageous when searching for correlations between these parameters. It is important to note that \suzaku\ is not a survey mission and therefore, the sample is not complete in terms of flux or volume. However, the NLS1 and BLS1 samples cover a similar range of X-ray luminosities and redshifts, which allows comparisons to be made between the X-ray properties of the two samples. The main results of this analysis are:

\begin{enumerate}[(i)]
	\item In agreement with previous works, the sample shows that NLS1 galaxies have steeper photon indices ($\Gamma$), higher X-ray Eddington ratios (L$_{x}$/L$_{\rm Edd}$) and stronger soft excesses (SE) than BLS1 galaxies.
	\item The luminosity of the narrow \feka\ core is lower on average in NLS1 galaxies. When we divide by the line luminosity by the continuum luminosity (i.e. a proxy for equivalent width) for each AGN (Fe/L$_x$), this ratio is also weaker for NLS1s.
	\item There is a significant correlation between L$_{x}$/L$_{\rm Edd}$ and $\Gamma$, and a weaker anti-correlation between L$_{x}$/L$_{\rm Edd}$ and Fe/L$_x$ (a result of the X-ray Baldwin effect). In the PCA, these three components make up PC1, with the strongest contribution from the X-ray Eddington ratio. 
	\item NLS1 galaxies, on average, have stronger hard excesses than BLS1 galaxies. For NLS1 galaxies, we also find a significant correlation between the soft and hard excesses, which is not found in the BLS1 galaxies. 
	\item When performing the PCA on the full sample, this SE and HE correlation appears as PC2. When the PCA is repeated for the individual samples, the NLS1s still show the soft and hard excess as PC2. However, for the BLS1s, PC2 shows contributions from the $\Gamma$ and the soft excess.
	\item The PCA shows large overlap between the NLS1 and BLS1 parameter space. However, the behaviours of the two samples are distinct. BLS1 galaxies appear to show an anti-correlation between PC1 and PC2. NLS1 galaxies are more spread out, occupying extreme regions of parameter space. 
\end{enumerate}

These results and possible physical interpretations are discussed in more detail in the following sections. 

\subsection{Iron line properties}
When considering the properties of the narrow \feka\ core in our sample, we find some evidence that on average BLS1 galaxies have higher luminosity iron lines than NLS1 galaxies. This is also the case when considering the ratio of iron line luminosity to X-ray luminosity (Fe/L$_x$; i.e. a proxy for equivalent width), implying that NLS1 iron lines are also weaker relative to the continuum. 

The narrow \feka\ emission likely originates in distant material like the torus or broad line region (e.g. \citealt{Nandra+2007}). The stronger iron lines in BLS1 galaxies could result from increased contributions from the torus. This could result in high-levels of absorption in the spectra of BLS1s, which is not evident.  Using ztbabs as a proxy for overall galaxy absorption, there appears to be little difference in the levels of intrinsic absorption in NLS1s and BLS1s (see panel (b) of Fig.~\ref{fig:hist}).

Previous works show differing interpretations on analyses of absorption in NLS1 galaxies: for example, \cite{Boller+1996}, find little evidence for large column densities in NLS1 galaxies observed with \rosat. However, some previous interpretations of the properties of NLS1 galaxies (e.g. \citealt{Mathur+2000,Williams+2002,Middleton+2007}), invoke increased levels of absorption to help explain observed UV and X-ray properties. More consideration to potential explanations of the iron line properties will be given in a future work. 

\subsection{The soft and hard excesses in Seyfert 1 AGN}
Previous works have shown that NLS1 galaxies have steeper and stronger soft excesses than BLS1s (e.g. \citealt{Boller+1996,Leighly1999,Vaughan+1999,Gliozzi+2020,Ohja+2020}), while less is known about the hard excess given the difficulty in detecting high energy emission. In our sample, NLS1 galaxies have stronger soft and hard excesses (SE and HE) than BLS1 galaxies (see Fig.~\ref{fig:hist}). The re-sampled KS test p-values demonstrate that this distinction can be made even when considering the errors on the parameters. The relationship between these parameters is shown in panel (d) of Fig.~\ref{fig:scatter}, where the NLS1s are shown in blue and BLS1s are shown in red. NLS1 galaxies show a significant correlation between SE and HE, which is not seen in the BLS1 sample. The positive correlation seen in the full sample (Fig.~\ref{fig:quilt}) is therefore driven mainly by the NLS1s. 

The positive correlation between SE and HE in the NLS1 sample may point to a common origin between these two components. In particular, this correlation can be understood and is even predicted from the blurred reflection model (e.g. \citealt{RossFabian+2005}; \citealt{Vasudevan+2014}). In this scenario, photons are emitted isotropically from the corona. While some are observed directly (forming the power law component), others are first reflected off of the innermost accretion disc (e.g. \citealt{RossFabian+2005}). This produces a multitude of fluorescent emission lines at low energies, which are relativistically blurred to form the soft excess. Higher energy photons penetrate further into the disc, where the material becomes more optically thick. These photons then undergo Compton scattering, forming the Compton hump peaking around $20\kev$. The ratio of reflected emission to continuum emission is then defined as the reflection fraction, $R$, which depends primarily upon the geometry of the corona. It is tempting to suggest that the distribution of NLS1 galaxies may be explained by different reflection fractions for each source. 

By contrast, the lack of correlation between the soft and hard excesses in the BLS1 galaxies may indicate that the origin of the soft and hard excesses in BLS1s may not necessarily require significant blurred reflection. \cite{Boissay+2016} argue that the positive correlation between SE and $\Gamma$  observed in a BAT sample study, and that is also seen in the BLS1s in this work, cannot be explained with a blurred reflection model. Separate mechanisms may instead be responsible for the production of the soft and hard excesses (e.g. partial covering or soft Comptonisation producing the soft excess, and reflection from neutral material which produces the Compton hump). The narrow range of observed soft excesses in the BLS1 sample is also of interest and worthy of further exploration.

Correlations between these parameters have been studied in previous works. \cite{Vasudevan+2014} present a similar positive correlation using data from \xmm\ and the BAT 58 month catalogue. This work uses a ratio of blackbody emission from $0.4-3\kev$ and power law emission from $1.5-6\kev$ to define the soft excess, and the reflection fraction $R$ from the {\sc pexrav} (\citealt{pexrav}) to define the strength of the Compton hump, differing slightly from our analysis. This work also cautions that non-simultaneous observations of the soft and hard energy bands are used, which may affect model parameters. \cite{Boissay+2016} use a sample of type 1 AGN observed with \xmm\ and \swift\ BAT and instead find an anti-correlation between the soft excess strength and the reflection fraction. This trend is also found when considering a sample of only 10 NLS1 galaxies. However, this work again uses different definitions for the soft excess and $R$, complicating a direct comparison of results.

\cite{Vasudevan+2014} and \cite{Boissay+2016} also present theoretical relationships between the hard excess ($R$) and soft excess based on blurred reflection simulations. In both works, these simulations show a positive correlation between the soft excess and $R$, akin to the one seen in our NLS1 sample. In particular, \cite{Vasudevan+2014} show that blurred reflection simulations can result in extremely strong soft excesses like the ones observed in some NLS1s in our sample. This lends further evidence that the soft and hard excesses in NLS1s may be produced through blurred reflection. By contrast, \cite{Vasudevan+2014} perform ionised absorption simulations and find no evidence for a correlation between the soft excess and $R$, similar to what we observe in our BLS1 sample. It is therefore plausible to suggest that ionised partial covering  (e.g. \citealt{Tanaka+2004,Chevallier+2006,Iso+2016}), or warm Comptonisation (e.g. \citealt{Petrucci+2018,Ballantyne+2020}), may be responsible for shaping some of the soft excesses observed in BLS1s.

The results of the PCA further support the distinct origins for the soft and hard excesses in the two samples. First, we note that the PCA performed on the full sample and on the NLS1s and BLS1s consistently reveals the X-ray Eddington ratio, L$_{x}$/L$_{\rm Edd}$, as the main contributor to PC1. However, the contributors to PC2 change depending on which sample is used. For the full sample, PC2 is dominated by changes in the soft and hard excesses. The SE and HE also form PC2 for the NLS1 sample, suggesting that these galaxies drive PC2 for the full sample. When only the BLS1 galaxies are used, PC2 instead becomes dominated by a combination of the soft excess and $\Gamma$. The PC2 results therefore appear to match the correlation analysis, and may demand a different origin for the soft excess in the two samples.

Another feature highlighting the differences between the two samples is the distribution of BLS1 galaxies in panel (b) of Fig.~\ref{fig:pca}. The BLS1 galaxies show a strong anti-correlation between PC1 and PC2, while no such correlation is seen in the NLS1 sample. This may be explained as a mathematical artefact of the PCA due to the large variance within the NLS1 sample. Such mathematical artefacts have been previously observed in PCA analysis; \cite{Gallant+2018} show that for spectral PCAs performed on a sample of blazars, PC3 sometimes appears as a distinctive bowed feature. Based on simulation work, this PC shape does not appear to represent a physical process, but rather a mathematical artefact resulting from the first two PCs. The anti-correlation observed here may have a similar origin. When the PCA is performed separately on the NLS1 and BLS1 samples, no such anti-correlation exists.

If this is indeed the case, this artefact may appear due to the much larger variance within the NLS1 sample. Several  NLS1 sources have extreme soft excesses (1H0707-495, PG1404+226, RE J1034+396), and others are radio loud and have high X-ray Eddington ratios (PKS 0558-504, RX J0134.2-4258, RX J1633.3+4718). These sources all appear as outliers in panel (b) of Fig.~\ref{fig:pca}. This again highlights the extreme behaviour of the NLS1 sample and further suggests intrinsic spectral differences between the two samples, and in particular, in the measured soft and hard excesses.

\section{Conclusions}
\label{sect:conclusion}

In this work, we present a sample of Seyfert 1 galaxies observed with \suzaku. We select all NLS1 and BLS1 galaxies with redshifts $<0.5$ and low host-galaxy column densities ($<10^{22}$\pscm) observed with \suzaku. This results in a sample of 69 AGN, of which 22 are NLS1s and 47 are BLS1s. These selection criteria allow for proper characterisation of the soft excess, often observed below $\sim2\kev$. The sample covers a similar range of redshifts and X-ray luminosities for both classes and is thus ideal for probing differences between the X-ray emission in NLS1 and BLS1 galaxies. 

In this work, we have focussed on measuring and comparing the broad band X-ray spectral properties of NLS1 and BLS1 galaxies using a toy model, which confirmed many previous results as well as presenting new findings on the soft and hard excesses. The acquired sample is suitable for the analysis of many additional properties of type-1 AGN, including further spectral and variability analysis.

This work has shown that the soft excess, and the relationship between the soft and hard excesses, are important characteristics in distinguishing the X-ray properties of Seyfert 1 galaxies. The work highlights the importance of simultaneous broadband X-ray spectroscopy to uncover the nature of the soft and hard excess in Seyfert 1 galaxies.

\section*{Acknowledgements}
We thank the referee for their helpful comments and suggestions which improved this manuscript. We would also like to thank Adam Gonzalez for helpful discussions. This research has made use of data obtained from the \suzaku\ satellite, a collaborative mission between the space agencies of Japan (JAXA) and the USA (NASA). The authors acknowledge the support of the Natural Sciences and Engineering Research Council of Canada (NSERC). LCG acknowledges financial support from the Canadian Space Agency (CSA).

\section*{Data Availability}
The data used in this work are publicly available in the \suzaku\ DARTS archive (https://darts.isas.jaxa.jp/astro/suzaku/data/). 

%

\bibliographystyle{mnras}
\bibliography{sept8_sample_arxiv_4pp}

\begin{thebibliography}{}
\makeatletter
\relax
\def\mn@urlcharsother{\let\do\@makeother \do\$\do\&\do\#\do\^\do\_\do\%\do\~}
\def\mn@doi{\begingroup\mn@urlcharsother \@ifnextchar [ {\mn@doi@}
  {\mn@doi@[]}}
\def\mn@doi@[#1]#2{\def\@tempa{#1}\ifx\@tempa\@empty \href
  {http://dx.doi.org/#2} {doi:#2}\else \href {http://dx.doi.org/#2} {#1}\fi
  \endgroup}
\def\mn@eprint#1#2{\mn@eprint@#1:#2::\@nil}
\def\mn@eprint@arXiv#1{\href {http://arxiv.org/abs/#1} {{\tt arXiv:#1}}}
\def\mn@eprint@dblp#1{\href {http://dblp.uni-trier.de/rec/bibtex/#1.xml}
  {dblp:#1}}
\def\mn@eprint@#1:#2:#3:#4\@nil{\def\@tempa {#1}\def\@tempb {#2}\def\@tempc
  {#3}\ifx \@tempc \@empty \let \@tempc \@tempb \let \@tempb \@tempa \fi \ifx
  \@tempb \@empty \def\@tempb {arXiv}\fi \@ifundefined
  {mn@eprint@\@tempb}{\@tempb:\@tempc}{\expandafter \expandafter \csname
  mn@eprint@\@tempb\endcsname \expandafter{\@tempc}}}

\bibitem[\protect\citeauthoryear{{Alston} et~al.,}{{Alston}
  et~al.}{2019}]{Alston+2019}
{Alston} W.~N.,  et~al., 2019, \mn@doi [MNRAS] {10.1093/mnras/sty2527}, \href
  {https://ui.adsabs.harvard.edu/abs/2019MNRAS.482.2088A} {482, 2088}

\bibitem[\protect\citeauthoryear{{Arnaud}}{{Arnaud}}{1996}]{xspec}
{Arnaud} K.~A.,  1996, in {Jacoby} G.~H.,  {Barnes} J.,  eds,  Astronomical
  Society of the Pacific Conference Series Vol. 101, Astronomical Data Analysis
  Software and Systems V. p.~17

\bibitem[\protect\citeauthoryear{{Ballantyne}}{{Ballantyne}}{2005}]{Bal+2005}
{Ballantyne} D.~R.,  2005, \mn@doi [\mnras] {10.1111/j.1365-2966.2005.09345.x},
  \href {https://ui.adsabs.harvard.edu/abs/2005MNRAS.362.1183B} {362, 1183}

\bibitem[\protect\citeauthoryear{{Ballantyne}}{{Ballantyne}}{2020}]{Ballantyne+2020}
{Ballantyne} D.~R.,  2020, \mn@doi [\mnras] {10.1093/mnras/stz3294}, \href
  {https://ui.adsabs.harvard.edu/abs/2020MNRAS.491.3553B} {491, 3553}

\bibitem[\protect\citeauthoryear{{Ballantyne}, {Ross}  \&
  {Fabian}}{{Ballantyne} et~al.}{2001}]{Ballantyne+2001}
{Ballantyne} D.~R.,  {Ross} R.~R.,   {Fabian} A.~C.,  2001, \mn@doi [\mnras]
  {10.1046/j.1365-8711.2001.04432.x}, \href
  {https://ui.adsabs.harvard.edu/abs/2001MNRAS.327...10B} {327, 10}

\bibitem[\protect\citeauthoryear{{Ballo}, {Braito}, {Reeves}, {Sambruna}  \&
  {Tombesi}}{{Ballo} et~al.}{2011}]{3c111}
{Ballo} L.,  {Braito} V.,  {Reeves} J.~N.,  {Sambruna} R.~M.,   {Tombesi} F.,
  2011, \mn@doi [\mnras] {10.1111/j.1365-2966.2011.19629.x}, \href
  {https://ui.adsabs.harvard.edu/abs/2011MNRAS.418.2367B} {418, 2367}

\bibitem[\protect\citeauthoryear{{Barth} et~al.,}{{Barth}
  et~al.}{2011}]{zw229hb}
{Barth} A.~J.,  et~al., 2011, \mn@doi [\apj] {10.1088/0004-637X/732/2/121},
  \href {https://ui.adsabs.harvard.edu/abs/2011ApJ...732..121B} {732, 121}

\bibitem[\protect\citeauthoryear{{Bentz} \& {Katz}}{{Bentz} \&
  {Katz}}{2015}]{AGNmass}
{Bentz} M.~C.,  {Katz} S.,  2015, \mn@doi [PASP] {10.1086/679601}, \href
  {https://ui.adsabs.harvard.edu/abs/2015PASP..127...67B} {127, 67}

\bibitem[\protect\citeauthoryear{{Bian} \& {Zhao}}{{Bian} \&
  {Zhao}}{2003}]{Bian+2003}
{Bian} W.,  {Zhao} Y.,  2003, \mn@doi [\mnras]
  {10.1046/j.1365-8711.2003.06650.x}, \href
  {https://ui.adsabs.harvard.edu/abs/2003MNRAS.343..164B} {343, 164}

\bibitem[\protect\citeauthoryear{{Bianchi}, {Guainazzi}, {Matt}  \& {Fonseca
  Bonilla}}{{Bianchi} et~al.}{2007}]{Bianchi+2007}
{Bianchi} S.,  {Guainazzi} M.,  {Matt} G.,   {Fonseca Bonilla} N.,  2007,
  \mn@doi [\aap] {10.1051/0004-6361:20077331}, \href
  {https://ui.adsabs.harvard.edu/abs/2007A&A...467L..19B} {467, L19}

\bibitem[\protect\citeauthoryear{{Boissay}, {Ricci}  \& {Paltani}}{{Boissay}
  et~al.}{2016}]{Boissay+2016}
{Boissay} R.,  {Ricci} C.,   {Paltani} S.,  2016, \mn@doi [\aap]
  {10.1051/0004-6361/201526982}, \href
  {https://ui.adsabs.harvard.edu/abs/2016A&A...588A..70B} {588, A70}

\bibitem[\protect\citeauthoryear{{Boller}, {Brandt}  \& {Fink}}{{Boller}
  et~al.}{1996}]{Boller+1996}
{Boller} T.,  {Brandt} W.~N.,   {Fink} H.,  1996, AAP, \href
  {https://ui.adsabs.harvard.edu/abs/1996A&A...305...53B} {305, 53}

\bibitem[\protect\citeauthoryear{{Boller} et~al.,}{{Boller}
  et~al.}{2002}]{Boller+2002}
{Boller} T.,  et~al., 2002, \mn@doi [\mnras]
  {10.1046/j.1365-8711.2002.05040.x}, \href
  {https://ui.adsabs.harvard.edu/abs/2002MNRAS.329L...1B} {329, L1}

\bibitem[\protect\citeauthoryear{{Boroson} \& {Green}}{{Boroson} \&
  {Green}}{1992}]{Boroson+1992}
{Boroson} T.~A.,  {Green} R.~F.,  1992, \mn@doi [ApJs] {10.1086/191661}, \href
  {https://ui.adsabs.harvard.edu/abs/1992ApJS...80..109B} {80, 109}

\bibitem[\protect\citeauthoryear{{Brandt}, {Mathur}  \& {Elvis}}{{Brandt}
  et~al.}{1997}]{Brandt+1997}
{Brandt} W.~N.,  {Mathur} S.,   {Elvis} M.,  1997, \mn@doi [MNRAS]
  {10.1093/mnras/285.3.L25}, \href
  {https://ui.adsabs.harvard.edu/abs/1997MNRAS.285L..25B} {285, L25}

\bibitem[\protect\citeauthoryear{{Brightman} et~al.,}{{Brightman}
  et~al.}{2013}]{Brightman+2013}
{Brightman} M.,  et~al., 2013, \mn@doi [\mnras] {10.1093/mnras/stt920}, \href
  {https://ui.adsabs.harvard.edu/abs/2013MNRAS.433.2485B} {433, 2485}

\bibitem[\protect\citeauthoryear{{Buhariwalla}, {Waddell}, {Gallo}, {Grupe}  \&
  {Komossa}}{{Buhariwalla} et~al.}{2020}]{m1239+2020}
{Buhariwalla} M.~Z.,  {Waddell} S. G.~H.,  {Gallo} L.~C.,  {Grupe} D.,
  {Komossa} S.,  2020, arXiv e-prints, \href
  {https://ui.adsabs.harvard.edu/abs/2020arXiv200808027B} {p. arXiv:2008.08027}

\bibitem[\protect\citeauthoryear{{Cash}}{{Cash}}{1979}]{Cash1979}
{Cash} W.,  1979, \mn@doi [ApJ] {10.1086/156922}, \href
  {https://ui.adsabs.harvard.edu/abs/1979ApJ...228..939C} {228, 939}

\bibitem[\protect\citeauthoryear{{Chevallier}, {Collin}, {Dumont}, {Czerny},
  {Mouchet}, {Gon{\c{c}}alves}  \& {Goosmann}}{{Chevallier}
  et~al.}{2006}]{Chevallier+2006}
{Chevallier} L.,  {Collin} S.,  {Dumont} A.~M.,  {Czerny} B.,  {Mouchet} M.,
  {Gon{\c{c}}alves} A.~C.,   {Goosmann} R.,  2006, \mn@doi [\aap]
  {10.1051/0004-6361:20053730}, \href
  {https://ui.adsabs.harvard.edu/abs/2006A&A...449..493C} {449, 493}

\bibitem[\protect\citeauthoryear{{Crummy}, {Fabian}, {Gallo}  \&
  {Ross}}{{Crummy} et~al.}{2006}]{Crummy+2006}
{Crummy} J.,  {Fabian} A.~C.,  {Gallo} L.,   {Ross} R.~R.,  2006, \mn@doi
  [\mnras] {10.1111/j.1365-2966.2005.09844.x}, \href
  {https://ui.adsabs.harvard.edu/abs/2006MNRAS.365.1067C} {365, 1067}

\bibitem[\protect\citeauthoryear{{Czerny} et~al.,}{{Czerny}
  et~al.}{2016}]{Czerny+2016}
{Czerny} B.,  et~al., 2016, \mn@doi [\aap] {10.1051/0004-6361/201628103}, \href
  {https://ui.adsabs.harvard.edu/abs/2016A&A...594A.102C} {594, A102}

\bibitem[\protect\citeauthoryear{{Done}, {Davis}, {Jin}, {Blaes}  \&
  {Ward}}{{Done} et~al.}{2012}]{Done+2012}
{Done} C.,  {Davis} S.~W.,  {Jin} C.,  {Blaes} O.,   {Ward} M.,  2012, \mn@doi
  [MNRAS] {10.1111/j.1365-2966.2011.19779.x}, \href
  {https://ui.adsabs.harvard.edu/abs/2012MNRAS.420.1848D} {420, 1848}

\bibitem[\protect\citeauthoryear{{Du} et~al.,}{{Du} et~al.}{2018}]{Du+2018}
{Du} P.,  et~al., 2018, \mn@doi [\apj] {10.3847/1538-4357/aaed2c}, \href
  {https://ui.adsabs.harvard.edu/abs/2018ApJ...869..142D} {869, 142}

\bibitem[\protect\citeauthoryear{{Fabian}, {Rees}, {Stella}  \&
  {White}}{{Fabian} et~al.}{1989}]{Fabian+1989}
{Fabian} A.~C.,  {Rees} M.~J.,  {Stella} L.,   {White} N.~E.,  1989, \mn@doi
  [MNRAS] {10.1093/mnras/238.3.729}, \href
  {https://ui.adsabs.harvard.edu/abs/1989MNRAS.238..729F} {238, 729}

\bibitem[\protect\citeauthoryear{{Fabian} et~al.,}{{Fabian}
  et~al.}{2012}]{Fabian+2012}
{Fabian} A.~C.,  et~al., 2012, \mn@doi [\mnras]
  {10.1111/j.1365-2966.2011.19676.x}, \href
  {https://ui.adsabs.harvard.edu/abs/2012MNRAS.419..116F} {419, 116}

\bibitem[\protect\citeauthoryear{{Fukazawa} et~al.,}{{Fukazawa}
  et~al.}{2009}]{Fukazawa+2009}
{Fukazawa} Y.,  et~al., 2009, \mn@doi [\pasj] {10.1093/pasj/61.sp1.S17}, \href
  {https://ui.adsabs.harvard.edu/abs/2009PASJ...61S..17F} {61, S17}

\bibitem[\protect\citeauthoryear{{Fukazawa} et~al.,}{{Fukazawa}
  et~al.}{2011}]{Fukazawa+2011}
{Fukazawa} Y.,  et~al., 2011, \mn@doi [ApJ] {10.1088/0004-637X/727/1/19}, \href
  {https://ui.adsabs.harvard.edu/abs/2011ApJ...727...19F} {727, 19}

\bibitem[\protect\citeauthoryear{{Fukazawa}, {Furui}, {Hayashi}, {Ohno},
  {Hiragi}  \& {Noda}}{{Fukazawa} et~al.}{2016}]{Fukazawa+2016}
{Fukazawa} Y.,  {Furui} S.,  {Hayashi} K.,  {Ohno} M.,  {Hiragi} K.,   {Noda}
  H.,  2016, \mn@doi [\apj] {10.3847/0004-637X/821/1/15}, \href
  {https://ui.adsabs.harvard.edu/abs/2016ApJ...821...15F} {821, 15}

\bibitem[\protect\citeauthoryear{{Gallant}, {Gallo}  \& {Parker}}{{Gallant}
  et~al.}{2018}]{Gallant+2018}
{Gallant} D.,  {Gallo} L.~C.,   {Parker} M.~L.,  2018, \mn@doi [MNRAS]
  {10.1093/mnras/sty1987}, \href
  {https://ui.adsabs.harvard.edu/abs/2018MNRAS.480.1999G} {480, 1999}

\bibitem[\protect\citeauthoryear{{Gallo}}{{Gallo}}{2006}]{Gallo2006}
{Gallo} L.~C.,  2006, \mn@doi [MNRAS] {10.1111/j.1365-2966.2006.10137.x}, \href
  {https://ui.adsabs.harvard.edu/abs/2006MNRAS.368..479G} {368, 479}

\bibitem[\protect\citeauthoryear{{Gallo}}{{Gallo}}{2018}]{Gallo+2018}
{Gallo} L.,  2018, in Revisiting Narrow-Line Seyfert 1 Galaxies and their Place
  in the Universe. p.~34 (\mn@eprint {arXiv} {1807.09838})

\bibitem[\protect\citeauthoryear{{Gallo}, {Boller}, {Brandt}, {Fabian}  \&
  {Grupe}}{{Gallo} et~al.}{2004}]{Gallo+2004}
{Gallo} L.~C.,  {Boller} T.,  {Brandt} W.~N.,  {Fabian} A.~C.,   {Grupe} D.,
  2004, \mn@doi [\mnras] {10.1111/j.1365-2966.2004.08003.x}, \href
  {https://ui.adsabs.harvard.edu/abs/2004MNRAS.352..744G} {352, 744}

\bibitem[\protect\citeauthoryear{{Gallo} et~al.,}{{Gallo}
  et~al.}{2019}]{Gallo+2019m335}
{Gallo} L.~C.,  et~al., 2019, \mn@doi [MNRAS] {10.1093/mnras/stz274}, \href
  {https://ui.adsabs.harvard.edu/abs/2019MNRAS.484.4287G} {484, 4287}

\bibitem[\protect\citeauthoryear{{Gierli{\'n}ski} \& {Done}}{{Gierli{\'n}ski}
  \& {Done}}{2004}]{Gierlinski+2004}
{Gierli{\'n}ski} M.,  {Done} C.,  2004, \mn@doi [\mnras]
  {10.1111/j.1365-2966.2004.07687.x}, \href
  {https://ui.adsabs.harvard.edu/abs/2004MNRAS.349L...7G} {349, L7}

\bibitem[\protect\citeauthoryear{{Gliozzi} \& {Williams}}{{Gliozzi} \&
  {Williams}}{2020}]{Gliozzi+2020}
{Gliozzi} M.,  {Williams} J.~K.,  2020, \mn@doi [\mnras]
  {10.1093/mnras/stz3005}, \href
  {https://ui.adsabs.harvard.edu/abs/2020MNRAS.491..532G} {491, 532}

\bibitem[\protect\citeauthoryear{{Gliozzi}, {Papadakis}  \&
  {Brinkmann}}{{Gliozzi} et~al.}{2007}]{Gliozzi+2007}
{Gliozzi} M.,  {Papadakis} I.~E.,   {Brinkmann} W.~P.,  2007, \mn@doi [\apj]
  {10.1086/510798}, \href
  {https://ui.adsabs.harvard.edu/abs/2007ApJ...656..691G} {656, 691}

\bibitem[\protect\citeauthoryear{{Gonz{\'a}lez-Mart{\'\i}n} \&
  {Vaughan}}{{Gonz{\'a}lez-Mart{\'\i}n} \& {Vaughan}}{2012}]{GM+2012}
{Gonz{\'a}lez-Mart{\'\i}n} O.,  {Vaughan} S.,  2012, \mn@doi [\aap]
  {10.1051/0004-6361/201219008}, \href
  {https://ui.adsabs.harvard.edu/abs/2012A&A...544A..80G} {544, A80}

\bibitem[\protect\citeauthoryear{{Goodrich}}{{Goodrich}}{1989}]{Goodrich+1989}
{Goodrich} R.~W.,  1989, \mn@doi [ApJ] {10.1086/167586}, \href
  {https://ui.adsabs.harvard.edu/abs/1989ApJ...342..224G} {342, 224}

\bibitem[\protect\citeauthoryear{{Grupe}}{{Grupe}}{1996}]{Grupe1996}
{Grupe} D.,  1996, PhD thesis, -

\bibitem[\protect\citeauthoryear{{Grupe}}{{Grupe}}{2004}]{Grupe+2004}
{Grupe} D.,  2004, \mn@doi [ApJ] {10.1086/382516}, \href
  {https://ui.adsabs.harvard.edu/abs/2004AJ....127.1799G} {127, 1799}

\bibitem[\protect\citeauthoryear{{Grupe}, {Leighly}, {Thomas}  \&
  {Laurent-Muehleisen}}{{Grupe} et~al.}{2000}]{Grupe+2000}
{Grupe} D.,  {Leighly} K.~M.,  {Thomas} H.~C.,   {Laurent-Muehleisen} S.~A.,
  2000, \aap, \href {https://ui.adsabs.harvard.edu/abs/2000A&A...356...11G}
  {356, 11}

\bibitem[\protect\citeauthoryear{{Grupe}, {Mathur}  \& {Komossa}}{{Grupe}
  et~al.}{2004}]{m1239}
{Grupe} D.,  {Mathur} S.,   {Komossa} S.,  2004, \mn@doi [ApJ]
  {10.1086/421002}, \href
  {https://ui.adsabs.harvard.edu/abs/2004AJ....127.3161G} {127, 3161}

\bibitem[\protect\citeauthoryear{{Grupe}, {Komossa}  \& {Leighly}}{{Grupe}
  et~al.}{2010}]{Grupe+2010}
{Grupe} D.,  {Komossa} S.,   {Leighly} K.~M.,  2010, in {Maraschi} L.,
  {Ghisellini} G.,  {Della Ceca} R.,   {Tavecchio} F.,  eds,  Astronomical
  Society of the Pacific Conference Series Vol. 427, Accretion and Ejection in
  AGN: a Global View. p.~86

\bibitem[\protect\citeauthoryear{{Haardt} \& {Maraschi}}{{Haardt} \&
  {Maraschi}}{1991}]{Haardt+1991}
{Haardt} F.,  {Maraschi} L.,  1991, \mn@doi [\apjl] {10.1086/186171}, \href
  {https://ui.adsabs.harvard.edu/abs/1991ApJ...380L..51H} {380, L51}

\bibitem[\protect\citeauthoryear{{Haardt} \& {Maraschi}}{{Haardt} \&
  {Maraschi}}{1993}]{Haardt+1993}
{Haardt} F.,  {Maraschi} L.,  1993, \mn@doi [\apj] {10.1086/173020}, \href
  {https://ui.adsabs.harvard.edu/abs/1993ApJ...413..507H} {413, 507}

\bibitem[\protect\citeauthoryear{{Ishida} et~al.,}{{Ishida}
  et~al.}{2011}]{Ishida+2011}
{Ishida} M.,  et~al., 2011, \mn@doi [\pasj] {10.1093/pasj/63.sp3.S657}, \href
  {https://ui.adsabs.harvard.edu/abs/2011PASJ...63S.657I} {63, S657}

\bibitem[\protect\citeauthoryear{{Iso}, {Ebisawa}, {Sameshima}, {Mizumoto},
  {Miyakawa}, {Inoue}  \& {Yamasaki}}{{Iso} et~al.}{2016}]{Iso+2016}
{Iso} N.,  {Ebisawa} K.,  {Sameshima} H.,  {Mizumoto} M.,  {Miyakawa} T.,
  {Inoue} H.,   {Yamasaki} H.,  2016, \mn@doi [\pasj] {10.1093/pasj/psw015},
  \href {https://ui.adsabs.harvard.edu/abs/2016PASJ...68S..27I} {68, S27}

\bibitem[\protect\citeauthoryear{{Iwasawa} \& {Taniguchi}}{{Iwasawa} \&
  {Taniguchi}}{1993}]{Baldwin}
{Iwasawa} K.,  {Taniguchi} Y.,  1993, \mn@doi [\apjl] {10.1086/186948}, \href
  {https://ui.adsabs.harvard.edu/abs/1993ApJ...413L..15I} {413, L15}

\bibitem[\protect\citeauthoryear{{Kaastra} \& {Bleeker}}{{Kaastra} \&
  {Bleeker}}{2016}]{optbin}
{Kaastra} J.~S.,  {Bleeker} J.~A.~M.,  2016, \mn@doi [AAP]
  {10.1051/0004-6361/201527395}, \href
  {https://ui.adsabs.harvard.edu/abs/2016A&A...587A.151K} {587, A151}

\bibitem[\protect\citeauthoryear{{Kettula}, {Nevalainen}  \&
  {Miller}}{{Kettula} et~al.}{2013}]{Kettula+2013}
{Kettula} K.,  {Nevalainen} J.,   {Miller} E.~D.,  2013, \mn@doi [\aap]
  {10.1051/0004-6361/201220408}, \href
  {https://ui.adsabs.harvard.edu/abs/2013A&A...552A..47K} {552, A47}

\bibitem[\protect\citeauthoryear{{Koss} et~al.,}{{Koss} et~al.}{2017}]{swift70}
{Koss} M.,  et~al., 2017, \mn@doi [\apj] {10.3847/1538-4357/aa8ec9}, \href
  {https://ui.adsabs.harvard.edu/abs/2017ApJ...850...74K} {850, 74}

\bibitem[\protect\citeauthoryear{{La Mura}, {Ciroi}, {Cracco}, {Ilic},
  {Popovic}  \& {Rafanelli}}{{La Mura} et~al.}{2011}]{LaMura+2011}
{La Mura} G.,  {Ciroi} S.,  {Cracco} V.,  {Ilic} D.,  {Popovic} L.,
  {Rafanelli} P.,  2011, in Narrow-Line Seyfert 1 Galaxies and their Place in
  the Universe. p.~56 (\mn@eprint {arXiv} {1106.2454})

\bibitem[\protect\citeauthoryear{{Landt} et~al.,}{{Landt}
  et~al.}{2017}]{Landt+2017}
{Landt} H.,  et~al., 2017, \mn@doi [\mnras] {10.1093/mnras/stw2447}, \href
  {https://ui.adsabs.harvard.edu/abs/2017MNRAS.464.2565L} {464, 2565}

\bibitem[\protect\citeauthoryear{{Leighly}}{{Leighly}}{1999}]{Leighly1999}
{Leighly} K.~M.,  1999, \mn@doi [ApJs] {10.1086/313287}, \href
  {https://ui.adsabs.harvard.edu/abs/1999ApJS..125..317L} {125, 317}

\bibitem[\protect\citeauthoryear{{Liebmann}, {Fabian}, {Tsuruta}, {Haba}  \&
  {Kunieda}}{{Liebmann} et~al.}{2018}]{Liebmann+2018}
{Liebmann} A.~C.,  {Fabian} A.~C.,  {Tsuruta} S.,  {Haba} Y.,   {Kunieda} H.,
  2018, \mn@doi [\apj] {10.3847/1538-4357/aae309}, \href
  {https://ui.adsabs.harvard.edu/abs/2018ApJ...868...11L} {868, 11}

\bibitem[\protect\citeauthoryear{{Lusso} et~al.,}{{Lusso}
  et~al.}{2010}]{Lusso+2010}
{Lusso} E.,  et~al., 2010, \mn@doi [\aap] {10.1051/0004-6361/200913298}, \href
  {https://ui.adsabs.harvard.edu/abs/2010A&A...512A..34L} {512, A34}

\bibitem[\protect\citeauthoryear{{Lusso} et~al.,}{{Lusso}
  et~al.}{2012}]{Lusso+2012}
{Lusso} E.,  et~al., 2012, \mn@doi [\mnras] {10.1111/j.1365-2966.2012.21513.x},
  \href {https://ui.adsabs.harvard.edu/abs/2012MNRAS.425..623L} {425, 623}

\bibitem[\protect\citeauthoryear{{Magdziarz} \& {Zdziarski}}{{Magdziarz} \&
  {Zdziarski}}{1995}]{pexrav}
{Magdziarz} P.,  {Zdziarski} A.~A.,  1995, \mn@doi [\mnras]
  {10.1093/mnras/273.3.837}, \href
  {https://ui.adsabs.harvard.edu/abs/1995MNRAS.273..837M} {273, 837}

\bibitem[\protect\citeauthoryear{{Magdziarz}, {Blaes}, {Zdziarski}, {Johnson}
  \& {Smith}}{{Magdziarz} et~al.}{1998}]{Magdziarz+1998}
{Magdziarz} P.,  {Blaes} O.~M.,  {Zdziarski} A.~A.,  {Johnson} W.~N.,   {Smith}
  D.~A.,  1998, \mn@doi [\mnras] {10.1046/j.1365-8711.1998.02015.x}, \href
  {https://ui.adsabs.harvard.edu/abs/1998MNRAS.301..179M} {301, 179}

\bibitem[\protect\citeauthoryear{{Malizia} et~al.,}{{Malizia}
  et~al.}{2008}]{Malizia+2008}
{Malizia} A.,  et~al., 2008, \mn@doi [MNRAS]
  {10.1111/j.1365-2966.2008.13657.x}, \href
  {https://ui.adsabs.harvard.edu/abs/2008MNRAS.389.1360M} {389, 1360}

\bibitem[\protect\citeauthoryear{{Mallick} \& {Dewangan}}{{Mallick} \&
  {Dewangan}}{2018}]{Mallick+2018}
{Mallick} L.,  {Dewangan} G.~C.,  2018, \mn@doi [\apj]
  {10.3847/1538-4357/aad193}, \href
  {https://ui.adsabs.harvard.edu/abs/2018ApJ...863..178M} {863, 178}

\bibitem[\protect\citeauthoryear{{Mantovani}, {Nandra}  \& {Ponti}}{{Mantovani}
  et~al.}{2016}]{Mantovani+2016}
{Mantovani} G.,  {Nandra} K.,   {Ponti} G.,  2016, \mn@doi [MNRAS]
  {10.1093/mnras/stw596}, \href
  {https://ui.adsabs.harvard.edu/abs/2016MNRAS.458.4198M} {458, 4198}

\bibitem[\protect\citeauthoryear{{Mateos} et~al.,}{{Mateos}
  et~al.}{2010}]{Mateo+2010}
{Mateos} S.,  et~al., 2010, \mn@doi [\aap] {10.1051/0004-6361/200913187}, \href
  {https://ui.adsabs.harvard.edu/abs/2010A&A...510A..35M} {510, A35}

\bibitem[\protect\citeauthoryear{{Mathur}}{{Mathur}}{2000}]{Mathur+2000}
{Mathur} S.,  2000, \mn@doi [\mnras] {10.1046/j.1365-8711.2000.03530.x}, \href
  {https://ui.adsabs.harvard.edu/abs/2000MNRAS.314L..17M} {314, L17}

\bibitem[\protect\citeauthoryear{{Mathur} et~al.,}{{Mathur}
  et~al.}{2018}]{m590}
{Mathur} S.,  et~al., 2018, \mn@doi [ApJ] {10.3847/1538-4357/aadd91}, \href
  {https://ui.adsabs.harvard.edu/abs/2018ApJ...866..123M} {866, 123}

\bibitem[\protect\citeauthoryear{{Middleton}, {Done}  \&
  {Gierli{\'n}ski}}{{Middleton} et~al.}{2007}]{Middleton+2007}
{Middleton} M.,  {Done} C.,   {Gierli{\'n}ski} M.,  2007, \mn@doi [\mnras]
  {10.1111/j.1365-2966.2007.12341.x}, \href
  {https://ui.adsabs.harvard.edu/abs/2007MNRAS.381.1426M} {381, 1426}

\bibitem[\protect\citeauthoryear{{Miniutti}, {Piconcelli}, {Bianchi}, {Vignali}
   \& {Bozzo}}{{Miniutti} et~al.}{2010}]{rbs1124}
{Miniutti} G.,  {Piconcelli} E.,  {Bianchi} S.,  {Vignali} C.,   {Bozzo} E.,
  2010, \mn@doi [\mnras] {10.1111/j.1365-2966.2009.15726.x}, \href
  {https://ui.adsabs.harvard.edu/abs/2010MNRAS.401.1315M} {401, 1315}

\bibitem[\protect\citeauthoryear{{Miniutti} et~al.,}{{Miniutti}
  et~al.}{2014}]{Miniutti+2014}
{Miniutti} G.,  et~al., 2014, \mn@doi [\mnras] {10.1093/mnras/stt2005}, \href
  {https://ui.adsabs.harvard.edu/abs/2014MNRAS.437.1776M} {437, 1776}

\bibitem[\protect\citeauthoryear{{Mitsuda} et~al.,}{{Mitsuda}
  et~al.}{2007}]{suzaku}
{Mitsuda} K.,  et~al., 2007, \mn@doi [PASJ] {10.1093/pasj/59.sp1.S1}, \href
  {https://ui.adsabs.harvard.edu/abs/2007PASJ...59S...1M} {59, S1}

\bibitem[\protect\citeauthoryear{{Miyaji}, {Ishisaki}, {Ueda}, {Ogasaka},
  {Awaki}  \& {Hayashida}}{{Miyaji} et~al.}{2003}]{kaz102}
{Miyaji} T.,  {Ishisaki} Y.,  {Ueda} Y.,  {Ogasaka} Y.,  {Awaki} H.,
  {Hayashida} K.,  2003, \mn@doi [\pasj] {10.1093/pasj/55.2.L11}, \href
  {https://ui.adsabs.harvard.edu/abs/2003PASJ...55L..11M} {55, L11}

\bibitem[\protect\citeauthoryear{{Miyazawa}, {Haba}  \& {Kunieda}}{{Miyazawa}
  et~al.}{2009}]{Miyazawa+2009}
{Miyazawa} T.,  {Haba} Y.,   {Kunieda} H.,  2009, \mn@doi [PASJ]
  {10.1093/pasj/61.6.1331}, \href
  {https://ui.adsabs.harvard.edu/abs/2009PASJ...61.1331M} {61, 1331}

\bibitem[\protect\citeauthoryear{{Nandra} \& {Pounds}}{{Nandra} \&
  {Pounds}}{1994}]{Nandra+1994}
{Nandra} K.,  {Pounds} K.~A.,  1994, \mn@doi [\mnras]
  {10.1093/mnras/268.2.405}, \href
  {https://ui.adsabs.harvard.edu/abs/1994MNRAS.268..405N} {268, 405}

\bibitem[\protect\citeauthoryear{{Nandra}, {O'Neill}, {George}  \&
  {Reeves}}{{Nandra} et~al.}{2007}]{Nandra+2007}
{Nandra} K.,  {O'Neill} P.~M.,  {George} I.~M.,   {Reeves} J.~N.,  2007,
  \mn@doi [MNRAS] {10.1111/j.1365-2966.2007.12331.x}, \href
  {https://ui.adsabs.harvard.edu/abs/2007MNRAS.382..194N} {382, 194}

\bibitem[\protect\citeauthoryear{{Nelson} \& {Whittle}}{{Nelson} \&
  {Whittle}}{1995}]{ngc7213}
{Nelson} C.~H.,  {Whittle} M.,  1995, \mn@doi [\apjs] {10.1086/192179}, \href
  {https://ui.adsabs.harvard.edu/abs/1995ApJS...99...67N} {99, 67}

\bibitem[\protect\citeauthoryear{{Niko{\l}ajuk}, {Czerny}  \&
  {Gurynowicz}}{{Niko{\l}ajuk} et~al.}{2009}]{Nik+2009}
{Niko{\l}ajuk} M.,  {Czerny} B.,   {Gurynowicz} P.,  2009, \mn@doi [MNRAS]
  {10.1111/j.1365-2966.2009.14478.x}, \href
  {https://ui.adsabs.harvard.edu/abs/2009MNRAS.394.2141N} {394, 2141}

\bibitem[\protect\citeauthoryear{{Noda} \& {Done}}{{Noda} \&
  {Done}}{2018}]{m1018}
{Noda} H.,  {Done} C.,  2018, \mn@doi [\mnras] {10.1093/mnras/sty2032}, \href
  {https://ui.adsabs.harvard.edu/abs/2018MNRAS.480.3898N} {480, 3898}

\bibitem[\protect\citeauthoryear{{Noda}, {Makishima}, {Nakazawa}, {Uchiyama},
  {Yamada}  \& {Sakurai}}{{Noda} et~al.}{2013}]{Noda+2013}
{Noda} H.,  {Makishima} K.,  {Nakazawa} K.,  {Uchiyama} H.,  {Yamada} S.,
  {Sakurai} S.,  2013, \mn@doi [\pasj] {10.1093/pasj/65.1.4}, \href
  {https://ui.adsabs.harvard.edu/abs/2013PASJ...65....4N} {65, 4}

\bibitem[\protect\citeauthoryear{{Nowak} et~al.,}{{Nowak}
  et~al.}{2011}]{Nowak+2011}
{Nowak} M.~A.,  et~al., 2011, \mn@doi [ApJ] {10.1088/0004-637X/728/1/13}, \href
  {https://ui.adsabs.harvard.edu/abs/2011ApJ...728...13N} {728, 13}

\bibitem[\protect\citeauthoryear{{Ojha}, {Chand}, {Dewangan}  \&
  {Rakshit}}{{Ojha} et~al.}{2020}]{Ohja+2020}
{Ojha} V.,  {Chand} H.,  {Dewangan} G.~C.,   {Rakshit} S.,  2020, arXiv
  e-prints, \href {https://ui.adsabs.harvard.edu/abs/2020arXiv200508352O} {p.
  arXiv:2005.08352}

\bibitem[\protect\citeauthoryear{{Osterbrock} \& {Pogge}}{{Osterbrock} \&
  {Pogge}}{1985}]{Osterbrock+1985}
{Osterbrock} D.~E.,  {Pogge} R.~W.,  1985, \mn@doi [ApJ] {10.1086/163513},
  \href {https://ui.adsabs.harvard.edu/abs/1985ApJ...297..166O} {297, 166}

\bibitem[\protect\citeauthoryear{{Paliya}, {Sahayanathan}, {Parker}, {Fabian},
  {Stalin}, {Anjum}  \& {Pand ey}}{{Paliya} et~al.}{2014}]{Paliya+2014}
{Paliya} V.~S.,  {Sahayanathan} S.,  {Parker} M.~L.,  {Fabian} A.~C.,  {Stalin}
  C.~S.,  {Anjum} A.,   {Pand ey} S.~B.,  2014, \mn@doi [\apj]
  {10.1088/0004-637X/789/2/143}, \href
  {https://ui.adsabs.harvard.edu/abs/2014ApJ...789..143P} {789, 143}

\bibitem[\protect\citeauthoryear{{Petrucci}, {Ursini}, {De Rosa}, {Bianchi},
  {Cappi}, {Matt}, {Dadina}  \& {Malzac}}{{Petrucci}
  et~al.}{2018}]{Petrucci+2018}
{Petrucci} P.~O.,  {Ursini} F.,  {De Rosa} A.,  {Bianchi} S.,  {Cappi} M.,
  {Matt} G.,  {Dadina} M.,   {Malzac} J.,  2018, \mn@doi [\aap]
  {10.1051/0004-6361/201731580}, \href
  {https://ui.adsabs.harvard.edu/abs/2018A&A...611A..59P} {611, A59}

\bibitem[\protect\citeauthoryear{{Ponti}, {Papadakis}, {Bianchi}, {Guainazzi},
  {Matt}, {Uttley}  \& {Bonilla}}{{Ponti} et~al.}{2012}]{Ponti+2012}
{Ponti} G.,  {Papadakis} I.,  {Bianchi} S.,  {Guainazzi} M.,  {Matt} G.,
  {Uttley} P.,   {Bonilla} N.~F.,  2012, \mn@doi [\aap]
  {10.1051/0004-6361/201118326}, \href
  {https://ui.adsabs.harvard.edu/abs/2012A&A...542A..83P} {542, A83}

\bibitem[\protect\citeauthoryear{{Porquet}, {Reeves}, {O'Brien}  \&
  {Brinkmann}}{{Porquet} et~al.}{2004}]{Porquet+2004}
{Porquet} D.,  {Reeves} J.~N.,  {O'Brien} P.,   {Brinkmann} W.,  2004, \mn@doi
  [AAP] {10.1051/0004-6361:20047108}, \href
  {https://ui.adsabs.harvard.edu/abs/2004A&A...422...85P} {422, 85}

\bibitem[\protect\citeauthoryear{{Pounds}, {Nandra}, {Stewart}, {George}  \&
  {Fabian}}{{Pounds} et~al.}{1990}]{Pounds+1990}
{Pounds} K.~A.,  {Nandra} K.,  {Stewart} G.~C.,  {George} I.~M.,   {Fabian}
  A.~C.,  1990, \mn@doi [\nat] {10.1038/344132a0}, \href
  {https://ui.adsabs.harvard.edu/abs/1990Natur.344..132P} {344, 132}

\bibitem[\protect\citeauthoryear{{Pounds}, {Done}  \& {Osborne}}{{Pounds}
  et~al.}{1995}]{Pounds+1995}
{Pounds} K.~A.,  {Done} C.,   {Osborne} J.~P.,  1995, \mn@doi [MNRAS]
  {10.1093/mnras/277.1.L5}, \href
  {https://ui.adsabs.harvard.edu/abs/1995MNRAS.277L...5P} {277, L5}

\bibitem[\protect\citeauthoryear{{Reeves} \& {Turner}}{{Reeves} \&
  {Turner}}{2000}]{Reeves+2000}
{Reeves} J.~N.,  {Turner} M.~J.~L.,  2000, \mn@doi [\mnras]
  {10.1046/j.1365-8711.2000.03510.x}, \href
  {https://ui.adsabs.harvard.edu/abs/2000MNRAS.316..234R} {316, 234}

\bibitem[\protect\citeauthoryear{{Reeves} et~al.,}{{Reeves}
  et~al.}{2006}]{Reeves+2006}
{Reeves} J.~N.,  et~al., 2006, \mn@doi [Astronomische Nachrichten]
  {10.1002/asna.200610696}, \href
  {https://ui.adsabs.harvard.edu/abs/2006AN....327.1079R} {327, 1079}

\bibitem[\protect\citeauthoryear{{Rivers}, {Markowitz}  \&
  {Rothschild}}{{Rivers} et~al.}{2011}]{Rivers+2011}
{Rivers} E.,  {Markowitz} A.,   {Rothschild} R.,  2011, \mn@doi [\apj]
  {10.1088/0004-637X/732/1/36}, \href
  {https://ui.adsabs.harvard.edu/abs/2011ApJ...732...36R} {732, 36}

\bibitem[\protect\citeauthoryear{{Ross} \& {Fabian}}{{Ross} \&
  {Fabian}}{2005}]{RossFabian+2005}
{Ross} R.~R.,  {Fabian} A.~C.,  2005, \mn@doi [MNRAS]
  {10.1111/j.1365-2966.2005.08797.x}, \href
  {https://ui.adsabs.harvard.edu/abs/2005MNRAS.358..211R} {358, 211}

\bibitem[\protect\citeauthoryear{{Schmid}, {Appenzeller}  \& {Burch}}{{Schmid}
  et~al.}{2003}]{Schmid+2003}
{Schmid} H.~M.,  {Appenzeller} I.,   {Burch} U.,  2003, \mn@doi [\aap]
  {10.1051/0004-6361:20030558}, \href
  {https://ui.adsabs.harvard.edu/abs/2003A&A...404..505S} {404, 505}

\bibitem[\protect\citeauthoryear{{Shemmer}, {Brandt}, {Netzer}, {Maiolino}  \&
  {Kaspi}}{{Shemmer} et~al.}{2008}]{Shemmer+2008}
{Shemmer} O.,  {Brandt} W.~N.,  {Netzer} H.,  {Maiolino} R.,   {Kaspi} S.,
  2008, \mn@doi [\apj] {10.1086/588776}, \href
  {https://ui.adsabs.harvard.edu/abs/2008ApJ...682...81S} {682, 81}

\bibitem[\protect\citeauthoryear{{Soldi} et~al.,}{{Soldi}
  et~al.}{2014}]{Soldi+2014}
{Soldi} S.,  et~al., 2014, \mn@doi [AAP] {10.1051/0004-6361/201322653}, \href
  {https://ui.adsabs.harvard.edu/abs/2014A&A...563A..57S} {563, A57}

\bibitem[\protect\citeauthoryear{{Tanaka}, {Boller}, {Gallo}, {Keil}  \&
  {Ueda}}{{Tanaka} et~al.}{2004}]{Tanaka+2004}
{Tanaka} Y.,  {Boller} T.,  {Gallo} L.,  {Keil} R.,   {Ueda} Y.,  2004, \mn@doi
  [PASJ] {10.1093/pasj/56.3.L9}, \href
  {https://ui.adsabs.harvard.edu/abs/2004PASJ...56L...9T} {56, L9}

\bibitem[\protect\citeauthoryear{{Tawa} et~al.,}{{Tawa} et~al.}{2008}]{suzbkg}
{Tawa} N.,  et~al., 2008, \mn@doi [\pasj] {10.1093/pasj/60.sp1.S11}, \href
  {https://ui.adsabs.harvard.edu/abs/2008PASJ...60S..11T} {60, S11}

\bibitem[\protect\citeauthoryear{{Tortosa}}{{Tortosa}}{2017}]{mcg08}
{Tortosa} A.,  2017, in {Ness} J.-U.,  {Migliari} S.,  eds, The X-ray Universe
  2017. p.~225

\bibitem[\protect\citeauthoryear{{Tripathi}, {Waddell}, {Gallo}, {Welsh}  \&
  {Chiang}}{{Tripathi} et~al.}{2019}]{zw229}
{Tripathi} S.,  {Waddell} S.~G.~H.,  {Gallo} L.~C.,  {Welsh} W.~F.,   {Chiang}
  C.~Y.,  2019, \mn@doi [\mnras] {10.1093/mnras/stz1988}, \href
  {https://ui.adsabs.harvard.edu/abs/2019MNRAS.488.4831T} {488, 4831}

\bibitem[\protect\citeauthoryear{{Vasudevan} \& {Fabian}}{{Vasudevan} \&
  {Fabian}}{2007}]{pg16}
{Vasudevan} R.~V.,  {Fabian} A.~C.,  2007, \mn@doi [\mnras]
  {10.1111/j.1365-2966.2007.12328.x}, \href
  {https://ui.adsabs.harvard.edu/abs/2007MNRAS.381.1235V} {381, 1235}

\bibitem[\protect\citeauthoryear{{Vasudevan} \& {Fabian}}{{Vasudevan} \&
  {Fabian}}{2009}]{Vasudevan+2009}
{Vasudevan} R.~V.,  {Fabian} A.~C.,  2009, \mn@doi [MNRAS]
  {10.1111/j.1365-2966.2008.14108.x}, \href
  {https://ui.adsabs.harvard.edu/abs/2009MNRAS.392.1124V} {392, 1124}

\bibitem[\protect\citeauthoryear{{Vasudevan}, {Mushotzky}, {Winter}  \&
  {Fabian}}{{Vasudevan} et~al.}{2009}]{swift9}
{Vasudevan} R.~V.,  {Mushotzky} R.~F.,  {Winter} L.~M.,   {Fabian} A.~C.,
  2009, \mn@doi [\mnras] {10.1111/j.1365-2966.2009.15371.x}, \href
  {https://ui.adsabs.harvard.edu/abs/2009MNRAS.399.1553V} {399, 1553}

\bibitem[\protect\citeauthoryear{{Vasudevan}, {Mushotzky}, {Reynolds},
  {Fabian}, {Lohfink}, {Zoghbi}, {Gallo}  \& {Walton}}{{Vasudevan}
  et~al.}{2014}]{Vasudevan+2014}
{Vasudevan} R.~V.,  {Mushotzky} R.~F.,  {Reynolds} C.~S.,  {Fabian} A.~C.,
  {Lohfink} A.~M.,  {Zoghbi} A.,  {Gallo} L.~C.,   {Walton} D.,  2014, \mn@doi
  [\apj] {10.1088/0004-637X/785/1/30}, \href
  {https://ui.adsabs.harvard.edu/abs/2014ApJ...785...30V} {785, 30}

\bibitem[\protect\citeauthoryear{{Vaughan}, {Reeves}, {Warwick}  \&
  {Edelson}}{{Vaughan} et~al.}{1999}]{Vaughan+1999}
{Vaughan} S.,  {Reeves} J.,  {Warwick} R.,   {Edelson} R.,  1999, \mn@doi
  [MNRAS] {10.1046/j.1365-8711.1999.02811.x}, \href
  {https://ui.adsabs.harvard.edu/abs/1999MNRAS.309..113V} {309, 113}

\bibitem[\protect\citeauthoryear{{Vestergaard}}{{Vestergaard}}{2002}]{m841}
{Vestergaard} M.,  2002, \mn@doi [\apj] {10.1086/340045}, \href
  {https://ui.adsabs.harvard.edu/abs/2002ApJ...571..733V} {571, 733}

\bibitem[\protect\citeauthoryear{{Walter} \& {Fink}}{{Walter} \&
  {Fink}}{1993}]{Walter+1993}
{Walter} R.,  {Fink} H.~H.,  1993, \aap, \href
  {https://ui.adsabs.harvard.edu/abs/1993A&A...274..105W} {274, 105}

\bibitem[\protect\citeauthoryear{{Walton}, {Nardini}, {Fabian}, {Gallo}  \&
  {Reis}}{{Walton} et~al.}{2013a}]{Walton+2013}
{Walton} D.~J.,  {Nardini} E.,  {Fabian} A.~C.,  {Gallo} L.~C.,   {Reis} R.~C.,
   2013a, \mn@doi [MNRAS] {10.1093/mnras/sts227}, \href
  {https://ui.adsabs.harvard.edu/abs/2013MNRAS.428.2901W} {428, 2901}

\bibitem[\protect\citeauthoryear{{Walton} et~al.,}{{Walton}
  et~al.}{2013b}]{ngc6814}
{Walton} D.~J.,  et~al., 2013b, \mn@doi [\apjl] {10.1088/2041-8205/777/2/L23},
  \href {https://ui.adsabs.harvard.edu/abs/2013ApJ...777L..23W} {777, L23}

\bibitem[\protect\citeauthoryear{{Wang} \& {Lu}}{{Wang} \&
  {Lu}}{2001}]{Wang+2001}
{Wang} T.,  {Lu} Y.,  2001, \mn@doi [\aap] {10.1051/0004-6361:20011071}, \href
  {https://ui.adsabs.harvard.edu/abs/2001A&A...377...52W} {377, 52}

\bibitem[\protect\citeauthoryear{{Wang}, {Brinkmann}  \& {Bergeron}}{{Wang}
  et~al.}{1996}]{Wang+1996}
{Wang} T.,  {Brinkmann} W.,   {Bergeron} J.,  1996, \aap, \href
  {https://ui.adsabs.harvard.edu/abs/1996A&A...309...81W} {309, 81}

\bibitem[\protect\citeauthoryear{{Wang}, {Mao}  \& {Wei}}{{Wang}
  et~al.}{2009}]{Wang+2009}
{Wang} J.,  {Mao} Y.~F.,   {Wei} J.~Y.,  2009, \mn@doi [\aj]
  {10.1088/0004-6256/137/2/3388}, \href
  {https://ui.adsabs.harvard.edu/abs/2009AJ....137.3388W} {137, 3388}

\bibitem[\protect\citeauthoryear{{Wilkins}, {Gallo}, {Grupe}, {Bonson},
  {Komossa}  \& {Fabian}}{{Wilkins} et~al.}{2015}]{Wilkins+2015}
{Wilkins} D.~R.,  {Gallo} L.~C.,  {Grupe} D.,  {Bonson} K.,  {Komossa} S.,
  {Fabian} A.~C.,  2015, \mn@doi [MNRAS] {10.1093/mnras/stv2130}, \href
  {https://ui.adsabs.harvard.edu/abs/2015MNRAS.454.4440W} {454, 4440}

\bibitem[\protect\citeauthoryear{{Williams}, {Pogge}  \& {Mathur}}{{Williams}
  et~al.}{2002}]{Williams+2002}
{Williams} R.~J.,  {Pogge} R.~W.,   {Mathur} S.,  2002, \mn@doi [\aj]
  {10.1086/344765}, \href
  {https://ui.adsabs.harvard.edu/abs/2002AJ....124.3042W} {124, 3042}

\bibitem[\protect\citeauthoryear{{Willingale}, {Starling}, {Beardmore},
  {Tanvir}  \& {O'Brien}}{{Willingale} et~al.}{2013}]{Willingale+2013}
{Willingale} R.,  {Starling} R.~L.~C.,  {Beardmore} A.~P.,  {Tanvir} N.~R.,
  {O'Brien} P.~T.,  2013, \mn@doi [MNRAS] {10.1093/mnras/stt175}, \href
  {https://ui.adsabs.harvard.edu/abs/2013MNRAS.431..394W} {431, 394}

\bibitem[\protect\citeauthoryear{{Wilms}, {Allen}  \& {McCray}}{{Wilms}
  et~al.}{2000}]{Wilms+2000}
{Wilms} J.,  {Allen} A.,   {McCray} R.,  2000, \mn@doi [ApJ] {10.1086/317016},
  \href {https://ui.adsabs.harvard.edu/abs/2000ApJ...542..914W} {542, 914}

\bibitem[\protect\citeauthoryear{{Winter}, {Lewis}, {Koss}, {Veilleux},
  {Keeney}  \& {Mushotzky}}{{Winter} et~al.}{2010}]{Winter+2010}
{Winter} L.~M.,  {Lewis} K.~T.,  {Koss} M.,  {Veilleux} S.,  {Keeney} B.,
  {Mushotzky} R.~F.,  2010, \mn@doi [\apj] {10.1088/0004-637X/710/1/503}, \href
  {https://ui.adsabs.harvard.edu/abs/2010ApJ...710..503W} {710, 503}

\bibitem[\protect\citeauthoryear{{Woo} \& {Urry}}{{Woo} \&
  {Urry}}{2002}]{Woo+2002}
{Woo} J.-H.,  {Urry} C.~M.,  2002, \mn@doi [\apj] {10.1086/342878}, \href
  {https://ui.adsabs.harvard.edu/abs/2002ApJ...579..530W} {579, 530}

\bibitem[\protect\citeauthoryear{{Woo}, {Kim}, {Imanishi}  \& {Park}}{{Woo}
  et~al.}{2012}]{m530}
{Woo} J.-H.,  {Kim} J.~H.,  {Imanishi} M.,   {Park} D.,  2012, \mn@doi [\aj]
  {10.1088/0004-6256/143/2/49}, \href
  {https://ui.adsabs.harvard.edu/abs/2012AJ....143...49W} {143, 49}

\bibitem[\protect\citeauthoryear{{You}, {Cao}  \& {Yuan}}{{You}
  et~al.}{2016}]{pg13}
{You} B.,  {Cao} X.-W.,   {Yuan} Y.-F.,  2016, \mn@doi [Research in Astronomy
  and Astrophysics] {10.1088/1674-4527/16/4/055}, \href
  {https://ui.adsabs.harvard.edu/abs/2016RAA....16...55Y} {16, 55}

\bibitem[\protect\citeauthoryear{{Yuan}, {Liu}, {Zhou}  \& {Wang}}{{Yuan}
  et~al.}{2010}]{Yuan+2010}
{Yuan} W.,  {Liu} B.~F.,  {Zhou} H.,   {Wang} T.~G.,  2010, \mn@doi [\apj]
  {10.1088/0004-637X/723/1/508}, \href
  {https://ui.adsabs.harvard.edu/abs/2010ApJ...723..508Y} {723, 508}

\makeatother
\end{thebibliography}

\newpage
\appendix

\section{Full sample spectra}

\onecolumn

\begin{figure*}
	\centering
	\includegraphics[width=75mm]{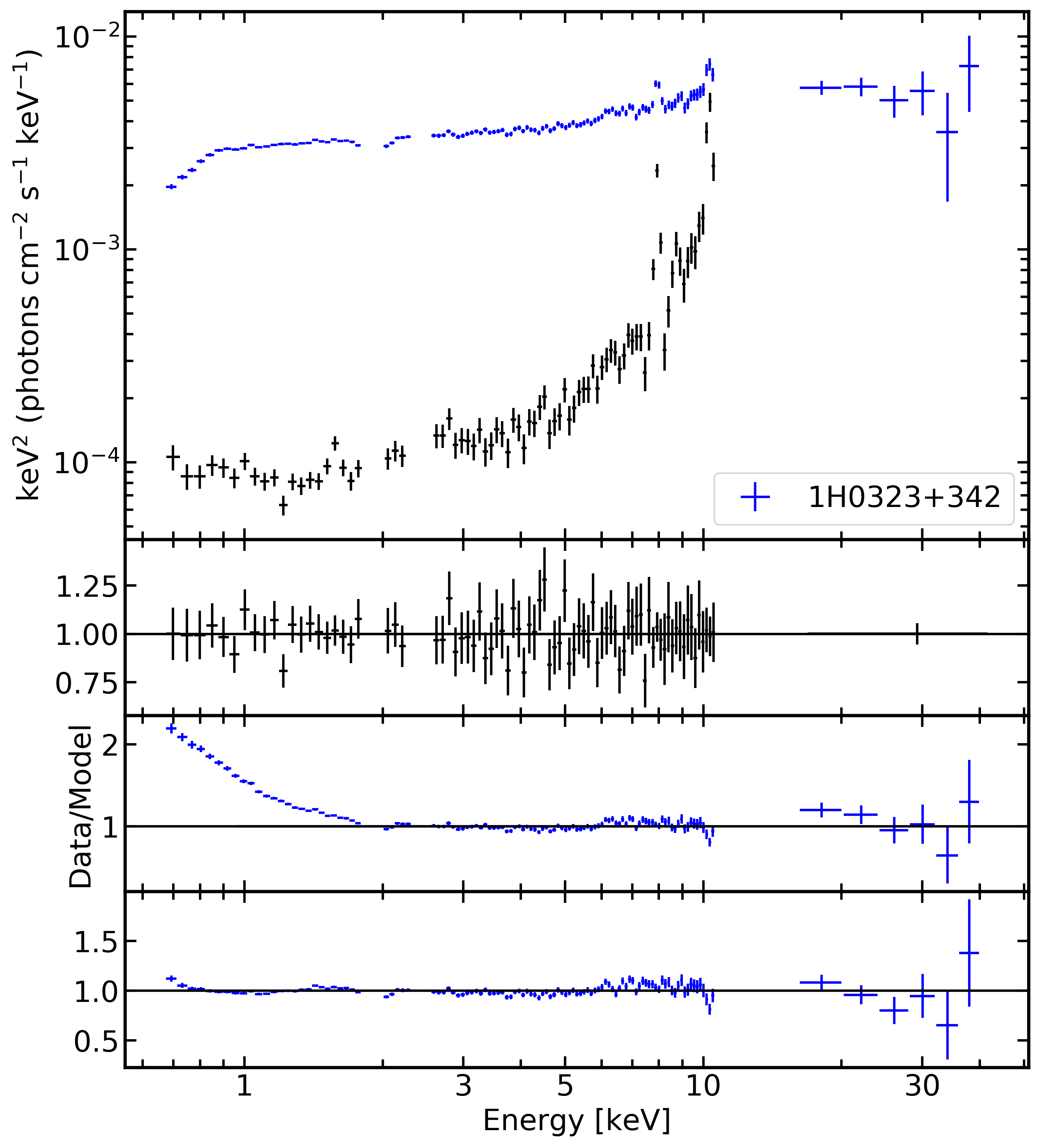}
	\hspace{7mm}
	\includegraphics[width=75mm]{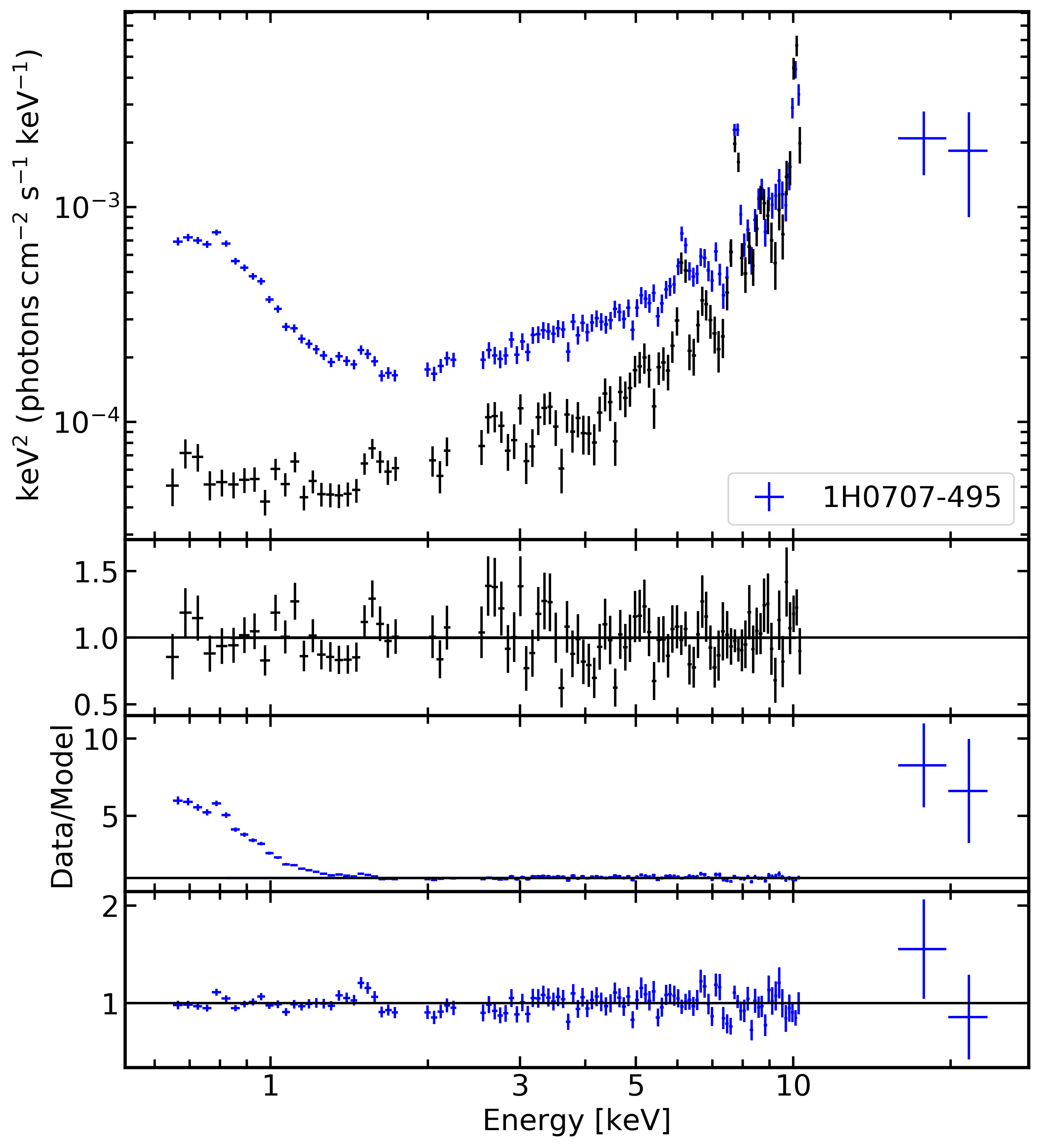}
	\hspace{0mm}
	\includegraphics[width=75mm]{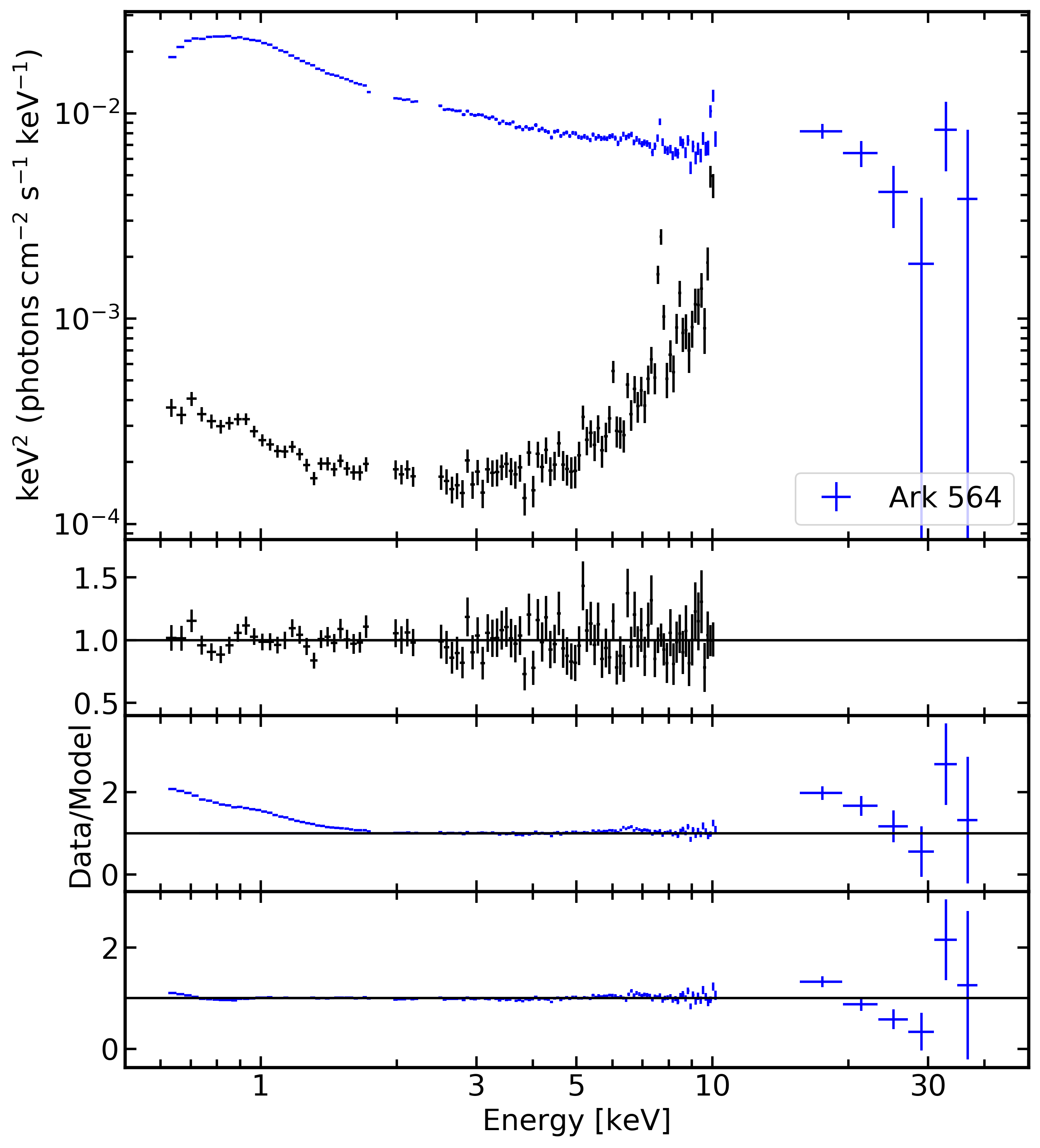}
	\hspace{7mm}
	\includegraphics[width=75mm]{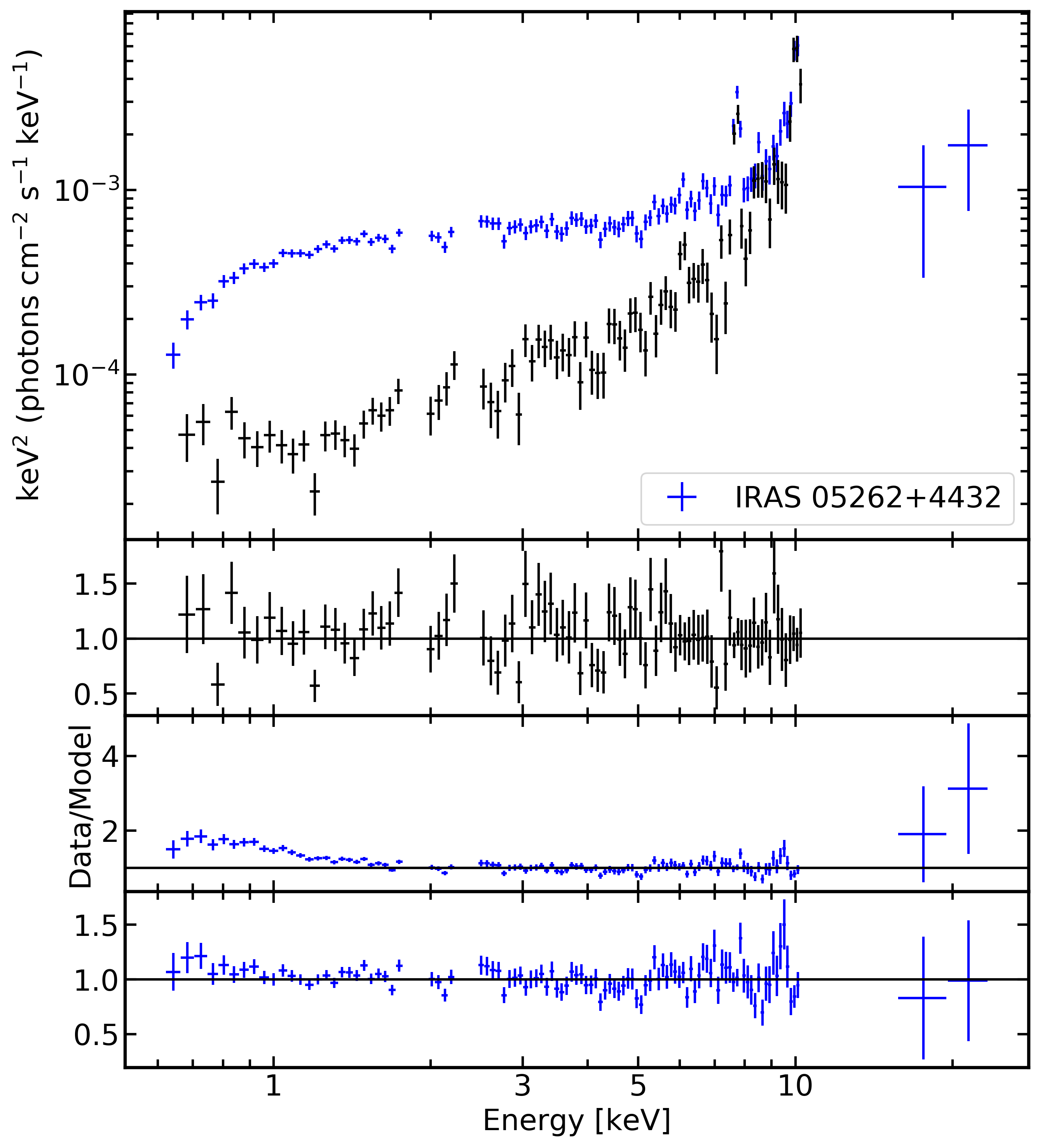}
	\caption{\label{fig:po} Modelling procedures shown for each AGN in the sample. NLS1 sources are shown in blue, and BLS1s are shown in red The top panel shows the source+background XIS and PIN data (blue/red) and background (black) unfolded against a power law with $\Gamma=0$. Object names are given in the bottom right corner. The second panel shows the residuals (data/model) for the background spectrum. The third panel shows the residuals for a power law fit from $2-4$ and $7-10~{\rm keV}$, extrapolated to the full energy range. The last panel shows the residuals for the broad band model. PIN data have been rebinned for clarity.}
\end{figure*}

\setcounter{figure}{0}

\begin{figure*}
	\centering
	\includegraphics[width=75mm]{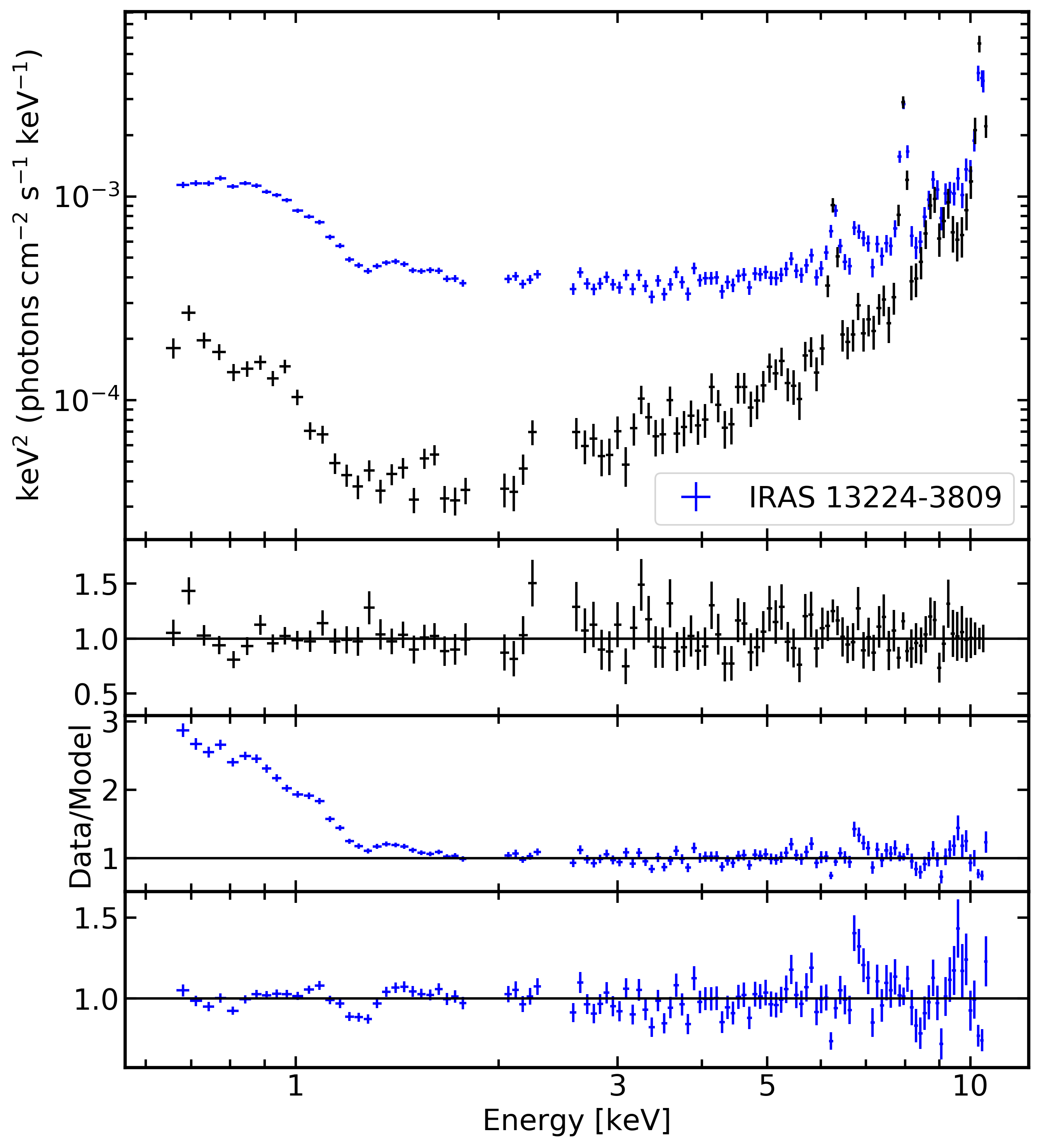}
	\hspace{7mm}
	\includegraphics[width=75mm]{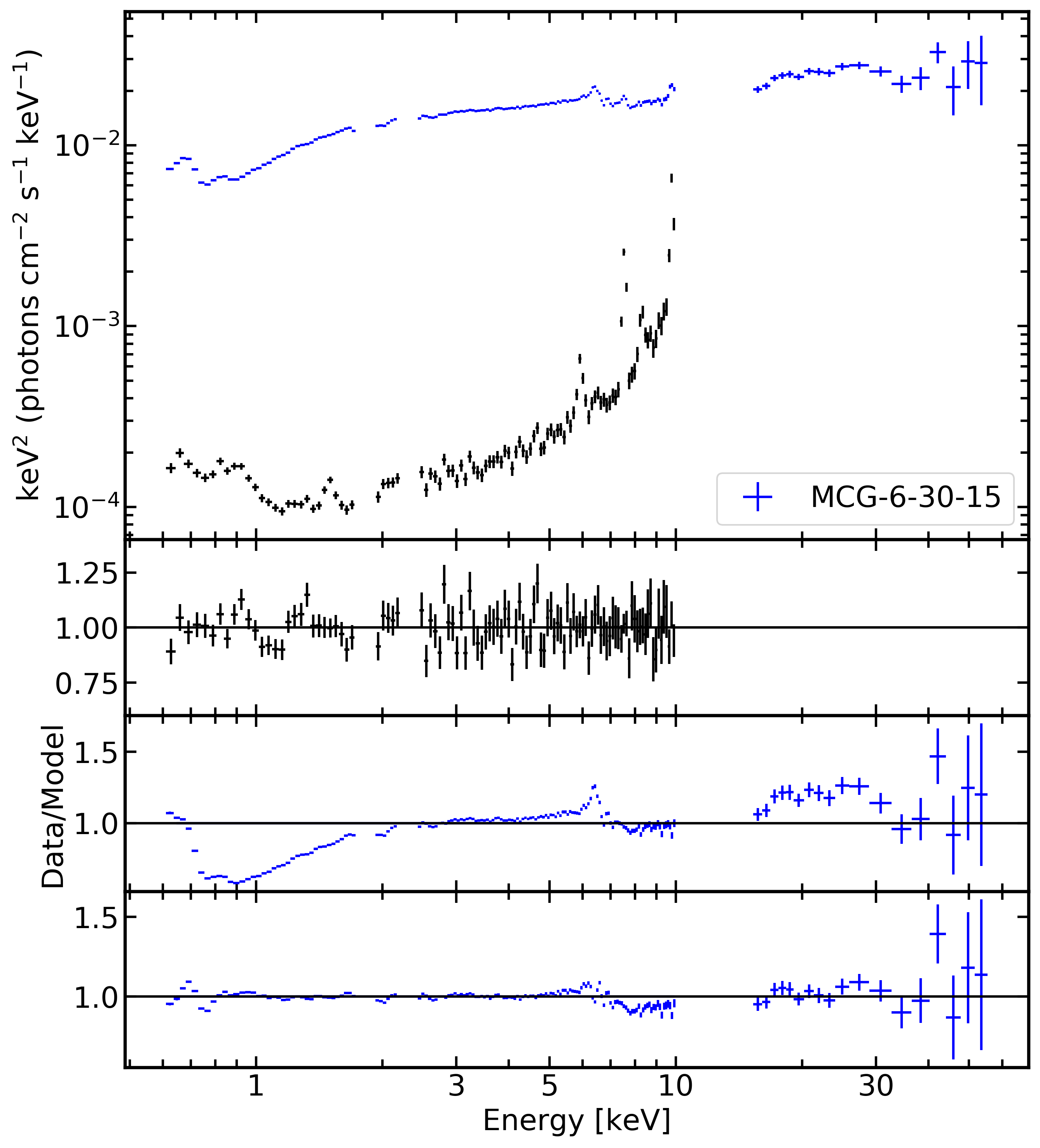}
	\hspace{0mm}
	\includegraphics[width=75mm]{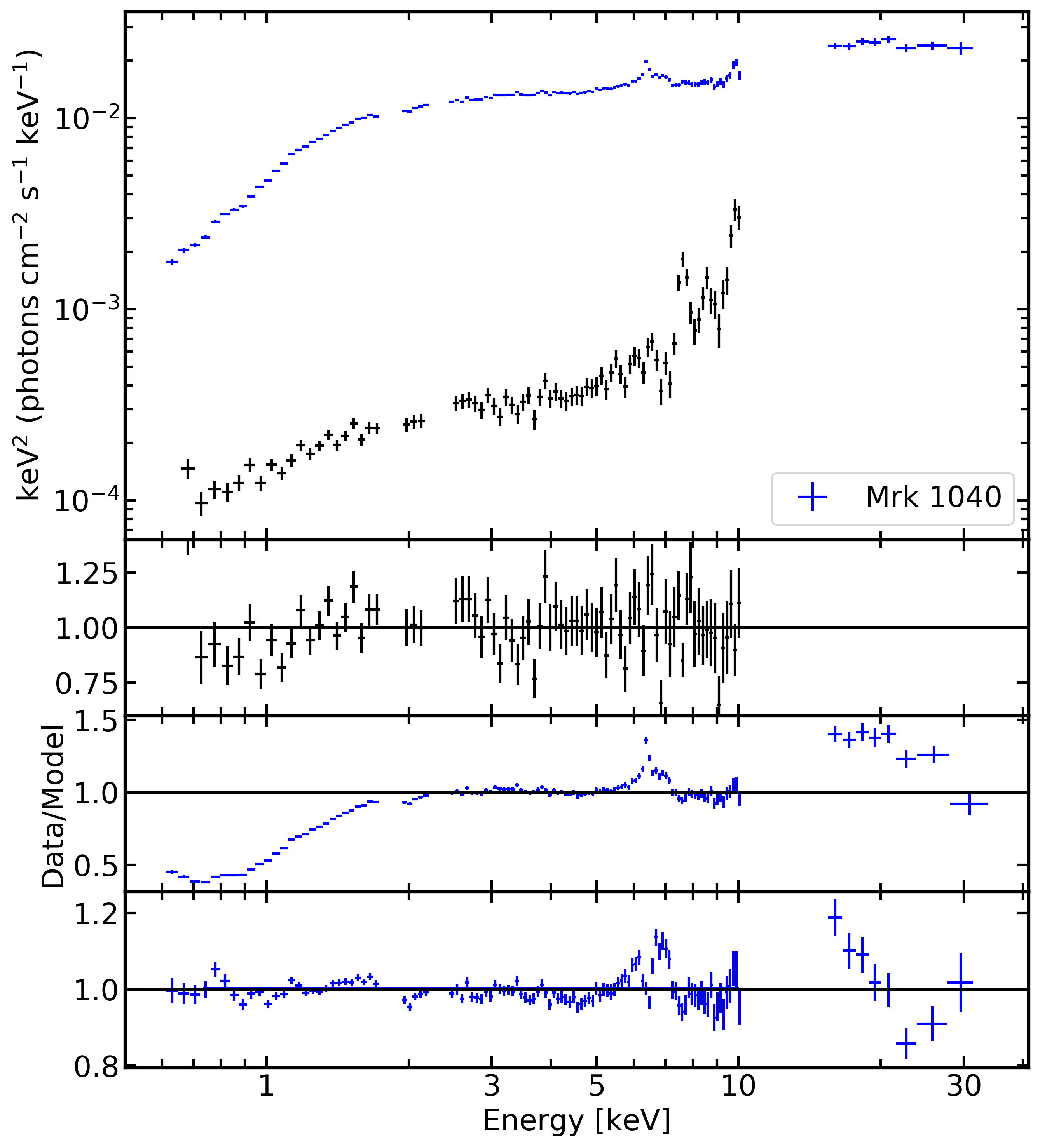}
	\hspace{7mm}
	\includegraphics[width=75mm]{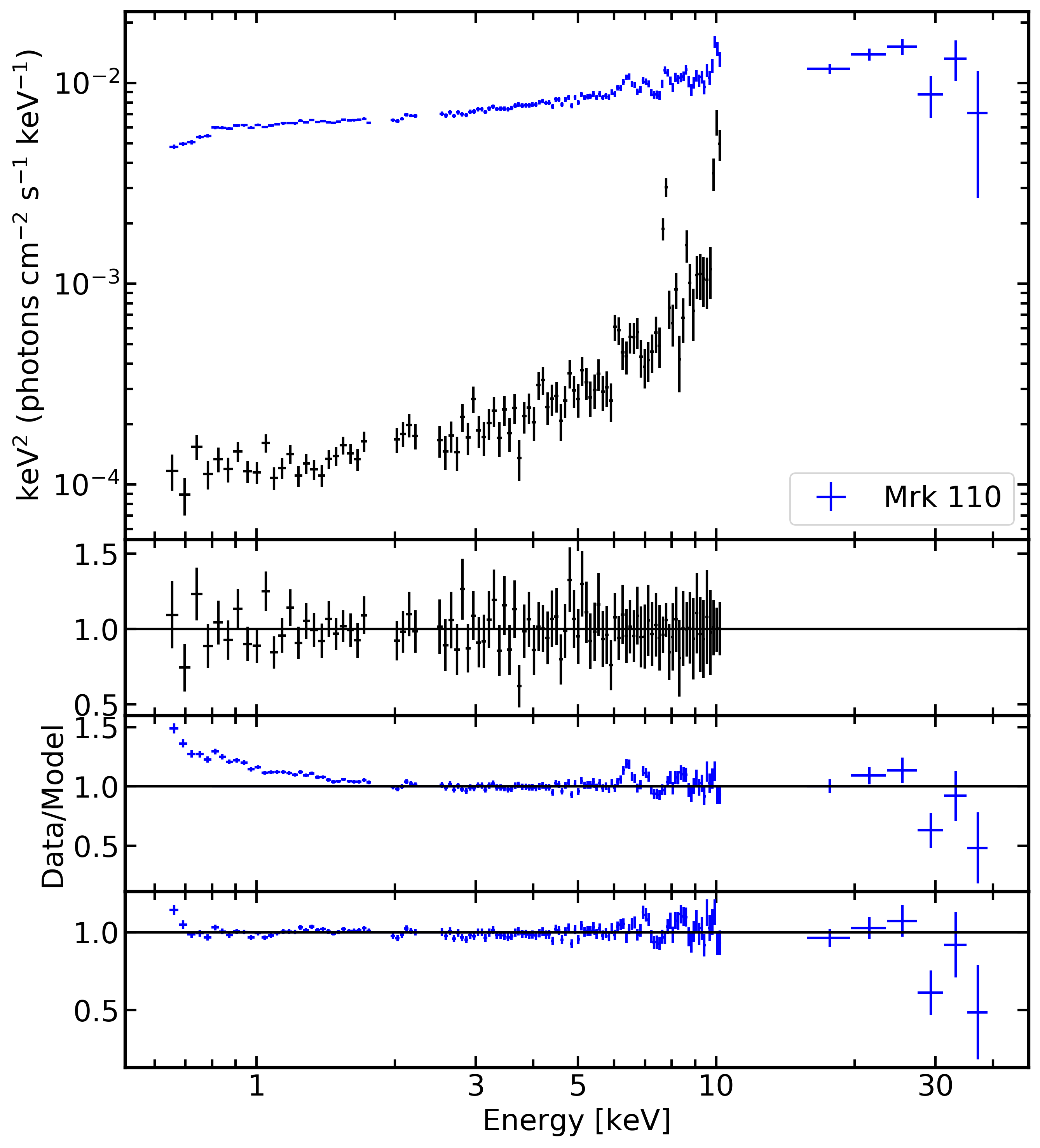}
	\caption{\label{fig:po}continued}
\end{figure*}

\setcounter{figure}{0}

\begin{figure*}
	\centering
	\includegraphics[width=75mm]{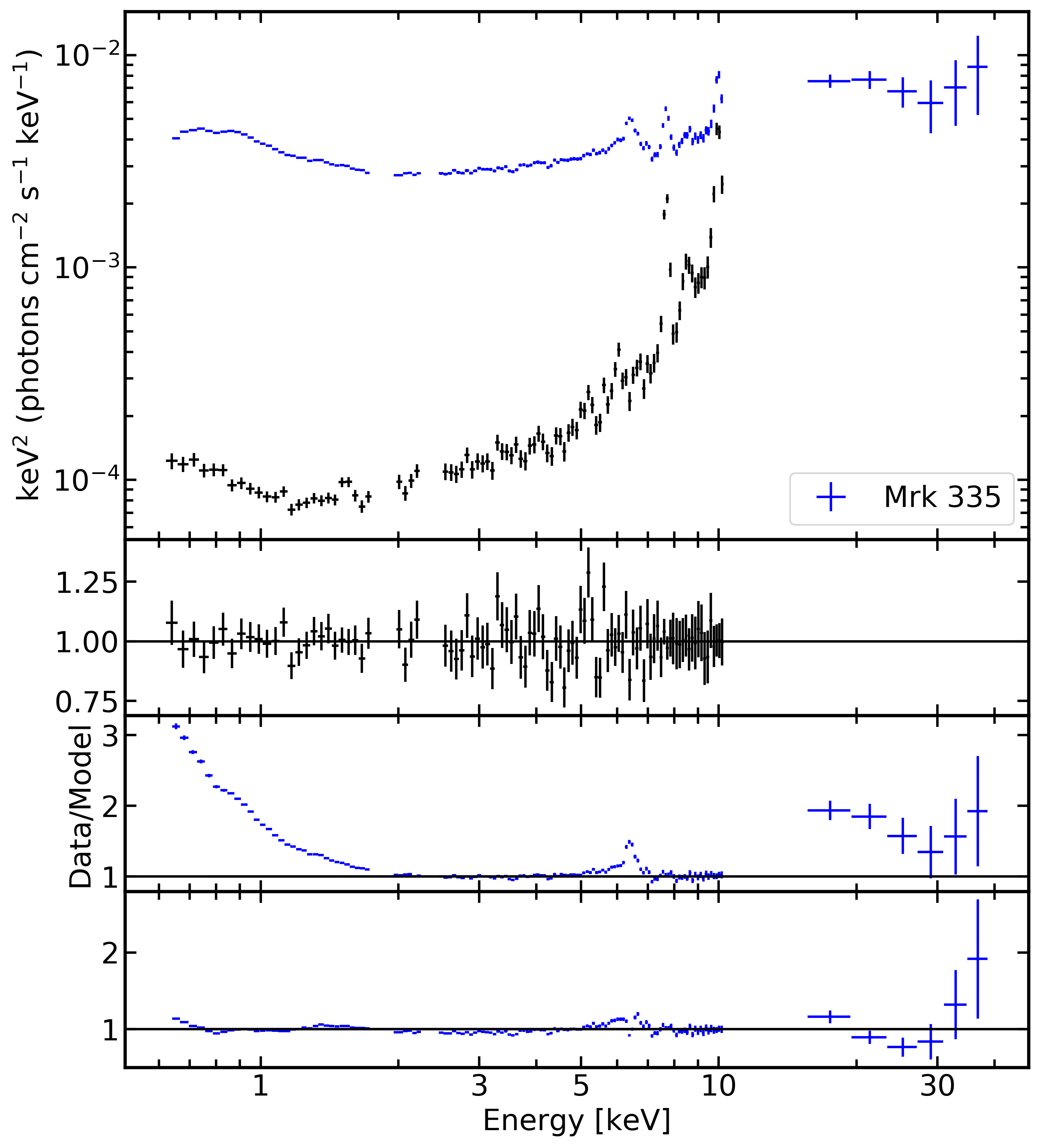}
	\hspace{7mm}
	\includegraphics[width=75mm]{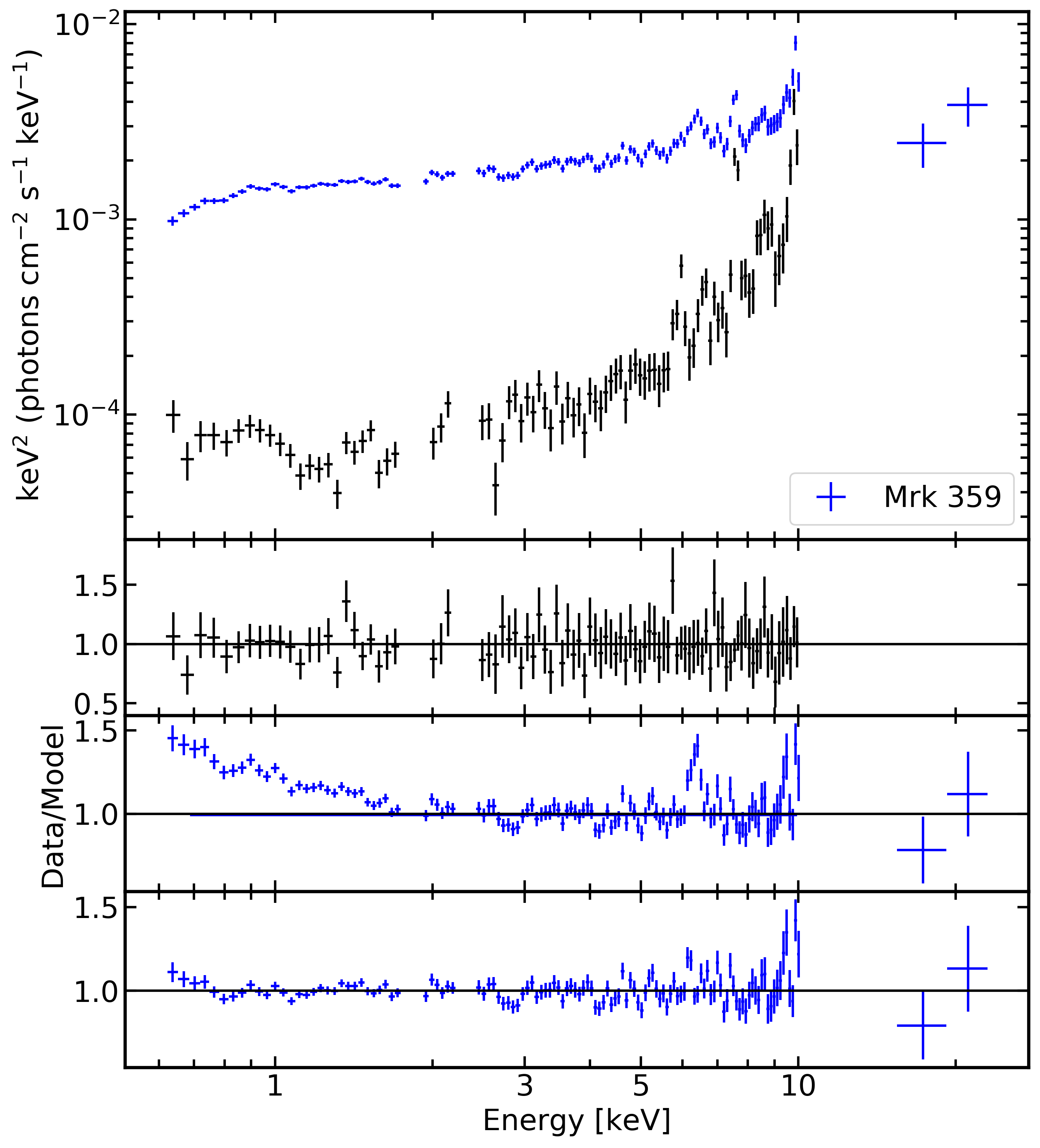}
	\hspace{0mm}
	\includegraphics[width=75mm]{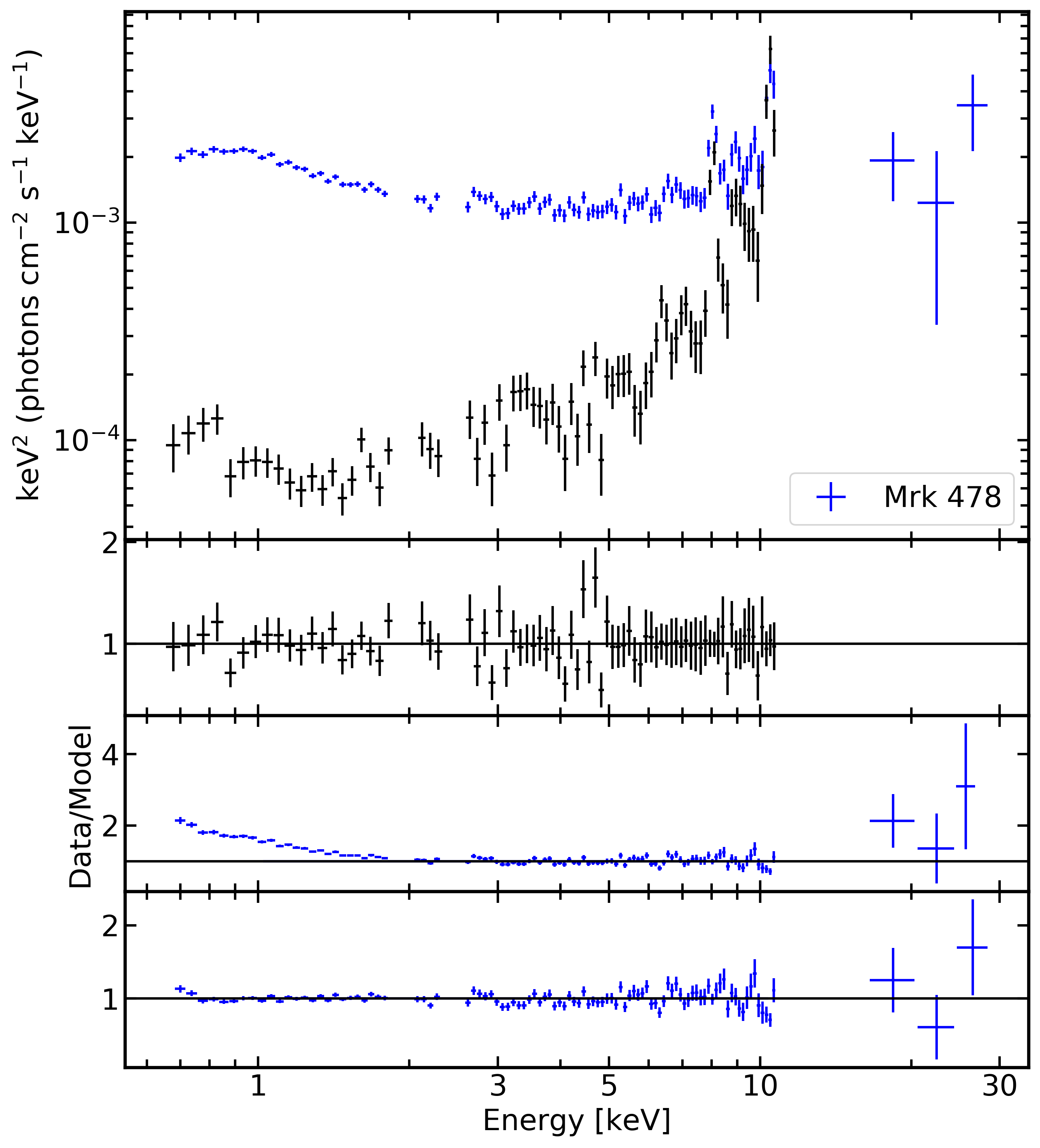}
	\hspace{7mm}
	\includegraphics[width=75mm]{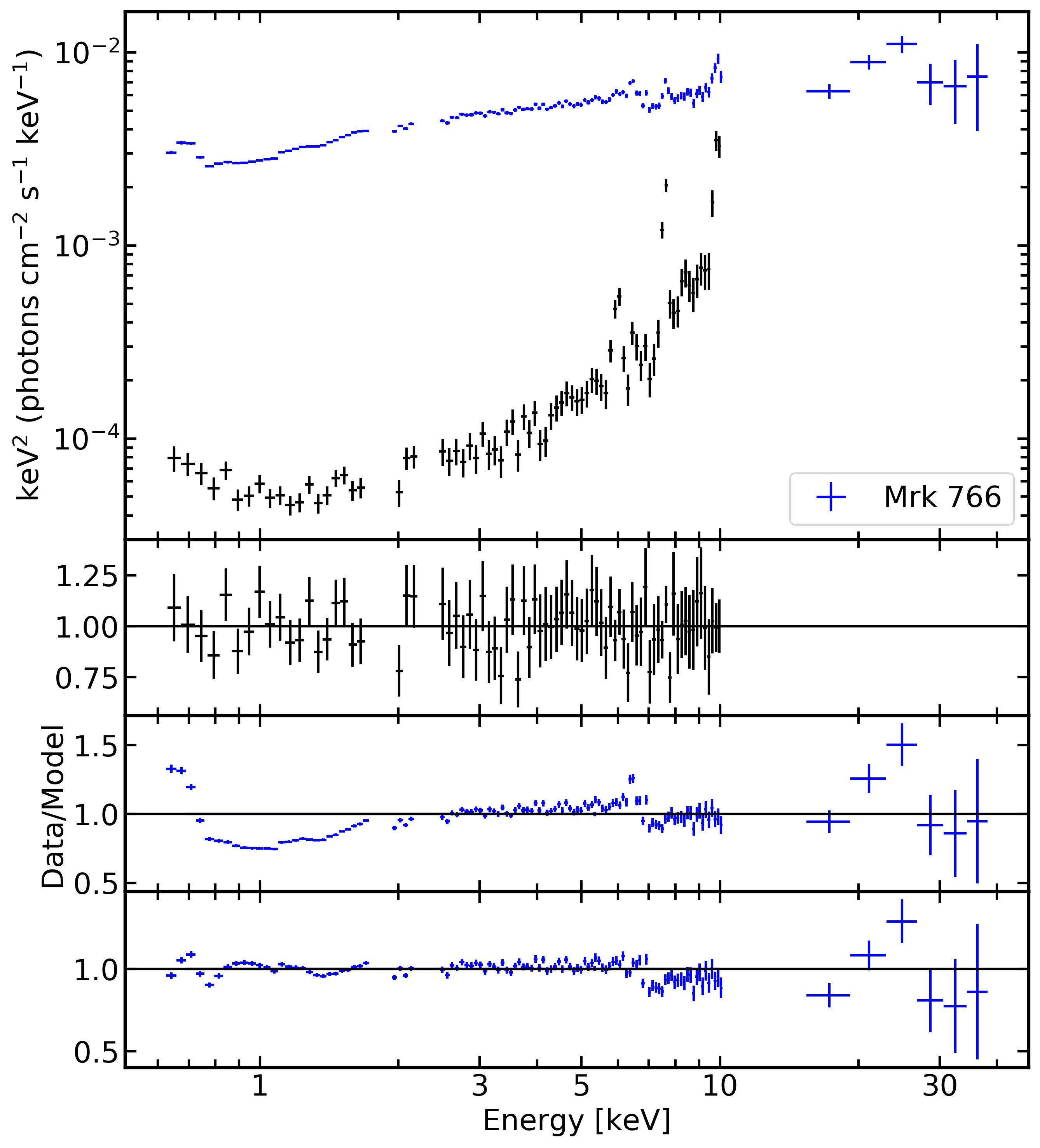}
	\caption{\label{fig:po}continued}
\end{figure*}

\setcounter{figure}{0}

\begin{figure*}
	\centering
	\includegraphics[width=75mm]{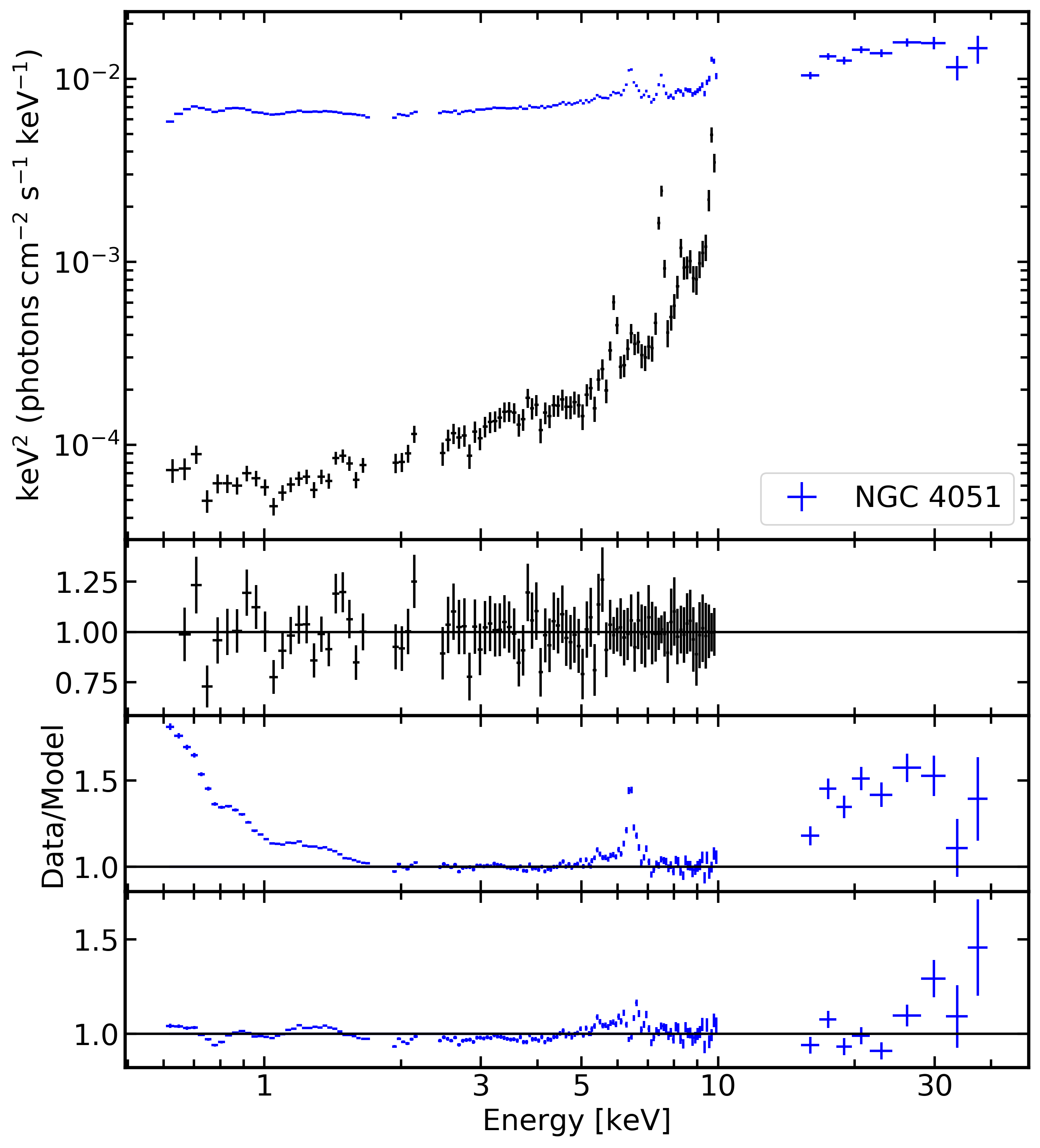}
	\hspace{7mm}
	\includegraphics[width=75mm]{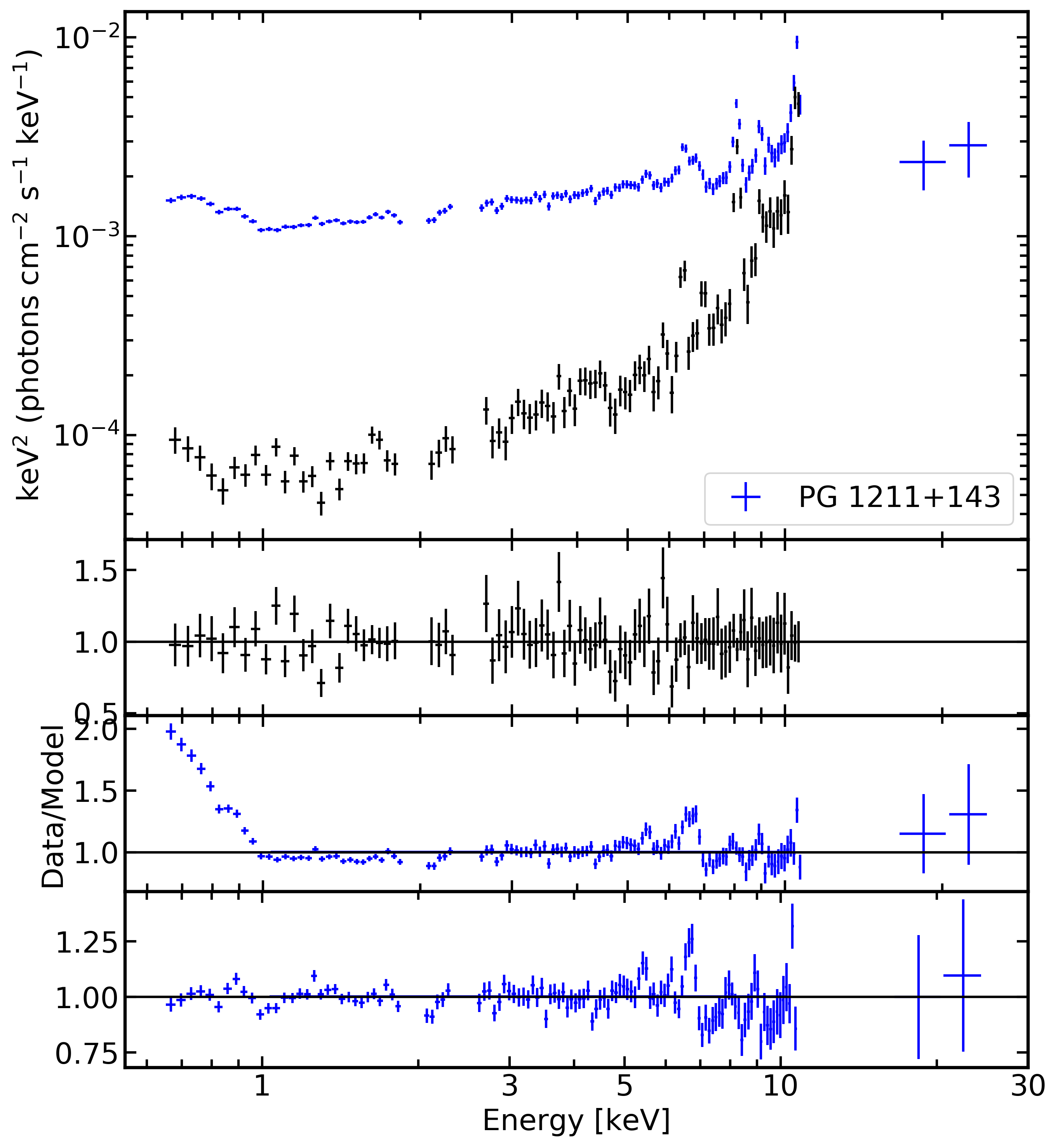}
	\hspace{0mm}
	\includegraphics[width=75mm]{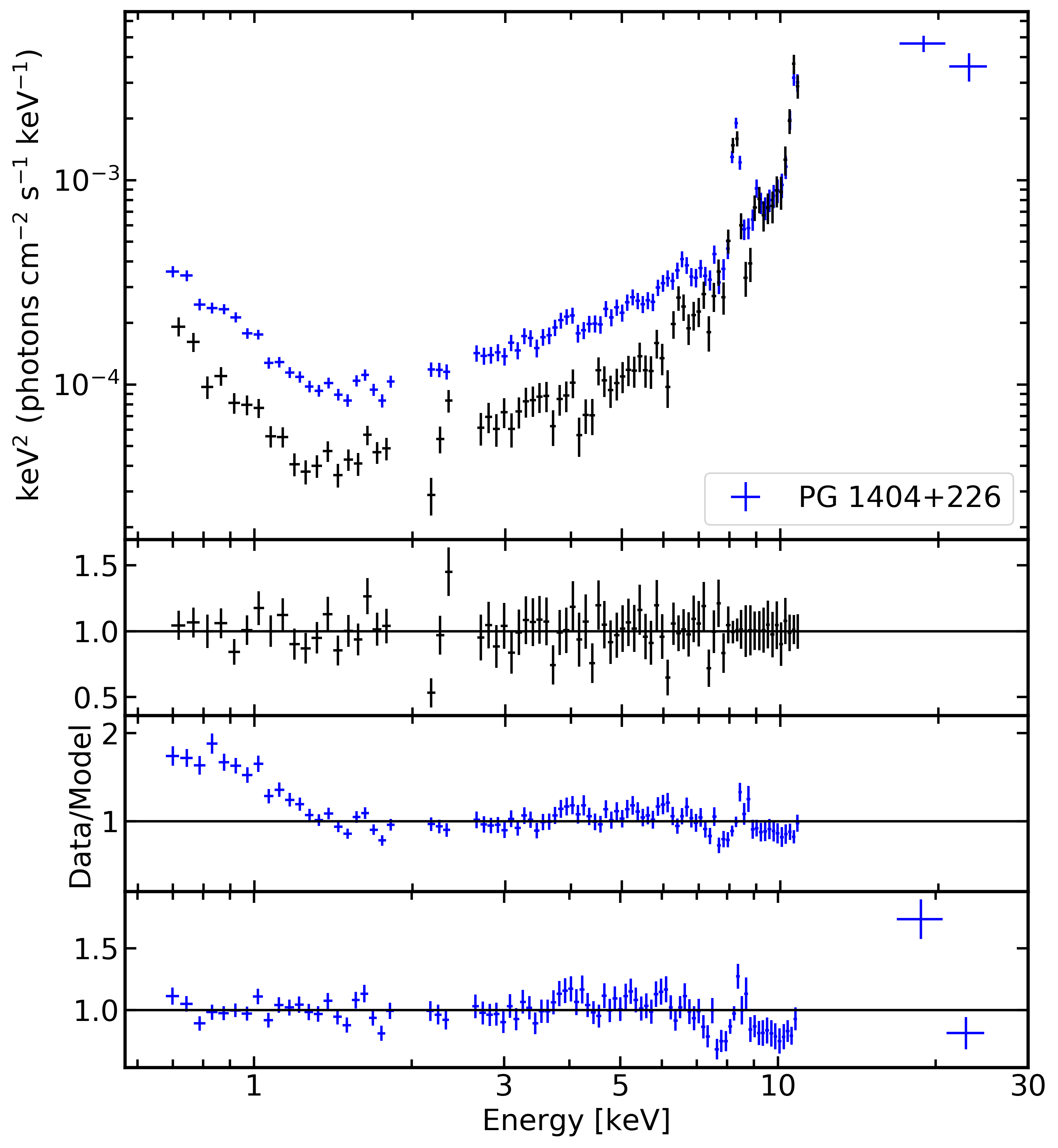}
	\hspace{7mm}
	\includegraphics[width=75mm]{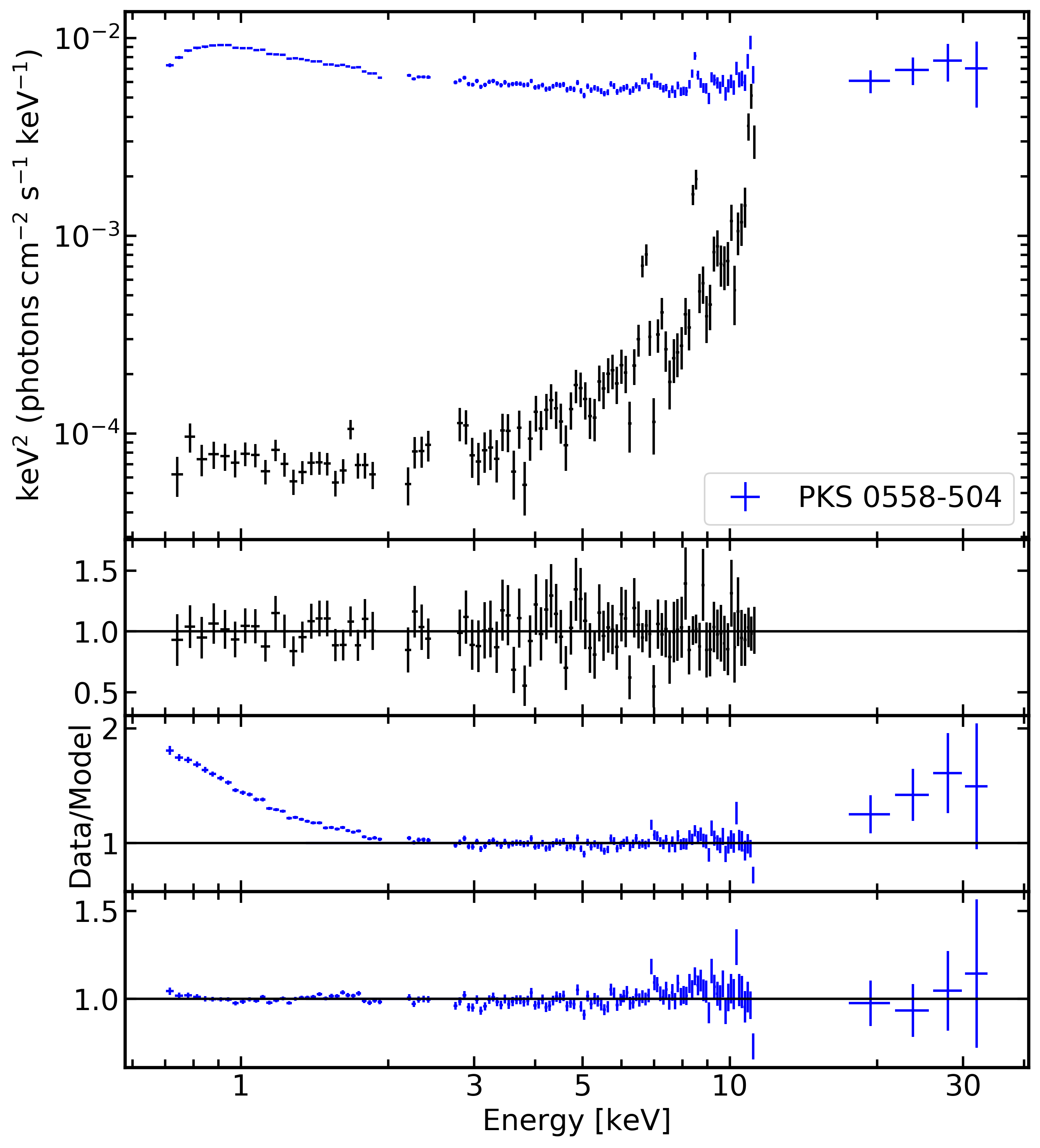}
	\caption{\label{fig:po}continued}
\end{figure*}

\setcounter{figure}{0}

\begin{figure*}
	\centering
	\includegraphics[width=75mm]{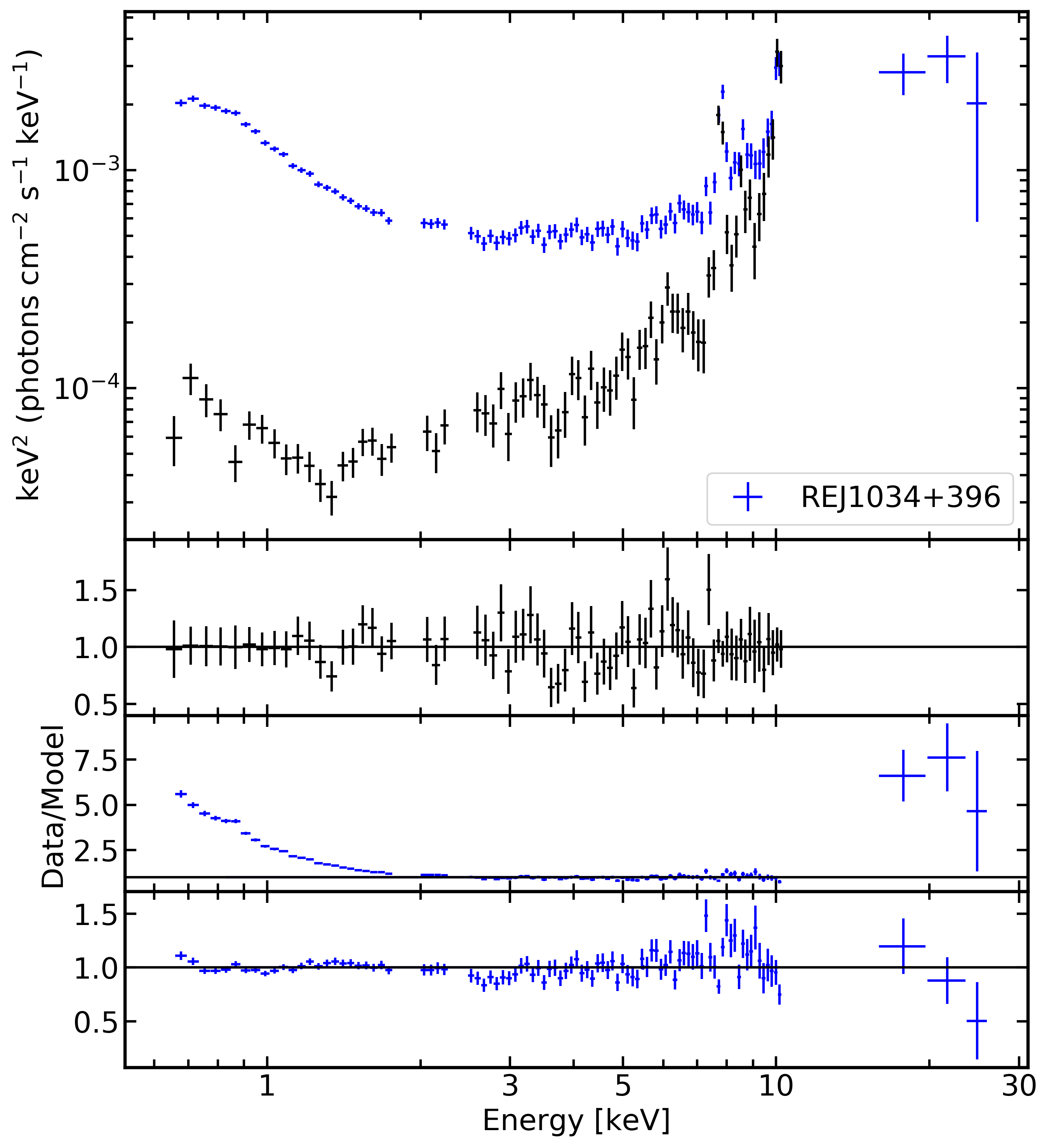}
	\hspace{7mm}
	\includegraphics[width=75mm]{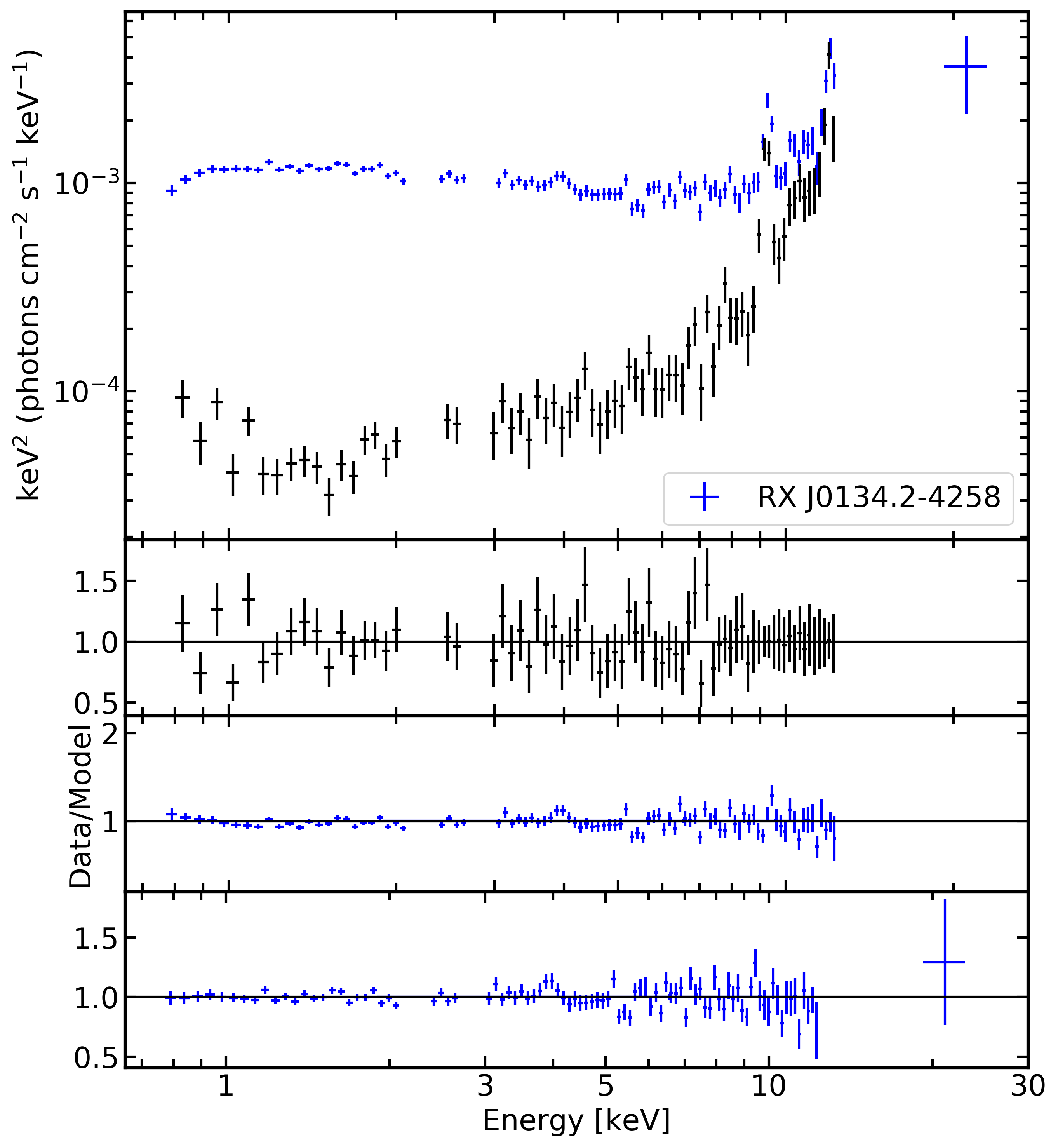}
	\hspace{0mm}
	\includegraphics[width=75mm]{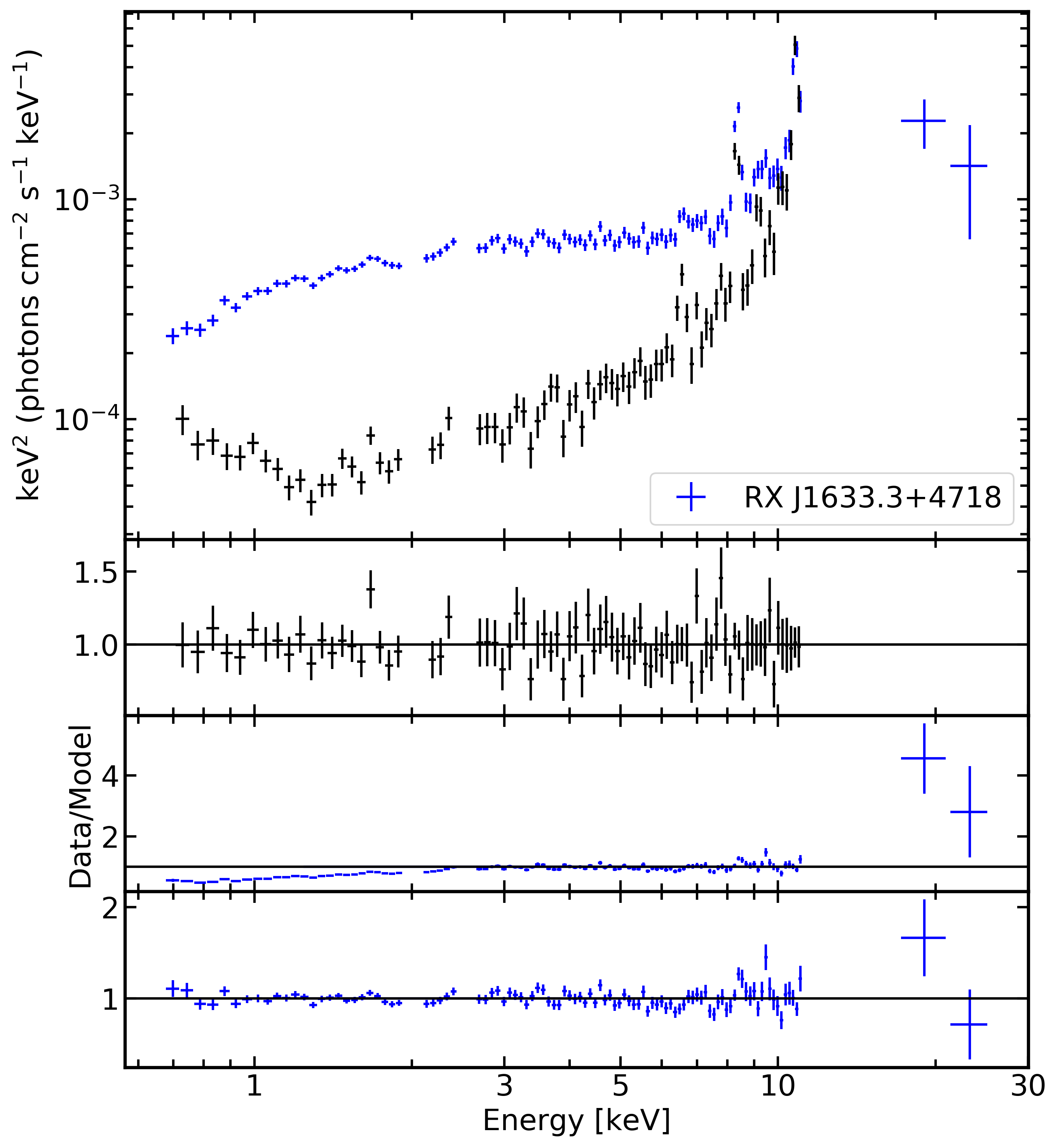}
	\hspace{7mm}
	\includegraphics[width=75mm]{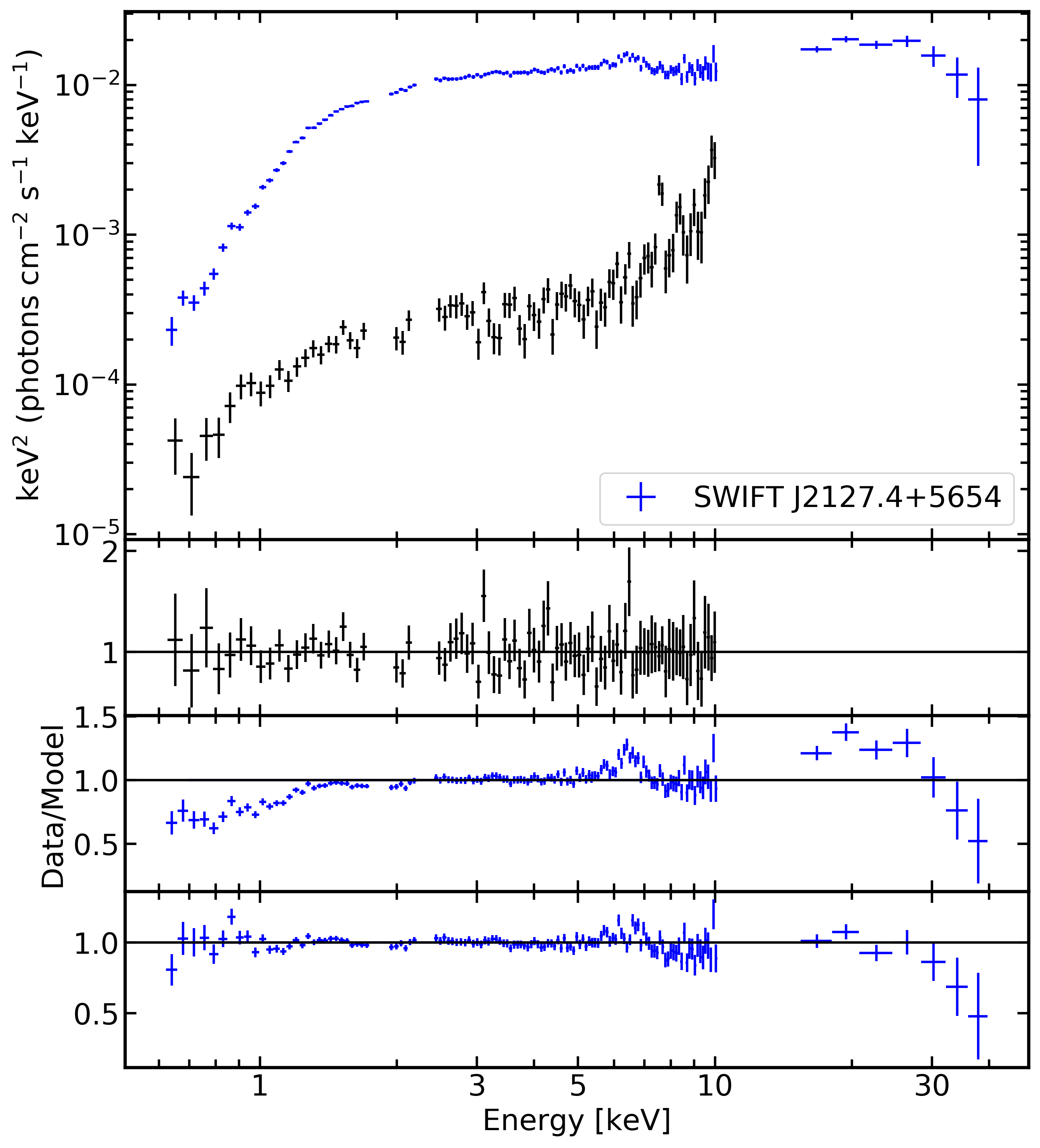}
	\caption{\label{fig:po}continued}
\end{figure*}

\setcounter{figure}{0}

\begin{figure*}
	\centering
	\includegraphics[width=75mm]{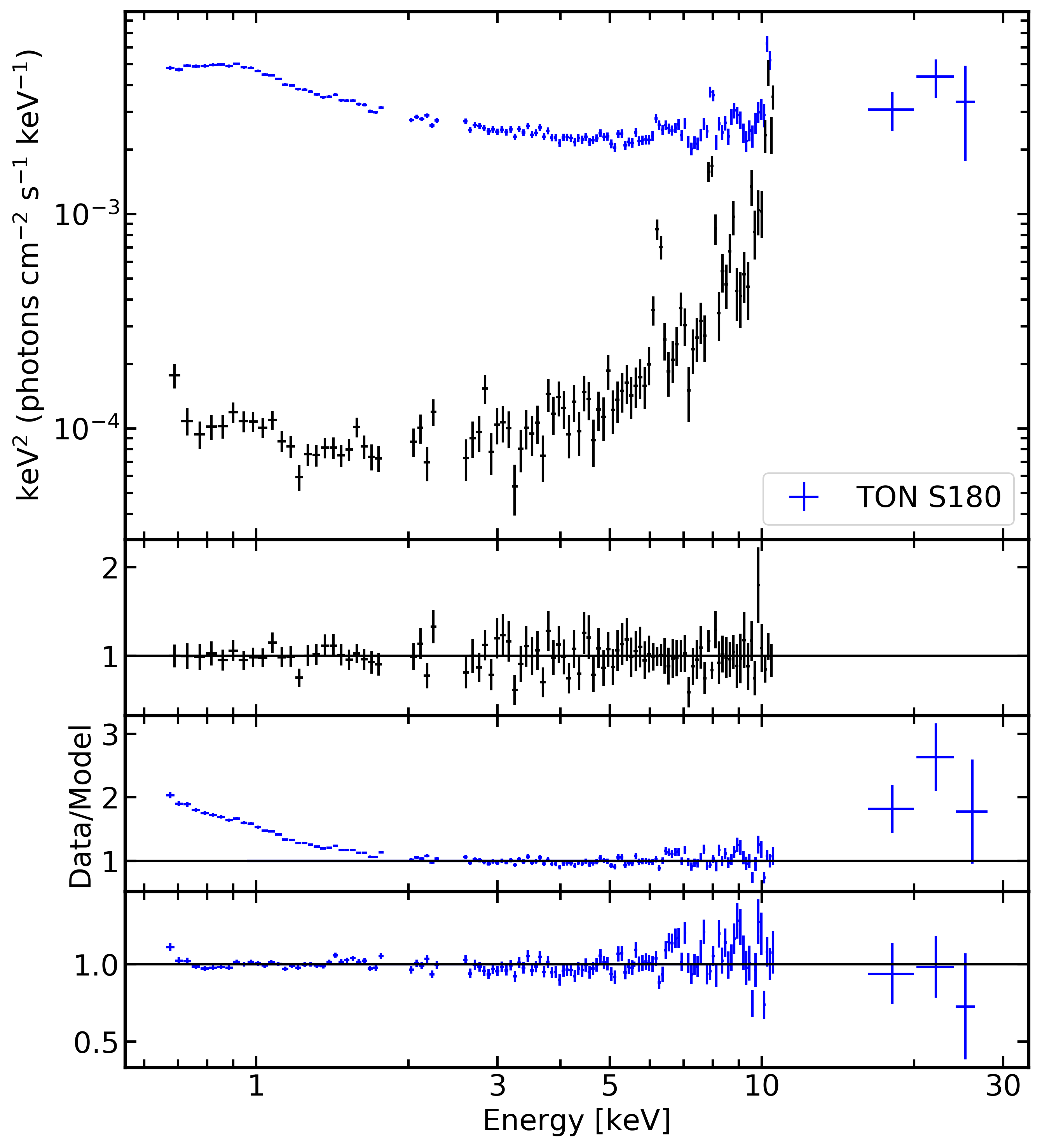}
	\hspace{7mm}
	\includegraphics[width=75mm]{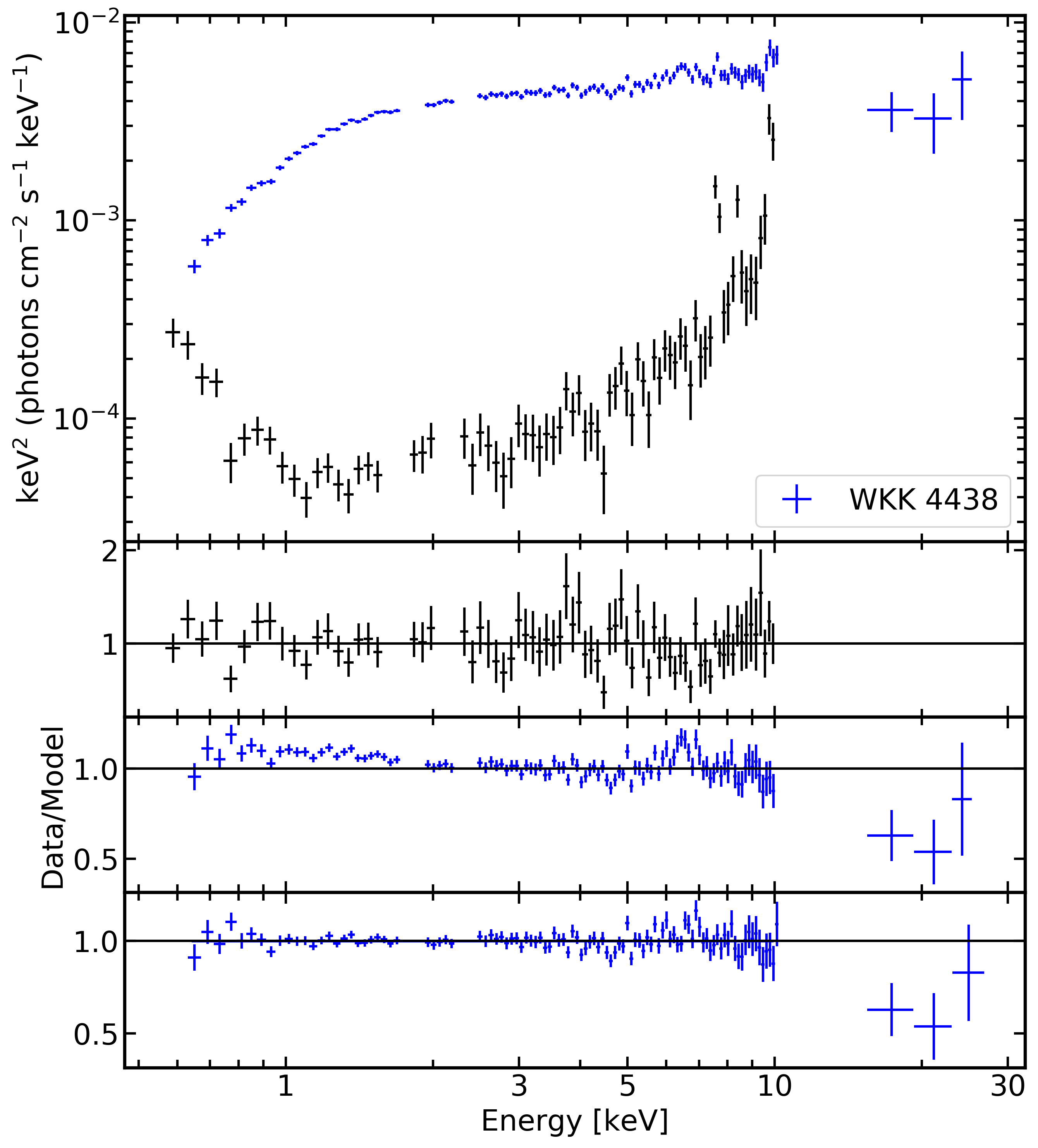}
	\hspace{0mm}
	\includegraphics[width=75mm]{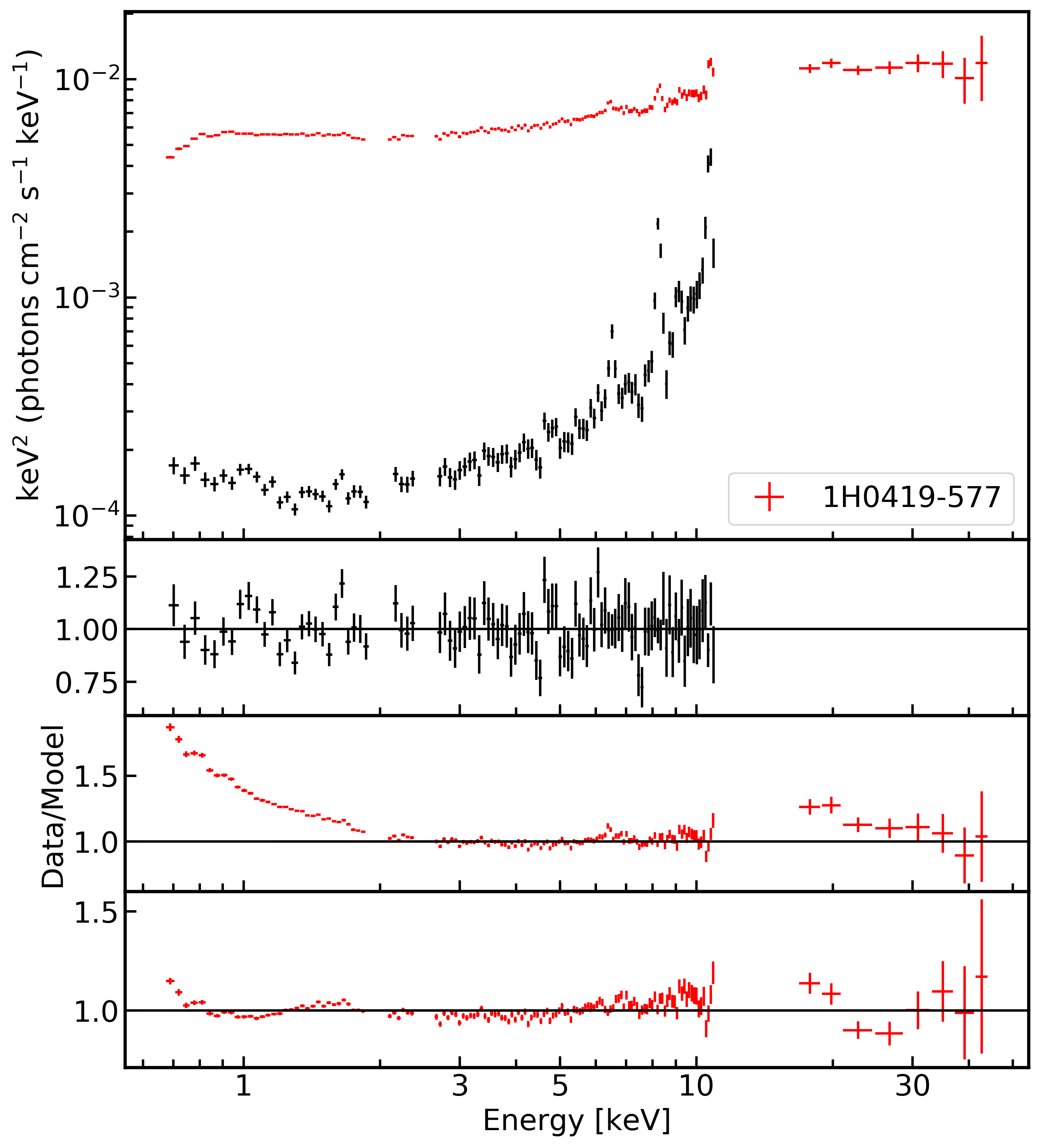}
	\hspace{7mm}
	\includegraphics[width=75mm]{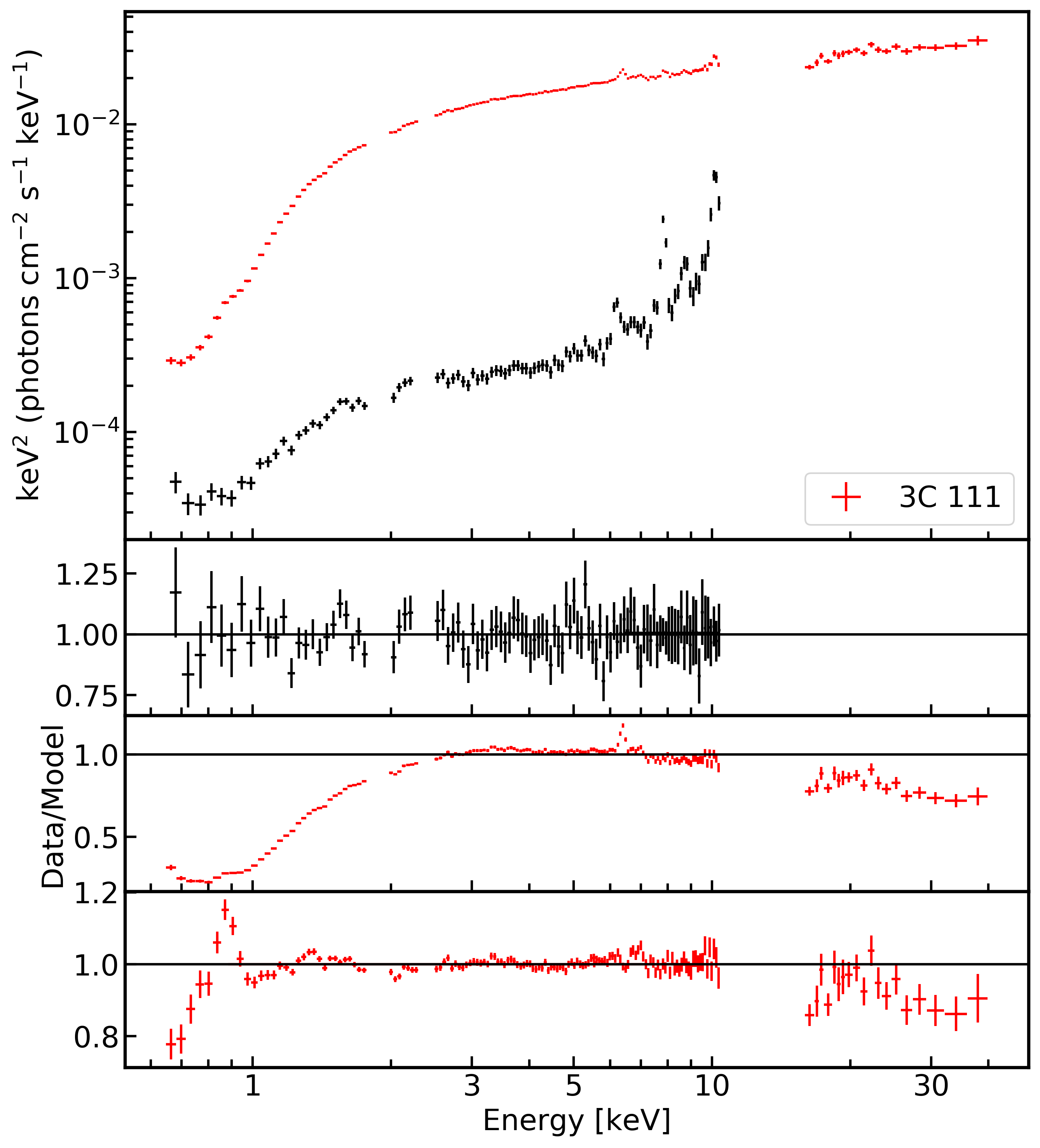}
	\caption{\label{fig:po}continued}
\end{figure*}

\setcounter{figure}{0}

\begin{figure*}
	\centering
	\includegraphics[width=75mm]{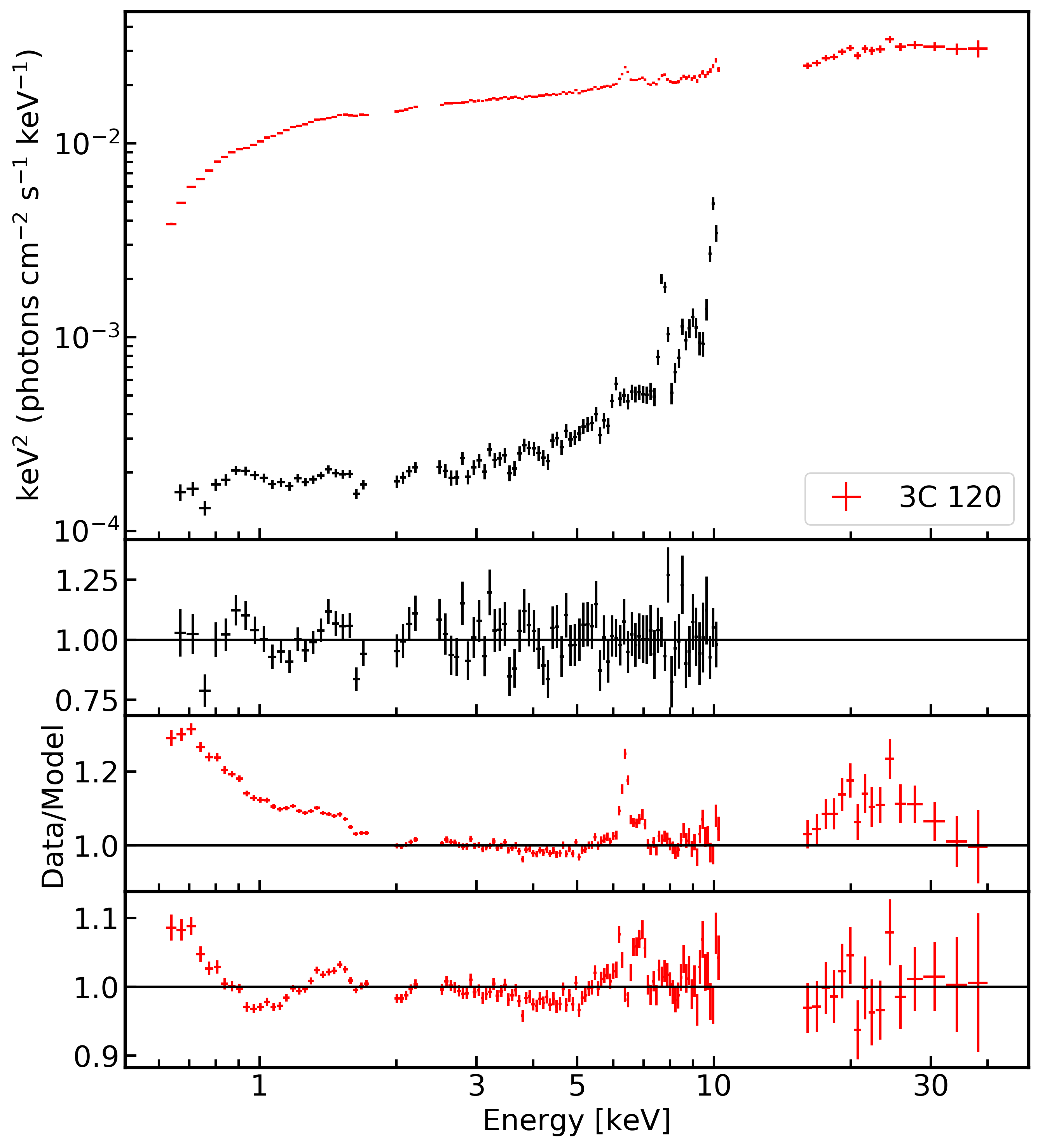}
	\hspace{7mm}
	\includegraphics[width=75mm]{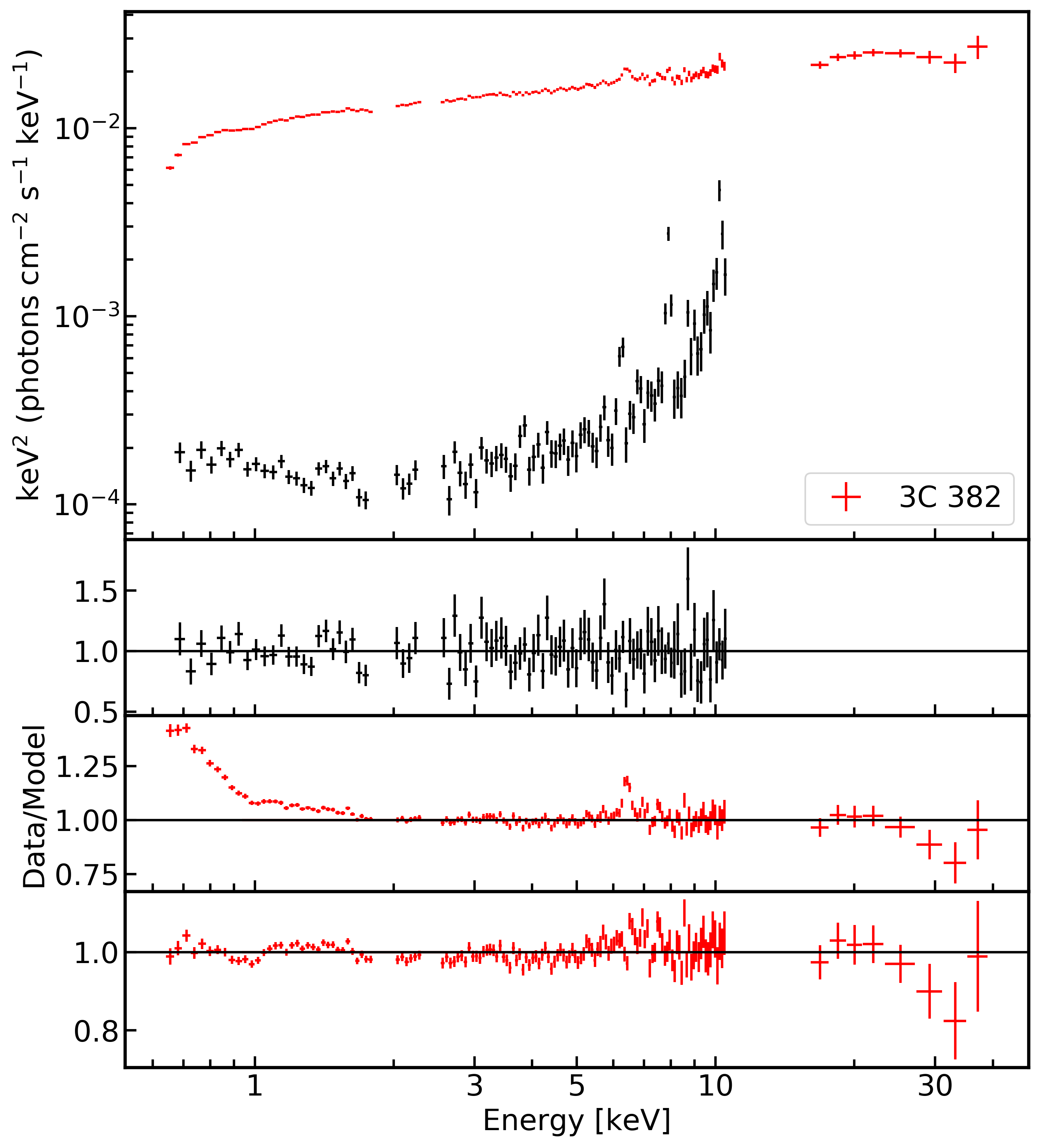}
	\hspace{0mm}
	\includegraphics[width=75mm]{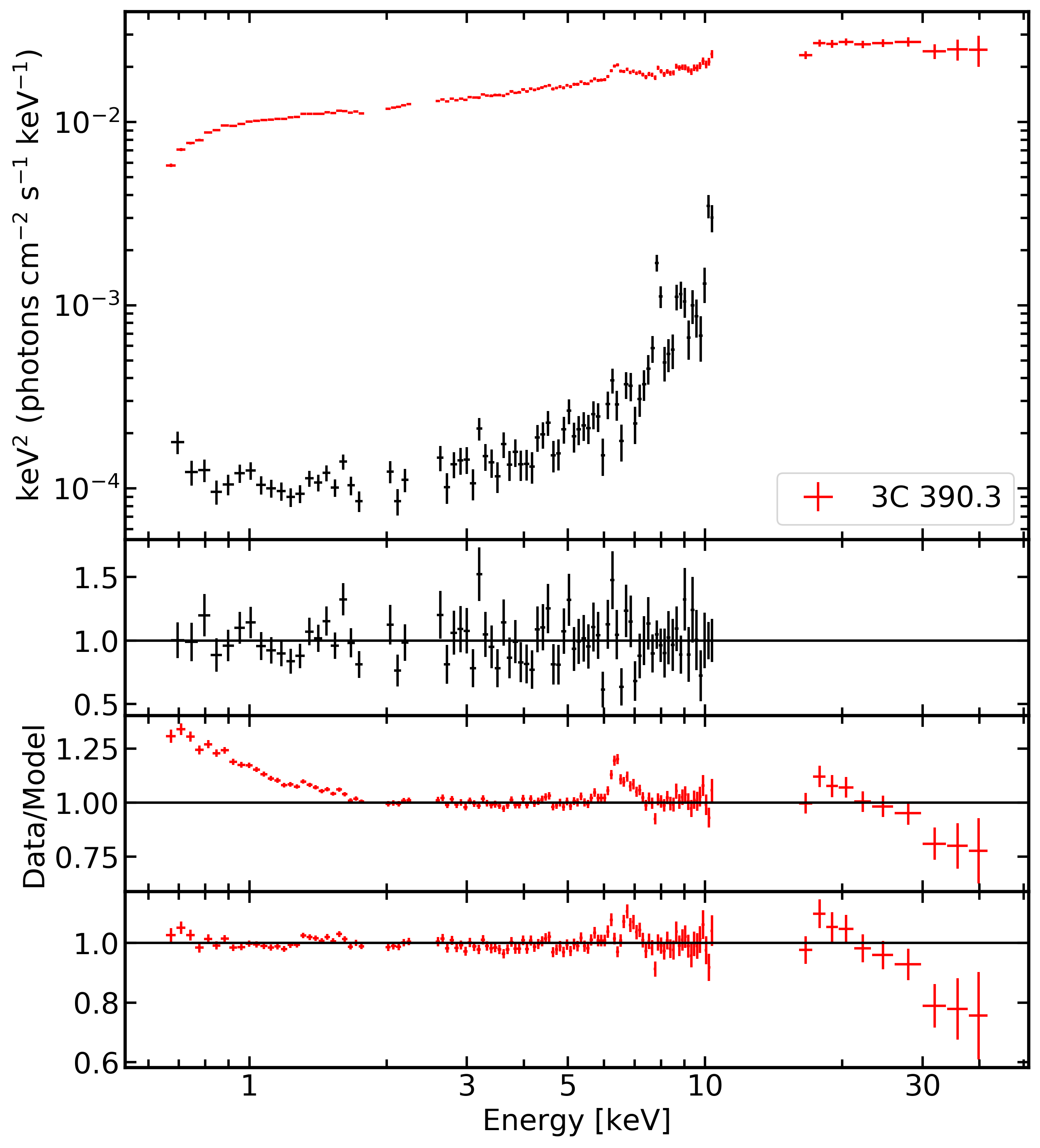}
	\hspace{7mm}
	\includegraphics[width=75mm]{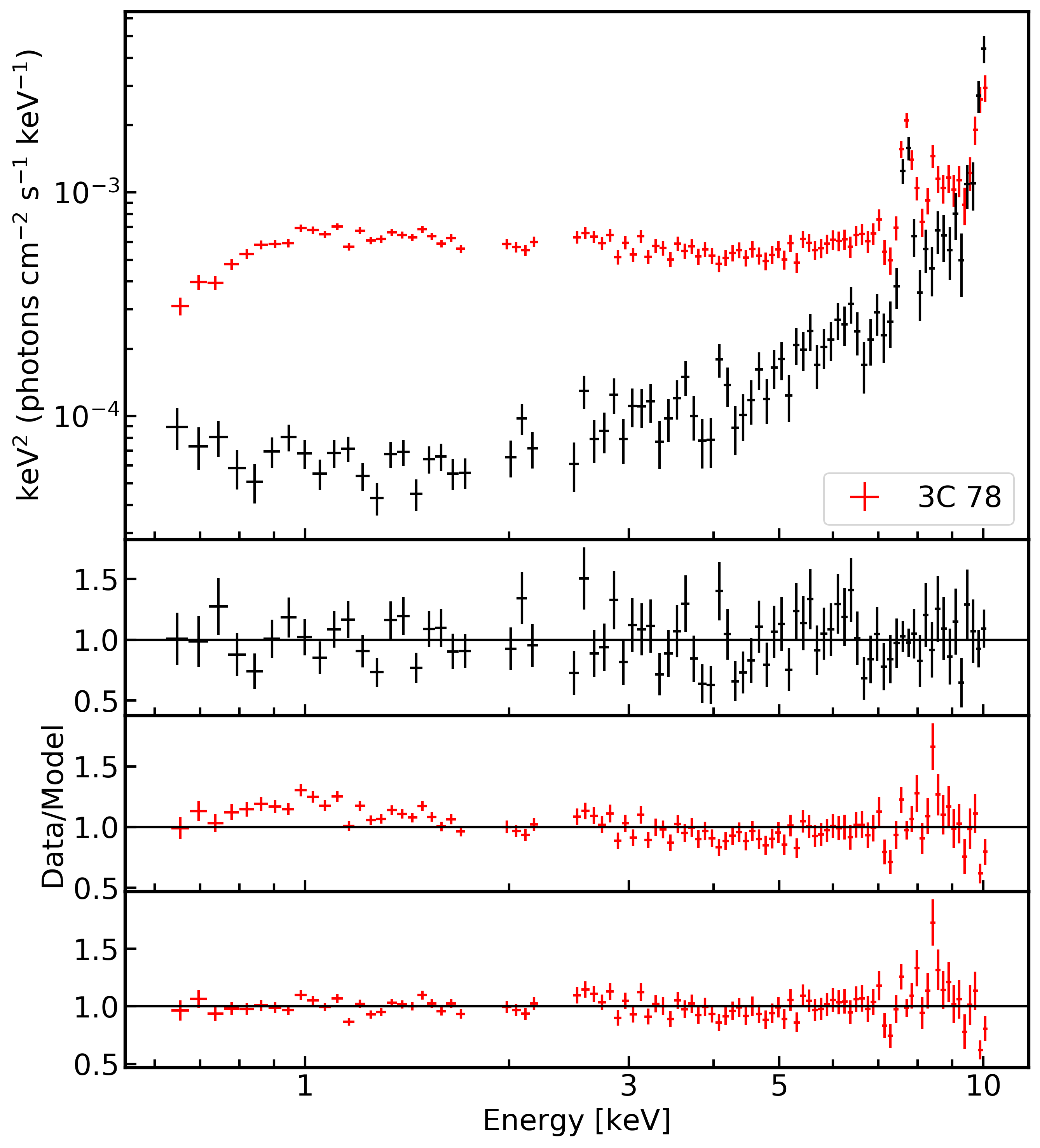}
	\caption{\label{fig:po}continued}
\end{figure*}

\setcounter{figure}{0}

\begin{figure*}
	\centering
	\includegraphics[width=75mm]{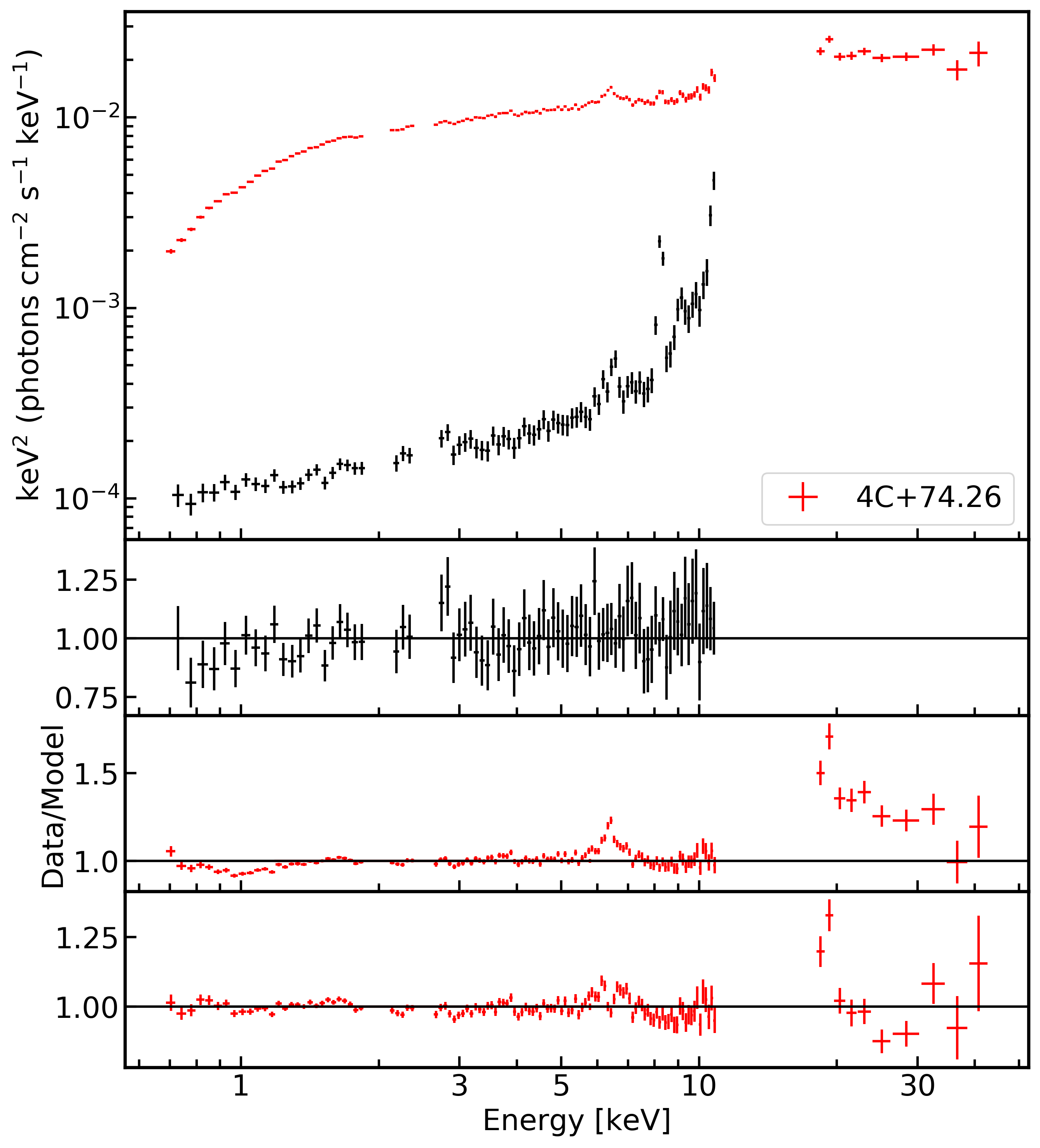}
	\hspace{7mm}
	\includegraphics[width=75mm]{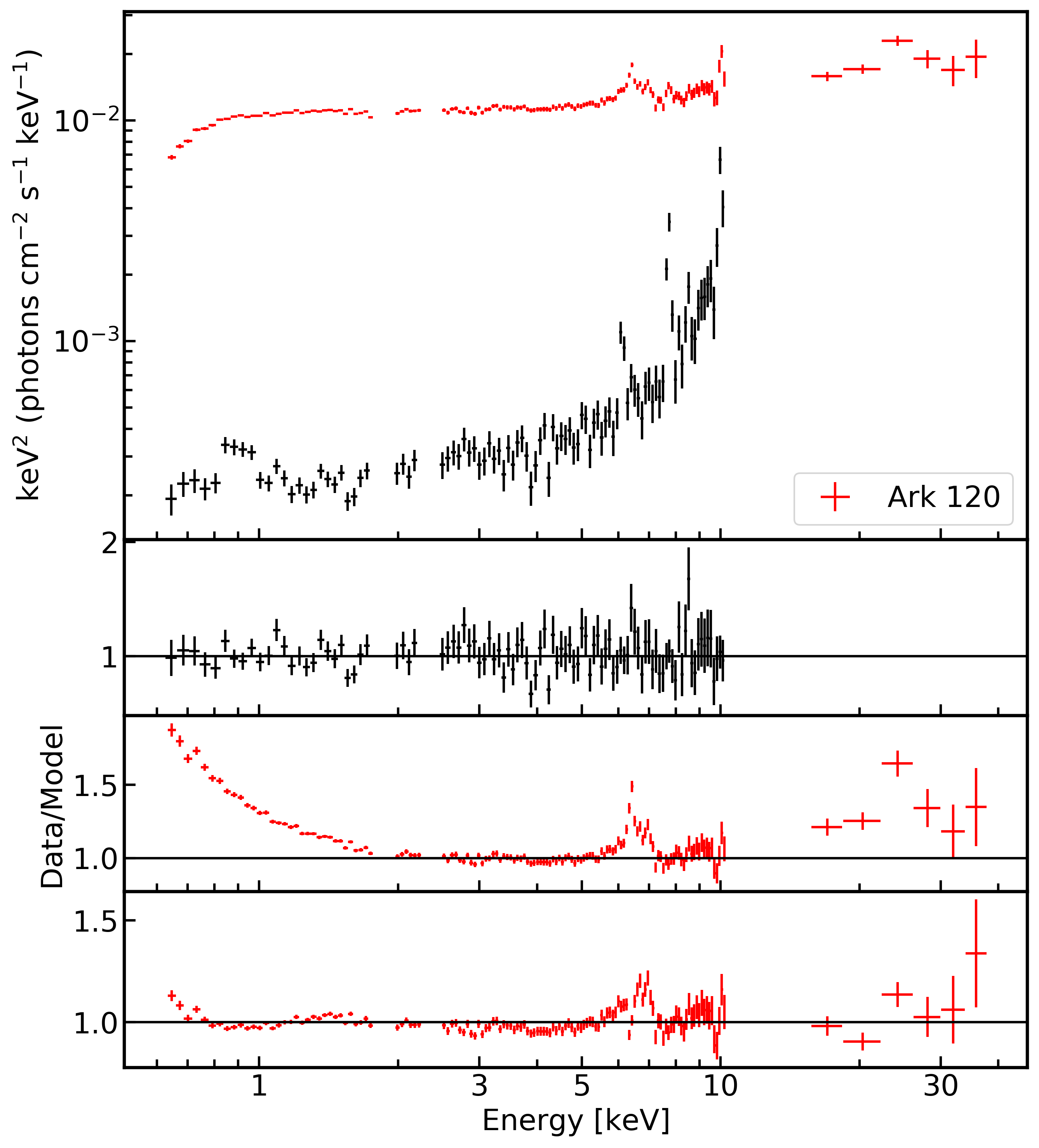}
	\hspace{0mm}
	\includegraphics[width=75mm]{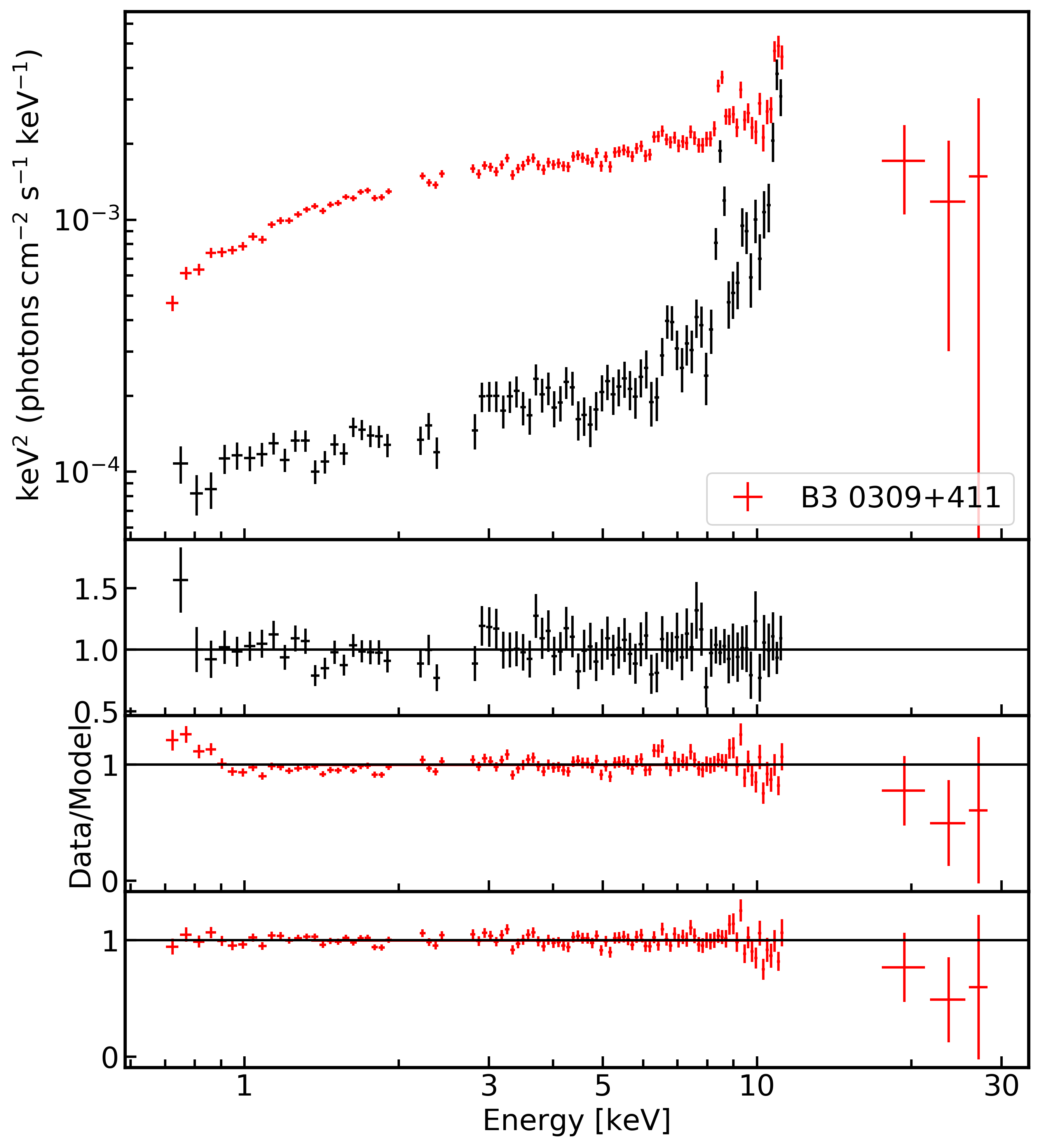}
	\hspace{7mm}
	\includegraphics[width=75mm]{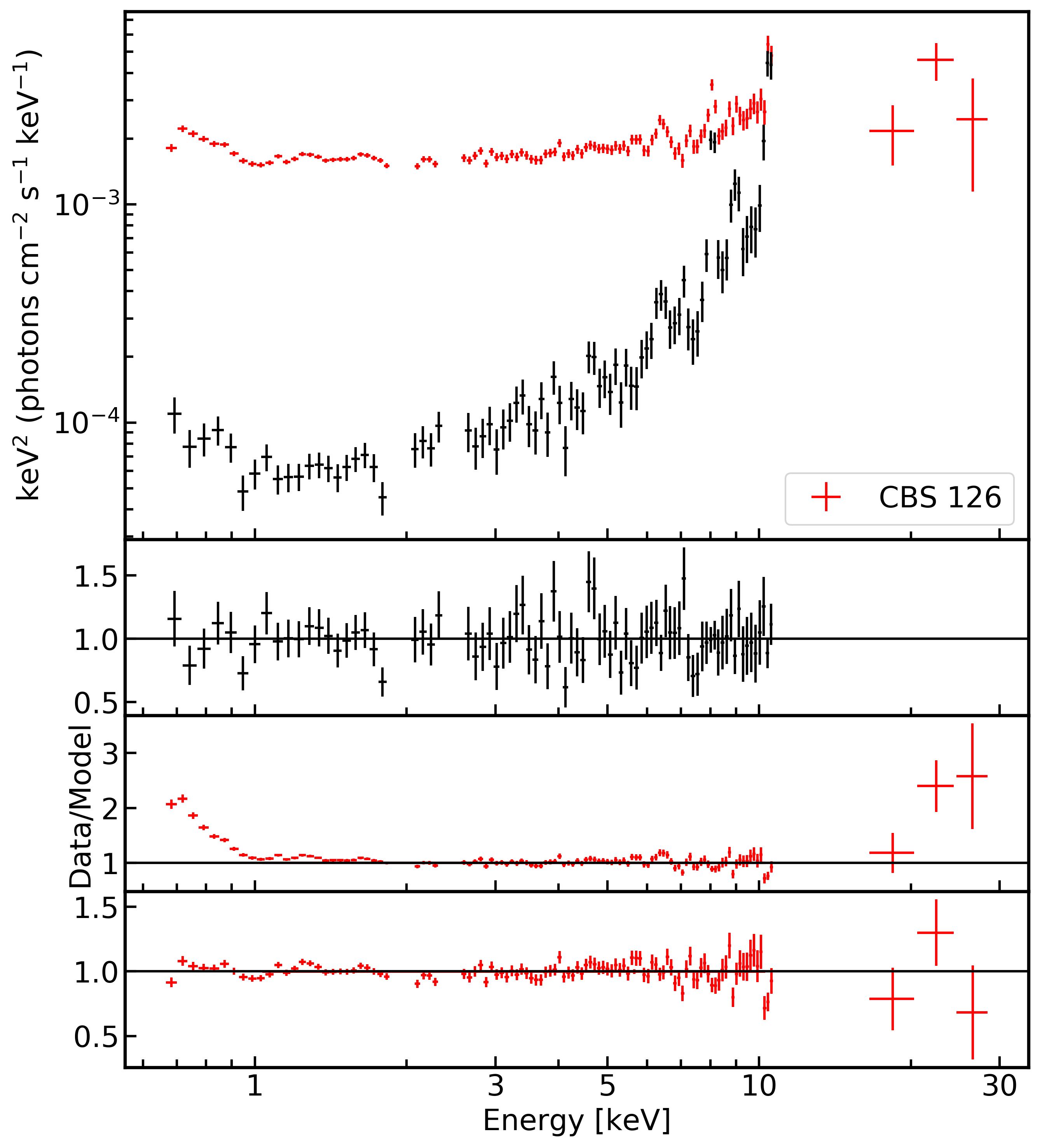}
	\caption{\label{fig:po}continued}
\end{figure*}

\begin{figure*}
	\centering	
	\includegraphics[width=75mm]{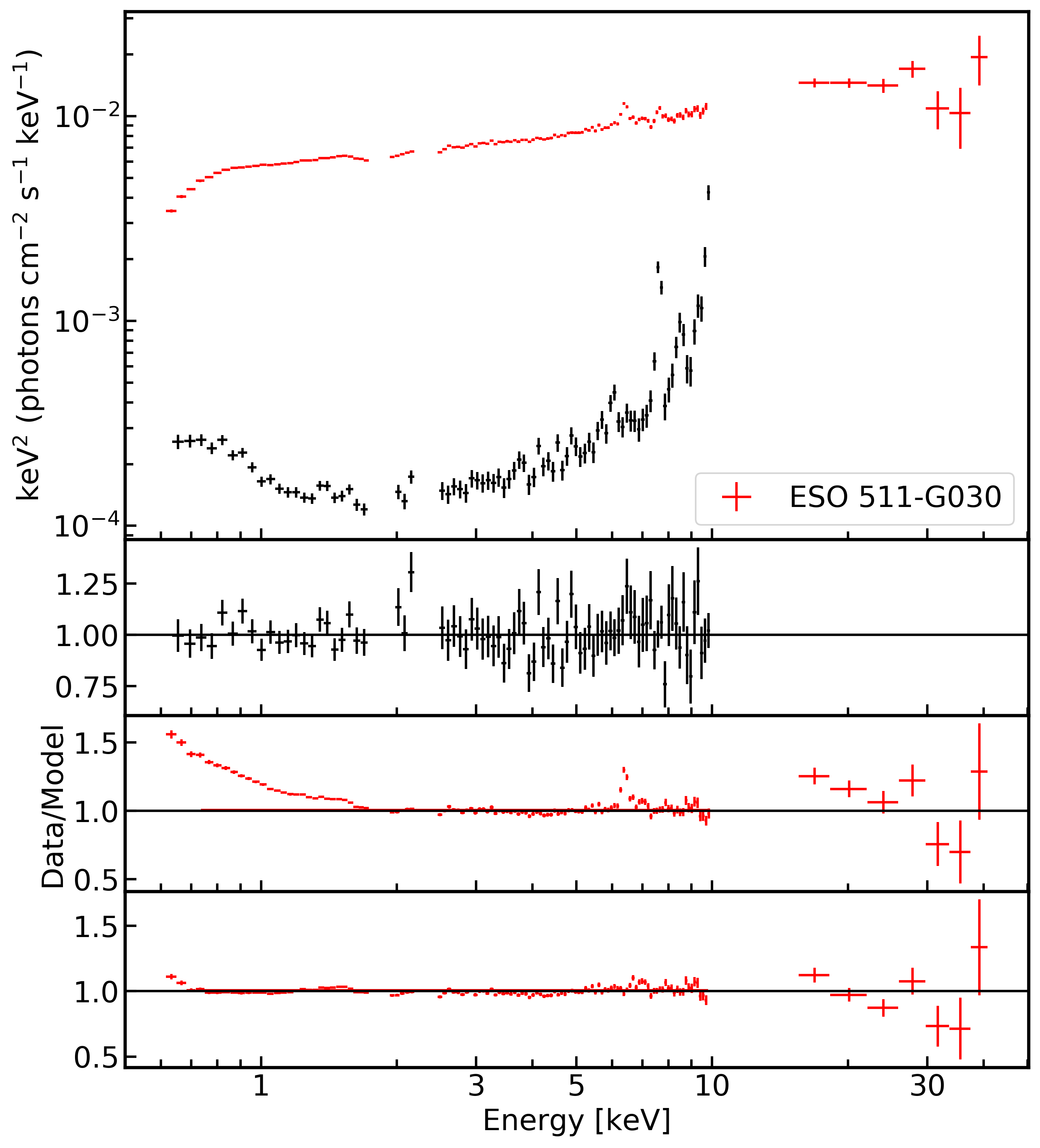}
	\hspace{7mm}
	\includegraphics[width=75mm]{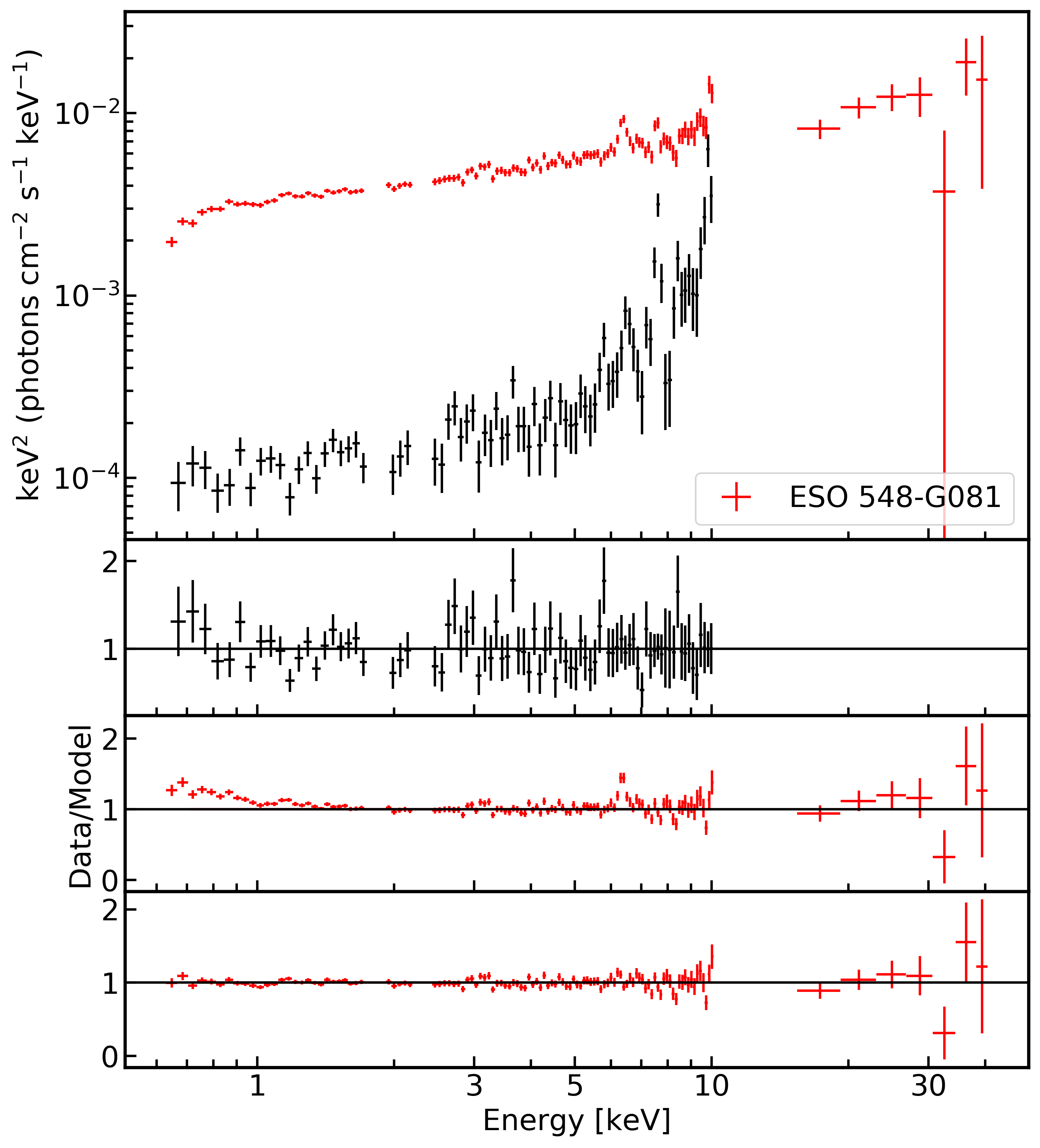}
	\hspace{0mm}
	\includegraphics[width=75mm]{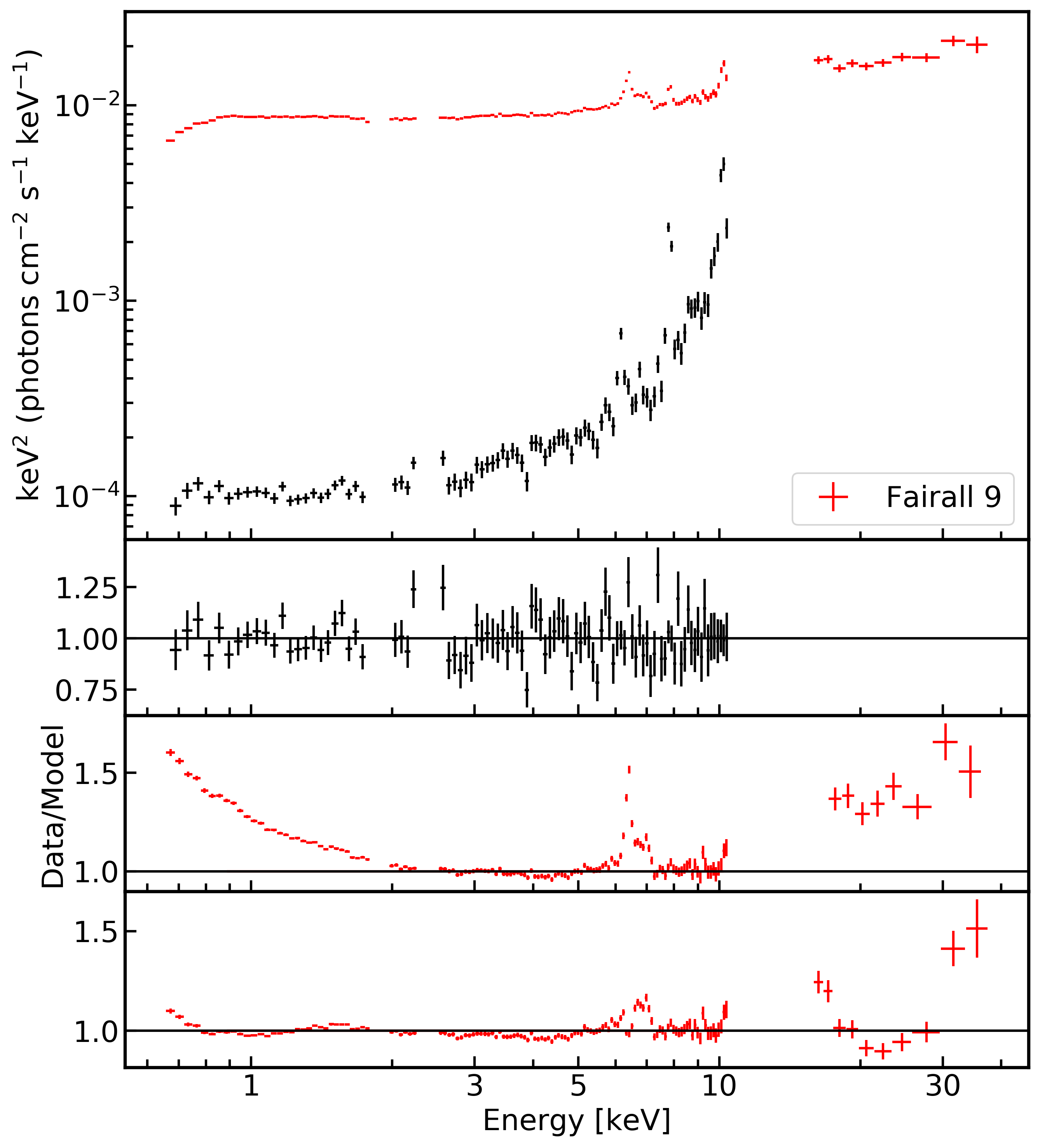}
	\hspace{7mm}
	\includegraphics[width=75mm]{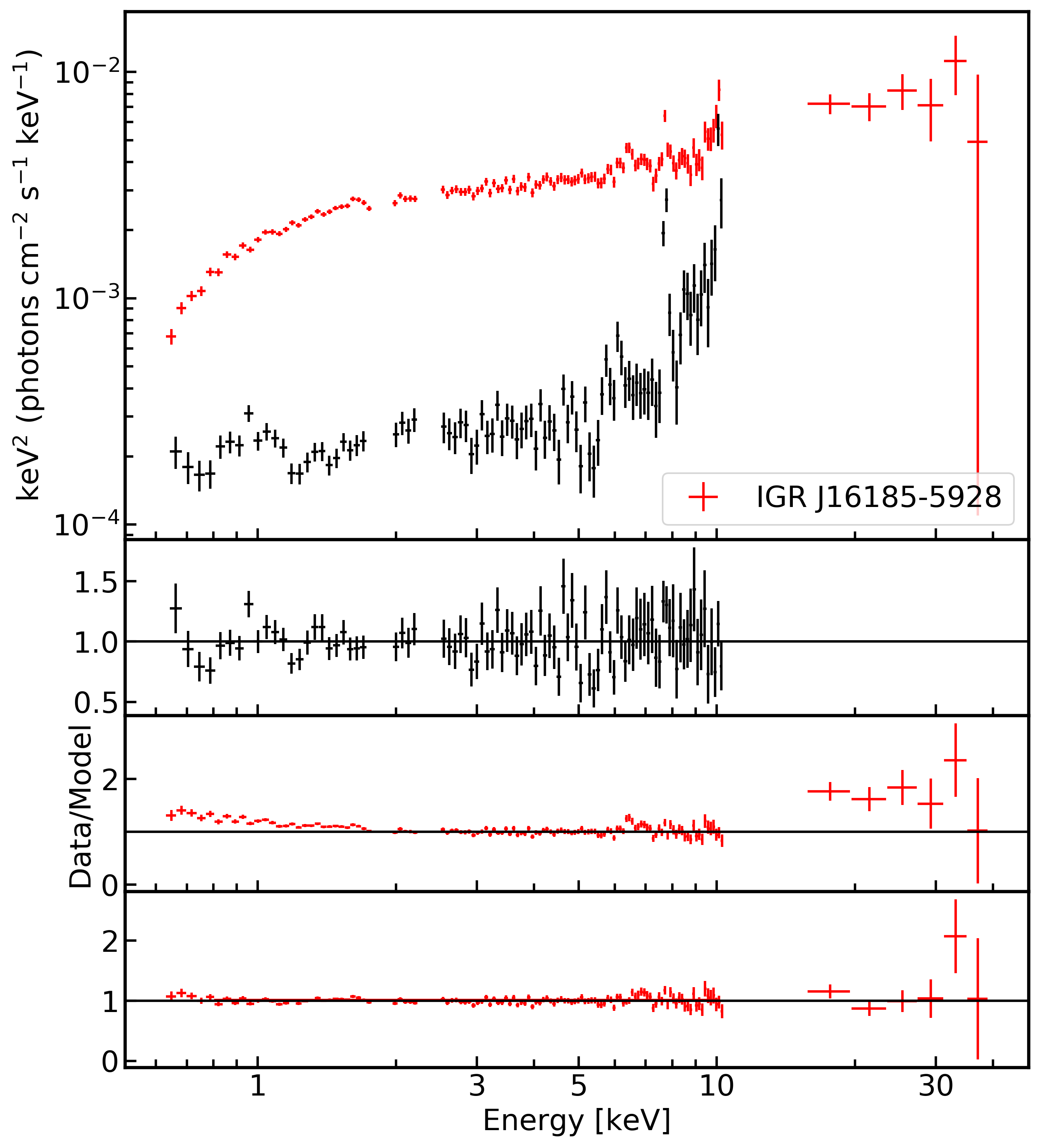}
	\caption{\label{fig:po}continued}
\end{figure*}

\setcounter{figure}{0}

\begin{figure*}
	\centering
	\includegraphics[width=75mm]{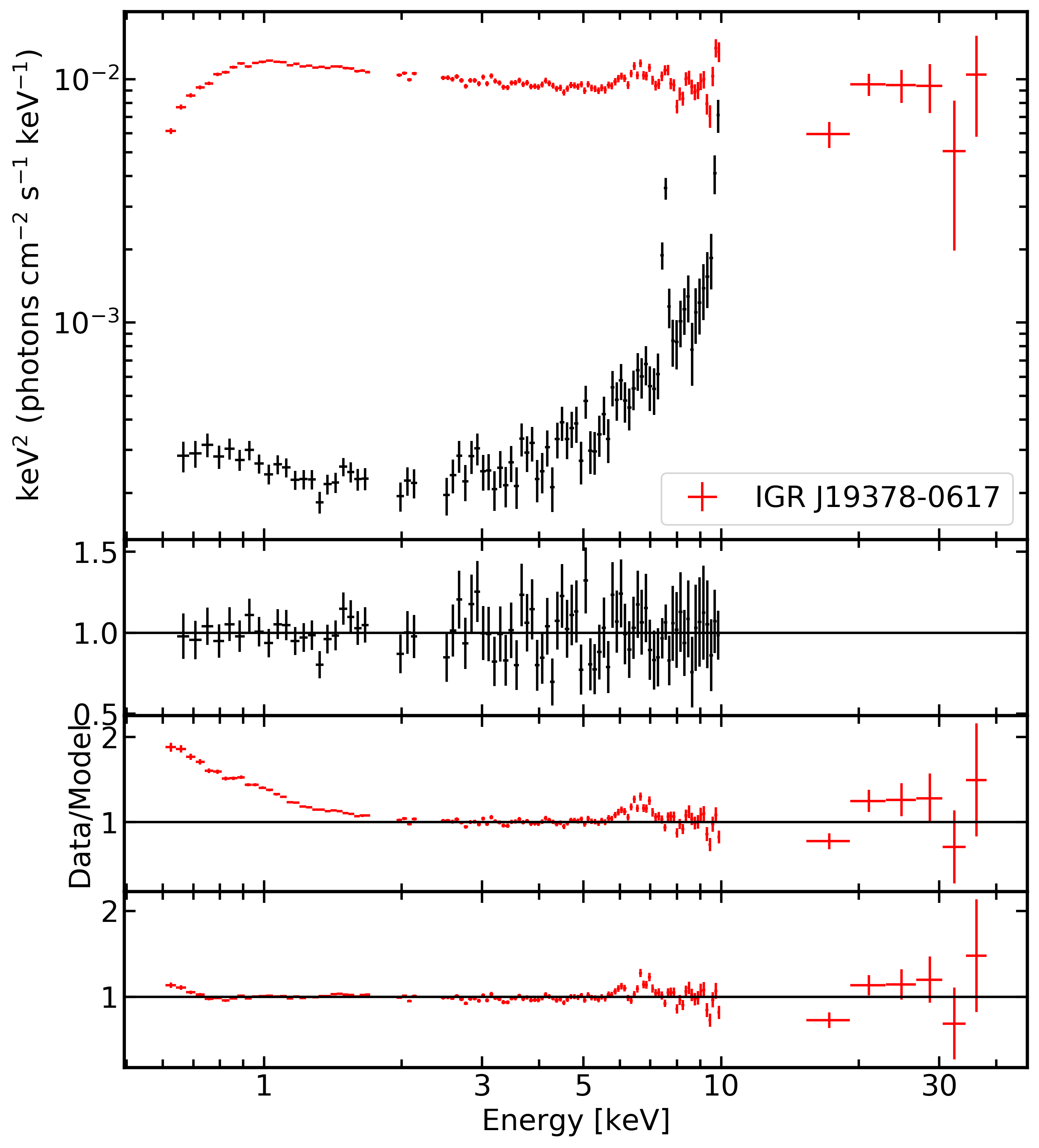}
	\hspace{7mm}
	\includegraphics[width=75mm]{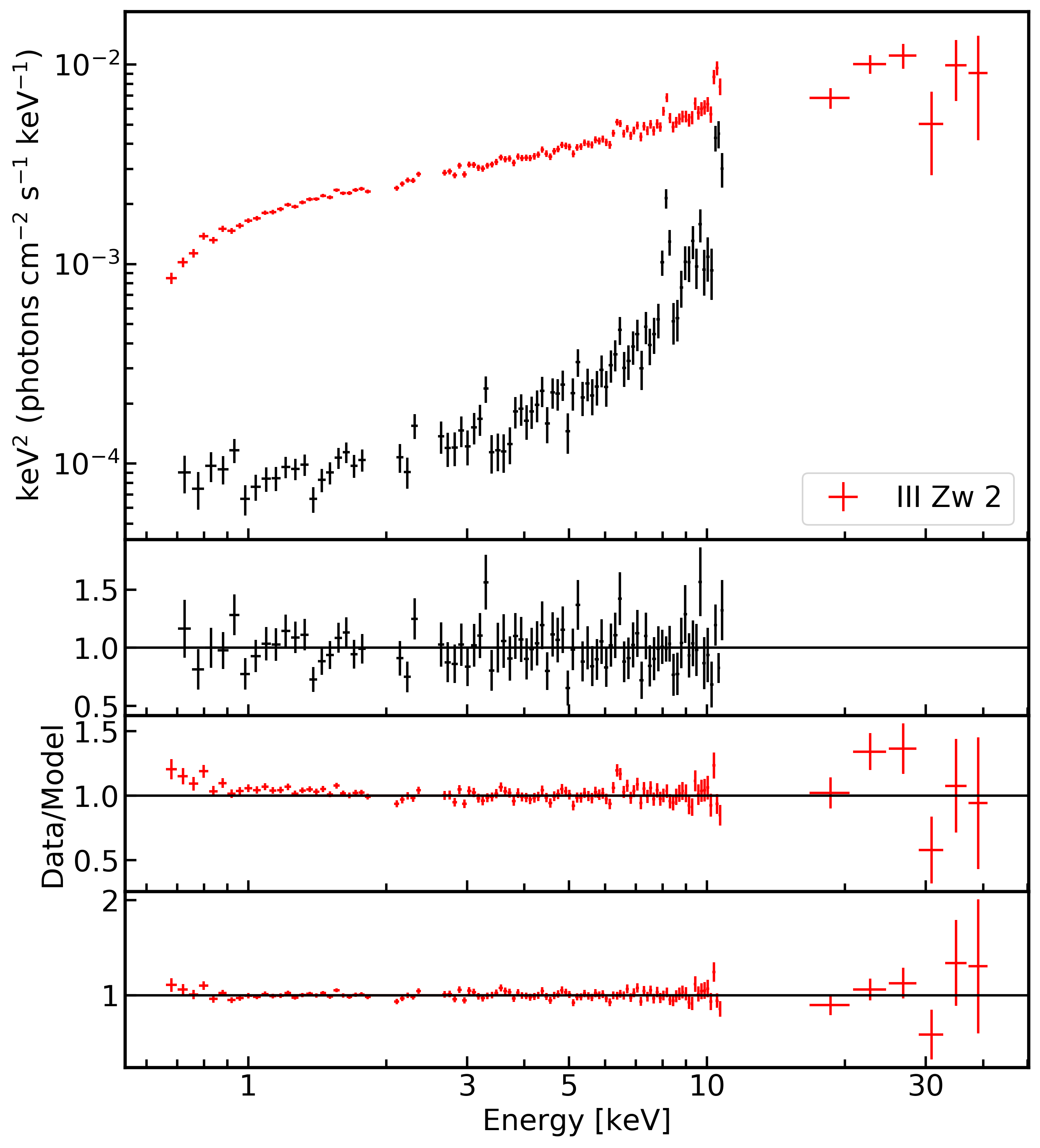}
	\hspace{0mm}
	\includegraphics[width=75mm]{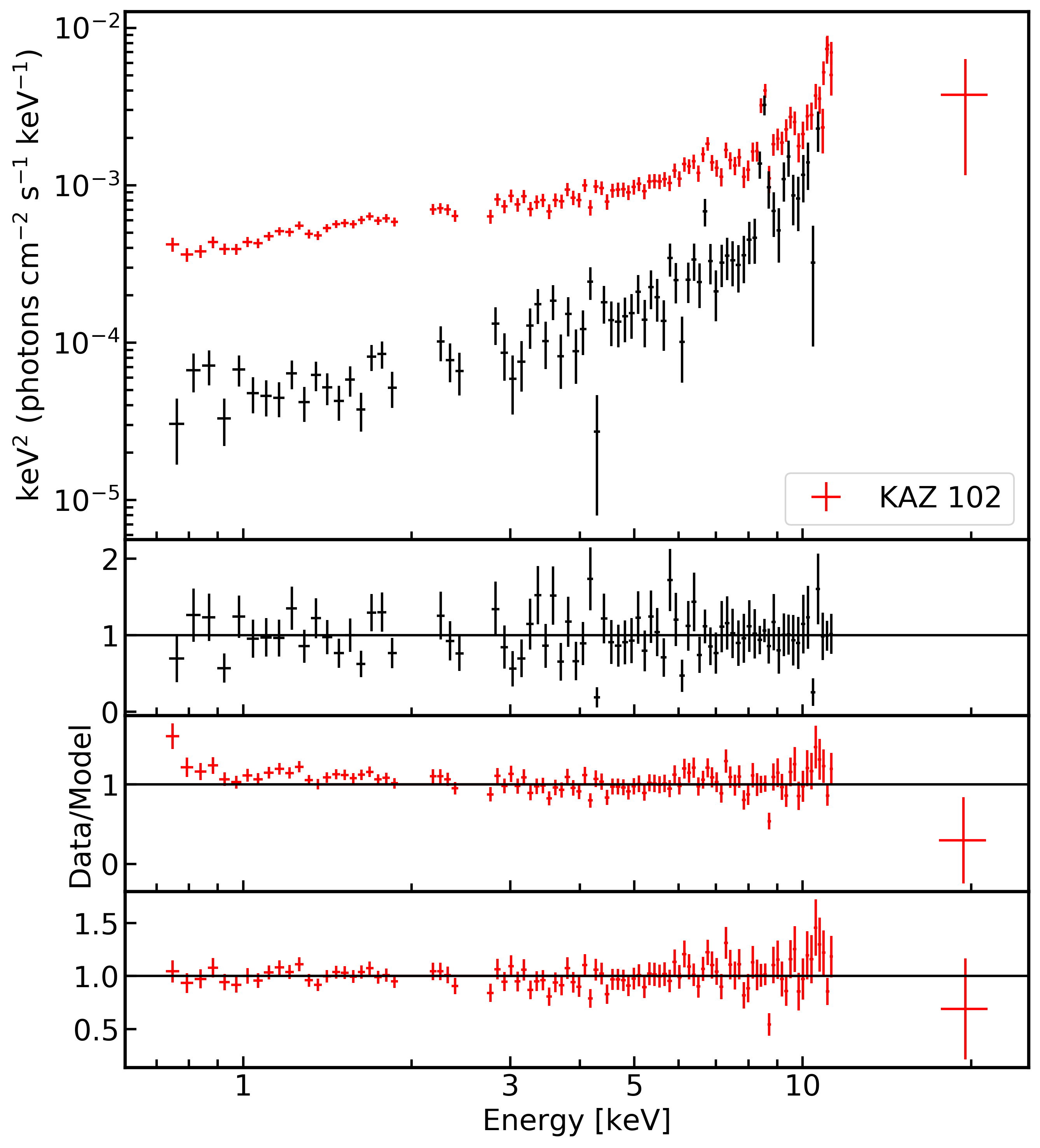}
	\hspace{7mm}
	\includegraphics[width=75mm]{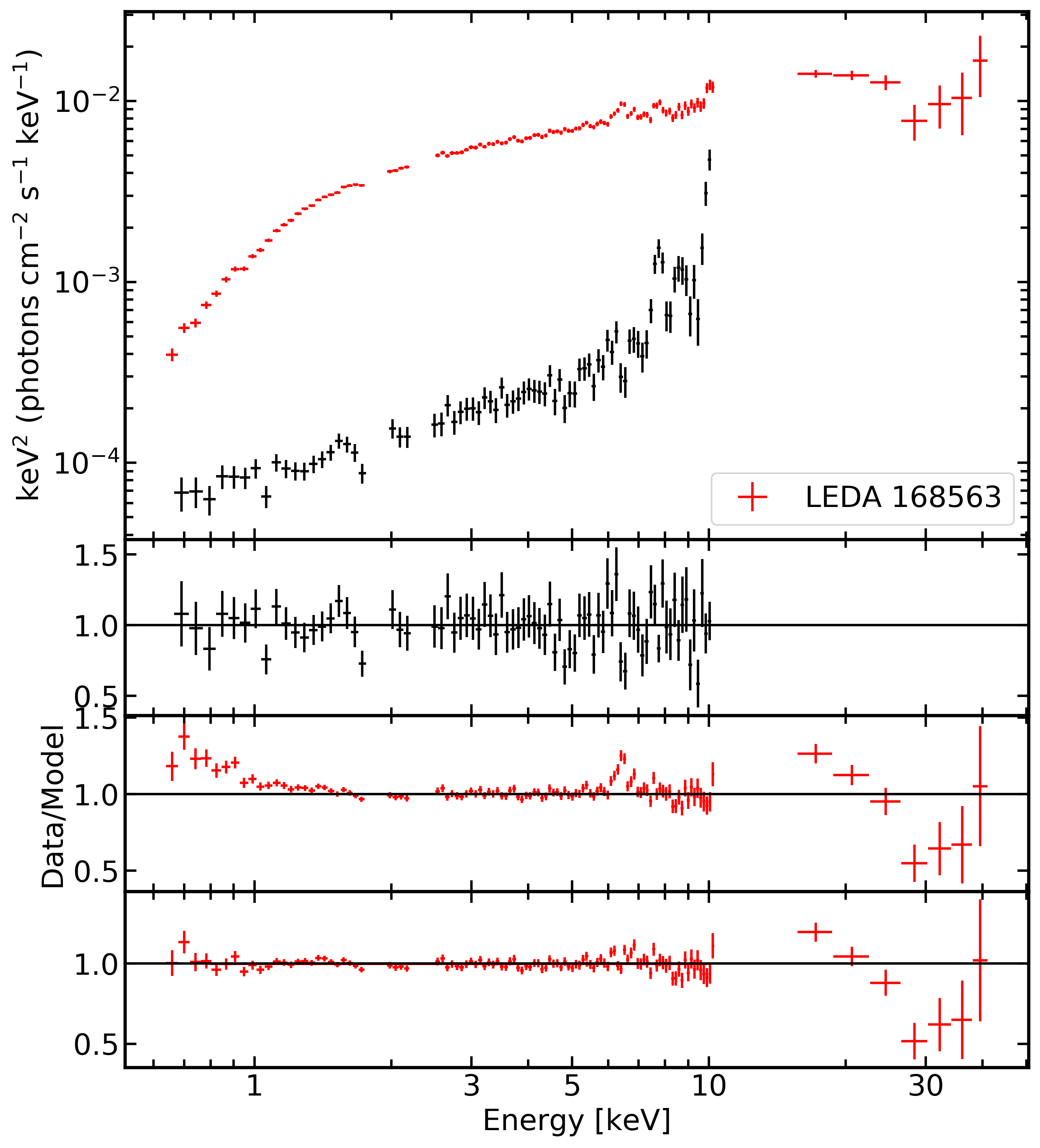}
	\caption{\label{fig:po}continued}
\end{figure*}

\setcounter{figure}{0}

\begin{figure*}
	\centering
	\includegraphics[width=75mm]{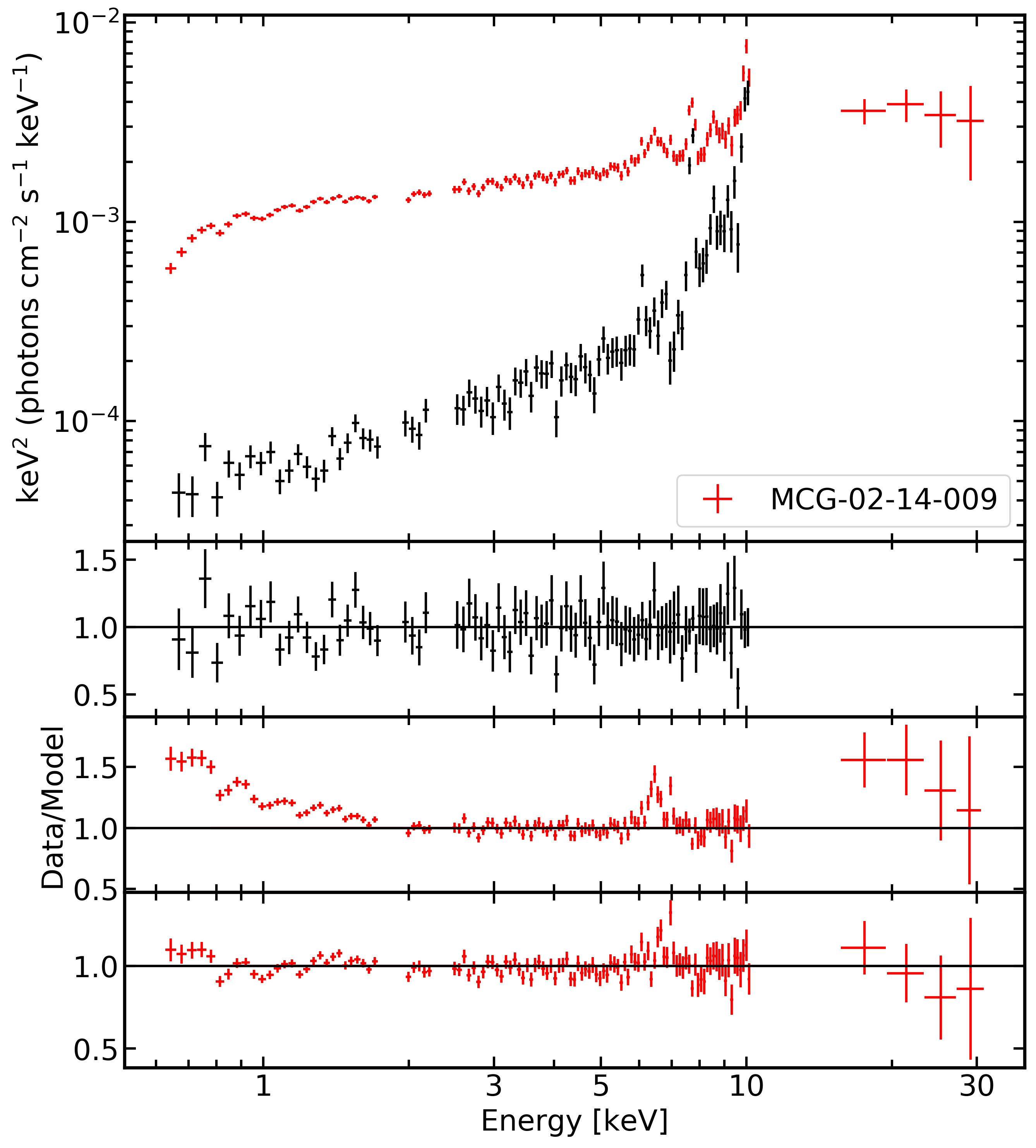}
	\hspace{7mm}
	\includegraphics[width=75mm]{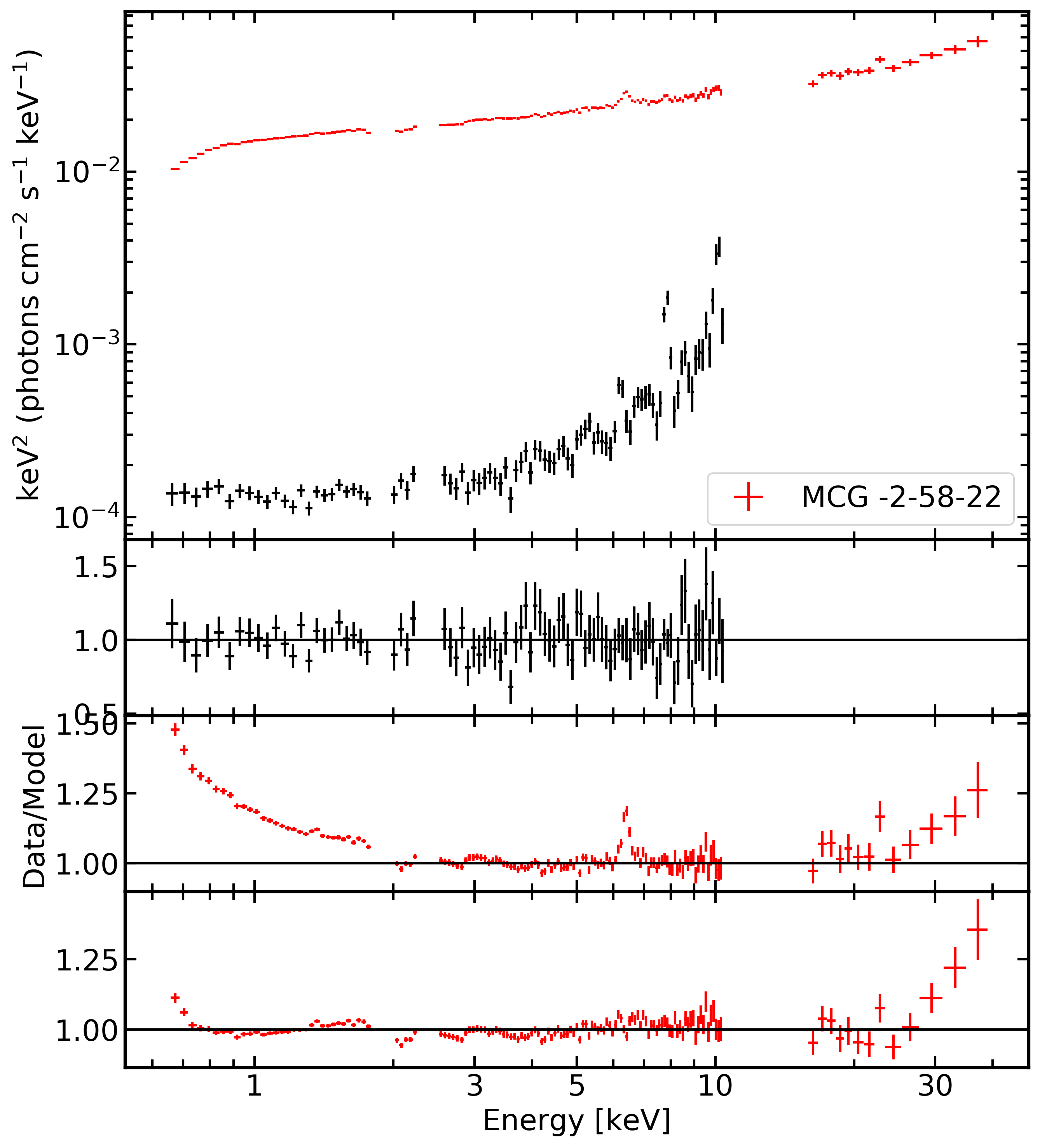}
	\hspace{0mm}
	\includegraphics[width=75mm]{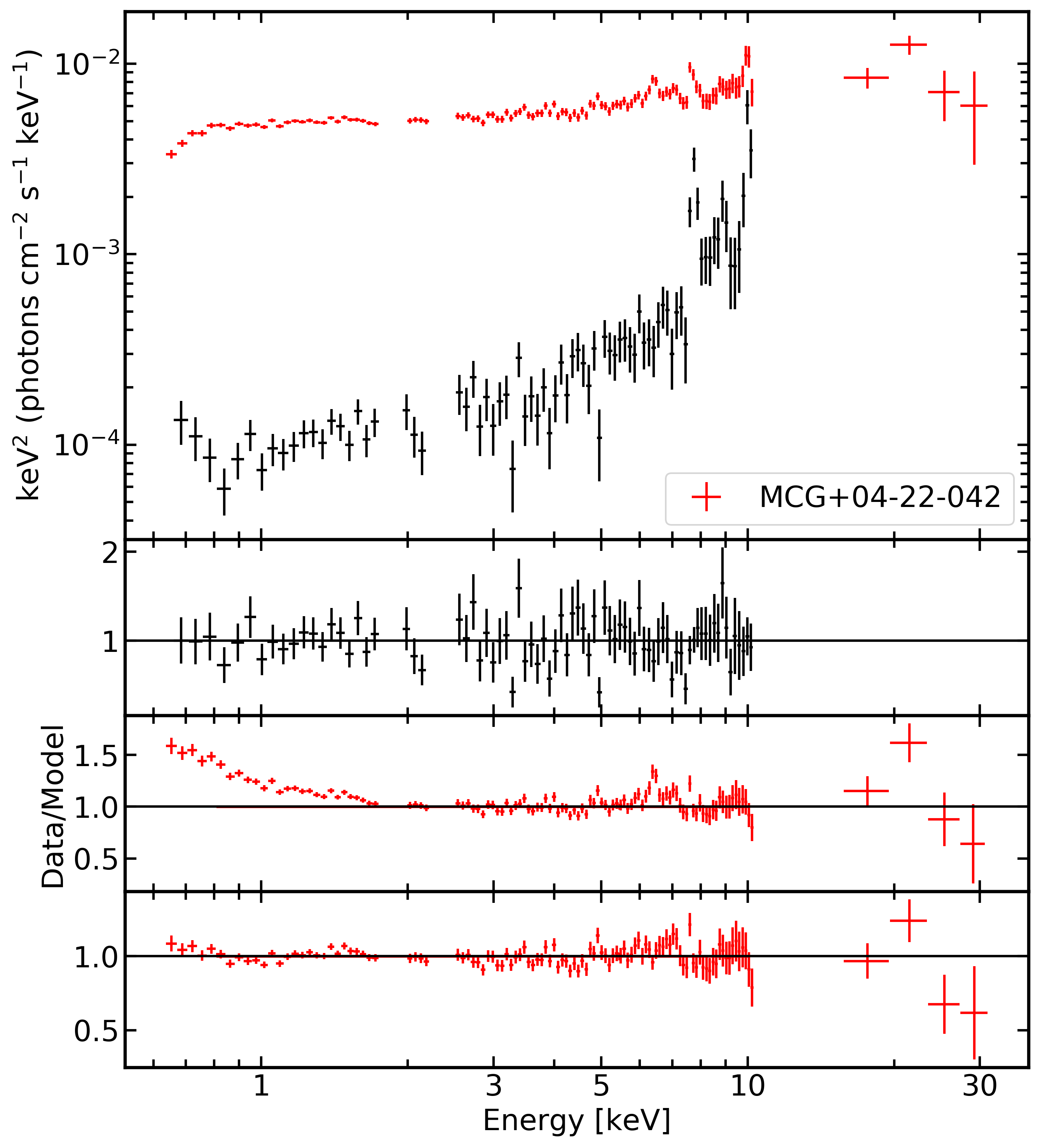}
	\hspace{7mm}
	\includegraphics[width=75mm]{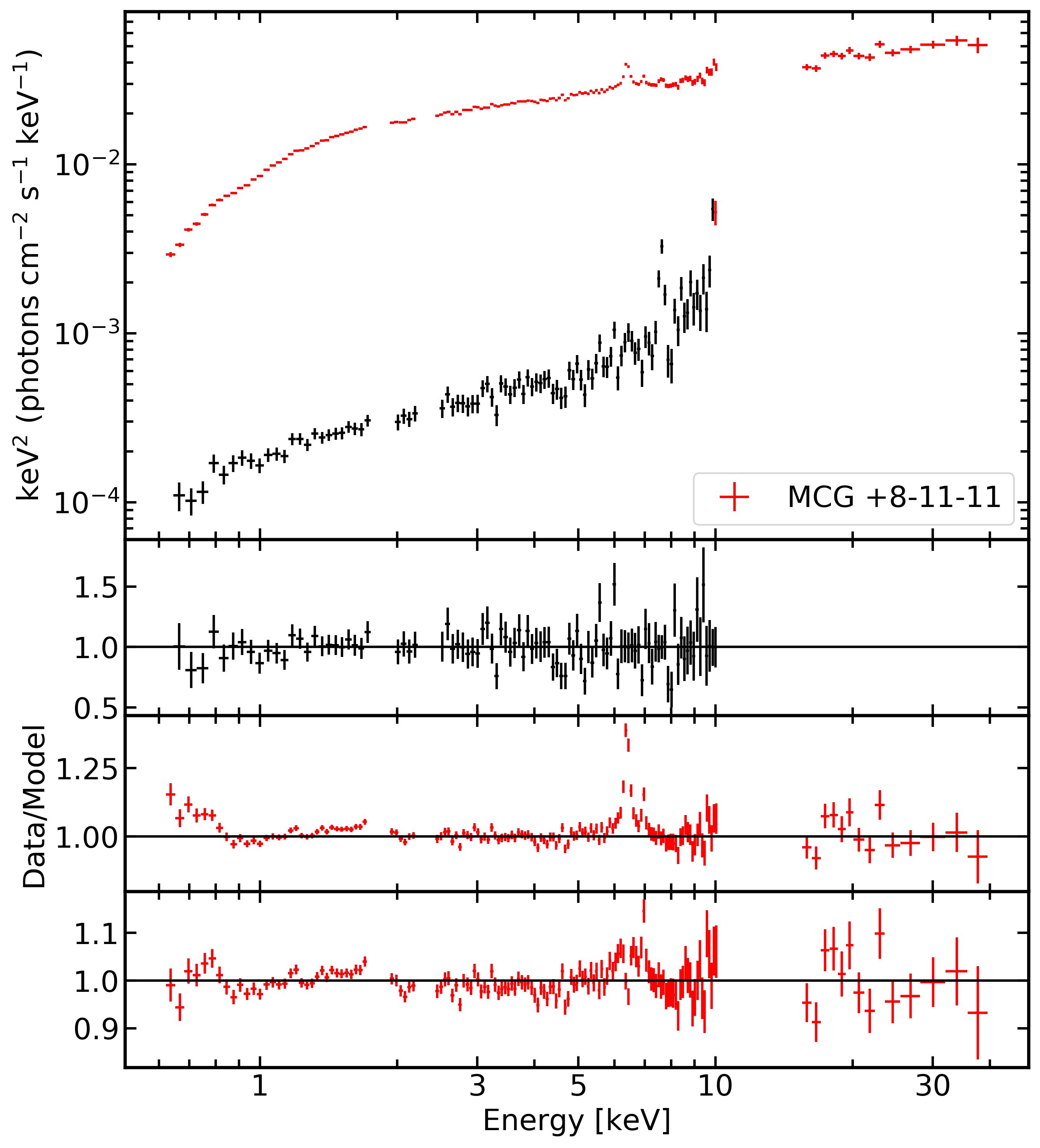}
	\caption{\label{fig:po}continued}
\end{figure*}	

\setcounter{figure}{0}

\begin{figure*}
	\centering	
	\includegraphics[width=75mm]{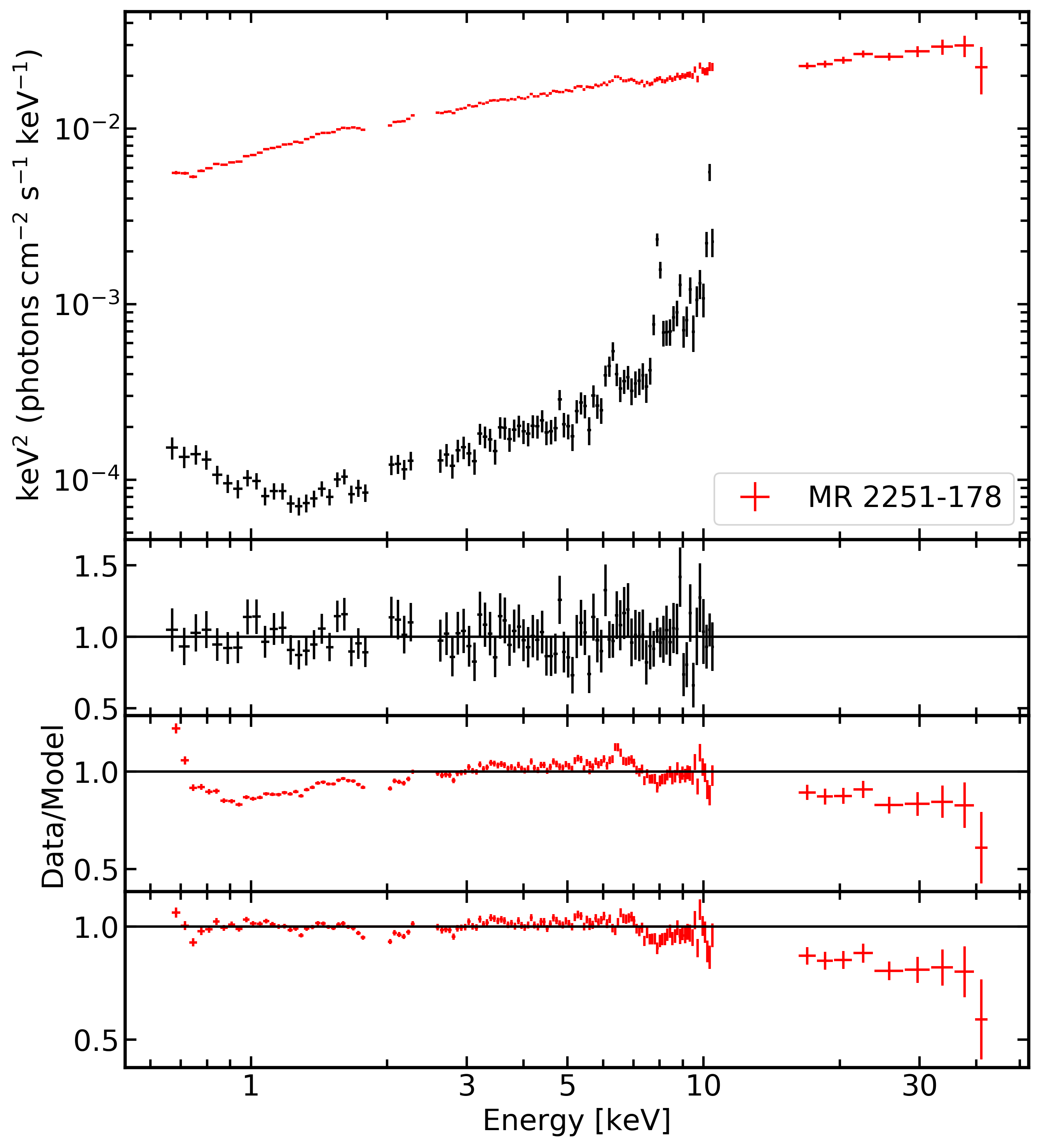}
	\hspace{7mm}
	\includegraphics[width=75mm]{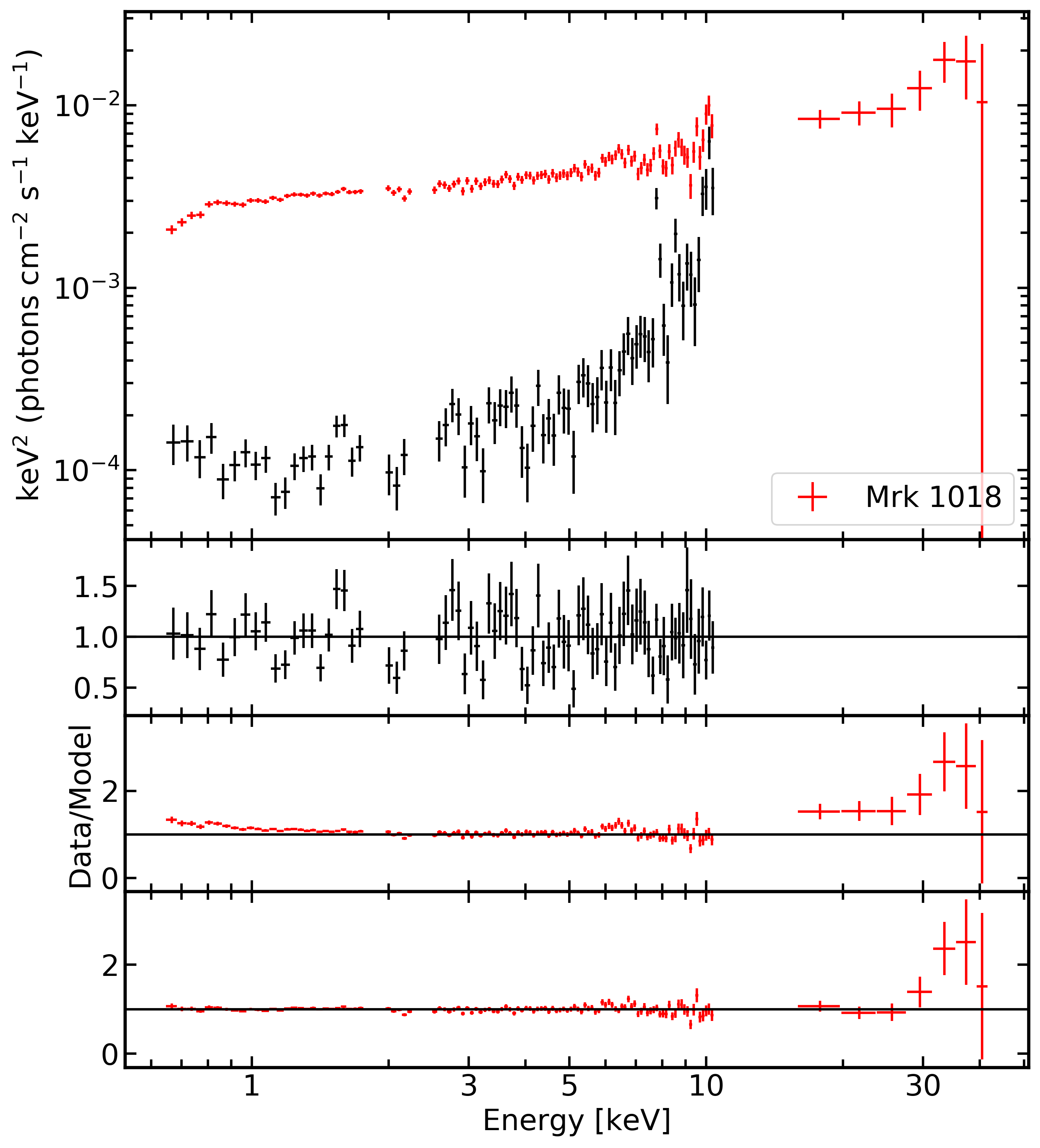}
	\hspace{0mm}
	\includegraphics[width=75mm]{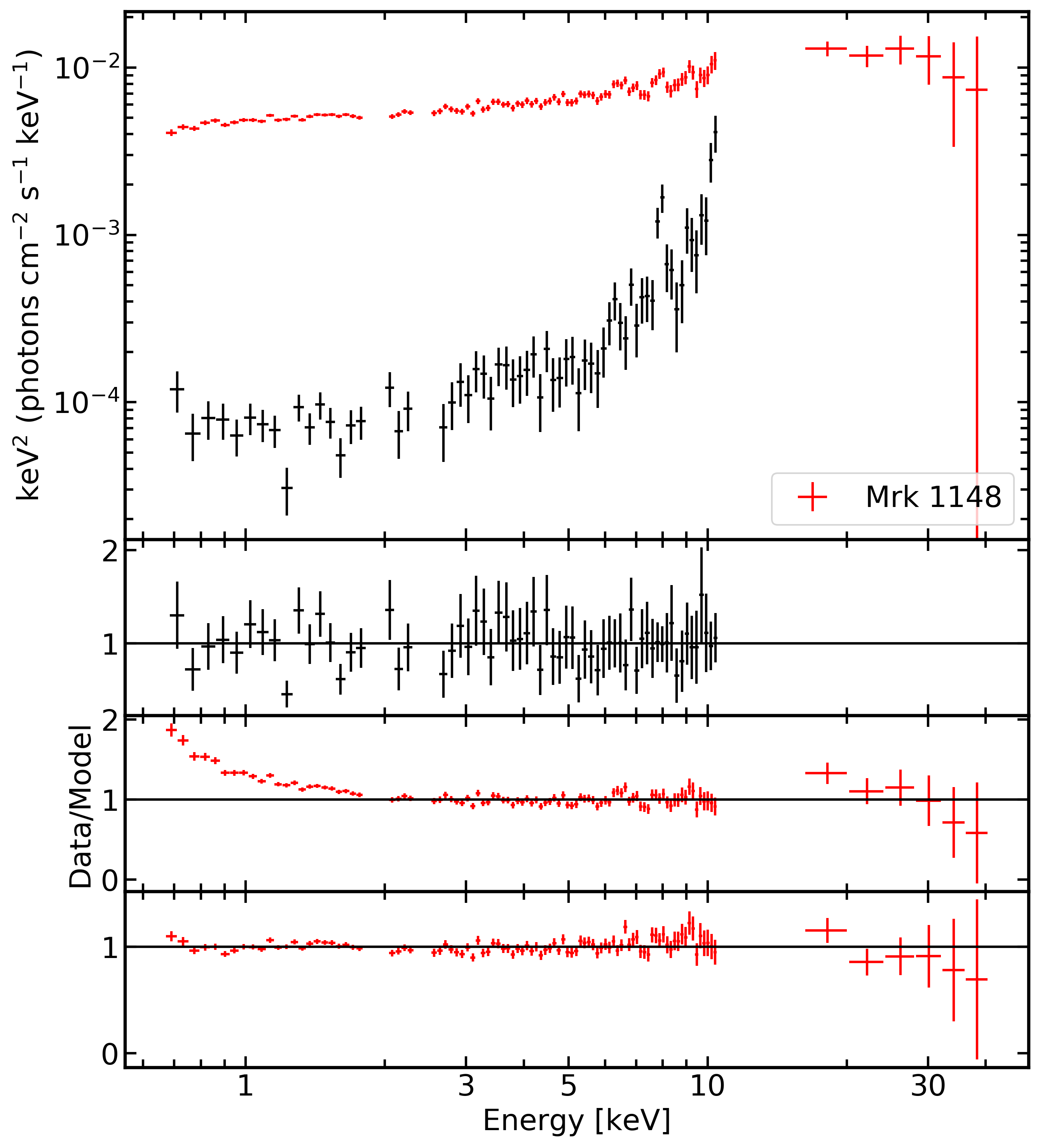}
	\hspace{7mm}
	\includegraphics[width=75mm]{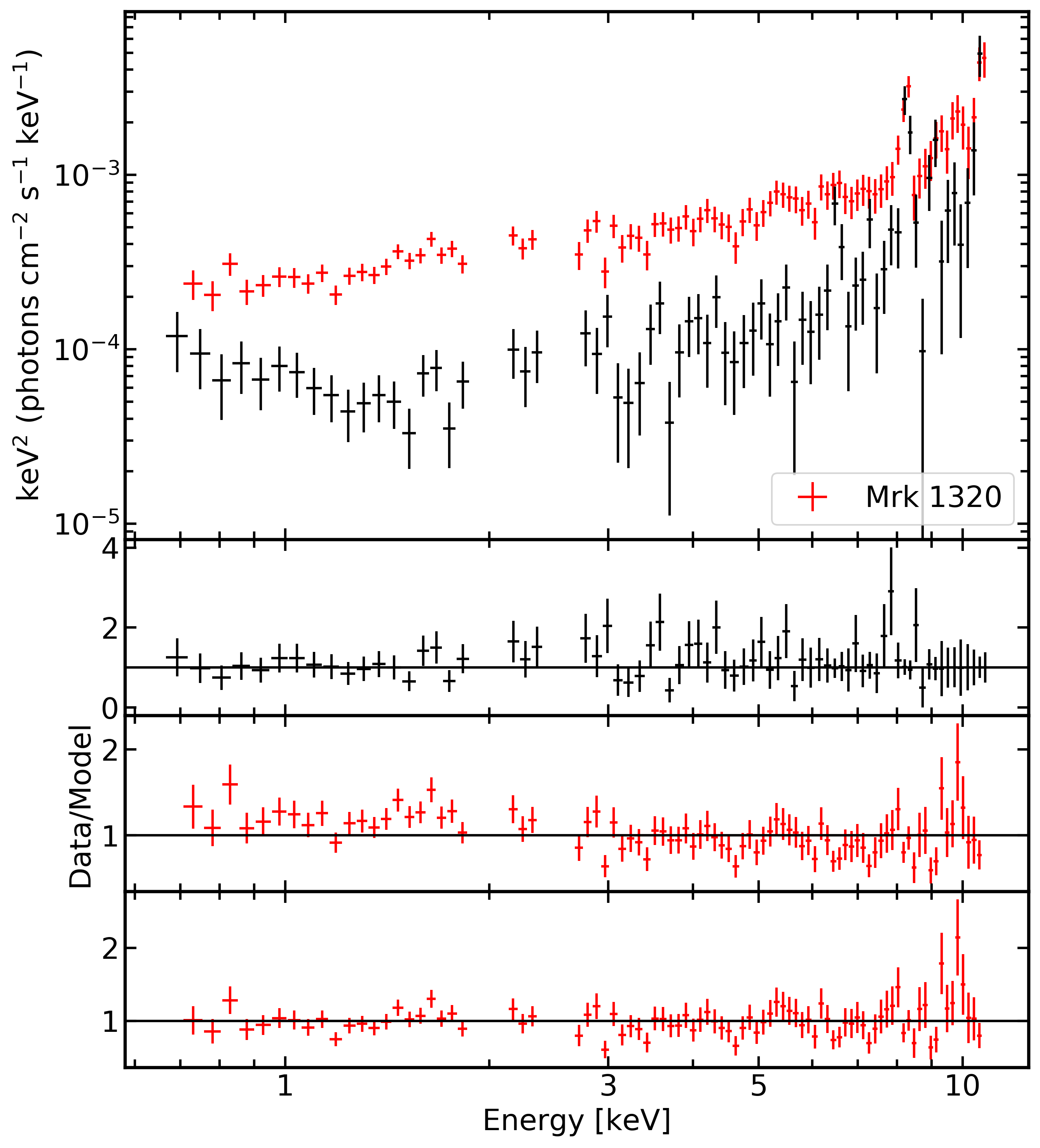}
	\caption{\label{fig:po}continued}
\end{figure*}

\setcounter{figure}{0}

\begin{figure*}
	\centering
	\includegraphics[width=75mm]{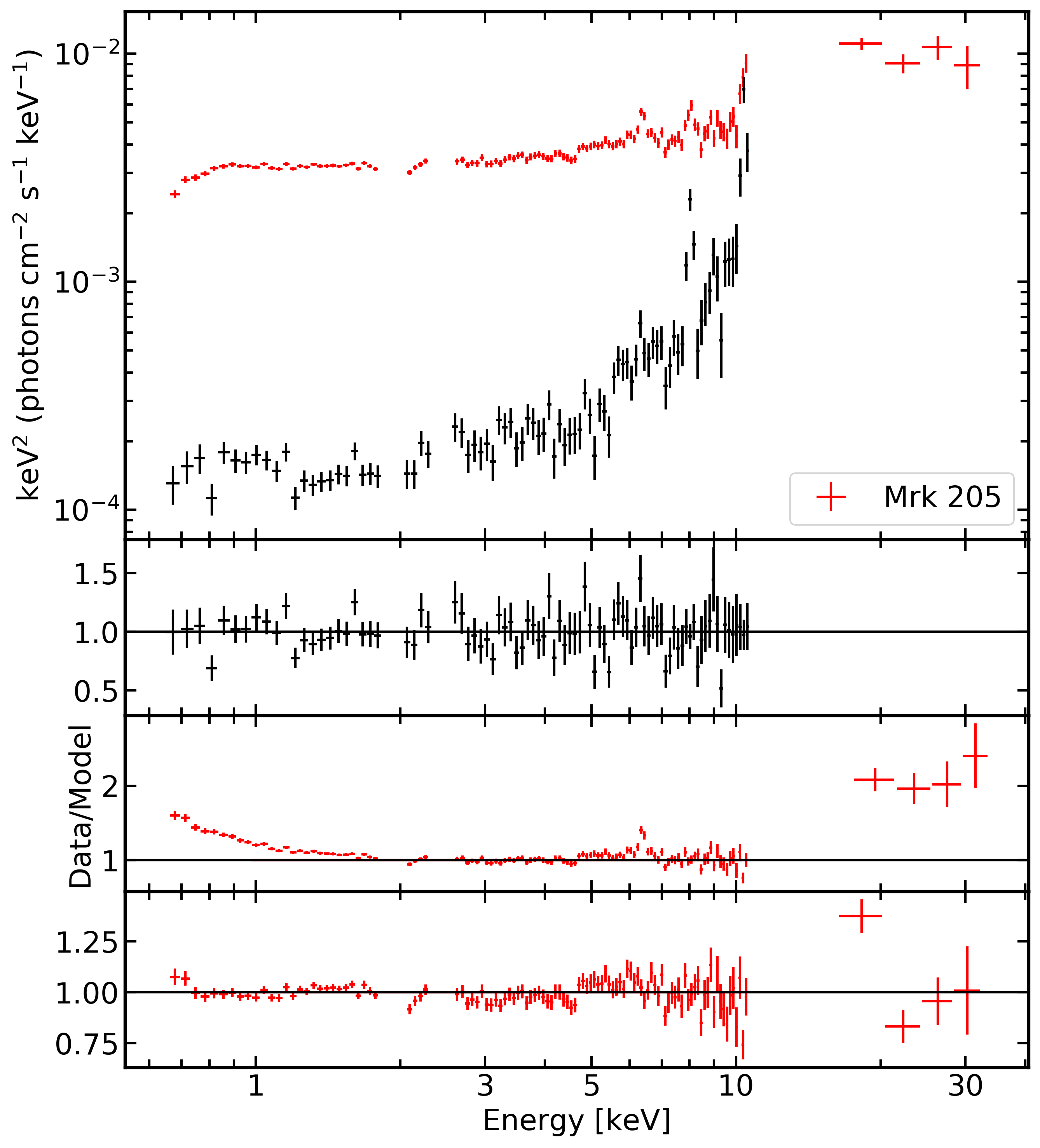}
	\hspace{7mm}
	\includegraphics[width=75mm]{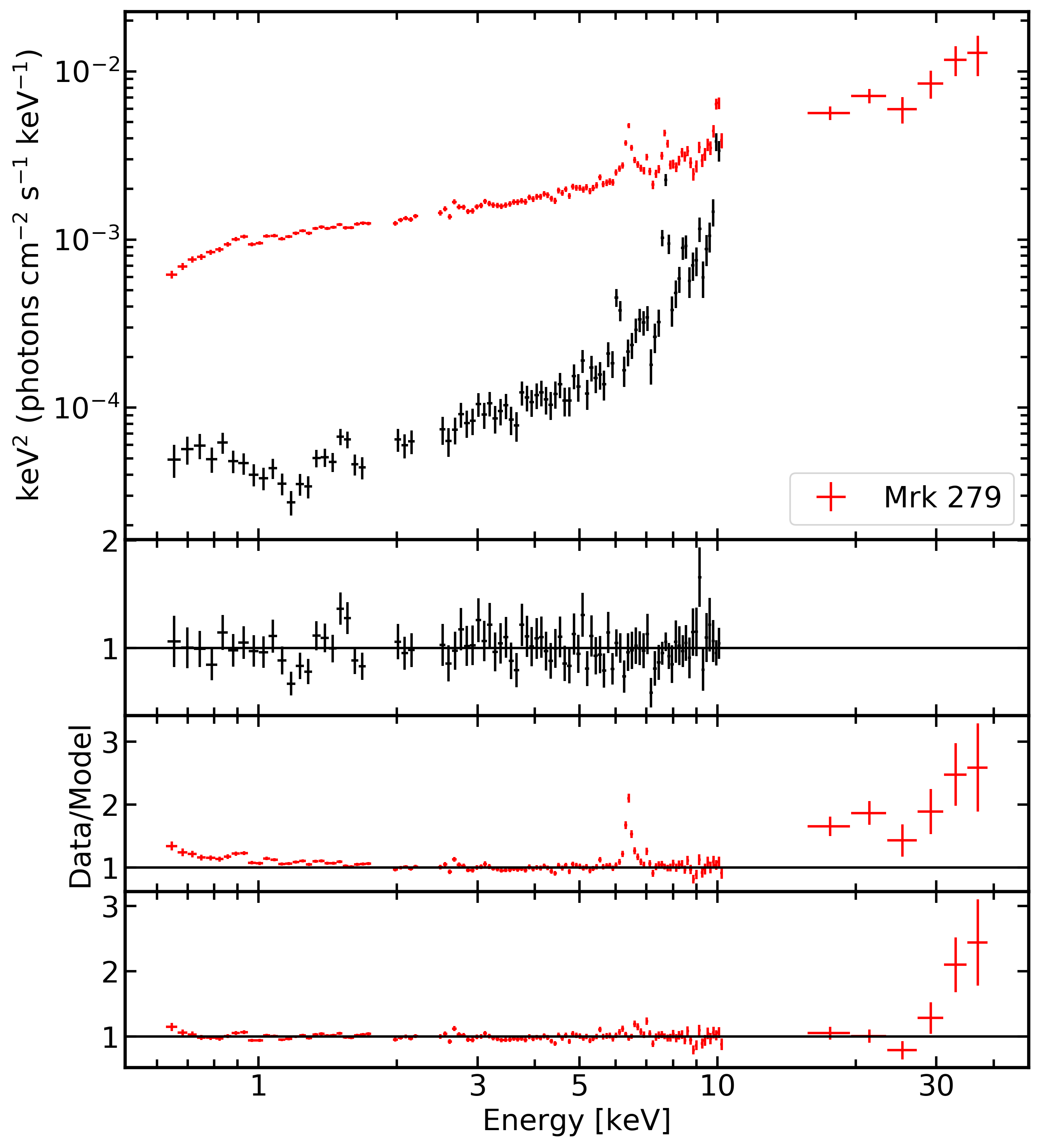}
	\hspace{0mm}
	\includegraphics[width=75mm]{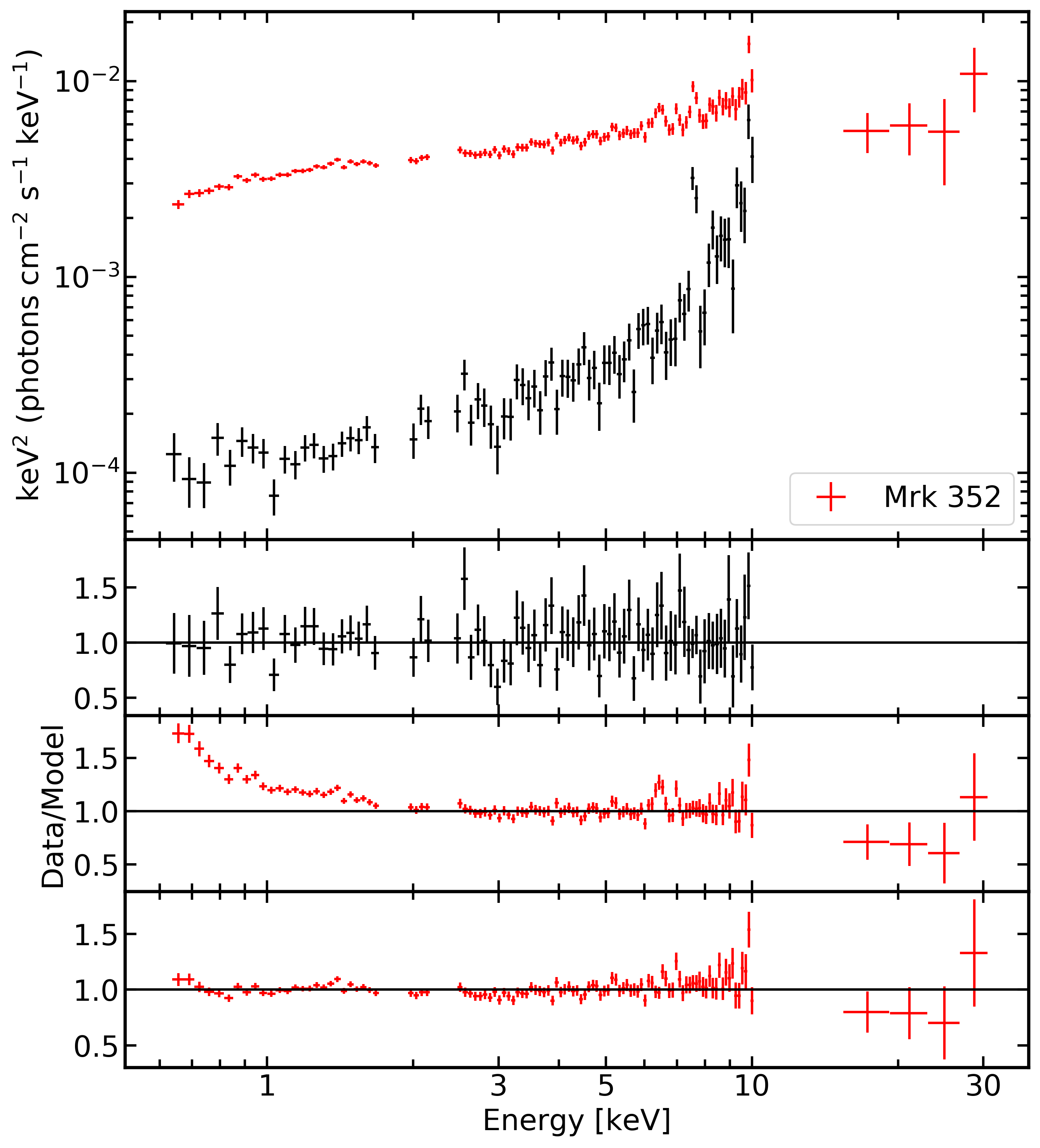}
	\hspace{7mm}
	\includegraphics[width=75mm]{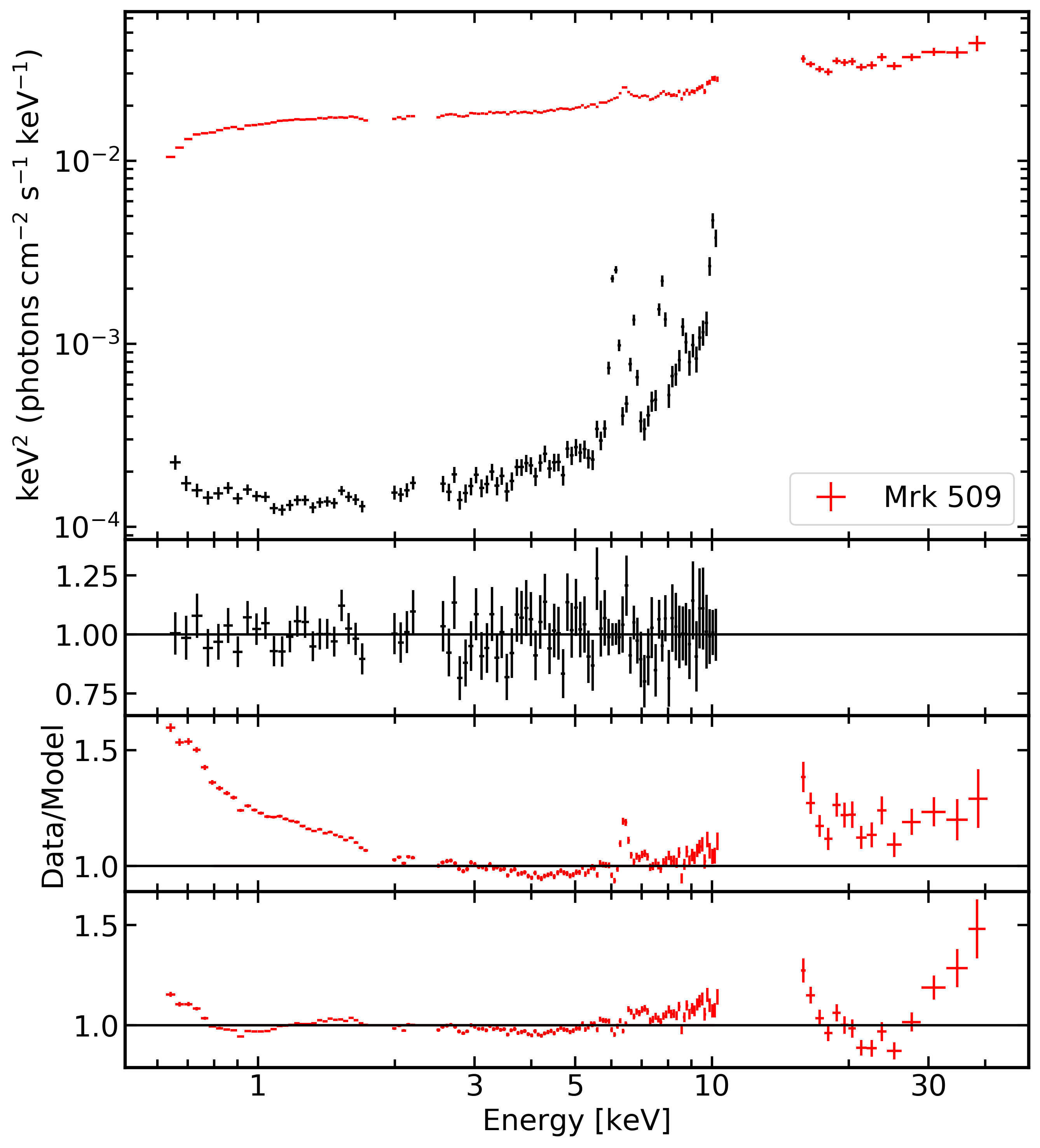}
	\caption{\label{fig:po}continued}
\end{figure*}

\setcounter{figure}{0}

\begin{figure*}
	\centering
	\includegraphics[width=75mm]{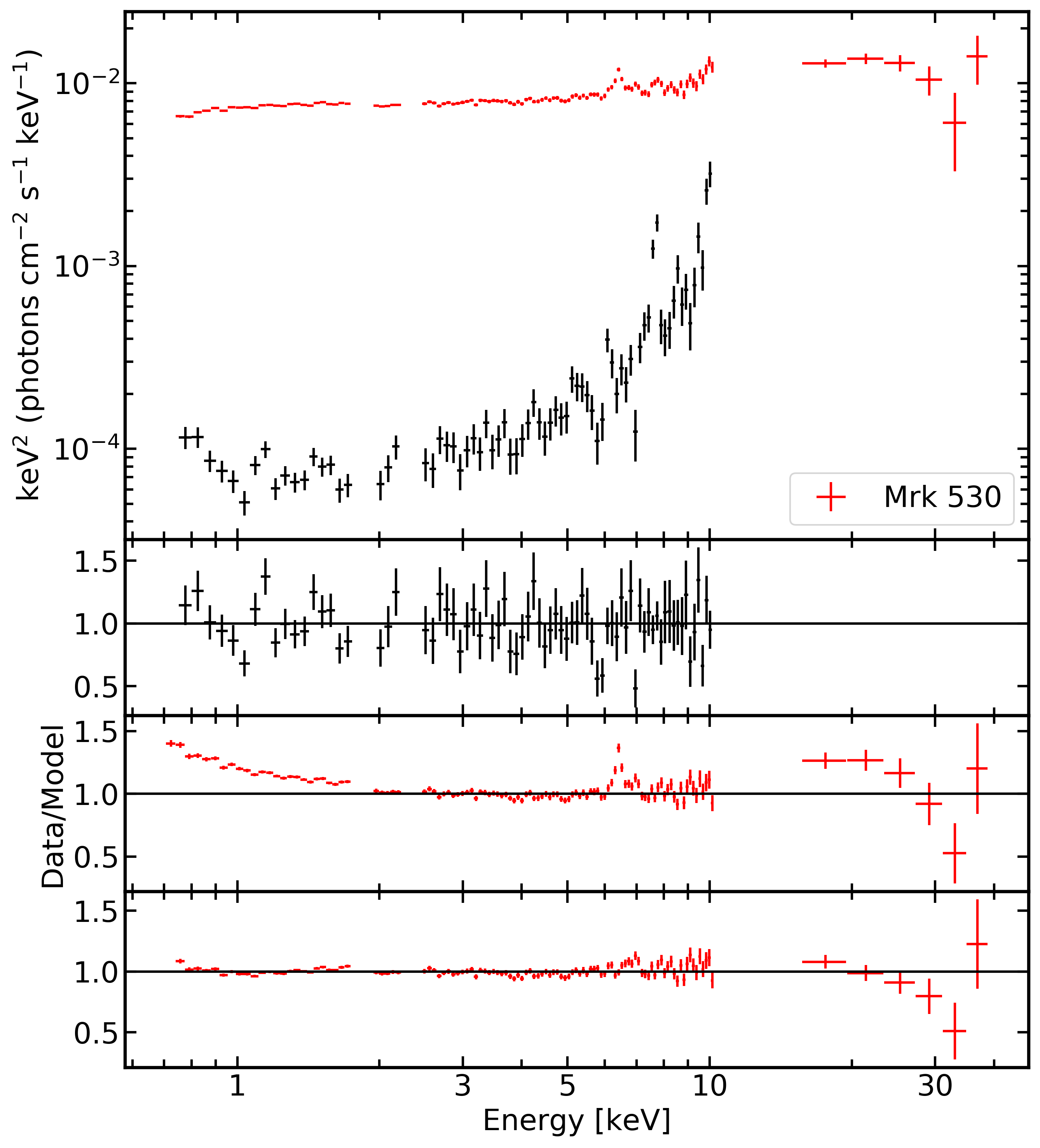}
	\hspace{7mm}
	\includegraphics[width=75mm]{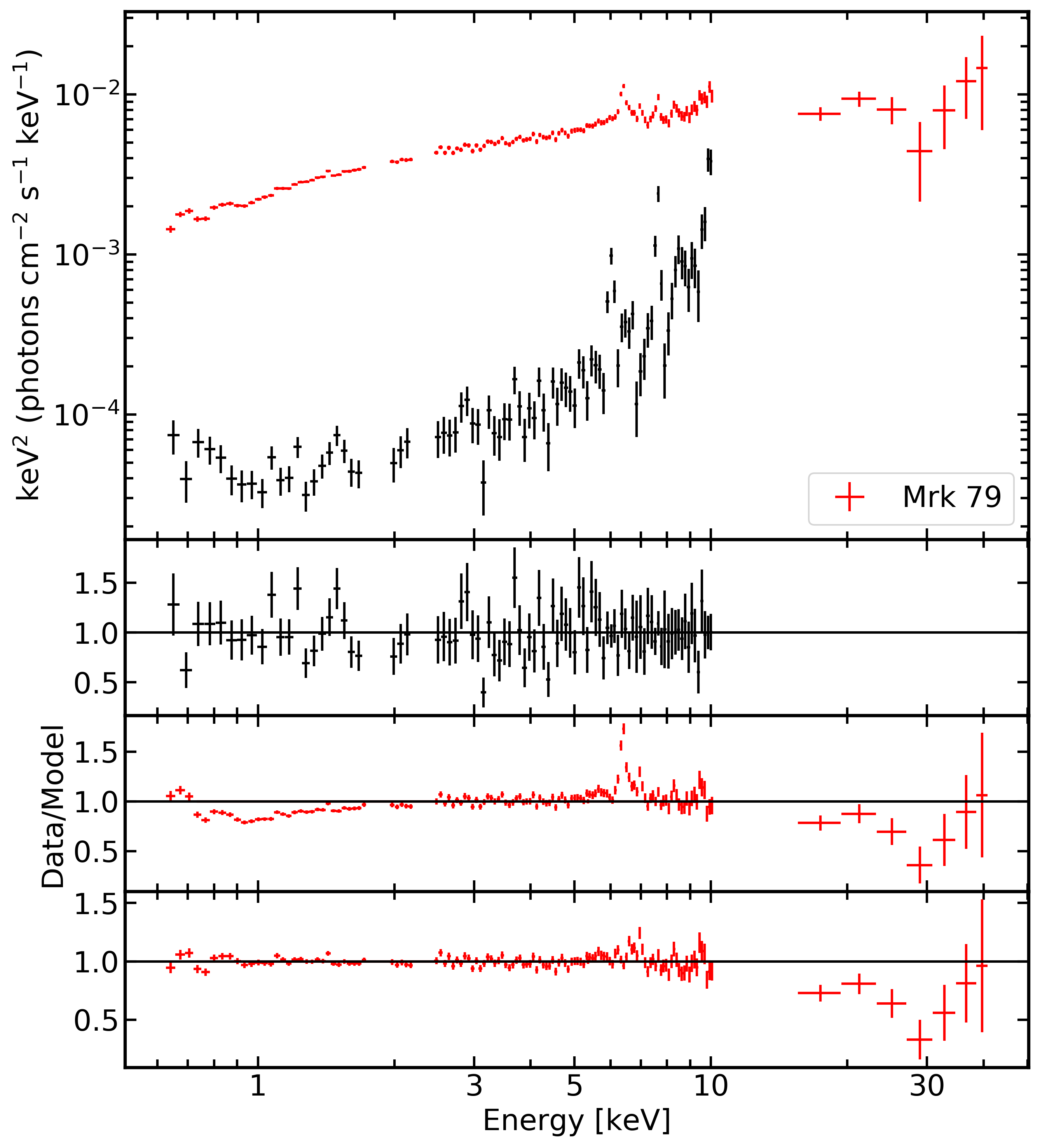}
	\hspace{0mm}
	\includegraphics[width=75mm]{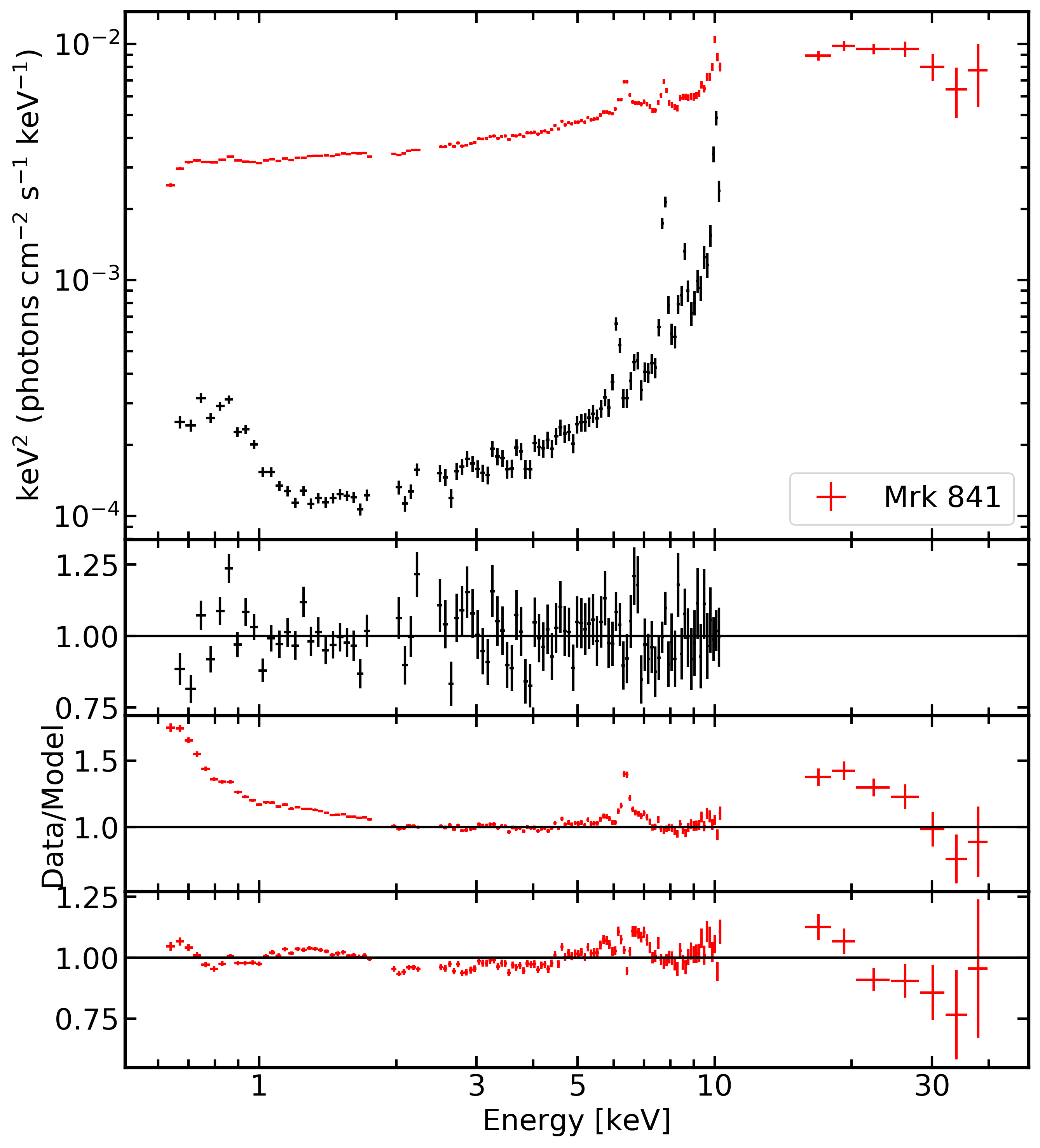}
	\hspace{7mm}
	\includegraphics[width=75mm]{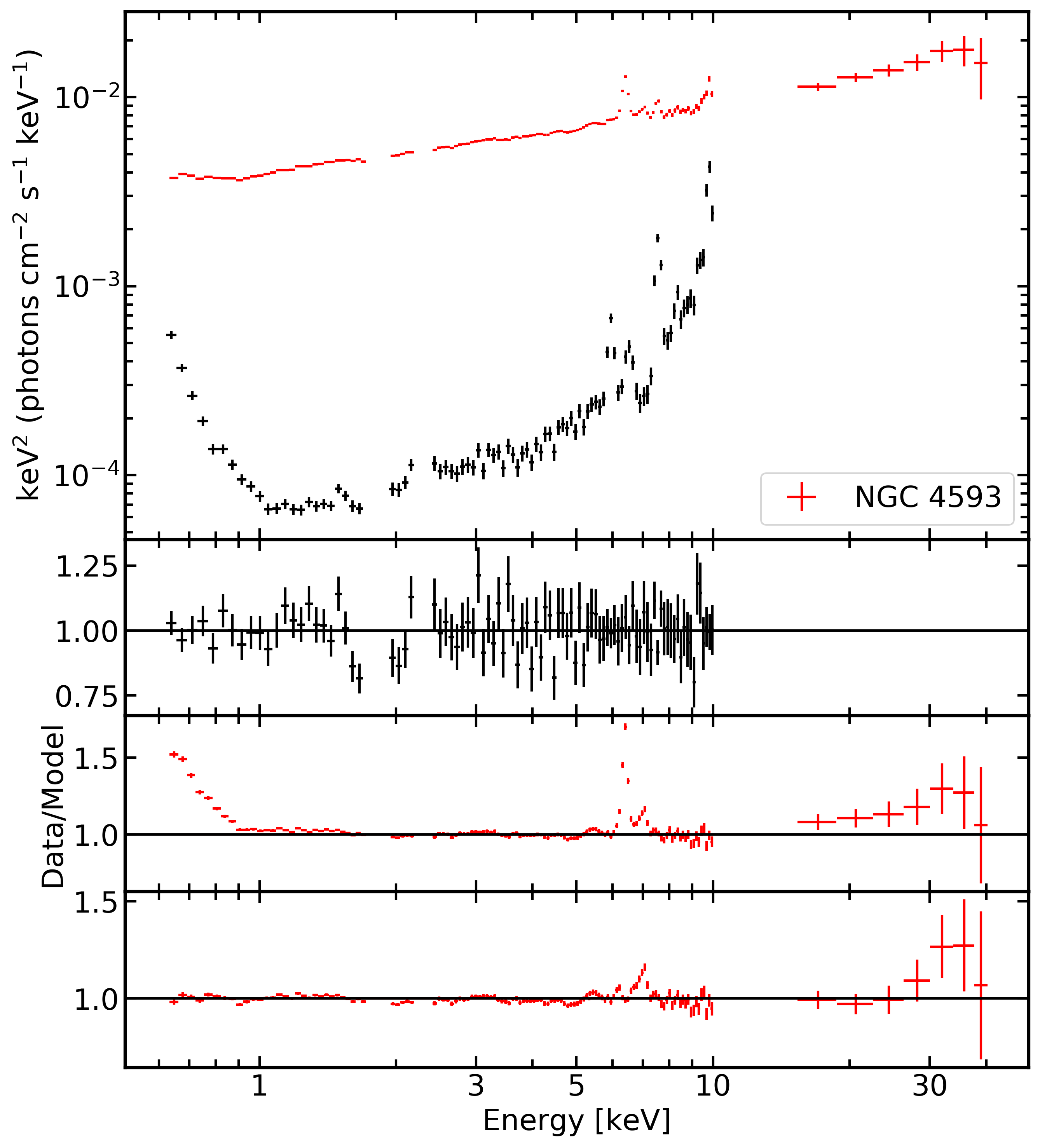}
	\caption{\label{fig:po}continued}
\end{figure*}

\setcounter{figure}{0}

\begin{figure*}
	\centering
	\includegraphics[width=75mm]{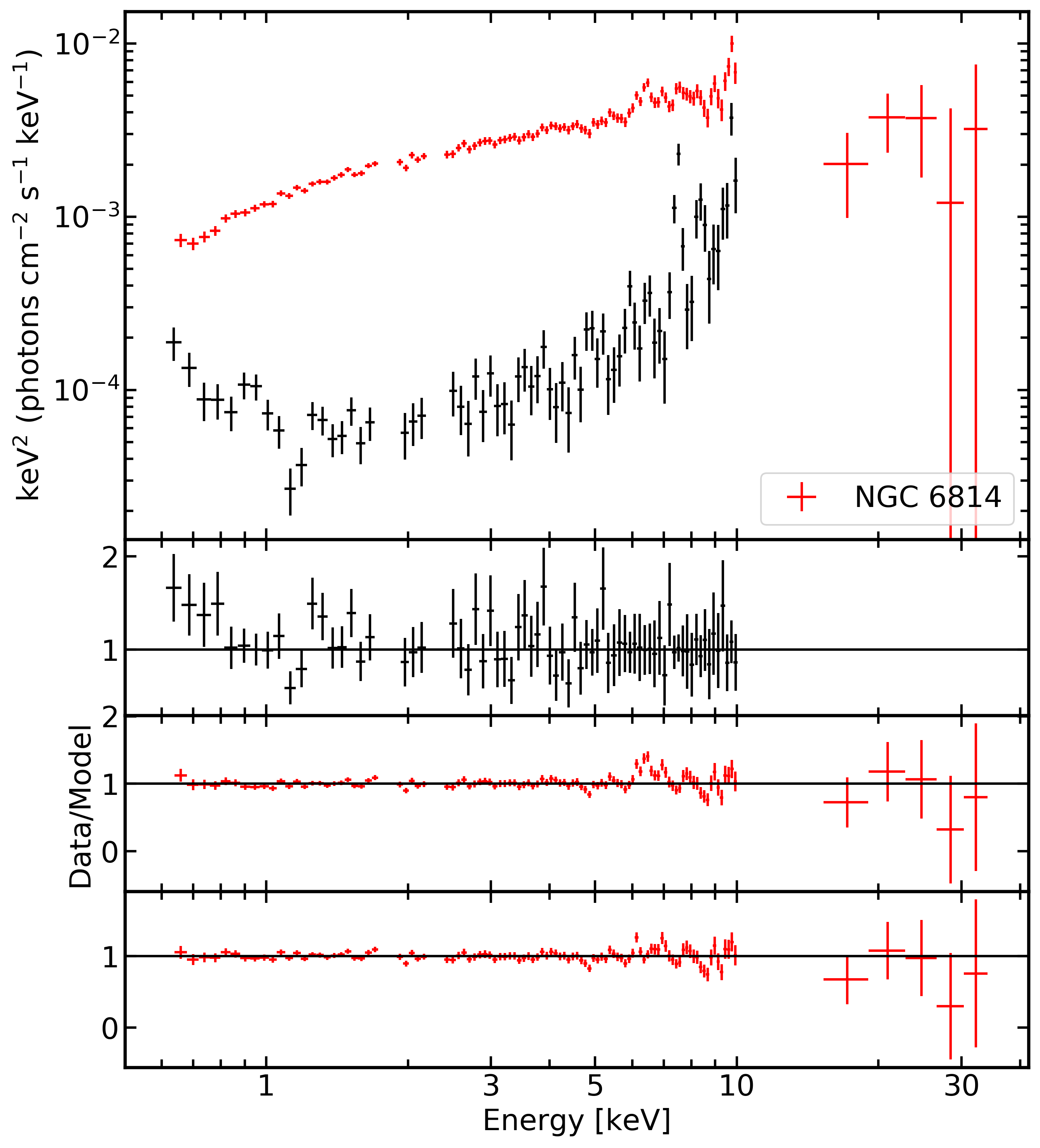}
	\hspace{7mm}
	\includegraphics[width=75mm]{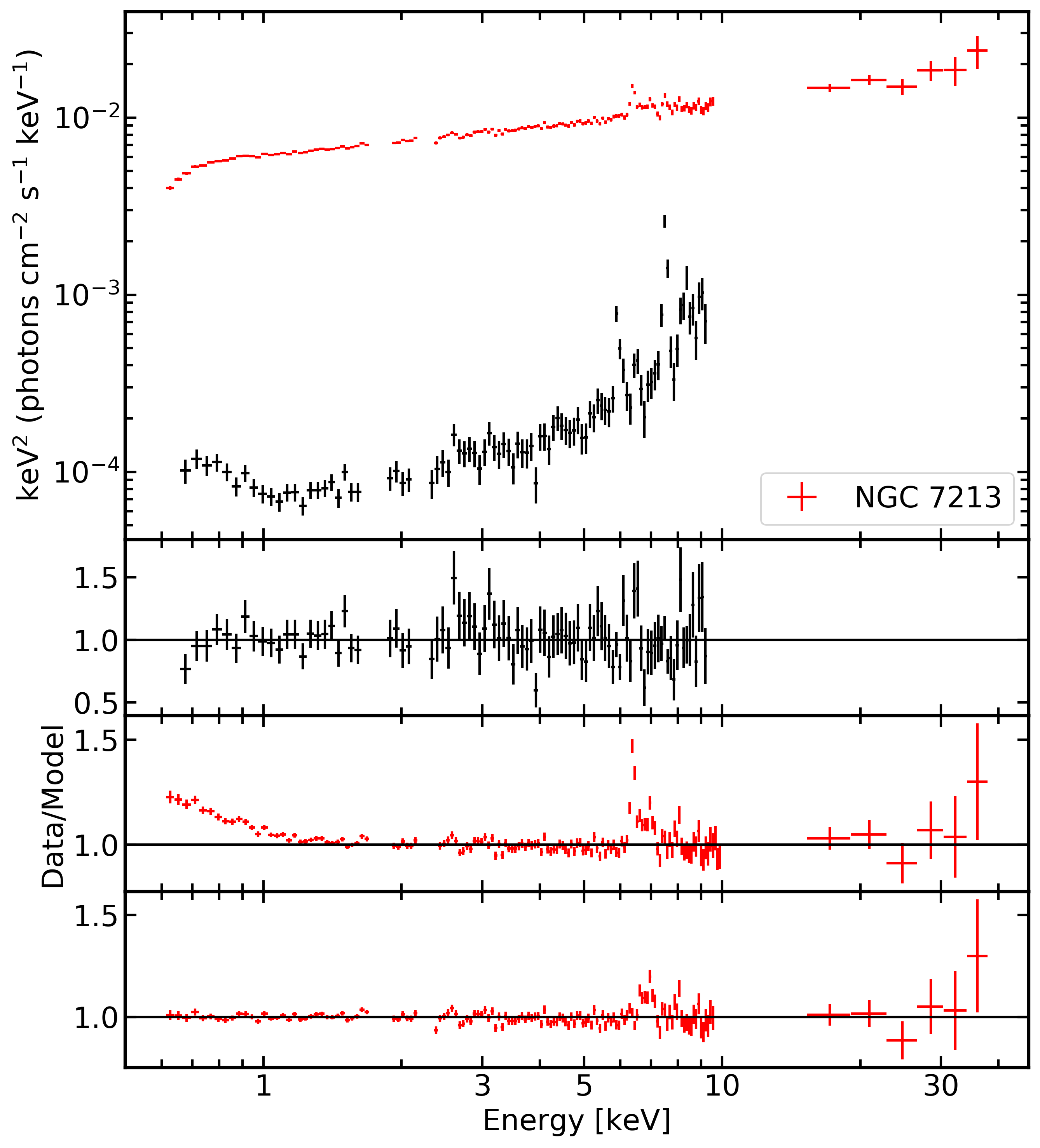}
	\hspace{0mm}
	\includegraphics[width=75mm]{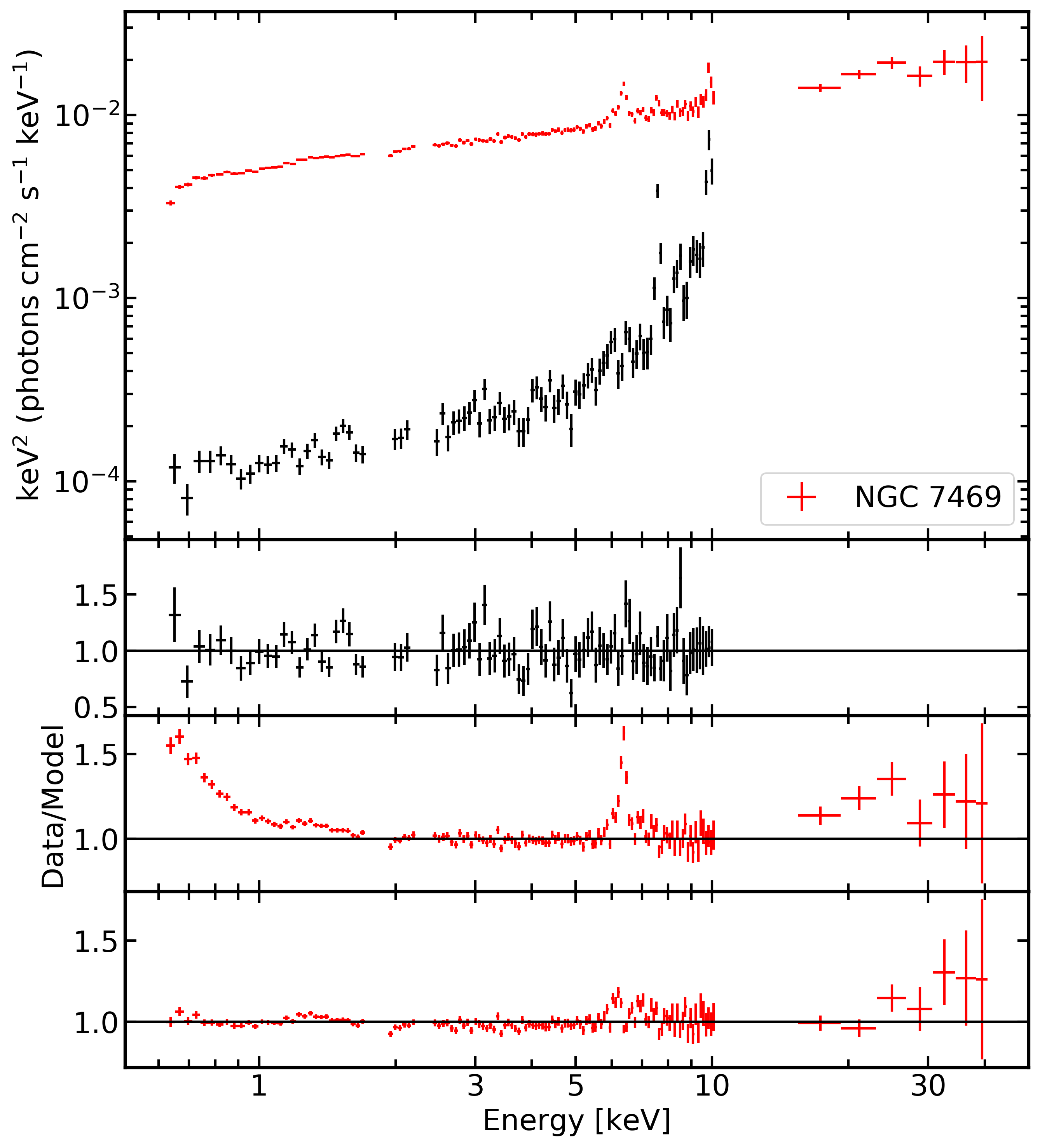}
	\hspace{7mm}
	\includegraphics[width=75mm]{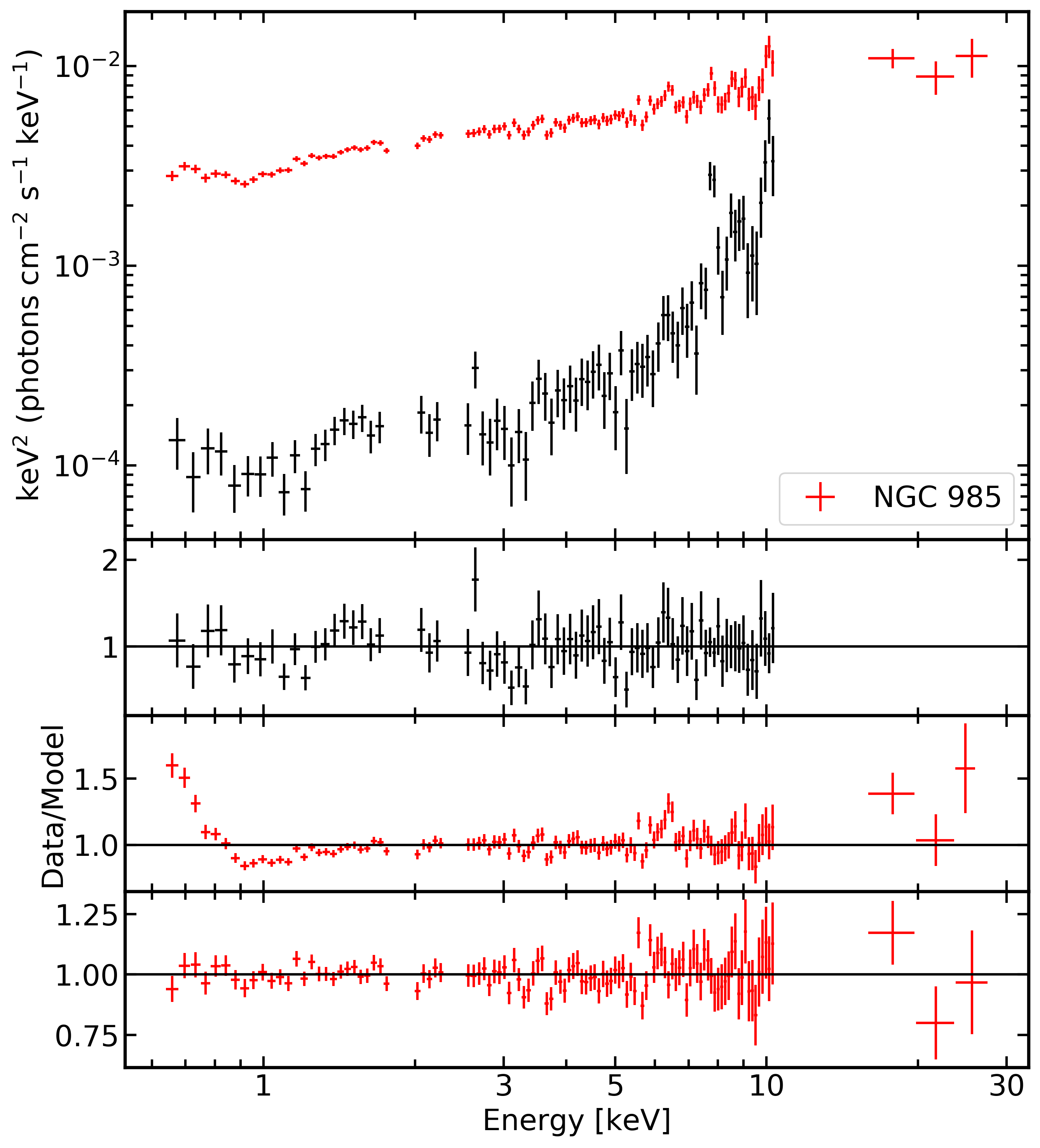}
	\caption{\label{fig:po}continued}
\end{figure*}

\setcounter{figure}{0}

\begin{figure*}
	\centering
	\includegraphics[width=75mm]{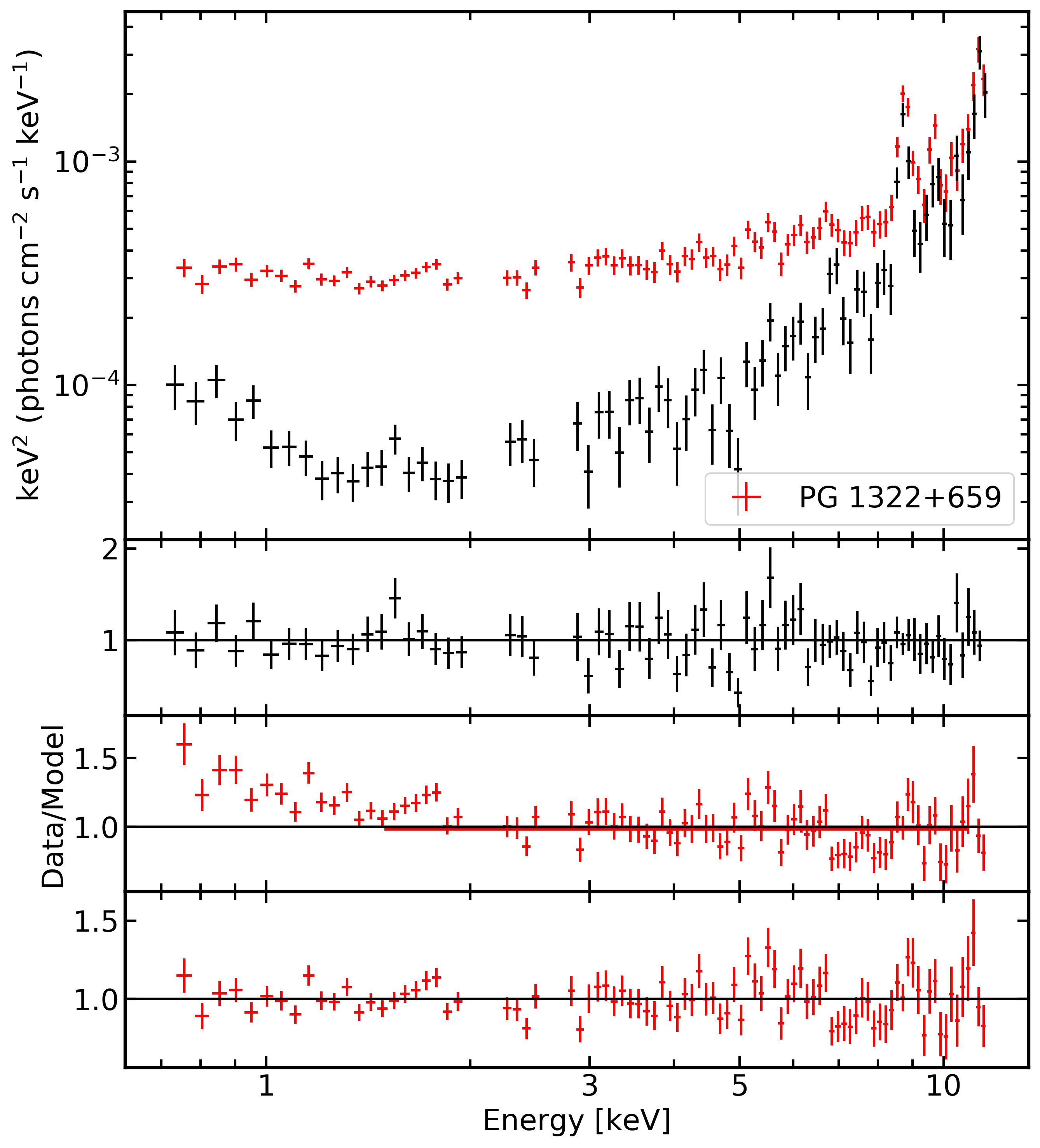}
	\hspace{7mm}
	\includegraphics[width=75mm]{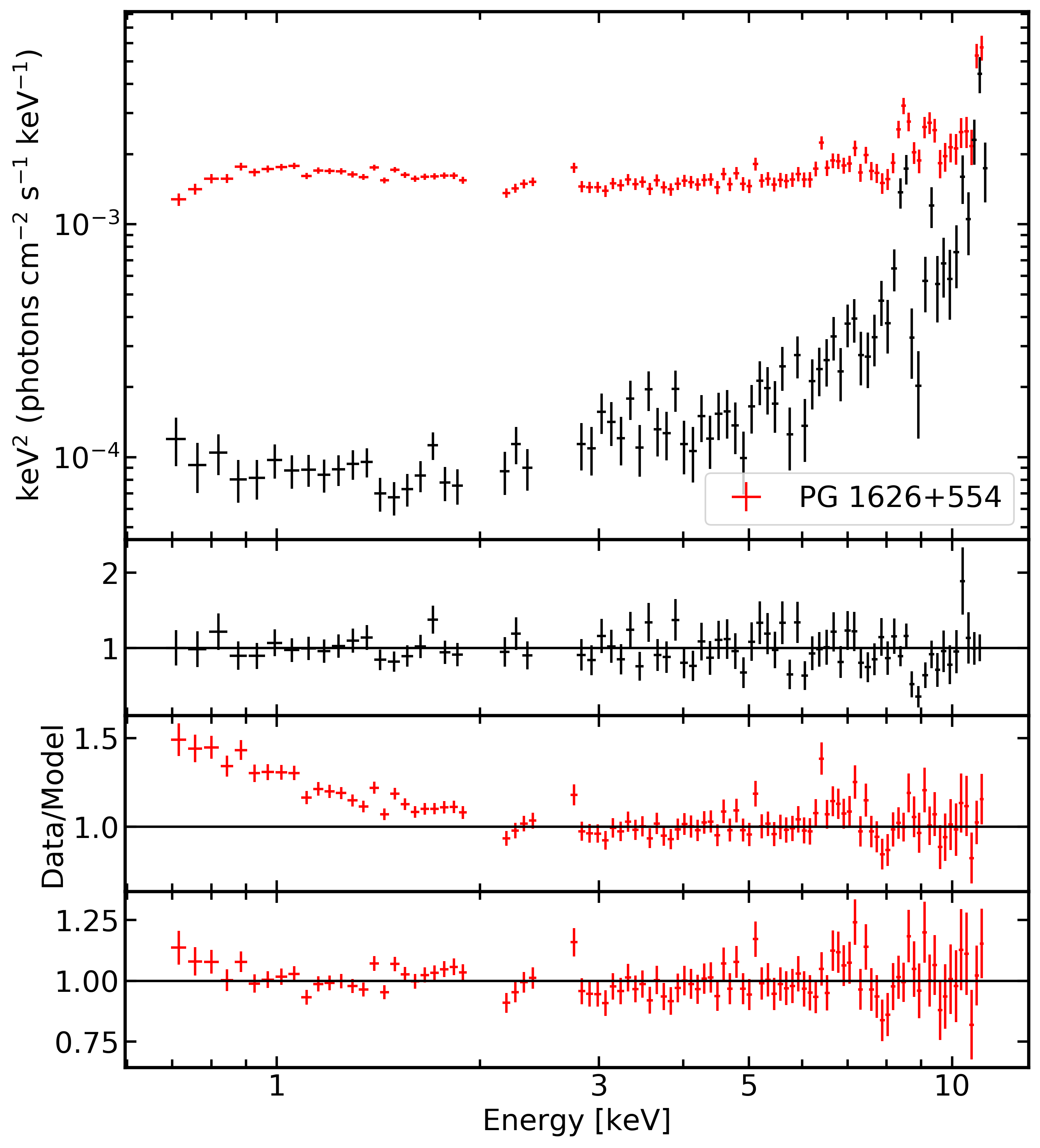}
	\hspace{0mm}
	\includegraphics[width=75mm]{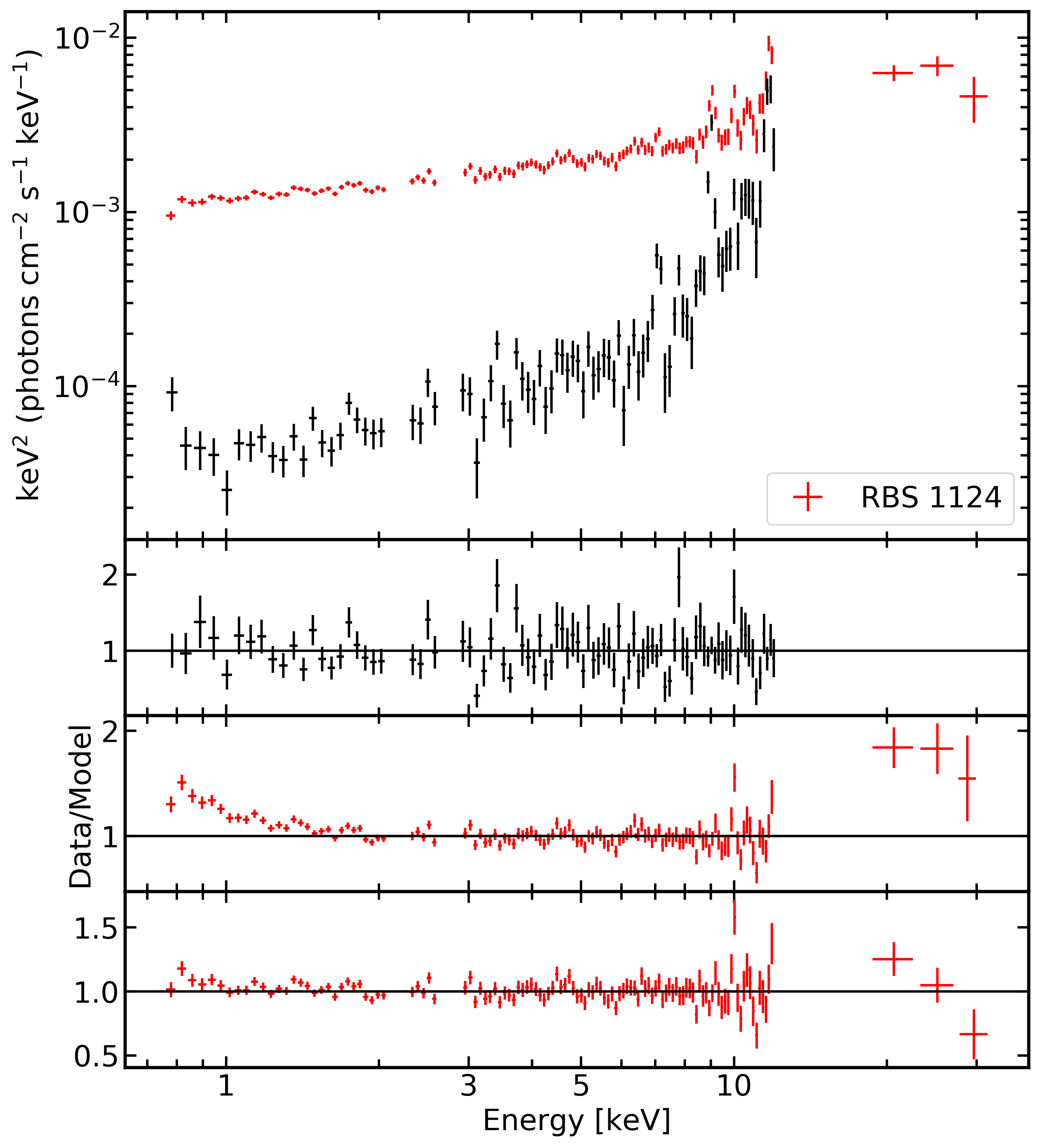}
	\hspace{7mm}
	\includegraphics[width=75mm]{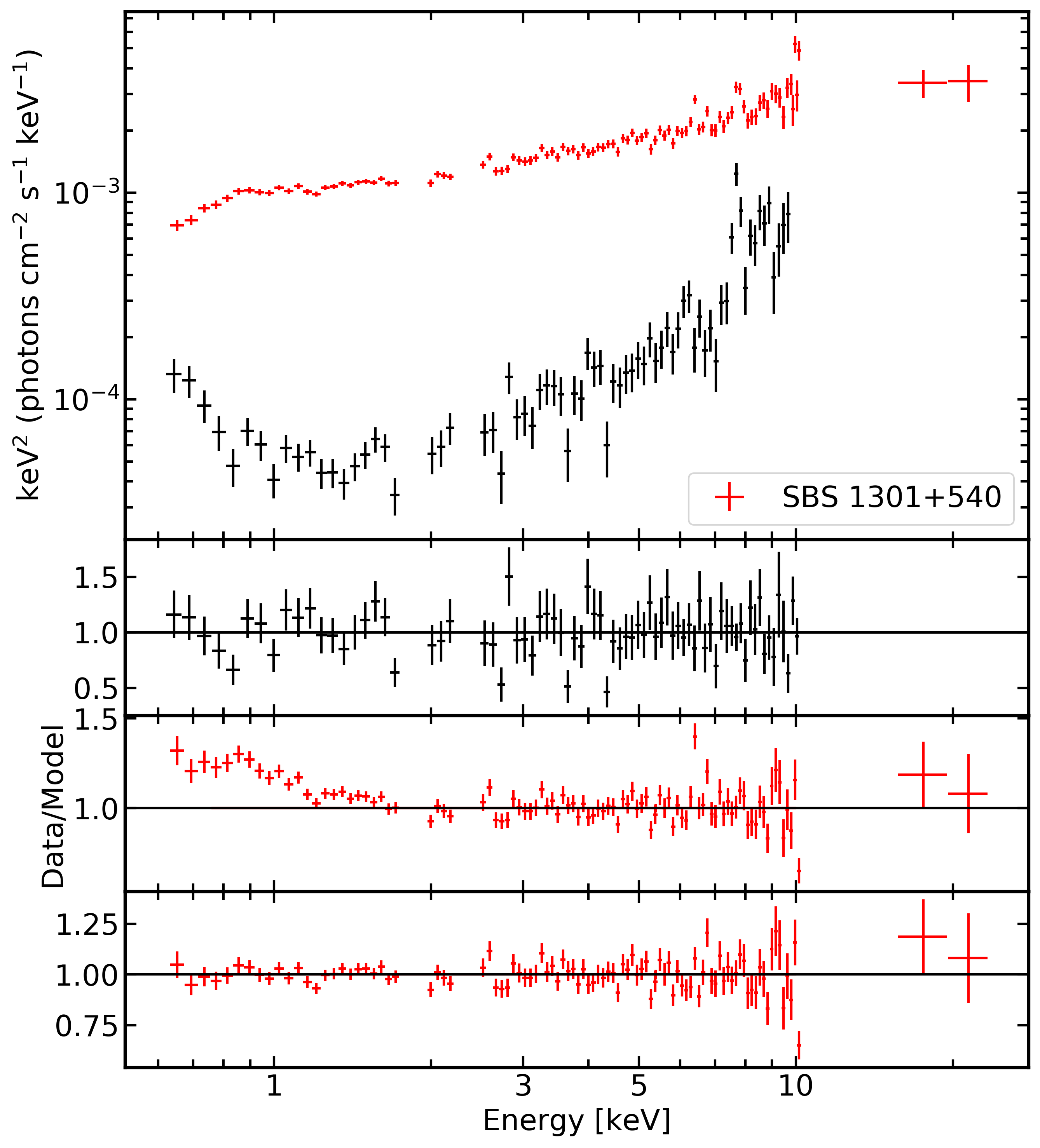}
	\caption{\label{fig:po}continued}
\end{figure*}

\setcounter{figure}{0}

\begin{figure*}
	\centering
	\includegraphics[width=75mm]{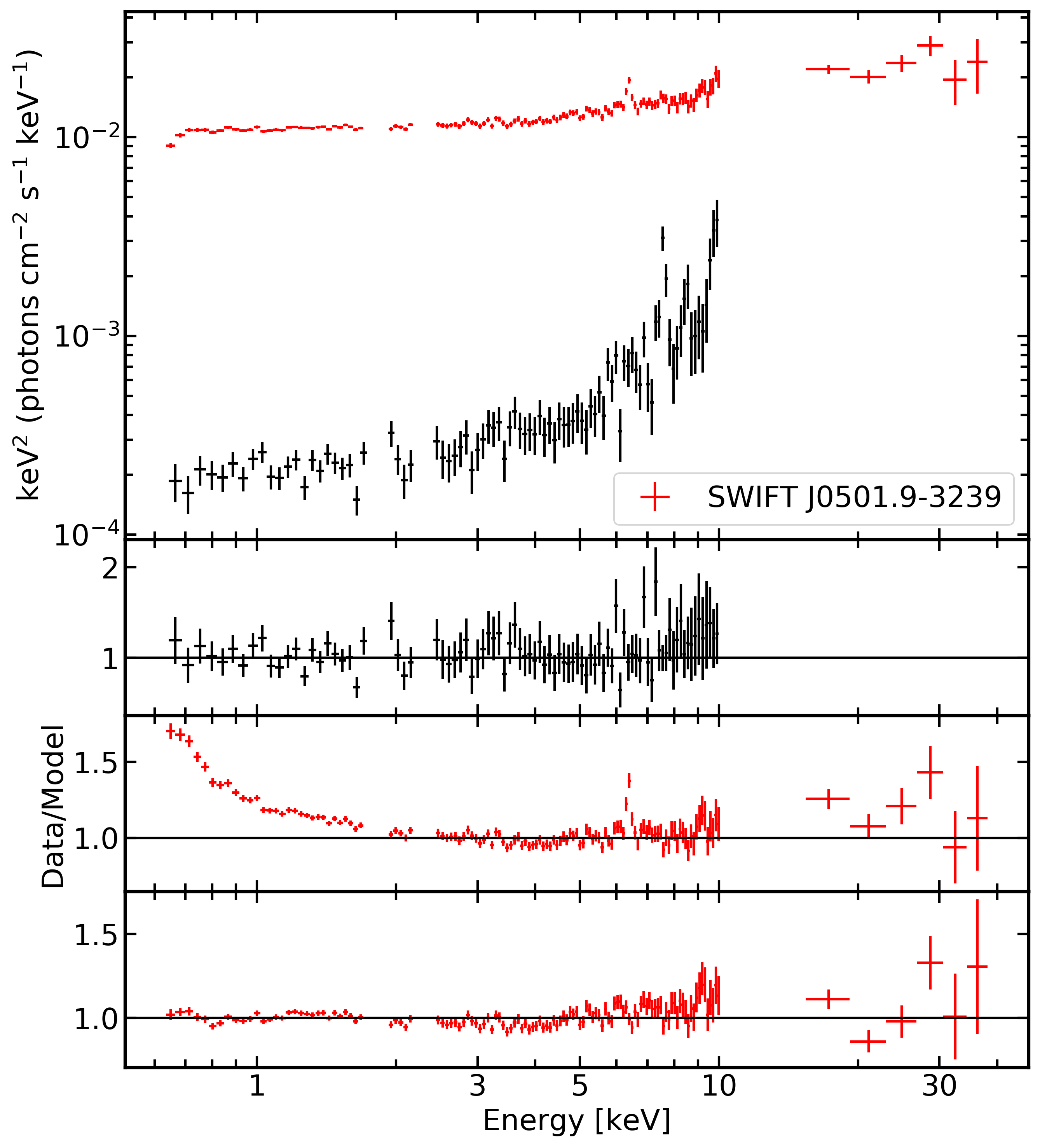}
	\hspace{7mm}
	\includegraphics[width=75mm]{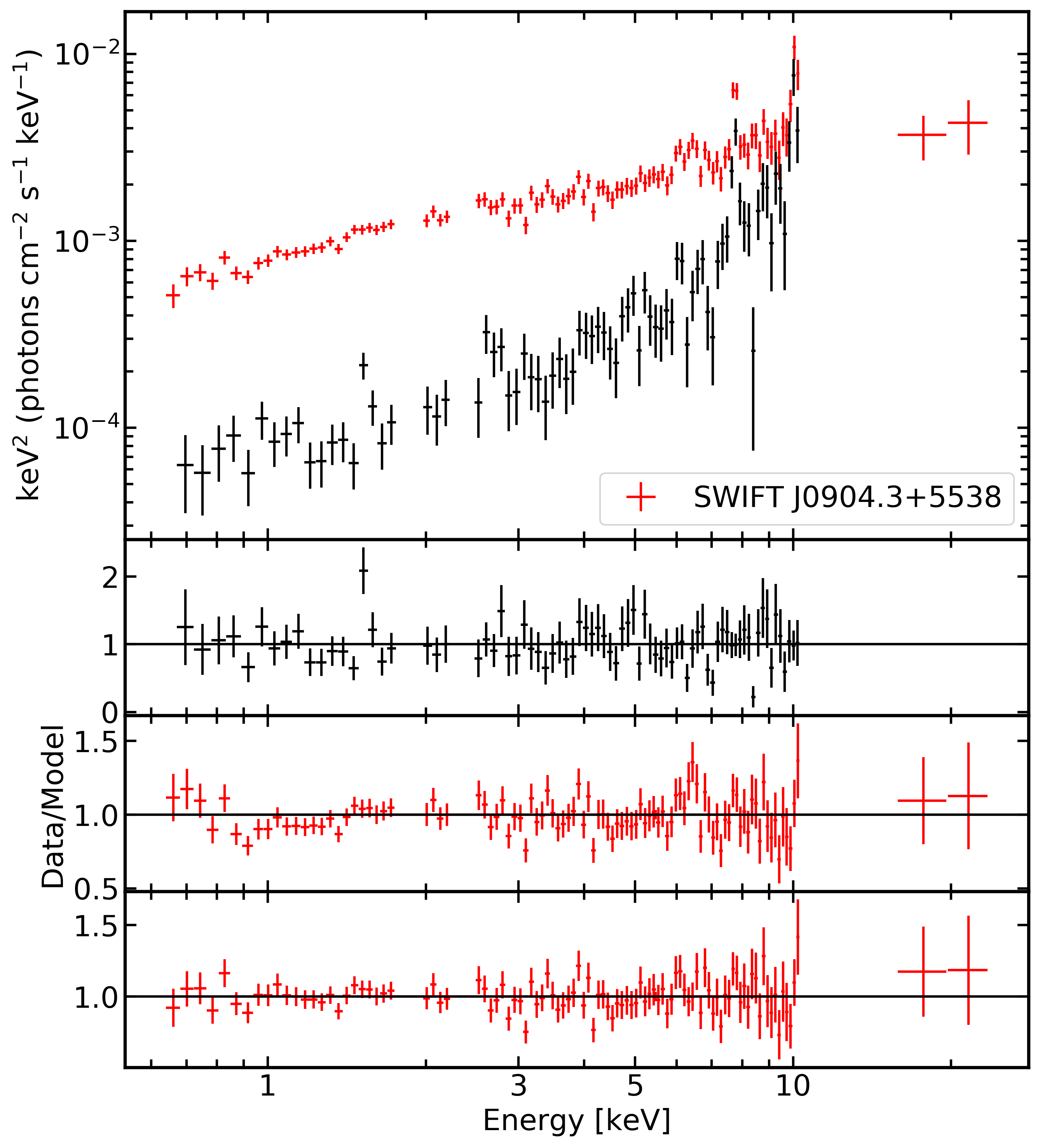}
	\hspace{0mm}
	\includegraphics[width=75mm]{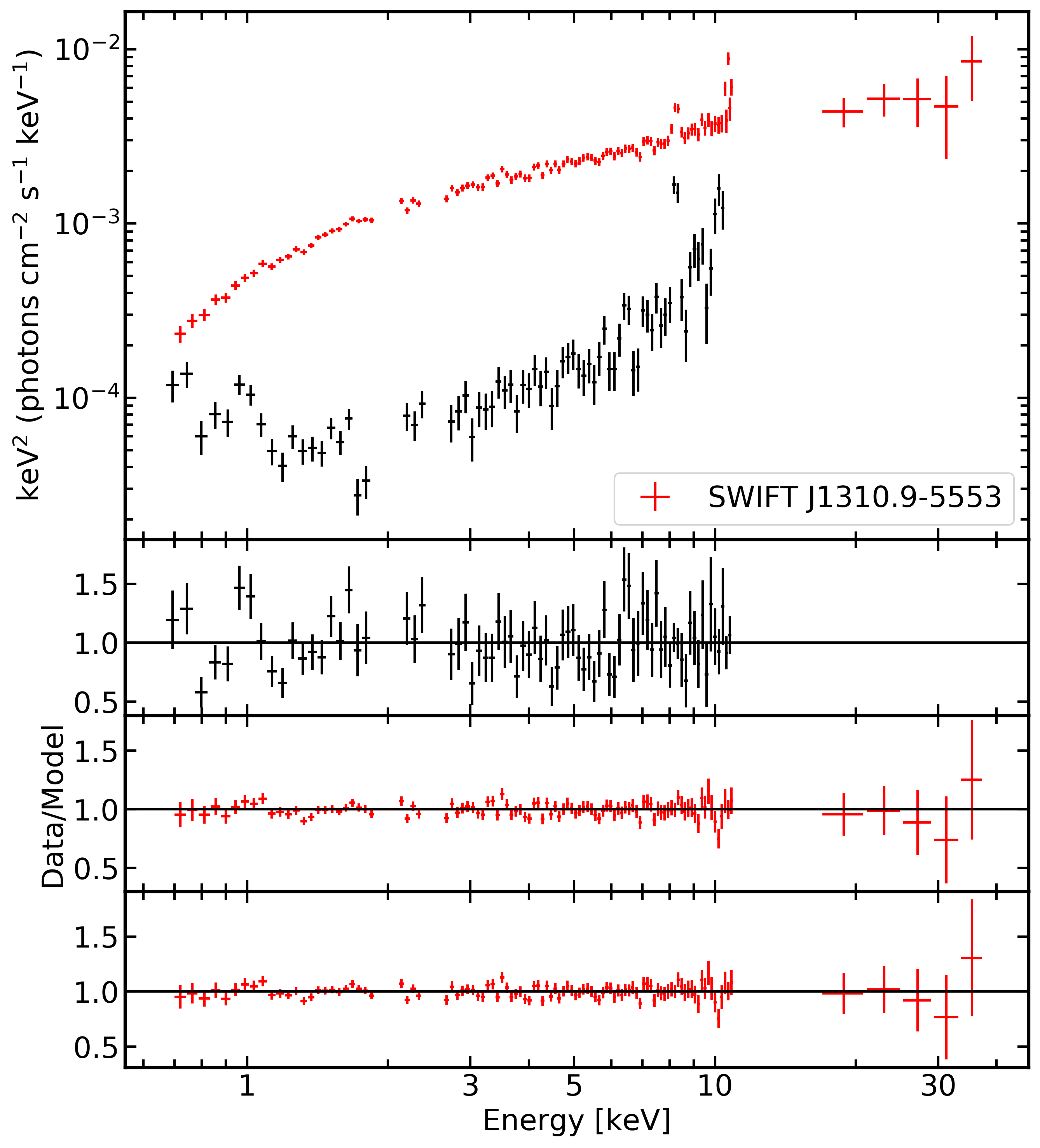}
	\caption{\label{fig:po}continued}
\end{figure*}

\setcounter{figure}{0}

\begin{figure*}
	\centering
	\includegraphics[width=75mm]{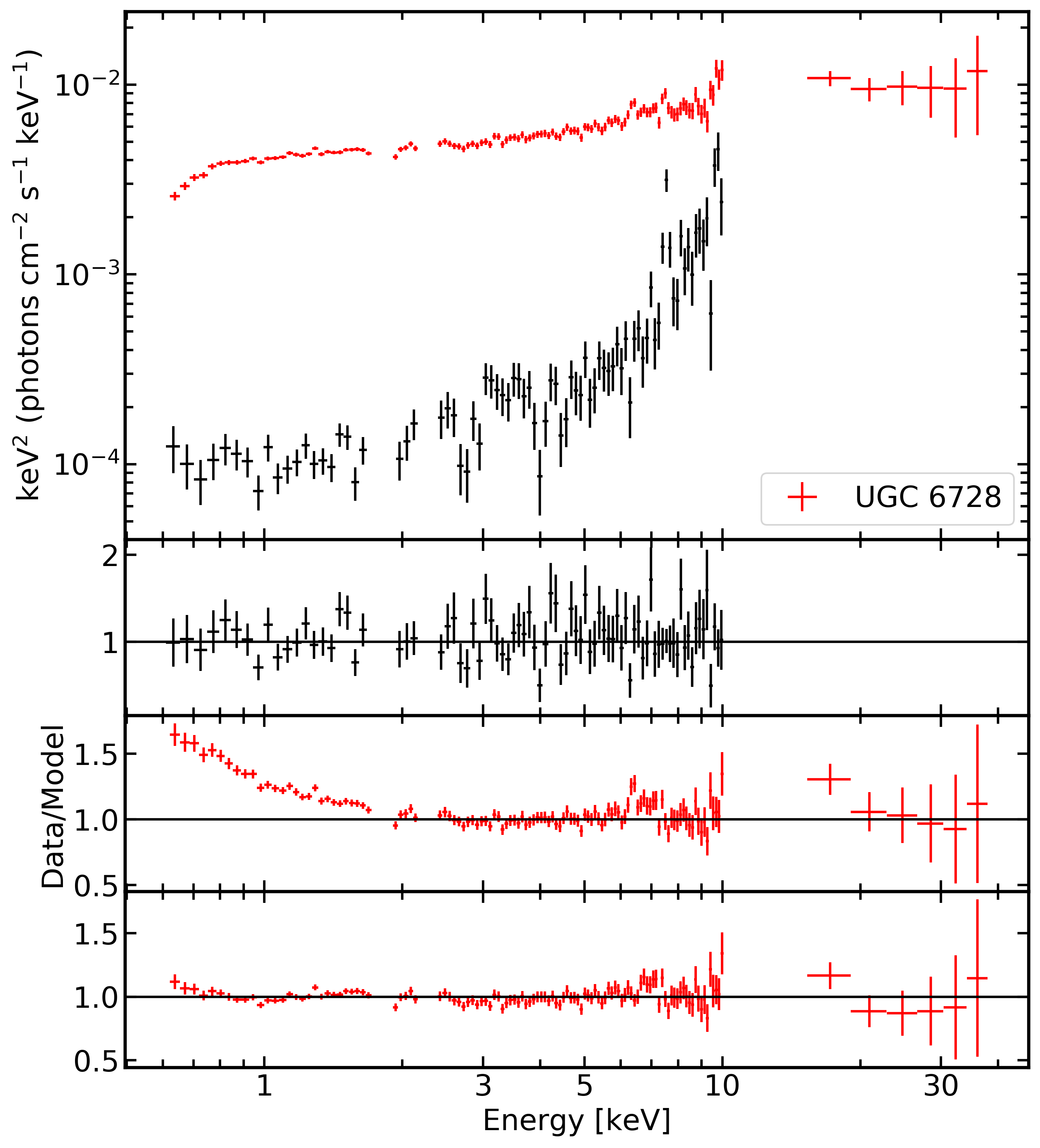}
	\hspace{7mm}
	\includegraphics[width=75mm]{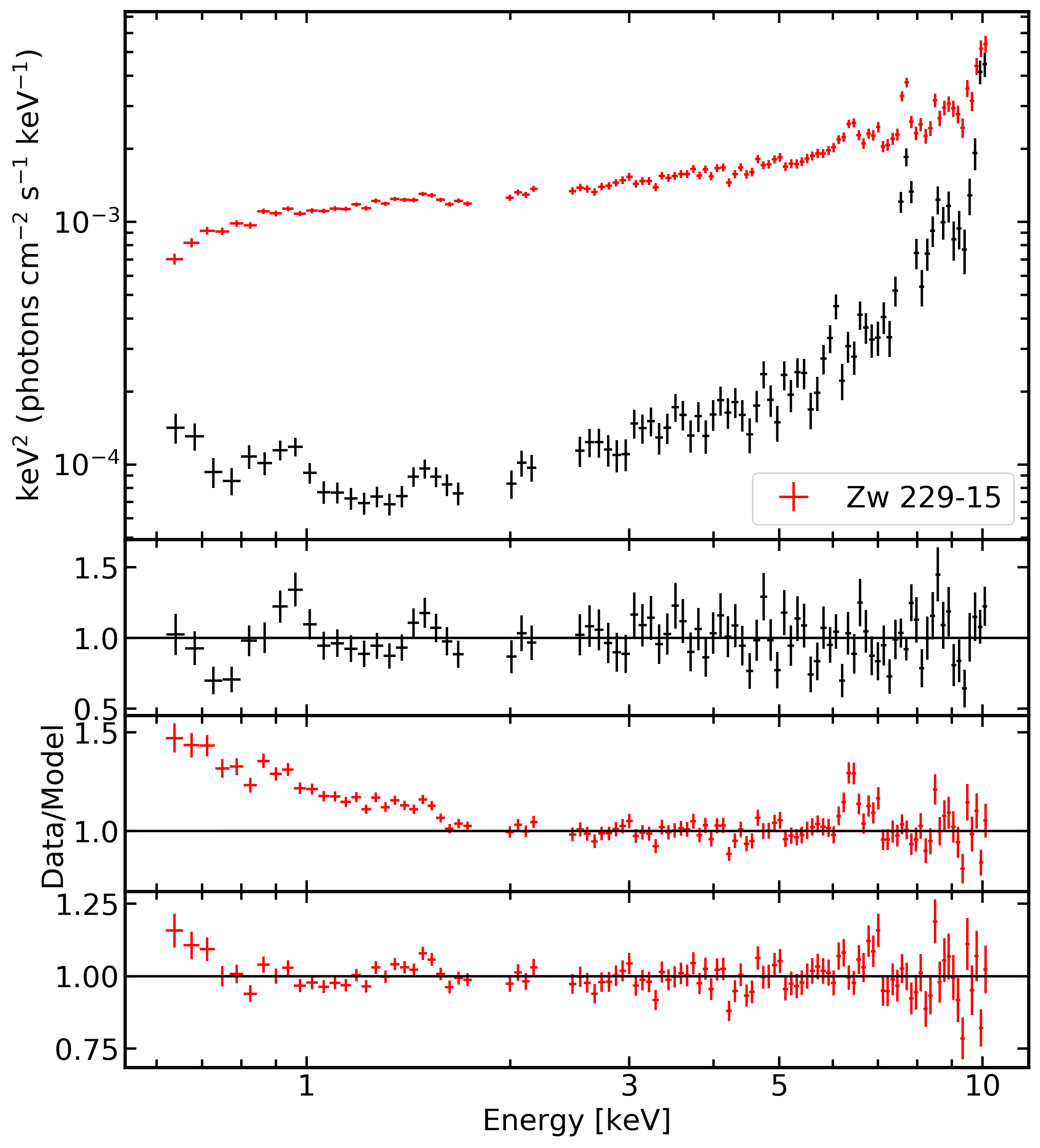}
	\caption{\label{fig:po}continued}
\end{figure*}



\bsp	
\label{lastpage}
\end{document}